\documentclass[longauth]{aa}

\usepackage{graphicx}
\usepackage{grffile} 
\usepackage{txfonts}
\usepackage{hyperref}
\usepackage{amsmath,amssymb}
\usepackage{mathrsfs}
\usepackage{ulem}
\makeatletter
\renewcommand*\maketitle{%
  \thispagestyle{firstpage}
\begingroup
    \if@wideboxfn
    \setlength\bibindent{1.4\parindent}
    \else
    \setlength\bibindent{\parindent}
    \fi
    \renewcommand*\thefootnote{\@fnsymbol\c@footnote}%
    \renewcommand\@makefntext[1]{%
    \ifaa@longfn\hsize\textwidth\fi
    \noindent
    \hb@xt@\bibindent{\hss\@makefnmark\enspace}##1}
  \ifaa@twocolumn
  \begingroup
    \begin{aa@strip}
          \aa@maketitle
    \end{aa@strip}
    \@thanks            
  \endgroup
  \else
    \begingroup
      \let\thanks\footnote
      \aa@maketitle
    \endgroup
  \fi
\endgroup
  \setcounter{footnote}{0}
}
\makeatother

\makeatletter
\renewcommand*\aa@pageof{, page \thepage{} of \pageref*{LastPage}}
\makeatother

\hypersetup{
    colorlinks=true,
    linkcolor=red,
    urlcolor=magenta,
}

\def\deg{\ensuremath{^\circ}}

\providecommand{\mas}{\ensuremath{\textrm{mas}}}

\providecommand{\deg}{\ensuremath{^\circ}}
\providecommand{\gmag}{\ensuremath{G}~}

\providecommand{\red}{\textcolor{red}}

\newcommand\gdr[1]{\gaia~DR#1}

\newcommand{\gaia}{\textit{Gaia}\xspace}
\renewcommand*\vec[1]{\ensuremath{\boldsymbol{#1}}}
\newcommand{\matfont}[1]{\ensuremath{\boldsymbol{\mathsf{#1}}}}
\newcommand{\mat}[1]{\matfont{#1}}
\newcommand{\orcit}[1]{\protect\href{https://orcid.org/#1}{\protect\includegraphics[width=8pt]{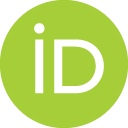}}}

\begin{document}

\title{\gaia Focused Product Release: A catalogue of sources around quasars to search for strongly lensed quasars.}

\author{
{\it Gaia} Collaboration
\and A.        ~Krone-Martins                 \orcit{0000-0002-2308-6623}\inst{\ref{inst:0001},\ref{inst:0002}}
\and C.        ~Ducourant                     \orcit{0000-0003-4843-8979}\inst{\ref{inst:0003}}
\and L.        ~Galluccio                     \orcit{0000-0002-8541-0476}\inst{\ref{inst:0004}}
\and L.        ~Delchambre                    \orcit{0000-0003-2559-408X}\inst{\ref{inst:0005}}
\and I.        ~Oreshina-Slezak               \inst{\ref{inst:0004}}
\and R.        ~Teixeira                      \orcit{0000-0002-6806-6626}\inst{\ref{inst:0007}}
\and J.        ~Braine                        \orcit{0000-0003-1740-1284}\inst{\ref{inst:0003}}
\and J.-F.     ~Le Campion                    \inst{\ref{inst:0003}}
\and F.        ~Mignard                       \inst{\ref{inst:0004}}
\and W.        ~Roux                          \orcit{0000-0002-7816-1950}\inst{\ref{inst:0011}}
\and A.        ~Blazere                       \inst{\ref{inst:0012}}
\and L.        ~Pegoraro                      \inst{\ref{inst:0011}}
\and A.G.A.    ~Brown                         \orcit{0000-0002-7419-9679}\inst{\ref{inst:0014}}
\and A.        ~Vallenari                     \orcit{0000-0003-0014-519X}\inst{\ref{inst:0015}}
\and T.        ~Prusti                        \orcit{0000-0003-3120-7867}\inst{\ref{inst:0016}}
\and J.H.J.    ~de Bruijne                    \orcit{0000-0001-6459-8599}\inst{\ref{inst:0016}}
\and F.        ~Arenou                        \orcit{0000-0003-2837-3899}\inst{\ref{inst:0018}}
\and C.        ~Babusiaux                     \orcit{0000-0002-7631-348X}\inst{\ref{inst:0019}}
\and A.        ~Barbier                       \orcit{0009-0004-0983-931X}\inst{\ref{inst:0011}}
\and M.        ~Biermann                      \orcit{0000-0002-5791-9056}\inst{\ref{inst:0021}}
\and O.L.      ~Creevey                       \orcit{0000-0003-1853-6631}\inst{\ref{inst:0004}}
\and D.W.      ~Evans                         \orcit{0000-0002-6685-5998}\inst{\ref{inst:0023}}
\and L.        ~Eyer                          \orcit{0000-0002-0182-8040}\inst{\ref{inst:0024}}
\and R.        ~Guerra                        \orcit{0000-0002-9850-8982}\inst{\ref{inst:0025}}
\and A.        ~Hutton                        \inst{\ref{inst:0026}}
\and C.        ~Jordi                         \orcit{0000-0001-5495-9602}\inst{\ref{inst:0027},\ref{inst:0028},\ref{inst:0029}}
\and S.A.      ~Klioner                       \orcit{0000-0003-4682-7831}\inst{\ref{inst:0030}}
\and U.        ~Lammers                       \orcit{0000-0001-8309-3801}\inst{\ref{inst:0025}}
\and L.        ~Lindegren                     \orcit{0000-0002-5443-3026}\inst{\ref{inst:0032}}
\and X.        ~Luri                          \orcit{0000-0001-5428-9397}\inst{\ref{inst:0027},\ref{inst:0028},\ref{inst:0029}}
\and S.        ~Randich                       \orcit{0000-0003-2438-0899}\inst{\ref{inst:0036}}
\and P.        ~Sartoretti                    \orcit{0000-0002-6574-7565}\inst{\ref{inst:0018}}
\and R.        ~Smiljanic                     \orcit{0000-0003-0942-7855}\inst{\ref{inst:0038}}
\and P.        ~Tanga                         \orcit{0000-0002-2718-997X}\inst{\ref{inst:0004}}
\and N.A.      ~Walton                        \orcit{0000-0003-3983-8778}\inst{\ref{inst:0023}}
\and C.A.L.    ~Bailer-Jones                  \inst{\ref{inst:0041}}
\and U.        ~Bastian                       \orcit{0000-0002-8667-1715}\inst{\ref{inst:0021}}
\and M.        ~Cropper                       \orcit{0000-0003-4571-9468}\inst{\ref{inst:0043}}
\and R.        ~Drimmel                       \orcit{0000-0002-1777-5502}\inst{\ref{inst:0044}}
\and D.        ~Katz                          \orcit{0000-0001-7986-3164}\inst{\ref{inst:0018}}
\and C.        ~Soubiran                      \orcit{0000-0003-3304-8134}\inst{\ref{inst:0003}}
\and F.        ~van Leeuwen                   \orcit{0000-0003-1781-4441}\inst{\ref{inst:0023}}
\and M.        ~Audard                        \orcit{0000-0003-4721-034X}\inst{\ref{inst:0024},\ref{inst:0049}}
\and J.        ~Bakker                        \inst{\ref{inst:0025}}
\and R.        ~Blomme                        \orcit{0000-0002-2526-346X}\inst{\ref{inst:0051}}
\and J.        ~Casta\~{n}eda                 \orcit{0000-0001-7820-946X}\inst{\ref{inst:0052},\ref{inst:0027},\ref{inst:0029}}
\and F.        ~De Angeli                     \orcit{0000-0003-1879-0488}\inst{\ref{inst:0023}}
\and C.        ~Fabricius                     \orcit{0000-0003-2639-1372}\inst{\ref{inst:0029},\ref{inst:0027},\ref{inst:0028}}
\and M.        ~Fouesneau                     \orcit{0000-0001-9256-5516}\inst{\ref{inst:0041}}
\and Y.        ~Fr\'{e}mat                    \orcit{0000-0002-4645-6017}\inst{\ref{inst:0051}}
\and A.        ~Guerrier                      \inst{\ref{inst:0011}}
\and E.        ~Masana                        \orcit{0000-0002-4819-329X}\inst{\ref{inst:0029},\ref{inst:0027},\ref{inst:0028}}
\and R.        ~Messineo                      \inst{\ref{inst:0065}}
\and C.        ~Nicolas                       \inst{\ref{inst:0011}}
\and K.        ~Nienartowicz                  \orcit{0000-0001-5415-0547}\inst{\ref{inst:0067},\ref{inst:0049}}
\and F.        ~Pailler                       \orcit{0000-0002-4834-481X}\inst{\ref{inst:0011}}
\and P.        ~Panuzzo                       \orcit{0000-0002-0016-8271}\inst{\ref{inst:0018}}
\and F.        ~Riclet                        \inst{\ref{inst:0011}}
\and G.M.      ~Seabroke                      \orcit{0000-0003-4072-9536}\inst{\ref{inst:0043}}
\and R.        ~Sordo                         \orcit{0000-0003-4979-0659}\inst{\ref{inst:0015}}
\and F.        ~Th\'{e}venin                  \orcit{0000-0002-5032-2476}\inst{\ref{inst:0004}}
\and G.        ~Gracia-Abril                  \inst{\ref{inst:0075},\ref{inst:0021}}
\and J.        ~Portell                       \orcit{0000-0002-8886-8925}\inst{\ref{inst:0027},\ref{inst:0028},\ref{inst:0029}}
\and D.        ~Teyssier                      \orcit{0000-0002-6261-5292}\inst{\ref{inst:0080}}
\and M.        ~Altmann                       \orcit{0000-0002-0530-0913}\inst{\ref{inst:0021},\ref{inst:0082}}
\and K.        ~Benson                        \inst{\ref{inst:0043}}
\and J.        ~Berthier                      \orcit{0000-0003-1846-6485}\inst{\ref{inst:0084}}
\and P.W.      ~Burgess                       \orcit{0009-0002-6668-4559}\inst{\ref{inst:0023}}
\and D.        ~Busonero                      \orcit{0000-0002-3903-7076}\inst{\ref{inst:0044}}
\and G.        ~Busso                         \orcit{0000-0003-0937-9849}\inst{\ref{inst:0023}}
\and H.        ~C\'{a}novas                   \orcit{0000-0001-7668-8022}\inst{\ref{inst:0080}}
\and B.        ~Carry                         \orcit{0000-0001-5242-3089}\inst{\ref{inst:0004}}
\and N.        ~Cheek                         \inst{\ref{inst:0090}}
\and G.        ~Clementini                    \orcit{0000-0001-9206-9723}\inst{\ref{inst:0091}}
\and Y.        ~Damerdji                      \orcit{0000-0002-3107-4024}\inst{\ref{inst:0005},\ref{inst:0093}}
\and M.        ~Davidson                      \orcit{0000-0001-9271-4411}\inst{\ref{inst:0094}}
\and P.        ~de Teodoro                    \inst{\ref{inst:0025}}
\and A.        ~Dell'Oro                      \orcit{0000-0003-1561-9685}\inst{\ref{inst:0036}}
\and E.        ~Fraile Garcia                 \orcit{0000-0001-7742-9663}\inst{\ref{inst:0097}}
\and D.        ~Garabato                      \orcit{0000-0002-7133-6623}\inst{\ref{inst:0098}}
\and P.        ~Garc\'{i}a-Lario              \orcit{0000-0003-4039-8212}\inst{\ref{inst:0025}}
\and N.        ~Garralda Torres               \inst{\ref{inst:0100}}
\and P.        ~Gavras                        \orcit{0000-0002-4383-4836}\inst{\ref{inst:0097}}
\and R.        ~Haigron                       \inst{\ref{inst:0018}}
\and N.C.      ~Hambly                        \orcit{0000-0002-9901-9064}\inst{\ref{inst:0094}}
\and D.L.      ~Harrison                      \orcit{0000-0001-8687-6588}\inst{\ref{inst:0023},\ref{inst:0105}}
\and D.        ~Hatzidimitriou                \orcit{0000-0002-5415-0464}\inst{\ref{inst:0106}}
\and J.        ~Hern\'{a}ndez                 \orcit{0000-0002-0361-4994}\inst{\ref{inst:0025}}
\and S.T.      ~Hodgkin                       \orcit{0000-0002-5470-3962}\inst{\ref{inst:0023}}
\and B.        ~Holl                          \orcit{0000-0001-6220-3266}\inst{\ref{inst:0024},\ref{inst:0049}}
\and S.        ~Jamal                         \orcit{0000-0002-3929-6668}\inst{\ref{inst:0041}}
\and S.        ~Jordan                        \orcit{0000-0001-6316-6831}\inst{\ref{inst:0021}}
\and A.C.      ~Lanzafame                     \orcit{0000-0002-2697-3607}\inst{\ref{inst:0113},\ref{inst:0114}}
\and W.        ~L\"{ o}ffler                  \inst{\ref{inst:0021}}
\and A.        ~Lorca                         \orcit{0000-0002-7985-250X}\inst{\ref{inst:0026}}
\and O.        ~Marchal                       \orcit{ 0000-0001-7461-892}\inst{\ref{inst:0117}}
\and P.M.      ~Marrese                       \orcit{0000-0002-8162-3810}\inst{\ref{inst:0118},\ref{inst:0119}}
\and A.        ~Moitinho                      \orcit{0000-0003-0822-5995}\inst{\ref{inst:0002}}
\and K.        ~Muinonen                      \orcit{0000-0001-8058-2642}\inst{\ref{inst:0121},\ref{inst:0122}}
\and M.        ~Nu\~{n}ez Campos              \inst{\ref{inst:0026}}
\and P.        ~Osborne                       \orcit{0000-0003-4482-3538}\inst{\ref{inst:0023}}
\and E.        ~Pancino                       \orcit{0000-0003-0788-5879}\inst{\ref{inst:0036},\ref{inst:0119}}
\and T.        ~Pauwels                       \inst{\ref{inst:0051}}
\and A.        ~Recio-Blanco                  \orcit{0000-0002-6550-7377}\inst{\ref{inst:0004}}
\and M.        ~Riello                        \orcit{0000-0002-3134-0935}\inst{\ref{inst:0023}}
\and L.        ~Rimoldini                     \orcit{0000-0002-0306-585X}\inst{\ref{inst:0049}}
\and A.C.      ~Robin                         \orcit{0000-0001-8654-9499}\inst{\ref{inst:0131}}
\and T.        ~Roegiers                      \orcit{0000-0002-1231-4440}\inst{\ref{inst:0132}}
\and L.M.      ~Sarro                         \orcit{0000-0002-5622-5191}\inst{\ref{inst:0133}}
\and M.        ~Schultheis                    \orcit{0000-0002-6590-1657}\inst{\ref{inst:0004}}
\and C.        ~Siopis                        \orcit{0000-0002-6267-2924}\inst{\ref{inst:0135}}
\and M.        ~Smith                         \inst{\ref{inst:0043}}
\and A.        ~Sozzetti                      \orcit{0000-0002-7504-365X}\inst{\ref{inst:0044}}
\and E.        ~Utrilla                       \inst{\ref{inst:0026}}
\and M.        ~van Leeuwen                   \orcit{0000-0001-9698-2392}\inst{\ref{inst:0023}}
\and K.        ~Weingrill                     \orcit{0000-0002-8163-2493}\inst{\ref{inst:0140}}
\and U.        ~Abbas                         \orcit{0000-0002-5076-766X}\inst{\ref{inst:0044}}
\and P.        ~\'{A}brah\'{a}m               \orcit{0000-0001-6015-646X}\inst{\ref{inst:0142},\ref{inst:0143}}
\and A.        ~Abreu Aramburu                \orcit{0000-0003-3959-0856}\inst{\ref{inst:0100}}
\and C.        ~Aerts                         \orcit{0000-0003-1822-7126}\inst{\ref{inst:0145},\ref{inst:0146},\ref{inst:0041}}
\and G.        ~Altavilla                     \orcit{0000-0002-9934-1352}\inst{\ref{inst:0118},\ref{inst:0119}}
\and M.A.      ~\'{A}lvarez                   \orcit{0000-0002-6786-2620}\inst{\ref{inst:0098}}
\and J.        ~Alves                         \orcit{0000-0002-4355-0921}\inst{\ref{inst:0151}}
\and R.I.      ~Anderson                      \orcit{0000-0001-8089-4419}\inst{\ref{inst:0152}}
\and T.        ~Antoja                        \orcit{0000-0003-2595-5148}\inst{\ref{inst:0027},\ref{inst:0028},\ref{inst:0029}}
\and D.        ~Baines                        \orcit{0000-0002-6923-3756}\inst{\ref{inst:0156}}
\and S.G.      ~Baker                         \orcit{0000-0002-6436-1257}\inst{\ref{inst:0043}}
\and Z.        ~Balog                         \orcit{0000-0003-1748-2926}\inst{\ref{inst:0021},\ref{inst:0041}}
\and C.        ~Barache                       \inst{\ref{inst:0082}}
\and D.        ~Barbato                       \inst{\ref{inst:0024},\ref{inst:0044}}
\and M.        ~Barros                        \orcit{0000-0002-9728-9618}\inst{\ref{inst:0163}}
\and M.A.      ~Barstow                       \orcit{0000-0002-7116-3259}\inst{\ref{inst:0164}}
\and S.        ~Bartolom\'{e}                 \orcit{0000-0002-6290-6030}\inst{\ref{inst:0029},\ref{inst:0027},\ref{inst:0028}}
\and D.        ~Bashi                         \orcit{0000-0002-9035-2645}\inst{\ref{inst:0168},\ref{inst:0169}}
\and N.        ~Bauchet                       \orcit{0000-0002-2307-8973}\inst{\ref{inst:0018}}
\and N.        ~Baudeau                       \inst{\ref{inst:0171}}
\and U.        ~Becciani                      \orcit{0000-0002-4389-8688}\inst{\ref{inst:0113}}
\and L.R.      ~Bedin                         \inst{\ref{inst:0015}}
\and I.        ~Bellas-Velidis                \inst{\ref{inst:0174}}
\and M.        ~Bellazzini                    \orcit{0000-0001-8200-810X}\inst{\ref{inst:0091}}
\and W.        ~Beordo                        \orcit{0000-0002-5094-1306}\inst{\ref{inst:0044},\ref{inst:0177}}
\and A.        ~Berihuete                     \orcit{0000-0002-8589-4423}\inst{\ref{inst:0178}}
\and M.        ~Bernet                        \orcit{0000-0001-7503-1010}\inst{\ref{inst:0027},\ref{inst:0028},\ref{inst:0029}}
\and C.        ~Bertolotto                    \inst{\ref{inst:0065}}
\and S.        ~Bertone                       \orcit{0000-0001-9885-8440}\inst{\ref{inst:0044}}
\and L.        ~Bianchi                       \orcit{0000-0002-7999-4372}\inst{\ref{inst:0184}}
\and A.        ~Binnenfeld                    \orcit{0000-0002-9319-3838}\inst{\ref{inst:0185}}
\and T.        ~Boch                          \orcit{0000-0001-5818-2781}\inst{\ref{inst:0117}}
\and A.        ~Bombrun                       \inst{\ref{inst:0187}}
\and S.        ~Bouquillon                    \inst{\ref{inst:0082},\ref{inst:0189}}
\and A.        ~Bragaglia                     \orcit{0000-0002-0338-7883}\inst{\ref{inst:0091}}
\and L.        ~Bramante                      \inst{\ref{inst:0065}}
\and E.        ~Breedt                        \orcit{0000-0001-6180-3438}\inst{\ref{inst:0023}}
\and A.        ~Bressan                       \orcit{0000-0002-7922-8440}\inst{\ref{inst:0193}}
\and N.        ~Brouillet                     \orcit{0000-0002-3274-7024}\inst{\ref{inst:0003}}
\and E.        ~Brugaletta                    \orcit{0000-0003-2598-6737}\inst{\ref{inst:0113}}
\and B.        ~Bucciarelli                   \orcit{0000-0002-5303-0268}\inst{\ref{inst:0044},\ref{inst:0177}}
\and A.G.      ~Butkevich                     \orcit{0000-0002-4098-3588}\inst{\ref{inst:0044}}
\and R.        ~Buzzi                         \orcit{0000-0001-9389-5701}\inst{\ref{inst:0044}}
\and E.        ~Caffau                        \orcit{0000-0001-6011-6134}\inst{\ref{inst:0018}}
\and R.        ~Cancelliere                   \orcit{0000-0002-9120-3799}\inst{\ref{inst:0201}}
\and S.        ~Cannizzo                      \inst{\ref{inst:0202}}
\and R.        ~Carballo                      \orcit{0000-0001-7412-2498}\inst{\ref{inst:0204}}
\and T.        ~Carlucci                      \inst{\ref{inst:0082}}
\and M.I.      ~Carnerero                     \orcit{0000-0001-5843-5515}\inst{\ref{inst:0044}}
\and J.M.      ~Carrasco                      \orcit{0000-0002-3029-5853}\inst{\ref{inst:0029},\ref{inst:0027},\ref{inst:0028}}
\and J.        ~Carretero                     \orcit{0000-0002-3130-0204}\inst{\ref{inst:0210},\ref{inst:0211}}
\and S.        ~Carton                        \inst{\ref{inst:0202}}
\and L.        ~Casamiquela                   \orcit{0000-0001-5238-8674}\inst{\ref{inst:0003},\ref{inst:0018}}
\and M.        ~Castellani                    \orcit{0000-0002-7650-7428}\inst{\ref{inst:0118}}
\and A.        ~Castro-Ginard                 \orcit{0000-0002-9419-3725}\inst{\ref{inst:0014}}
\and V.        ~Cesare                        \orcit{0000-0003-1119-4237}\inst{\ref{inst:0113}}
\and P.        ~Charlot                       \orcit{0000-0002-9142-716X}\inst{\ref{inst:0003}}
\and L.        ~Chemin                        \orcit{0000-0002-3834-7937}\inst{\ref{inst:0219}}
\and V.        ~Chiaramida                    \inst{\ref{inst:0065}}
\and A.        ~Chiavassa                     \orcit{0000-0003-3891-7554}\inst{\ref{inst:0004}}
\and N.        ~Chornay                       \orcit{0000-0002-8767-3907}\inst{\ref{inst:0023},\ref{inst:0049}}
\and R.        ~Collins                       \orcit{0000-0001-8437-1703}\inst{\ref{inst:0094}}
\and G.        ~Contursi                      \orcit{0000-0001-5370-1511}\inst{\ref{inst:0004}}
\and W.J.      ~Cooper                        \orcit{0000-0003-3501-8967}\inst{\ref{inst:0226},\ref{inst:0044}}
\and T.        ~Cornez                        \inst{\ref{inst:0202}}
\and M.        ~Crosta                        \orcit{0000-0003-4369-3786}\inst{\ref{inst:0044},\ref{inst:0230}}
\and C.        ~Crowley                       \orcit{0000-0002-9391-9360}\inst{\ref{inst:0187}}
\and C.        ~Dafonte                       \orcit{0000-0003-4693-7555}\inst{\ref{inst:0098}}
\and P.        ~de Laverny                    \orcit{0000-0002-2817-4104}\inst{\ref{inst:0004}}
\and F.        ~De Luise                      \orcit{0000-0002-6570-8208}\inst{\ref{inst:0235}}
\and R.        ~De March                      \orcit{0000-0003-0567-842X}\inst{\ref{inst:0065}}
\and R.        ~de Souza                      \orcit{0009-0007-7669-0254}\inst{\ref{inst:0007}}
\and A.        ~de Torres                     \inst{\ref{inst:0187}}
\and E.F.      ~del Peloso                    \inst{\ref{inst:0021}}
\and M.        ~Delbo                         \orcit{0000-0002-8963-2404}\inst{\ref{inst:0004}}
\and A.        ~Delgado                       \inst{\ref{inst:0097}}
\and T.E.      ~Dharmawardena                 \orcit{0000-0002-9583-5216}\inst{\ref{inst:0041}}
\and S.        ~Diakite                       \inst{\ref{inst:0243}}
\and C.        ~Diener                        \inst{\ref{inst:0023}}
\and E.        ~Distefano                     \orcit{0000-0002-2448-2513}\inst{\ref{inst:0113}}
\and C.        ~Dolding                       \inst{\ref{inst:0043}}
\and K.        ~Dsilva                        \orcit{0000-0002-1476-9772}\inst{\ref{inst:0135}}
\and J.        ~Dur\'{a}n                     \inst{\ref{inst:0097}}
\and H.        ~Enke                          \orcit{0000-0002-2366-8316}\inst{\ref{inst:0140}}
\and P.        ~Esquej                        \orcit{0000-0001-8195-628X}\inst{\ref{inst:0097}}
\and C.        ~Fabre                         \inst{\ref{inst:0012}}
\and M.        ~Fabrizio                      \orcit{0000-0001-5829-111X}\inst{\ref{inst:0118},\ref{inst:0119}}
\and S.        ~Faigler                       \orcit{0000-0002-8368-5724}\inst{\ref{inst:0168}}
\and M.        ~Fatovi\'{c}                   \orcit{0000-0003-1911-4326}\inst{\ref{inst:0255}}
\and G.        ~Fedorets                      \orcit{0000-0002-8418-4809}\inst{\ref{inst:0121},\ref{inst:0257}}
\and J.        ~Fern\'{a}ndez-Hern\'{a}ndez   \inst{\ref{inst:0097}}
\and P.        ~Fernique                      \orcit{0000-0002-3304-2923}\inst{\ref{inst:0117}}
\and F.        ~Figueras                      \orcit{0000-0002-3393-0007}\inst{\ref{inst:0027},\ref{inst:0028},\ref{inst:0029}}
\and Y.        ~Fournier                      \orcit{0000-0002-6633-9088}\inst{\ref{inst:0140}}
\and C.        ~Fouron                        \inst{\ref{inst:0171}}
\and M.        ~Gai                           \orcit{0000-0001-9008-134X}\inst{\ref{inst:0044}}
\and M.        ~Galinier                      \orcit{0000-0001-7920-0133}\inst{\ref{inst:0004}}
\and A.        ~Garcia-Gutierrez              \inst{\ref{inst:0029},\ref{inst:0027},\ref{inst:0028}}
\and M.        ~Garc\'{i}a-Torres             \orcit{0000-0002-6867-7080}\inst{\ref{inst:0270}}
\and A.        ~Garofalo                      \orcit{0000-0002-5907-0375}\inst{\ref{inst:0091}}
\and E.        ~Gerlach                       \orcit{0000-0002-9533-2168}\inst{\ref{inst:0030}}
\and R.        ~Geyer                         \orcit{0000-0001-6967-8707}\inst{\ref{inst:0030}}
\and P.        ~Giacobbe                      \orcit{0000-0001-7034-7024}\inst{\ref{inst:0044}}
\and G.        ~Gilmore                       \orcit{0000-0003-4632-0213}\inst{\ref{inst:0023},\ref{inst:0276}}
\and S.        ~Girona                        \orcit{0000-0002-1975-1918}\inst{\ref{inst:0277}}
\and G.        ~Giuffrida                     \orcit{0000-0002-8979-4614}\inst{\ref{inst:0118}}
\and R.        ~Gomel                         \inst{\ref{inst:0168}}
\and A.        ~Gomez                         \orcit{0000-0002-3796-3690}\inst{\ref{inst:0098}}
\and J.        ~Gonz\'{a}lez-N\'{u}\~{n}ez    \orcit{0000-0001-5311-5555}\inst{\ref{inst:0281}}
\and I.        ~Gonz\'{a}lez-Santamar\'{i}a   \orcit{0000-0002-8537-9384}\inst{\ref{inst:0098}}
\and E.        ~Gosset                        \inst{\ref{inst:0005},\ref{inst:0284}}
\and M.        ~Granvik                       \orcit{0000-0002-5624-1888}\inst{\ref{inst:0121},\ref{inst:0286}}
\and V.        ~Gregori Barrera               \inst{\ref{inst:0029},\ref{inst:0027},\ref{inst:0028}}
\and R.        ~Guti\'{e}rrez-S\'{a}nchez     \orcit{0009-0003-1500-4733}\inst{\ref{inst:0080}}
\and M.        ~Haywood                       \orcit{0000-0003-0434-0400}\inst{\ref{inst:0018}}
\and A.        ~Helmer                        \inst{\ref{inst:0202}}
\and A.        ~Helmi                         \orcit{0000-0003-3937-7641}\inst{\ref{inst:0293}}
\and K.        ~Henares                       \inst{\ref{inst:0156}}
\and S.L.      ~Hidalgo                       \orcit{0000-0002-0002-9298}\inst{\ref{inst:0295},\ref{inst:0296}}
\and T.        ~Hilger                        \orcit{0000-0003-1646-0063}\inst{\ref{inst:0030}}
\and D.        ~Hobbs                         \orcit{0000-0002-2696-1366}\inst{\ref{inst:0032}}
\and C.        ~Hottier                       \orcit{0000-0002-3498-3944}\inst{\ref{inst:0018}}
\and H.E.      ~Huckle                        \inst{\ref{inst:0043}}
\and M.        ~Jab\l{}o\'{n}ska              \orcit{0000-0001-6962-4979}\inst{\ref{inst:0301},\ref{inst:0302}}
\and F.        ~Jansen                        \inst{\ref{inst:0303}}
\and \'{O}.    ~Jim\'{e}nez-Arranz            \orcit{0000-0001-7434-5165}\inst{\ref{inst:0027},\ref{inst:0028},\ref{inst:0029}}
\and J.        ~Juaristi Campillo             \inst{\ref{inst:0021}}
\and S.        ~Khanna                        \orcit{0000-0002-2604-4277}\inst{\ref{inst:0044},\ref{inst:0293}}
\and G.        ~Kordopatis                    \orcit{0000-0002-9035-3920}\inst{\ref{inst:0004}}
\and \'{A}     ~K\'{o}sp\'{a}l                \orcit{0000-0001-7157-6275}\inst{\ref{inst:0142},\ref{inst:0041},\ref{inst:0143}}
\and Z.        ~Kostrzewa-Rutkowska           \inst{\ref{inst:0014}}
\and M.        ~Kun                           \orcit{0000-0002-7538-5166}\inst{\ref{inst:0142}}
\and S.        ~Lambert                       \orcit{0000-0001-6759-5502}\inst{\ref{inst:0082}}
\and A.F.      ~Lanza                         \orcit{0000-0001-5928-7251}\inst{\ref{inst:0113}}
\and Y.        ~Lebreton                      \orcit{0000-0002-4834-2144}\inst{\ref{inst:0318},\ref{inst:0319}}
\and T.        ~Lebzelter                     \orcit{0000-0002-0702-7551}\inst{\ref{inst:0151}}
\and S.        ~Leccia                        \orcit{0000-0001-5685-6930}\inst{\ref{inst:0321}}
\and I.        ~Lecoeur-Taibi                 \orcit{0000-0003-0029-8575}\inst{\ref{inst:0049}}
\and G.        ~Lecoutre                      \inst{\ref{inst:0131}}
\and S.        ~Liao                          \orcit{0000-0002-9346-0211}\inst{\ref{inst:0324},\ref{inst:0044},\ref{inst:0326}}
\and L.        ~Liberato                      \orcit{0000-0003-3433-6269}\inst{\ref{inst:0004},\ref{inst:0328}}
\and E.        ~Licata                        \orcit{0000-0002-5203-0135}\inst{\ref{inst:0044}}
\and H.E.P.    ~Lindstr{\o}m                  \orcit{0009-0004-8864-5459}\inst{\ref{inst:0044},\ref{inst:0331},\ref{inst:0332}}
\and T.A.      ~Lister                        \orcit{0000-0002-3818-7769}\inst{\ref{inst:0333}}
\and E.        ~Livanou                       \orcit{0000-0003-0628-2347}\inst{\ref{inst:0106}}
\and A.        ~Lobel                         \orcit{0000-0001-5030-019X}\inst{\ref{inst:0051}}
\and C.        ~Loup                          \inst{\ref{inst:0117}}
\and L.        ~Mahy                          \orcit{0000-0003-0688-7987}\inst{\ref{inst:0051}}
\and R.G.      ~Mann                          \orcit{0000-0002-0194-325X}\inst{\ref{inst:0094}}
\and M.        ~Manteiga                      \orcit{0000-0002-7711-5581}\inst{\ref{inst:0339}}
\and J.M.      ~Marchant                      \orcit{0000-0002-3678-3145}\inst{\ref{inst:0340}}
\and M.        ~Marconi                       \orcit{0000-0002-1330-2927}\inst{\ref{inst:0321}}
\and D.        ~Mar\'{i}n Pina                \orcit{0000-0001-6482-1842}\inst{\ref{inst:0027},\ref{inst:0028},\ref{inst:0029}}
\and S.        ~Marinoni                      \orcit{0000-0001-7990-6849}\inst{\ref{inst:0118},\ref{inst:0119}}
\and D.J.      ~Marshall                      \orcit{0000-0003-3956-3524}\inst{\ref{inst:0347}}
\and J.        ~Mart\'{i}n Lozano             \orcit{0009-0001-2435-6680}\inst{\ref{inst:0090}}
\and J.M.      ~Mart\'{i}n-Fleitas            \orcit{0000-0002-8594-569X}\inst{\ref{inst:0026}}
\and G.        ~Marton                        \orcit{0000-0002-1326-1686}\inst{\ref{inst:0142}}
\and N.        ~Mary                          \inst{\ref{inst:0202}}
\and A.        ~Masip                         \orcit{0000-0003-1419-0020}\inst{\ref{inst:0029},\ref{inst:0027},\ref{inst:0028}}
\and D.        ~Massari                       \orcit{0000-0001-8892-4301}\inst{\ref{inst:0091}}
\and A.        ~Mastrobuono-Battisti          \orcit{0000-0002-2386-9142}\inst{\ref{inst:0018}}
\and T.        ~Mazeh                         \orcit{0000-0002-3569-3391}\inst{\ref{inst:0168}}
\and P.J.      ~McMillan                      \orcit{0000-0002-8861-2620}\inst{\ref{inst:0032}}
\and J.        ~Meichsner                     \orcit{0000-0002-9900-7864}\inst{\ref{inst:0030}}
\and S.        ~Messina                       \orcit{0000-0002-2851-2468}\inst{\ref{inst:0113}}
\and D.        ~Michalik                      \orcit{0000-0002-7618-6556}\inst{\ref{inst:0016}}
\and N.R.      ~Millar                        \inst{\ref{inst:0023}}
\and A.        ~Mints                         \orcit{0000-0002-8440-1455}\inst{\ref{inst:0140}}
\and D.        ~Molina                        \orcit{0000-0003-4814-0275}\inst{\ref{inst:0028},\ref{inst:0027},\ref{inst:0029}}
\and R.        ~Molinaro                      \orcit{0000-0003-3055-6002}\inst{\ref{inst:0321}}
\and L.        ~Moln\'{a}r                    \orcit{0000-0002-8159-1599}\inst{\ref{inst:0142},\ref{inst:0369},\ref{inst:0143}}
\and G.        ~Monari                        \orcit{0000-0002-6863-0661}\inst{\ref{inst:0117}}
\and M.        ~Mongui\'{o}                   \orcit{0000-0002-4519-6700}\inst{\ref{inst:0027},\ref{inst:0028},\ref{inst:0029}}
\and P.        ~Montegriffo                   \orcit{0000-0001-5013-5948}\inst{\ref{inst:0091}}
\and A.        ~Montero                       \inst{\ref{inst:0090}}
\and R.        ~Mor                           \orcit{0000-0002-8179-6527}\inst{\ref{inst:0377},\ref{inst:0028},\ref{inst:0029}}
\and A.        ~Mora                          \inst{\ref{inst:0026}}
\and R.        ~Morbidelli                    \orcit{0000-0001-7627-4946}\inst{\ref{inst:0044}}
\and T.        ~Morel                         \orcit{0000-0002-8176-4816}\inst{\ref{inst:0005}}
\and D.        ~Morris                        \orcit{0000-0002-1952-6251}\inst{\ref{inst:0094}}
\and N.        ~Mowlavi                       \orcit{0000-0003-1578-6993}\inst{\ref{inst:0024}}
\and D.        ~Munoz                         \inst{\ref{inst:0202}}
\and T.        ~Muraveva                      \orcit{0000-0002-0969-1915}\inst{\ref{inst:0091}}
\and C.P.      ~Murphy                        \inst{\ref{inst:0025}}
\and I.        ~Musella                       \orcit{0000-0001-5909-6615}\inst{\ref{inst:0321}}
\and Z.        ~Nagy                          \orcit{0000-0002-3632-1194}\inst{\ref{inst:0142}}
\and S.        ~Nieto                         \inst{\ref{inst:0097}}
\and L.        ~Noval                         \inst{\ref{inst:0202}}
\and A.        ~Ogden                         \inst{\ref{inst:0023}}
\and C.        ~Ordenovic                     \inst{\ref{inst:0004}}
\and C.        ~Pagani                        \orcit{0000-0001-5477-4720}\inst{\ref{inst:0394}}
\and I.        ~Pagano                        \orcit{0000-0001-9573-4928}\inst{\ref{inst:0113}}
\and L.        ~Palaversa                     \orcit{0000-0003-3710-0331}\inst{\ref{inst:0255}}
\and P.A.      ~Palicio                       \orcit{0000-0002-7432-8709}\inst{\ref{inst:0004}}
\and L.        ~Pallas-Quintela               \orcit{0000-0001-9296-3100}\inst{\ref{inst:0098}}
\and A.        ~Panahi                        \orcit{0000-0001-5850-4373}\inst{\ref{inst:0168}}
\and C.        ~Panem                         \inst{\ref{inst:0011}}
\and S.        ~Payne-Wardenaar               \inst{\ref{inst:0021}}
\and A.        ~Penttil\"{ a}                 \orcit{0000-0001-7403-1721}\inst{\ref{inst:0121}}
\and P.        ~Pesciullesi                   \inst{\ref{inst:0097}}
\and A.M.      ~Piersimoni                    \orcit{0000-0002-8019-3708}\inst{\ref{inst:0235}}
\and M.        ~Pinamonti                     \orcit{0000-0002-4445-1845}\inst{\ref{inst:0044}}
\and F.-X.     ~Pineau                        \orcit{0000-0002-2335-4499}\inst{\ref{inst:0117}}
\and E.        ~Plachy                        \orcit{0000-0002-5481-3352}\inst{\ref{inst:0142},\ref{inst:0369},\ref{inst:0143}}
\and G.        ~Plum                          \inst{\ref{inst:0018}}
\and E.        ~Poggio                        \orcit{0000-0003-3793-8505}\inst{\ref{inst:0004},\ref{inst:0044}}
\and D.        ~Pourbaix$^\dagger$            \orcit{0000-0002-3020-1837}\inst{\ref{inst:0135},\ref{inst:0284}}
\and A.        ~Pr\v{s}a                      \orcit{0000-0002-1913-0281}\inst{\ref{inst:0415}}
\and L.        ~Pulone                        \orcit{0000-0002-5285-998X}\inst{\ref{inst:0118}}
\and E.        ~Racero                        \orcit{0000-0002-6101-9050}\inst{\ref{inst:0090},\ref{inst:0418}}
\and M.        ~Rainer                        \orcit{0000-0002-8786-2572}\inst{\ref{inst:0036},\ref{inst:0420}}
\and C.M.      ~Raiteri                       \orcit{0000-0003-1784-2784}\inst{\ref{inst:0044}}
\and P.        ~Ramos                         \orcit{0000-0002-5080-7027}\inst{\ref{inst:0422},\ref{inst:0027},\ref{inst:0029}}
\and M.        ~Ramos-Lerate                  \orcit{0009-0005-4677-8031}\inst{\ref{inst:0080}}
\and M.        ~Ratajczak                     \orcit{0000-0002-3218-2684}\inst{\ref{inst:0301}}
\and P.        ~Re Fiorentin                  \orcit{0000-0002-4995-0475}\inst{\ref{inst:0044}}
\and S.        ~Regibo                        \orcit{0000-0001-7227-9563}\inst{\ref{inst:0145}}
\and C.        ~Reyl\'{e}                     \orcit{0000-0003-2258-2403}\inst{\ref{inst:0131}}
\and V.        ~Ripepi                        \orcit{0000-0003-1801-426X}\inst{\ref{inst:0321}}
\and A.        ~Riva                          \orcit{0000-0002-6928-8589}\inst{\ref{inst:0044}}
\and H.-W.     ~Rix                           \orcit{0000-0003-4996-9069}\inst{\ref{inst:0041}}
\and G.        ~Rixon                         \orcit{0000-0003-4399-6568}\inst{\ref{inst:0023}}
\and N.        ~Robichon                      \orcit{0000-0003-4545-7517}\inst{\ref{inst:0018}}
\and C.        ~Robin                         \inst{\ref{inst:0202}}
\and M.        ~Romero-G\'{o}mez              \orcit{0000-0003-3936-1025}\inst{\ref{inst:0027},\ref{inst:0028},\ref{inst:0029}}
\and N.        ~Rowell                        \orcit{0000-0003-3809-1895}\inst{\ref{inst:0094}}
\and F.        ~Royer                         \orcit{0000-0002-9374-8645}\inst{\ref{inst:0018}}
\and D.        ~Ruz Mieres                    \orcit{0000-0002-9455-157X}\inst{\ref{inst:0023}}
\and K.A.      ~Rybicki                       \orcit{0000-0002-9326-9329}\inst{\ref{inst:0442}}
\and G.        ~Sadowski                      \orcit{0000-0002-3411-1003}\inst{\ref{inst:0135}}
\and A.        ~S\'{a}ez N\'{u}\~{n}ez        \orcit{0009-0001-6078-0868}\inst{\ref{inst:0029},\ref{inst:0027},\ref{inst:0028}}
\and A.        ~Sagrist\`{a} Sell\'{e}s       \orcit{0000-0001-6191-2028}\inst{\ref{inst:0021}}
\and J.        ~Sahlmann                      \orcit{0000-0001-9525-3673}\inst{\ref{inst:0097}}
\and V.        ~Sanchez Gimenez               \orcit{0000-0003-1797-3557}\inst{\ref{inst:0029},\ref{inst:0027},\ref{inst:0028}}
\and N.        ~Sanna                         \orcit{0000-0001-9275-9492}\inst{\ref{inst:0036}}
\and R.        ~Santove\~{n}a                 \orcit{0000-0002-9257-2131}\inst{\ref{inst:0098}}
\and M.        ~Sarasso                       \orcit{0000-0001-5121-0727}\inst{\ref{inst:0044}}
\and C.        ~Sarrate Riera                 \inst{\ref{inst:0052},\ref{inst:0027},\ref{inst:0029}}
\and E.        ~Sciacca                       \orcit{0000-0002-5574-2787}\inst{\ref{inst:0113}}
\and J.C.      ~Segovia                       \inst{\ref{inst:0090}}
\and D.        ~S\'{e}gransan                 \orcit{0000-0003-2355-8034}\inst{\ref{inst:0024}}
\and S.        ~Shahaf                        \orcit{0000-0001-9298-8068}\inst{\ref{inst:0442}}
\and A.        ~Siebert                       \orcit{0000-0001-8059-2840}\inst{\ref{inst:0117},\ref{inst:0463}}
\and L.        ~Siltala                       \orcit{0000-0002-6938-794X}\inst{\ref{inst:0121}}
\and E.        ~Slezak                        \inst{\ref{inst:0004}}
\and R.L.      ~Smart                         \orcit{0000-0002-4424-4766}\inst{\ref{inst:0044},\ref{inst:0226}}
\and O.N.      ~Snaith                        \orcit{0000-0003-1414-1296}\inst{\ref{inst:0018},\ref{inst:0469}}
\and E.        ~Solano                        \orcit{0000-0003-1885-5130}\inst{\ref{inst:0470}}
\and F.        ~Solitro                       \inst{\ref{inst:0065}}
\and D.        ~Souami                        \orcit{0000-0003-4058-0815}\inst{\ref{inst:0318},\ref{inst:0473}}
\and J.        ~Souchay                       \inst{\ref{inst:0082}}
\and L.        ~Spina                         \orcit{0000-0002-9760-6249}\inst{\ref{inst:0015}}
\and E.        ~Spitoni                       \orcit{0000-0001-9715-5727}\inst{\ref{inst:0004},\ref{inst:0477}}
\and F.        ~Spoto                         \orcit{0000-0001-7319-5847}\inst{\ref{inst:0478}}
\and L.A.      ~Squillante                    \inst{\ref{inst:0065}}
\and I.A.      ~Steele                        \orcit{0000-0001-8397-5759}\inst{\ref{inst:0340}}
\and H.        ~Steidelm\"{ u}ller            \inst{\ref{inst:0030}}
\and J.        ~Surdej                        \orcit{0000-0002-7005-1976}\inst{\ref{inst:0005}}
\and L.        ~Szabados                      \orcit{0000-0002-2046-4131}\inst{\ref{inst:0142}}
\and F.        ~Taris                         \inst{\ref{inst:0082}}
\and M.B.      ~Taylor                        \orcit{0000-0002-4209-1479}\inst{\ref{inst:0485}}
\and K.        ~Tisani\'{c}                   \orcit{0000-0001-6382-4937}\inst{\ref{inst:0255}}
\and L.        ~Tolomei                       \orcit{0000-0002-3541-3230}\inst{\ref{inst:0065}}
\and F.        ~Torra                         \orcit{0000-0002-8429-299X}\inst{\ref{inst:0052},\ref{inst:0027},\ref{inst:0029}}
\and G.        ~Torralba Elipe                \orcit{0000-0001-8738-194X}\inst{\ref{inst:0098},\ref{inst:0492},\ref{inst:0493}}
\and M.        ~Trabucchi                     \orcit{0000-0002-1429-2388}\inst{\ref{inst:0494},\ref{inst:0024}}
\and M.        ~Tsantaki                      \orcit{0000-0002-0552-2313}\inst{\ref{inst:0036}}
\and A.        ~Ulla                          \orcit{0000-0001-6424-5005}\inst{\ref{inst:0497},\ref{inst:0498}}
\and N.        ~Unger                         \orcit{0000-0003-3993-7127}\inst{\ref{inst:0024}}
\and O.        ~Vanel                         \orcit{0000-0002-7898-0454}\inst{\ref{inst:0018}}
\and A.        ~Vecchiato                     \orcit{0000-0003-1399-5556}\inst{\ref{inst:0044}}
\and D.        ~Vicente                       \orcit{0000-0002-1584-1182}\inst{\ref{inst:0277}}
\and S.        ~Voutsinas                     \inst{\ref{inst:0094}}
\and M.        ~Weiler                        \inst{\ref{inst:0029},\ref{inst:0027},\ref{inst:0028}}
\and \L{}.     ~Wyrzykowski                   \orcit{0000-0002-9658-6151}\inst{\ref{inst:0301}}
\and H.        ~Zhao                          \orcit{0000-0003-2645-6869}\inst{\ref{inst:0004},\ref{inst:0509}}
\and J.        ~Zorec                         \orcit{0000-0003-1257-6915}\inst{\ref{inst:0510}}
\and T.        ~Zwitter                       \orcit{0000-0002-2325-8763}\inst{\ref{inst:0511}}
\and L.        ~Balaguer-N\'{u}\~{n}ez      \orcit{0000-0001-9789-7069}\inst{\ref{inst:0029},\ref{inst:0027},\ref{inst:0028}}
\and N.        ~Leclerc                      \orcit{0009-0001-5569-6098}\inst{\ref{inst:0018}}
\and S.        ~Morgenthaler               \orcit{0009-0005-6349-3716}\inst{\ref{inst:0555}}
\and G.        ~Robert                         \inst{\ref{inst:0202}}
\and S.        ~Zucker                    \orcit{0000-0003-3173-3138}\inst{\ref{inst:0185}}
}
\institute{
     Donald Bren School of Information and Computer Sciences, University of California, Irvine, CA 92697, USA\relax                                                                                                                                                                                                                                                                                                  \label{inst:0001}
\and CENTRA, Faculdade de Ci\^{e}ncias, Universidade de Lisboa, Edif. C8, Campo Grande, 1749-016 Lisboa, Portugal\relax                                                                                                                                                                                                                                                                                              \label{inst:0002}
\and Laboratoire d'astrophysique de Bordeaux, Univ. Bordeaux, CNRS, B18N, all{\'e}e Geoffroy Saint-Hilaire, 33615 Pessac, France\relax                                                                                                                                                                                                                                                                               \label{inst:0003}
\and Universit\'{e} C\^{o}te d'Azur, Observatoire de la C\^{o}te d'Azur, CNRS, Laboratoire Lagrange, Bd de l'Observatoire, CS 34229, 06304 Nice Cedex 4, France\relax                                                                                                                                                                                                                                                \label{inst:0004}
\and Institut d'Astrophysique et de G\'{e}ophysique, Universit\'{e} de Li\`{e}ge, 19c, All\'{e}e du 6 Ao\^{u}t, B-4000 Li\`{e}ge, Belgium\relax                                                                                                                                                                                                                                                                      \label{inst:0005}
\and Instituto de Astronomia, Geof\`{i}sica e Ci\^{e}ncias Atmosf\'{e}ricas, Universidade de S\~{a}o Paulo, Rua do Mat\~{a}o, 1226, Cidade Universitaria, 05508-900 S\~{a}o Paulo, SP, Brazil\relax                                                                                                                                                                                                                  \label{inst:0007}
\and CNES Centre Spatial de Toulouse, 18 avenue Edouard Belin, 31401 Toulouse Cedex 9, France\relax                                                                                                                                                                                                                                                                                                                  \label{inst:0011}
\and ATOS for CNES Centre Spatial de Toulouse, 18 avenue Edouard Belin, 31401 Toulouse Cedex 9, France\relax                                                                                                                                                                                                                                                                                                         \label{inst:0012}
\and Leiden Observatory, Leiden University, Niels Bohrweg 2, 2333 CA Leiden, The Netherlands\relax                                                                                                                                                                                                                                                                                                                   \label{inst:0014}
\and INAF - Osservatorio astronomico di Padova, Vicolo Osservatorio 5, 35122 Padova, Italy\relax                                                                                                                                                                                                                                                                                                                     \label{inst:0015}
\and European Space Agency (ESA), European Space Research and Technology Centre (ESTEC), Keplerlaan 1, 2201AZ, Noordwijk, The Netherlands\relax                                                                                                                                                                                                                                                                      \label{inst:0016}
\and GEPI, Observatoire de Paris, Universit\'{e} PSL, CNRS, 5 Place Jules Janssen, 92190 Meudon, France\relax                                                                                                                                                                                                                                                                                                        \label{inst:0018}
\and Univ. Grenoble Alpes, CNRS, IPAG, 38000 Grenoble, France\relax                                                                                                                                                                                                                                                                                                                                                  \label{inst:0019}
\and Astronomisches Rechen-Institut, Zentrum f\"{ u}r Astronomie der Universit\"{ a}t Heidelberg, M\"{ o}nchhofstr. 12-14, 69120 Heidelberg, Germany\relax                                                                                                                                                                                                                                                           \label{inst:0021}
\and Institute of Astronomy, University of Cambridge, Madingley Road, Cambridge CB3 0HA, United Kingdom\relax                                                                                                                                                                                                                                                                                                        \label{inst:0023}
\and Department of Astronomy, University of Geneva, Chemin Pegasi 51, 1290 Versoix, Switzerland\relax                                                                                                                                                                                                                                                                                                                \label{inst:0024}
\and European Space Agency (ESA), European Space Astronomy Centre (ESAC), Camino bajo del Castillo, s/n, Urbanizaci\'{o}n Villafranca del Castillo, Villanueva de la Ca\~{n}ada, 28692 Madrid, Spain\relax                                                                                                                                                                                                           \label{inst:0025}
\and Aurora Technology for European Space Agency (ESA), Camino bajo del Castillo, s/n, Urbanizaci\'{o}n Villafranca del Castillo, Villanueva de la Ca\~{n}ada, 28692 Madrid, Spain\relax                                                                                                                                                                                                                             \label{inst:0026}
\and Institut de Ci\`{e}ncies del Cosmos (ICCUB), Universitat  de  Barcelona  (UB), Mart\'{i} i  Franqu\`{e}s  1, 08028 Barcelona, Spain\relax                                                                                                                                                                                                                                                                       \label{inst:0027}
\and Departament de F\'{i}sica Qu\`{a}ntica i Astrof\'{i}sica (FQA), Universitat de Barcelona (UB), c. Mart\'{i} i Franqu\`{e}s 1, 08028 Barcelona, Spain\relax                                                                                                                                                                                                                                                      \label{inst:0028}
\and Institut d'Estudis Espacials de Catalunya (IEEC), c. Gran Capit\`{a}, 2-4, 08034 Barcelona, Spain\relax                                                                                                                                                                                                                                                                                                         \label{inst:0029}
\and Lohrmann Observatory, Technische Universit\"{ a}t Dresden, Mommsenstra{\ss}e 13, 01062 Dresden, Germany\relax                                                                                                                                                                                                                                                                                                   \label{inst:0030}
\and Lund Observatory, Division of Astrophysics, Department of Physics, Lund University, Box 43, 22100 Lund, Sweden\relax                                                                                                                                                                                                                                                                                            \label{inst:0032}
\and INAF - Osservatorio Astrofisico di Arcetri, Largo Enrico Fermi 5, 50125 Firenze, Italy\relax                                                                                                                                                                                                                                                                                                                    \label{inst:0036}
\and Nicolaus Copernicus Astronomical Center, Polish Academy of Sciences, ul. Bartycka 18, 00-716 Warsaw, Poland\relax                                                                                                                                                                                                                                                                                               \label{inst:0038}
\and Max Planck Institute for Astronomy, K\"{ o}nigstuhl 17, 69117 Heidelberg, Germany\relax                                                                                                                                                                                                                                                                                                                         \label{inst:0041}
\and Mullard Space Science Laboratory, University College London, Holmbury St Mary, Dorking, Surrey RH5 6NT, United Kingdom\relax                                                                                                                                                                                                                                                                                    \label{inst:0043}
\and INAF - Osservatorio Astrofisico di Torino, via Osservatorio 20, 10025 Pino Torinese (TO), Italy\relax                                                                                                                                                                                                                                                                                                           \label{inst:0044}
\and Department of Astronomy, University of Geneva, Chemin d'Ecogia 16, 1290 Versoix, Switzerland\relax                                                                                                                                                                                                                                                                                                              \label{inst:0049}
\and Royal Observatory of Belgium, Ringlaan 3, 1180 Brussels, Belgium\relax                                                                                                                                                                                                                                                                                                                                          \label{inst:0051}
\and DAPCOM Data Services, c. dels Vilabella, 5-7, 80500 Vic, Barcelona, Spain\relax                                                                                                                                                                                                                                                                                                                                 \label{inst:0052}
\and ALTEC S.p.a, Corso Marche, 79,10146 Torino, Italy\relax                                                                                                                                                                                                                                                                                                                                                         \label{inst:0065}
\and Sednai S\`{a}rl, Geneva, Switzerland\relax                                                                                                                                                                                                                                                                                                                                                                      \label{inst:0067}
\and Gaia DPAC Project Office, ESAC, Camino bajo del Castillo, s/n, Urbanizaci\'{o}n Villafranca del Castillo, Villanueva de la Ca\~{n}ada, 28692 Madrid, Spain\relax                                                                                                                                                                                                                                                \label{inst:0075}
\and Telespazio UK S.L. for European Space Agency (ESA), Camino bajo del Castillo, s/n, Urbanizaci\'{o}n Villafranca del Castillo, Villanueva de la Ca\~{n}ada, 28692 Madrid, Spain\relax                                                                                                                                                                                                                            \label{inst:0080}
\and SYRTE, Observatoire de Paris, Universit\'{e} PSL, CNRS, Sorbonne Universit\'{e}, LNE, 61 avenue de l'Observatoire 75014 Paris, France\relax                                                                                                                                                                                                                                                                     \label{inst:0082}
\and IMCCE, Observatoire de Paris, Universit\'{e} PSL, CNRS, Sorbonne Universit{\'e}, Univ. Lille, 77 av. Denfert-Rochereau, 75014 Paris, France\relax                                                                                                                                                                                                                                                               \label{inst:0084}
\and Serco Gesti\'{o}n de Negocios for European Space Agency (ESA), Camino bajo del Castillo, s/n, Urbanizaci\'{o}n Villafranca del Castillo, Villanueva de la Ca\~{n}ada, 28692 Madrid, Spain\relax                                                                                                                                                                                                                 \label{inst:0090}
\and INAF - Osservatorio di Astrofisica e Scienza dello Spazio di Bologna, via Piero Gobetti 93/3, 40129 Bologna, Italy\relax                                                                                                                                                                                                                                                                                        \label{inst:0091}
\and CRAAG - Centre de Recherche en Astronomie, Astrophysique et G\'{e}ophysique, Route de l'Observatoire Bp 63 Bouzareah 16340 Algiers, Algeria\relax                                                                                                                                                                                                                                                               \label{inst:0093}
\and Institute for Astronomy, University of Edinburgh, Royal Observatory, Blackford Hill, Edinburgh EH9 3HJ, United Kingdom\relax                                                                                                                                                                                                                                                                                    \label{inst:0094}
\and RHEA for European Space Agency (ESA), Camino bajo del Castillo, s/n, Urbanizaci\'{o}n Villafranca del Castillo, Villanueva de la Ca\~{n}ada, 28692 Madrid, Spain\relax                                                                                                                                                                                                                                          \label{inst:0097}
\and CIGUS CITIC - Department of Computer Science and Information Technologies, University of A Coru\~{n}a, Campus de Elvi\~{n}a s/n, A Coru\~{n}a, 15071, Spain\relax                                                                                                                                                                                                                                               \label{inst:0098}
\and ATG Europe for European Space Agency (ESA), Camino bajo del Castillo, s/n, Urbanizaci\'{o}n Villafranca del Castillo, Villanueva de la Ca\~{n}ada, 28692 Madrid, Spain\relax                                                                                                                                                                                                                                    \label{inst:0100}
\and Kavli Institute for Cosmology Cambridge, Institute of Astronomy, Madingley Road, Cambridge, CB3 0HA\relax                                                                                                                                                                                                                                                                                                       \label{inst:0105}
\and Department of Astrophysics, Astronomy and Mechanics, National and Kapodistrian University of Athens, Panepistimiopolis, Zografos, 15783 Athens, Greece\relax                                                                                                                                                                                                                                                    \label{inst:0106}
\and INAF - Osservatorio Astrofisico di Catania, via S. Sofia 78, 95123 Catania, Italy\relax                                                                                                                                                                                                                                                                                                                         \label{inst:0113}
\and Dipartimento di Fisica e Astronomia ""Ettore Majorana"", Universit\`{a} di Catania, Via S. Sofia 64, 95123 Catania, Italy\relax                                                                                                                                                                                                                                                                                 \label{inst:0114}
\and Universit\'{e} de Strasbourg, CNRS, Observatoire astronomique de Strasbourg, UMR 7550,  11 rue de l'Universit\'{e}, 67000 Strasbourg, France\relax                                                                                                                                                                                                                                                              \label{inst:0117}
\and INAF - Osservatorio Astronomico di Roma, Via Frascati 33, 00078 Monte Porzio Catone (Roma), Italy\relax                                                                                                                                                                                                                                                                                                         \label{inst:0118}
\and Space Science Data Center - ASI, Via del Politecnico SNC, 00133 Roma, Italy\relax                                                                                                                                                                                                                                                                                                                               \label{inst:0119}
\and Department of Physics, University of Helsinki, P.O. Box 64, 00014 Helsinki, Finland\relax                                                                                                                                                                                                                                                                                                                       \label{inst:0121}
\and Finnish Geospatial Research Institute FGI, Vuorimiehentie 5, 02150 Espoo, Finland\relax                                                                                                                                                                                                                                                                                                                         \label{inst:0122}
\and Institut UTINAM CNRS UMR6213, Universit\'{e} de Franche-Comt\'{e}, OSU THETA Franche-Comt\'{e} Bourgogne, Observatoire de Besan\c{c}on, BP1615, 25010 Besan\c{c}on Cedex, France\relax                                                                                                                                                                                                                          \label{inst:0131}
\and HE Space Operations BV for European Space Agency (ESA), Keplerlaan 1, 2201AZ, Noordwijk, The Netherlands\relax                                                                                                                                                                                                                                                                                                  \label{inst:0132}
\and Dpto. de Inteligencia Artificial, UNED, c/ Juan del Rosal 16, 28040 Madrid, Spain\relax                                                                                                                                                                                                                                                                                                                         \label{inst:0133}
\and Institut d'Astronomie et d'Astrophysique, Universit\'{e} Libre de Bruxelles CP 226, Boulevard du Triomphe, 1050 Brussels, Belgium\relax                                                                                                                                                                                                                                                                         \label{inst:0135}
\and Leibniz Institute for Astrophysics Potsdam (AIP), An der Sternwarte 16, 14482 Potsdam, Germany\relax                                                                                                                                                                                                                                                                                                            \label{inst:0140}
\and Konkoly Observatory, Research Centre for Astronomy and Earth Sciences, E\"{ o}tv\"{ o}s Lor{\'a}nd Research Network (ELKH), MTA Centre of Excellence, Konkoly Thege Mikl\'{o}s \'{u}t 15-17, 1121 Budapest, Hungary\relax                                                                                                                                                                                       \label{inst:0142}
\and ELTE E\"{ o}tv\"{ o}s Lor\'{a}nd University, Institute of Physics, 1117, P\'{a}zm\'{a}ny P\'{e}ter s\'{e}t\'{a}ny 1A, Budapest, Hungary\relax                                                                                                                                                                                                                                                                   \label{inst:0143}
\and Instituut voor Sterrenkunde, KU Leuven, Celestijnenlaan 200D, 3001 Leuven, Belgium\relax                                                                                                                                                                                                                                                                                                                        \label{inst:0145}
\and Department of Astrophysics/IMAPP, Radboud University, P.O.Box 9010, 6500 GL Nijmegen, The Netherlands\relax                                                                                                                                                                                                                                                                                                     \label{inst:0146}
\and University of Vienna, Department of Astrophysics, T\"{ u}rkenschanzstra{\ss}e 17, A1180 Vienna, Austria\relax                                                                                                                                                                                                                                                                                                   \label{inst:0151}
\and Institute of Physics, Ecole Polytechnique F\'ed\'erale de Lausanne (EPFL), Observatoire de Sauverny, 1290 Versoix, Switzerland\relax                                                                                                                                                                                                                                                                            \label{inst:0152}
\and Quasar Science Resources for European Space Agency (ESA), Camino bajo del Castillo, s/n, Urbanizaci\'{o}n Villafranca del Castillo, Villanueva de la Ca\~{n}ada, 28692 Madrid, Spain\relax                                                                                                                                                                                                                      \label{inst:0156}
\and LASIGE, Faculdade de Ci\^{e}ncias, Universidade de Lisboa, Edif. C6, Campo Grande, 1749-016 Lisboa, Portugal\relax                                                                                                                                                                                                                                                                                              \label{inst:0163}
\and School of Physics and Astronomy , University of Leicester, University Road, Leicester LE1 7RH, United Kingdom\relax                                                                                                                                                                                                                                                                                             \label{inst:0164}
\and School of Physics and Astronomy, Tel Aviv University, Tel Aviv 6997801, Israel\relax                                                                                                                                                                                                                                                                                                                            \label{inst:0168}
\and Cavendish Laboratory, JJ Thomson Avenue, Cambridge CB3 0HE, United Kingdom\relax                                                                                                                                                                                                                                                                                                                                \label{inst:0169}
\and Telespazio for CNES Centre Spatial de Toulouse, 18 avenue Edouard Belin, 31401 Toulouse Cedex 9, France\relax                                                                                                                                                                                                                                                                                                   \label{inst:0171}
\and National Observatory of Athens, I. Metaxa and Vas. Pavlou, Palaia Penteli, 15236 Athens, Greece\relax                                                                                                                                                                                                                                                                                                           \label{inst:0174}
\and University of Turin, Department of Physics, Via Pietro Giuria 1, 10125 Torino, Italy\relax                                                                                                                                                                                                                                                                                                                      \label{inst:0177}
\and Depto. Estad\'istica e Investigaci\'on Operativa. Universidad de C\'adiz, Avda. Rep\'ublica Saharaui s/n, 11510 Puerto Real, C\'adiz, Spain\relax                                                                                                                                                                                                                                                               \label{inst:0178}
\and EURIX S.r.l., Corso Vittorio Emanuele II 61, 10128, Torino, Italy\relax                                                                                                                                                                                                                                                                                                                                         \label{inst:0184}
\and Porter School of the Environment and Earth Sciences, Tel Aviv University, Tel Aviv 6997801, Israel\relax                                                                                                                                                                                                                                                                                                        \label{inst:0185}
\and HE Space Operations BV for European Space Agency (ESA), Camino bajo del Castillo, s/n, Urbanizaci\'{o}n Villafranca del Castillo, Villanueva de la Ca\~{n}ada, 28692 Madrid, Spain\relax                                                                                                                                                                                                                        \label{inst:0187}
\and LFCA/DAS,Universidad de Chile,CNRS,Casilla 36-D, Santiago, Chile\relax                                                                                                                                                                                                                                                                                                                                          \label{inst:0189}
\and SISSA - Scuola Internazionale Superiore di Studi Avanzati, via Bonomea 265, 34136 Trieste, Italy\relax                                                                                                                                                                                                                                                                                                          \label{inst:0193}
\and University of Turin, Department of Computer Sciences, Corso Svizzera 185, 10149 Torino, Italy\relax                                                                                                                                                                                                                                                                                                             \label{inst:0201}
\and Thales Services for CNES Centre Spatial de Toulouse, 18 avenue Edouard Belin, 31401 Toulouse Cedex 9, France\relax                                                                                                                                                                                                                                                                                              \label{inst:0202}
\and Dpto. de Matem\'{a}tica Aplicada y Ciencias de la Computaci\'{o}n, Univ. de Cantabria, ETS Ingenieros de Caminos, Canales y Puertos, Avda. de los Castros s/n, 39005 Santander, Spain\relax                                                                                                                                                                                                                     \label{inst:0204}
\and Institut de F\'{i}sica d'Altes Energies (IFAE), The Barcelona Institute of Science and Technology, Campus UAB, 08193 Bellaterra (Barcelona), Spain\relax                                                                                                                                                                                                                                                        \label{inst:0210}
\and Port d'Informaci\'{o} Cient\'{i}fica (PIC), Campus UAB, C. Albareda s/n, 08193 Bellaterra (Barcelona), Spain\relax                                                                                                                                                                                                                                                                                              \label{inst:0211}
\and Instituto de Astrof\'{i}sica, Universidad Andres Bello, Fernandez Concha 700, Las Condes, Santiago RM, Chile\relax                                                                                                                                                                                                                                                                                              \label{inst:0219}
\and Centre for Astrophysics Research, University of Hertfordshire, College Lane, AL10 9AB, Hatfield, United Kingdom\relax                                                                                                                                                                                                                                                                                           \label{inst:0226}
\and University of Turin, Mathematical Department ""G.Peano"", Via Carlo Alberto 10, 10123 Torino, Italy\relax                                                                                                                                                                                                                                                                                                       \label{inst:0230}
\and INAF - Osservatorio Astronomico d'Abruzzo, Via Mentore Maggini, 64100 Teramo, Italy\relax                                                                                                                                                                                                                                                                                                                       \label{inst:0235}
\and M\'{e}socentre de calcul de Franche-Comt\'{e}, Universit\'{e} de Franche-Comt\'{e}, 16 route de Gray, 25030 Besan\c{c}on Cedex, France\relax                                                                                                                                                                                                                                                                    \label{inst:0243}
\and Ru{\dj}er Bo\v{s}kovi\'{c} Institute, Bijeni\v{c}ka cesta 54, 10000 Zagreb, Croatia\relax                                                                                                                                                                                                                                                                                                                       \label{inst:0255}
\and Astrophysics Research Centre, School of Mathematics and Physics, Queen's University Belfast, Belfast BT7 1NN, UK\relax                                                                                                                                                                                                                                                                                          \label{inst:0257}
\and Data Science and Big Data Lab, Pablo de Olavide University, 41013, Seville, Spain\relax                                                                                                                                                                                                                                                                                                                         \label{inst:0270}
\and Institute of Astrophysics, FORTH, Crete, Greece\relax                                                                                                                                                                                                                                                                                                                                                           \label{inst:0276}
\and Barcelona Supercomputing Center (BSC), Pla\c{c}a Eusebi G\"{ u}ell 1-3, 08034-Barcelona, Spain\relax                                                                                                                                                                                                                                                                                                            \label{inst:0277}
\and ETSE Telecomunicaci\'{o}n, Universidade de Vigo, Campus Lagoas-Marcosende, 36310 Vigo, Galicia, Spain\relax                                                                                                                                                                                                                                                                                                     \label{inst:0281}
\and F.R.S.-FNRS, Rue d'Egmont 5, 1000 Brussels, Belgium\relax                                                                                                                                                                                                                                                                                                                                                       \label{inst:0284}
\and Asteroid Engineering Laboratory, Lule\aa{} University of Technology, Box 848, S-981 28 Kiruna, Sweden\relax                                                                                                                                                                                                                                                                                                     \label{inst:0286}
\and Kapteyn Astronomical Institute, University of Groningen, Landleven 12, 9747 AD Groningen, The Netherlands\relax                                                                                                                                                                                                                                                                                                 \label{inst:0293}
\and IAC - Instituto de Astrofisica de Canarias, Via L\'{a}ctea s/n, 38200 La Laguna S.C., Tenerife, Spain\relax                                                                                                                                                                                                                                                                                                     \label{inst:0295}
\and Department of Astrophysics, University of La Laguna, Via L\'{a}ctea s/n, 38200 La Laguna S.C., Tenerife, Spain\relax                                                                                                                                                                                                                                                                                            \label{inst:0296}
\and Astronomical Observatory, University of Warsaw,  Al. Ujazdowskie 4, 00-478 Warszawa, Poland\relax                                                                                                                                                                                                                                                                                                               \label{inst:0301}
\and Research School of Astronomy \& Astrophysics, Australian National University, Cotter Road, Weston, ACT 2611, Australia\relax                                                                                                                                                                                                                                                                                     \label{inst:0302}
\and European Space Agency (ESA, retired), European Space Research and Technology Centre (ESTEC), Keplerlaan 1, 2201AZ, Noordwijk, The Netherlands\relax                                                                                                                                                                                                                                                             \label{inst:0303}
\and LESIA, Observatoire de Paris, Universit\'{e} PSL, CNRS, Sorbonne Universit\'{e}, Universit\'{e} de Paris, 5 Place Jules Janssen, 92190 Meudon, France\relax                                                                                                                                                                                                                                                     \label{inst:0318}
\and Universit\'{e} Rennes, CNRS, IPR (Institut de Physique de Rennes) - UMR 6251, 35000 Rennes, France\relax                                                                                                                                                                                                                                                                                                        \label{inst:0319}
\and INAF - Osservatorio Astronomico di Capodimonte, Via Moiariello 16, 80131, Napoli, Italy\relax                                                                                                                                                                                                                                                                                                                   \label{inst:0321}
\and Shanghai Astronomical Observatory, Chinese Academy of Sciences, 80 Nandan Road, Shanghai 200030, People's Republic of China\relax                                                                                                                                                                                                                                                                               \label{inst:0324}
\and University of Chinese Academy of Sciences, No.19(A) Yuquan Road, Shijingshan District, Beijing 100049, People's Republic of China\relax                                                                                                                                                                                                                                                                         \label{inst:0326}
\and S\~{a}o Paulo State University, Grupo de Din\^{a}mica Orbital e Planetologia, CEP 12516-410, Guaratinguet\'{a}, SP, Brazil\relax                                                                                                                                                                                                                                                                                \label{inst:0328}
\and Niels Bohr Institute, University of Copenhagen, Juliane Maries Vej 30, 2100 Copenhagen {\O}, Denmark\relax                                                                                                                                                                                                                                                                                                      \label{inst:0331}
\and DXC Technology, Retortvej 8, 2500 Valby, Denmark\relax                                                                                                                                                                                                                                                                                                                                                          \label{inst:0332}
\and Las Cumbres Observatory, 6740 Cortona Drive Suite 102, Goleta, CA 93117, USA\relax                                                                                                                                                                                                                                                                                                                              \label{inst:0333}
\and CIGUS CITIC, Department of Nautical Sciences and Marine Engineering, University of A Coru\~{n}a, Paseo de Ronda 51, 15071, A Coru\~{n}a, Spain\relax                                                                                                                                                                                                                                                            \label{inst:0339}
\and Astrophysics Research Institute, Liverpool John Moores University, 146 Brownlow Hill, Liverpool L3 5RF, United Kingdom\relax                                                                                                                                                                                                                                                                                    \label{inst:0340}
\and IRAP, Universit\'{e} de Toulouse, CNRS, UPS, CNES, 9 Av. colonel Roche, BP 44346, 31028 Toulouse Cedex 4, France\relax                                                                                                                                                                                                                                                                                          \label{inst:0347}
\and MTA CSFK Lend\"{ u}let Near-Field Cosmology Research Group, Konkoly Observatory, MTA Research Centre for Astronomy and Earth Sciences, Konkoly Thege Mikl\'{o}s \'{u}t 15-17, 1121 Budapest, Hungary\relax                                                                                                                                                                                                      \label{inst:0369}
\and Pervasive Technologies s.l., c. Saragossa 118, 08006 Barcelona, Spain\relax                                                                                                                                                                                                                                                                                                                                     \label{inst:0377}
\and School of Physics and Astronomy, University of Leicester, University Road, Leicester LE1 7RH, United Kingdom\relax                                                                                                                                                                                                                                                                                              \label{inst:0394}
\and Villanova University, Department of Astrophysics and Planetary Science, 800 E Lancaster Avenue, Villanova PA 19085, USA\relax                                                                                                                                                                                                                                                                                   \label{inst:0415}
\and Departmento de F\'{i}sica de la Tierra y Astrof\'{i}sica, Universidad Complutense de Madrid, 28040 Madrid, Spain\relax                                                                                                                                                                                                                                                                                          \label{inst:0418}
\and INAF - Osservatorio Astronomico di Brera, via E. Bianchi, 46, 23807 Merate (LC), Italy\relax                                                                                                                                                                                                                                                                                                                    \label{inst:0420}
\and National Astronomical Observatory of Japan, 2-21-1 Osawa, Mitaka, Tokyo 181-8588, Japan\relax                                                                                                                                                                                                                                                                                                                   \label{inst:0422}
\and Department of Particle Physics and Astrophysics, Weizmann Institute of Science, Rehovot 7610001, Israel\relax                                                                                                                                                                                                                                                                                                   \label{inst:0442}
\and Centre de Donn\'{e}es Astronomique de Strasbourg, Strasbourg, France\relax                                                                                                                                                                                                                                                                                                                                      \label{inst:0463}
\and University of Exeter, School of Physics and Astronomy, Stocker Road, Exeter, EX2 7SJ, United Kingdom\relax                                                                                                                                                                                                                                                                                                      \label{inst:0469}
\and Departamento de Astrof\'{i}sica, Centro de Astrobiolog\'{i}a (CSIC-INTA), ESA-ESAC. Camino Bajo del Castillo s/n. 28692 Villanueva de la Ca\~{n}ada, Madrid, Spain\relax                                                                                                                                                                                                                                        \label{inst:0470}
\and naXys, Department of Mathematics, University of Namur, Rue de Bruxelles 61, 5000 Namur, Belgium\relax                                                                                                                                                                                                                                                                                                           \label{inst:0473}
\and INAF. Osservatorio Astronomico di Trieste, via G.B. Tiepolo 11, 34131, Trieste, Italy\relax                                                                                                                                                                                                                                                                                                                     \label{inst:0477}
\and Harvard-Smithsonian Center for Astrophysics, 60 Garden St., MS 15, Cambridge, MA 02138, USA\relax                                                                                                                                                                                                                                                                                                               \label{inst:0478}
\and H H Wills Physics Laboratory, University of Bristol, Tyndall Avenue, Bristol BS8 1TL, United Kingdom\relax                                                                                                                                                                                                                                                                                                      \label{inst:0485}
\and Escuela de Arquitectura y Polit\'{e}cnica - Universidad Europea de Valencia, Spain\relax                                                                                                                                                                                                                                                                                                                        \label{inst:0492}
\and Escuela Superior de Ingenier\'{i}a y Tecnolog\'{i}a - Universidad Internacional de la Rioja, Spain\relax                                                                                                                                                                                                                                                                                                        \label{inst:0493}
\and Department of Physics and Astronomy G. Galilei, University of Padova, Vicolo dell'Osservatorio 3, 35122, Padova, Italy\relax                                                                                                                                                                                                                                                                                    \label{inst:0494}
\and Applied Physics Department, Universidade de Vigo, 36310 Vigo, Spain\relax                                                                                                                                                                                                                                                                                                                                       \label{inst:0497}
\and Instituto de F{'i}sica e Ciencias Aeroespaciais (IFCAE), Universidade de Vigo‚ \'{A} Campus de As Lagoas, 32004 Ourense, Spain\relax                                                                                                                                                                                                                                                                          \label{inst:0498}
\and Purple Mountain Observatory, Chinese Academy of Sciences, Nanjing 210023, China\relax                                                                                                                                                                                                                                                                                                                           \label{inst:0509}
\and Sorbonne Universit\'{e}, CNRS, UMR7095, Institut d'Astrophysique de Paris, 98bis bd. Arago, 75014 Paris, France\relax                                                                                                                                                                                                                                                                                           \label{inst:0510}
\and Faculty of Mathematics and Physics, University of Ljubljana, Jadranska ulica 19, 1000 Ljubljana, Slovenia\relax                                                                                                                                                                                                                                                                                                 \label{inst:0511}
\and Institute of Mathematics, Ecole Polytechnique F\'ed\'erale de Lausanne (EPFL), Switzerland\relax
\label{inst:0555}
}

\date{Received 23 June 2023 / Accepted 09 October 2023}
\abstract
   {Strongly lensed quasars are fundamental sources for cosmology. The \gaia space mission covers the entire sky with the unprecedented resolution of $0.18$" in the optical, making it an ideal instrument to search for gravitational lenses down to the limiting magnitude of 21. Nevertheless, the previous \gaia Data Releases are known to be incomplete for small angular separations such as those expected for most lenses.}
   {We present the Data Processing and Analysis Consortium GravLens pipeline, which was built to analyse all \gaia detections around quasars and to cluster them into sources, thus producing a catalogue of secondary sources around each quasar. We analysed the resulting catalogue to produce scores that indicate source configurations that are compatible with strongly lensed quasars.} 
   {GravLens uses the DBSCAN unsupervised clustering algorithm to detect sources around quasars. The resulting catalogue of multiplets is then analysed with several methods to identify potential gravitational lenses. We developed and applied an outlier scoring method, a comparison between the average BP and RP spectra of the components, and we also used an extremely randomised tree algorithm. These methods produce scores to identify the most probable configurations and to establish a list of lens candidates.}
   {We analysed the environment of 3\,760\,032 quasars. A total of 4\,760\,920 sources, including the quasars, were found within 6\arcsec of the quasar positions. This list is given in the \gaia archive. In 87\% of cases, the quasar remains a single source, and in 501\,385 cases neighbouring sources were detected. We propose a list of 381 lensed candidates, of which we identified 49 as the most promising. Beyond these candidates, the associate tables in this Focused Product Release allow the entire community to explore the unique \gaia data for strong lensing studies further.}
{}
\keywords{gravitational lensing: strong, quasars: general, methods: data analysis, catalogues, surveys}

\titlerunning{A catalogue of sources around quasars to search for strongly lensed quasars}
\authorrunning{Gaia collaboration, A. Krone-Martins et al.}
\maketitle

\section{Introduction}\label{intro}
An extensive census of quasars (QSOs) is fundamental to many cosmological studies. Specifically, in cases where these quasars exhibit multiple images, strongly lensed quasars enable the estimation of the Hubble constant $H_0$ directly from time delay measurements~\citep[e.g.][]{1964MNRAS.128..307R,2005IAUS..225..297C, 2020A&A...640A.105M, 2017MNRAS.468.2590S, 2019MNRAS.490.1743C, 2020MNRAS.498.1420W}. In addition to many other astrophysical and cosmological applications, these sources are also used for detailed studies of dark matter halos and their substructures~\citep[e.g.][]{2017MNRAS.471.2224N, 2017MNRAS.468.1426L, 2018PhRvD..97b3001D, 2020MNRAS.492L..12G, 2020MNRAS.492.5314N, 2020MNRAS.491.6077G, 2021MNRAS.507.2432G, 2021MNRAS.507.1202M}, 
and for constraining the dark energy equation of state \citep[e.g.][]{2004PhRvD..70d3534L, 2011PhRvD..84l3529L, 2012AJ....143..120O, 2017ApJ...834...75X, 2019MNRAS.487.1980L, 2022MNRAS.514.1433W}.  All lensed quasars discussed in this work are strongly lensed by definition.

Detecting lenses has been historically challenging because (i) most lensed quasars are distant and hence faint and (ii) broad sky coverage and high angular resolution are required as the angular separations between the lensed images are small. Moreover, chance projection can create astrometric configurations and even broad-band photometry similar to those expected from lensed quasars such that a reliable confirmation requires spectroscopic measurements of the quasar images and the lensing galaxy. Due to these difficulties, the number of confirmed lensed quasars has been low for decades. The ESA/\gaia mission~\citep{2016Prusti} and its data releases, particularly since Data Release 2~\citep{2018A&A...616A...1G,2021A&A...649A...1G,2022arXiv220800211G}, have been dramatically improving this situation. 

Although \gaia was primarily designed to study the Milky Way through astrometry at the micro-arcsecond level, it produces an all-sky survey including millions of galaxies and QSOs~\citep{2012A&A...543A.100R, 2013A&A...556A.102K,2014A&A...568A.124D, 2015A&A...576A..74D, 2022arXiv220605681G, 2022Ducourant, KMFlisvos}. In their work, \citet{2014A&A...568A.124D} and \citet{2015A&A...576A..74D} define downlink criteria. The sources can be used to create a homogeneous, magnitude-limited survey of lensed quasars down to image separations of $\sim0.18$\arcsec. This is comparable to the angular resolution of the NASA/ESA Hubble Space Telescope, but \gaia measurements have all-sky coverage. During this decade, the recently launched ESA/NASA Euclid space mission \citep{EuclidM} will also be surveying ${\sim}14\,000$ square degrees at comparable angular resolutions, providing deep and high-resolution images in multiple bands for the first time at such large scales \citep{ECSurveyI2022}. This is expected to revolutionise strong lensing studies by enabling almost direct confirmation of thousands of lensed quasars \citep[e.g.][]{2022A&ARv..30....8T}, and complementing \gaia's $\mu as$ astrometry with deep and precise photometry.

Conservative estimates of the number of lensed QSOs detectable by \gaia in standard $\Lambda$CDM ($\Lambda$ cold dark matter) ) cosmology ($\Lambda=0.7$, $\Omega_m=0.3$) suggest that $\sim3\,000$ multiply imaged QSOs could be detected by \gaia of which $> 250$ would have $\ge3$ images and the rest doubles~\citep{2002GaiaWGRel, 2016A&A...590A..42F}. \gaia could then lead to a ten-fold increase in the number of lensed QSOs resulting in a homogeneous survey providing precise astrometry for all lensed images. Such a census is being built by the astronomical community, which continuously scans the \gaia data releases for follow-up spectroscopic confirmation~\citep[e.g.][]{2018Krone-Martins, 2018Ducourant,2018MNRAS.479.4345A,2018MNRAS.479.5060L,2019A&A...622A.165D,2019arXiv191208977K,2019MNRAS.483.4242L,2021Stern,2022arXiv220607714L,2022MNRAS.509..738D}. This endeavour is already resulting in a unique and statistically significant sample of lenses that will be used to study the evolution of the population of the deflecting galaxies and to constrain cosmological parameters, including the value of $H_0$ that is currently under significant tension \citep[e.g.][]{2019NatAs...3..891V,2021CQGra..38o3001D}.

Current \gaia Data Releases are still incomplete at the lowest angular separations~\citep[e.g.][]{2017A&A...599A..50A,2018A&A...616A..17A,2021A&A...649A...5F,2021Torra}, as expected for early mission products. This has led some known lensed images to lack \gaia counterparts~\citep[see, e.g.][]{2018Ducourant}. And this has slowed the identification of new lenses since most lensed QSOs that have yet to be discovered are probably angularly small. Thus, within the \gaia Data Processing and Analysis Consortium, we developed a dedicated processing chain to analyse the environment of quasars and produce a catalogue of sources and clusters, including new lens candidates for further studies and confirmation by the community. The major goals of this work are to make this focused \gaia data available to the community, to present the new data, and to call the attention of the community to the possibility of using this data to study currently known lensed quasars and to discover new lensed quasars while also providing a first, non-exhaustive, candidate list.

Finally, new astrophysical cases can emerge from the large lens samples expected. For instance, because the quasar is unique and point-like, but the images are seen through different parts of the lensing object, we can use the colors and magnitudes to study the obscuring dust in the line of sight. Cosmic dust reddens light but not always in the same way. Let $A_v$ designate the V band absorption and $B-V$ the color of an object. When there is reddening, we define the excess $B-V$ as $E(B-V)=(B-V) -(B-V)_0$ where $(B-V)_0$ is the intrinsic color of the object. Milky Way dust shows a pattern where $A_v\approx 3.1 E(B-V)$ but the dust absorption spectrum is not identical for all galaxies~\citep[e.g.][]{2003ApJ...594..279G}.  \gaia provides much of this information, so using a large sample, as should come out of this work, we can reconstruct and study dust at cosmological distances via the $A_v/E(B-V)$ ratio, possibly inferring the existence of major dust features in the Universe~\citep[e.g. as done in][for features in our galaxy]{2018A&A...616A.132L,2019ApJ...887...93G,2021NatAs...5..832L,2022arXiv220411715L}, a new method to the best of our knowledge.

This paper is organized as follows. Sect.~\ref{inputlist} presents the list of quasars and quasar candidates used as inputs to GravLens. Sect.~\ref{environment} describes how the GravLens algorithm clusters transits (individual detections) mapping transits to sources along with some remaining issues. Sect.~\ref{catalogue} describes the resulting catalog, the field contents, and the new sources. Afterward, in Sect.~\ref{search}, we present the methods developed to create lens scores and a list of candidates. Finally, we present our conclusions in Sect.~\ref{conclusion}. 

\section{The list of quasars}\label{inputlist}
Our processing starts from an input list of quasars. Since gravitational lenses are rare, with less than $\sim80$ lensed quasars with 4 images --\textit{quads} hereafter-- known today \citep{2021Stern}, we want to make this input list of quasars for the GravLens processing as complete as possible to maximise our chances of detecting new lenses. Therefore we tolerate moderate stellar or galaxy contamination in this input list.

\begin{figure*}[ht]
    \includegraphics[width=1\textwidth]{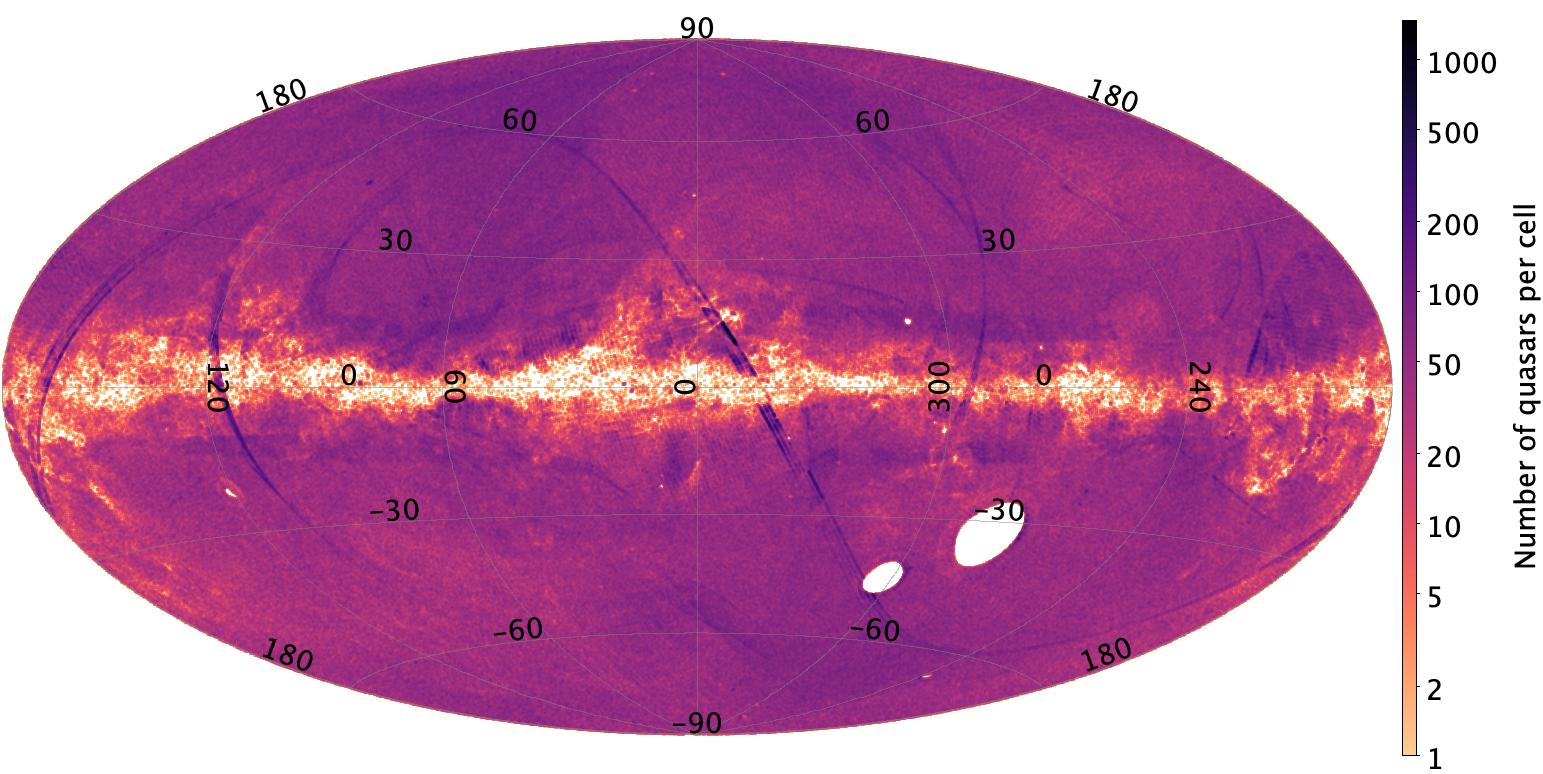} 
    \caption{Sky distribution in galactic coordinates of the quasars included in the input list. The cell of this map is approximately 0.2 deg$^2$, and the colour indicates the number of sources in each cell on a logarithmic scale.}
    \label{qso_lb}
\end{figure*}

\begin{figure*}
\centering
\begin{minipage}{0.49\textwidth} \centering \small
\includegraphics[width=0.98\textwidth]{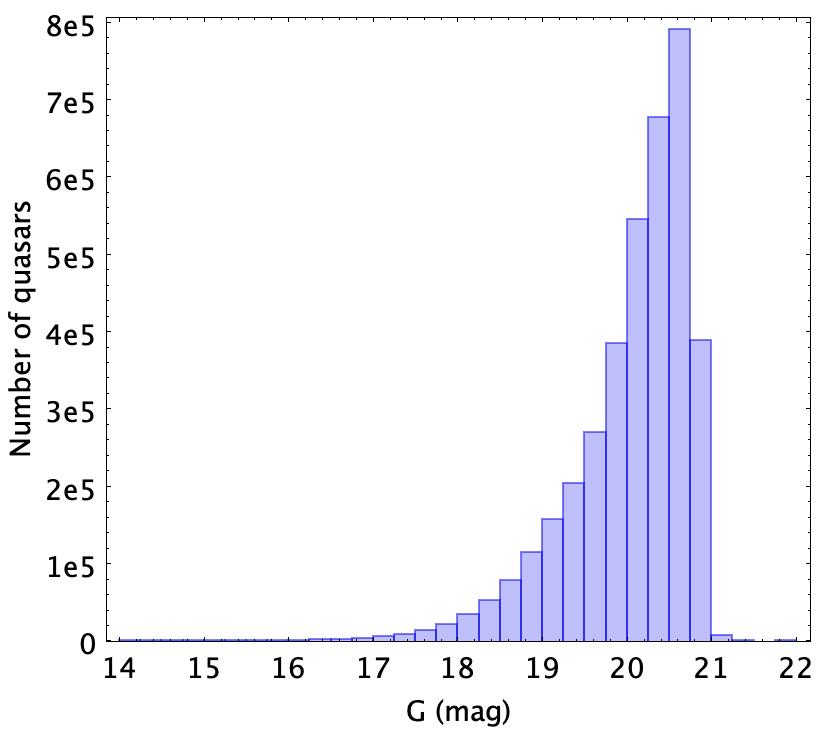} \\
(a)
\end{minipage}
\begin{minipage}{0.49\textwidth} \centering \small
\includegraphics[width=0.98\textwidth]{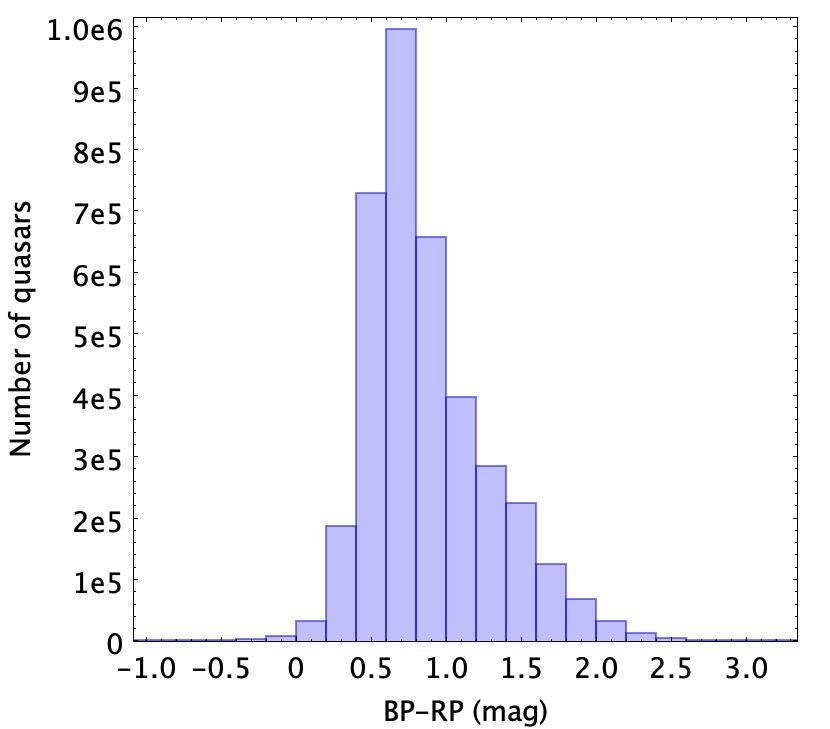} \\
(b)
\end{minipage} \\[0.5cm]

\begin{minipage}{0.49\textwidth} \centering \small
\includegraphics[width=0.98\textwidth]{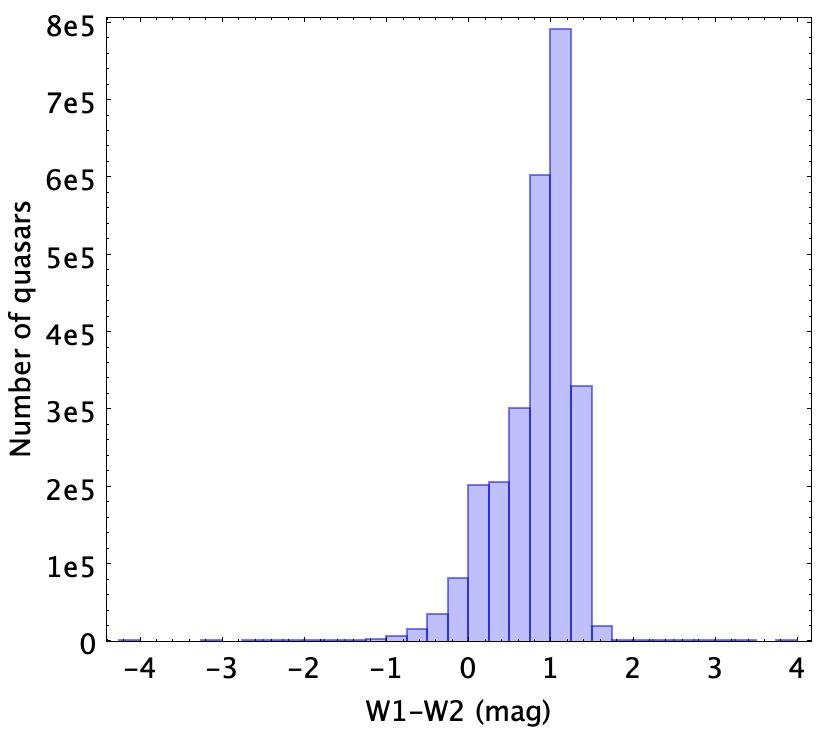} \\
(c)
\end{minipage}
\begin{minipage}{0.49\textwidth} \centering \small
\includegraphics[width=0.98\textwidth]{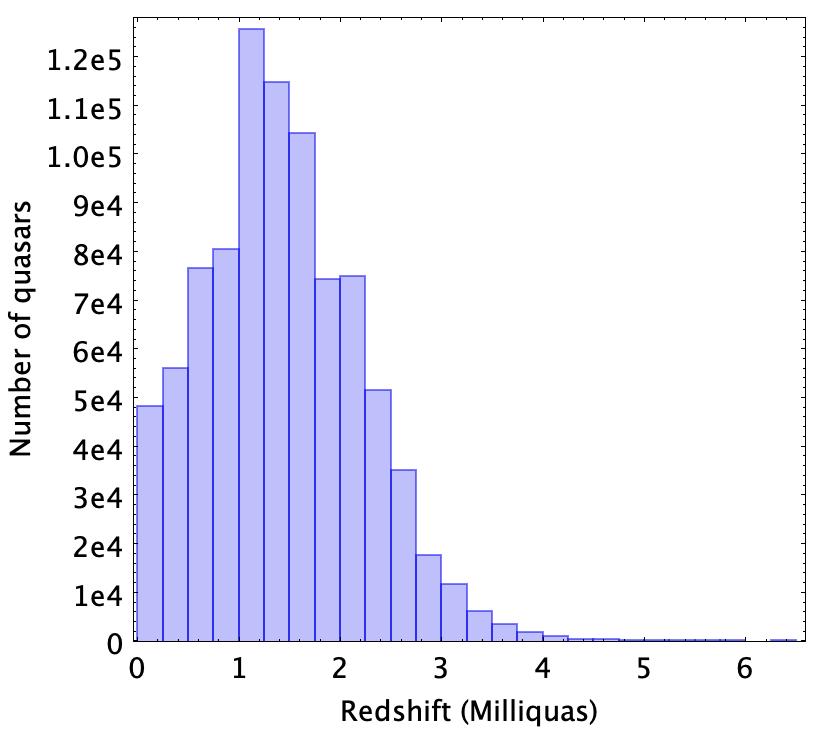} \\
(d)
\end{minipage}
\caption{Distributions of
(a): \gaia $G$ magnitudes (phot\_g\_mean\_mag) from the \gdr3 gaia\_source table, (b): $G_\mathrm{BP}-G_\mathrm{RP}$ colours (phot\_bp\_mean\_mag -- phot\_rp\_mean\_mag), (c): W1-W2  colours (from catWISE), (d): redshifts (from Milliquas) of the quasars and candidates from the input list.}
\label{qso_G}
\end{figure*}

We merged some major catalogues of quasars and candidate AGNs published before 2022. These include the data releases 6.4, 7.0, 7.1b, 7.4c, 7.5, 7.5b of the Milliquas catalogue \citep{2021Flesch, 2019Flesch}, the R90 and C75 selections of the AllWISE catalogue \citep{2018Assef}, the catalogue of AGN candidates from \citep{2019Shu}, a selection of sources from Klioner \textit{et al.} 2021 (private communication), a subset of the \gdr3 {\tt quasar\_candidates} table \citep{2022arXiv220605681G}, and additional quasars whose morphology was analysed by \cite{2022Ducourant}. Most of these catalogues contain stellar contaminants. 

The catalogues were cross-matched using a search radius of 3\arcsec \, and we only kept a single entry. The data priority follows the order listed above such that if a source is found in Milliquas 7.4c and 7.1b, only Milliquas 7.4c will be recorded. The compilation contains $\sim 24$ million total sources, of which $\sim5$ million are matched to a \gdr{3} source. 

Sources that are clearly stellar were eliminated by applying a weak astrometric filter rejecting proper motions larger than 14 mas/yr or parallaxes larger than 6 mas. This filter was derived from the astrometric properties of the multiply imaged quasars by gravitational lensing \citep[]{2018Ducourant} and is intentionally not severe because most quasars are in the faint luminosity regime of \gaia where the astrometry is less accurate and the potential presence of a surrounding host galaxy can perturb the astrometry of the central nucleus. We also filtered out objects brighter than $G=14$ magnitudes and excluded sources with colours compatible with stellar sources: 
$(G_\mathrm{BP}-G)>1$ \& $(G-G_\mathrm{RP})>0.8$ \& $G<20$ 
\citep[see][ Fig. 37]{2022arXiv220605681G}. We thus discarded $\sim\,21\,000$ sources judged stellar. The small number of discarded sources indicates that possibly some sort of astrometric filtering had already been performed in the construction of the original catalogues. Finally, sources in the direction of the Magellanic clouds, of other large galaxies, or of major globular clusters \citep{2010Harris} were removed. 

The final list of quasars and candidates contains 3\,760\,480 
sources with an entry in \gdr{3} and we refer to the list as the quasars or the quasar catalogue. The original catalogue name (e.g. Milliquas 7.4c) for each source is stored in \texttt{lens\_catalogue\_name}.

The sky distribution of the quasars in our input list is shown in Fig.~\ref{qso_lb} in galactic coordinates. The sky coverage of each of the merged catalogues is heterogeneous, as is the resulting compiled list. Most (81\%) of the sources have a \gmag  magnitude fainter than 19.5 mag, as seen in Fig.~\ref{qso_G}. Fig.~\ref{qso_G} also shows the \gaia colour 
$G_\mathrm{BP}-G_\mathrm{RP}$, the W1-W2 colour from catWISE \citep{2020catWISE}, and the redshift distribution when available from Milliquas ($\sim$900\,000 sources). The redshift distribution peaks at $z\sim1.4$  and extends to $z\sim6$ for a small number of sources.

\section{How GravLens searches for sources near quasars}\label{environment}
 
The all-sky coverage and $\sim 180$ mas angular resolution  make \gaia an exceptional instrument to search for lenses. Most currently known lenses have image separations $\gg$1\arcsec \, \citep[e.g.][]{2018Ducourant}\footnote{See also \url{https://research.ast.cam.ac.uk/lensedquasars/}}. Nevertheless, the expected distribution of lenses should peak at smaller separations, $\la$1\arcsec, making most of them quasi-undetectable from the ground \citep[e.g.][]{2016A&A...590A..42F}. Unfortunately, the \gdr2 and \gdr3 are incomplete at separations $\le$ 2\arcsec \, \citep[e.g.][]{2017A&A...599A..50A,2018A&A...616A..17A,2021A&A...649A...5F,2021Torra}. This results from a severe selection on the astrometric and photometric quality indicators of the sources that are published in these Data Releases. 

The primary goal of GravLens is to detect secondary sources near QSOs and QSO candidates, and derive their mean astrometry and raw photometry. In this context, each source that is detected in a field centred in the coordinates of a quasar is named \textit{component}. The ensemble of all the components in a field is named \textit{multiplets}. The field centred on a quasar is designated by \textit{quasar}. 

\subsection{The adopted \gaia data}
The instruments and focal plane of the satellite are well-described in \citep{2016Prusti}. Unlike most \gaia data processing chains that assign transits to a single source exploiting the \gaia cross-match \citep{2021Torra}, GravLens manipulates upstream data and allows a finer clustering to separate adjacent sources.
The data adopted by GravLens comes from the pre-processing step of \gaia treatment which is referred to as Image Parameter Determination~\citep[IPD,][]{Fabricius2016A&A...595A...3F}. The purpose of IPD is to transform the raw spacecraft telemetry into basic astrometric and photometric measures for the
Sky Mapper (SM) and Astrometric Fields (AF) windows. 
Our input data is the \gaia DR3 IPD outputs (flux and positions). These epoch positions are not the high-precision one-dimensional \gaia astrometry, but approximate 2D positions with a resolution of about one CCD pixel. GravLens uses the positions (right ascension and declination) of each transit, the fluxes in the $G$-band measured in SM and AF windows, and a rough on-board estimation of the $G$ magnitude done in the \gaia Video Processing Unit 
~\citep[VPU,][]{2015A&A...576A..74D,Fabricius2016A&A...595A...3F}\footnote{The resolution of the onboard estimation of the $G$ magnitude is 0.015625 mag.}. GravLens identifies by itself all \gaia transits within 6\arcsec of each quasar, without relying on the \gaia standard cross-matching since at this stage of the data processing, the \gaia cross-match is not yet known. The \gaia cross-match might subsequently identify more additional sources in the field at the later processing stages, but this is not included in GravLens.

\subsection{The GravLens clustering algorithm}\label{clustering}
GravLens uses the Density-Based Spatial Clustering of Applications with Noise (DBSCAN) algorithm \citep{1996Ester} for unsupervised clustering\footnote{We used the software implementation from the Apache Commons Math library v.3.6.1.}. Without indicating the number of clusters, as required, for example, for K-Means algorithms, it identifies groups of connected points and outliers. 
The principle of DBSCAN is to build a neighbourhood graph by connecting points (which here are individual detections in right ascension and declination) if their distance is smaller than a certain $\epsilon$. Here, adopt $\epsilon=100$ mas, a value chosen empirically that is within the PSF width of \gaia, thus corresponding to an angular distance that the instrument cannot physically resolve individual sources. The angular distances between the points are calculated using the haversine formula \citep[e.g.][]{de1795memoria}.

While there are non-connected points within $\epsilon$, the algorithm tries to connect them, and thus the graph grows. Otherwise, the set of connected points remains as is. When at least $minPts = 3$ (empirically chosen) points are connected, a cluster is formed (called a component); otherwise, the points are considered outliers. All \gaia transits associated with an entry of the quasar catalogue are then either outliers or within components (clusters). Once the clustering is complete, a sigma-clipping filter based on the positions and the magnitude is applied to the components, using $3\sigma$ as the threshold. The GravLens processing of the quasars has produced a catalogue of $\sim$4.7 million components.

\begin{figure}
    \includegraphics[width=0.45\textwidth]{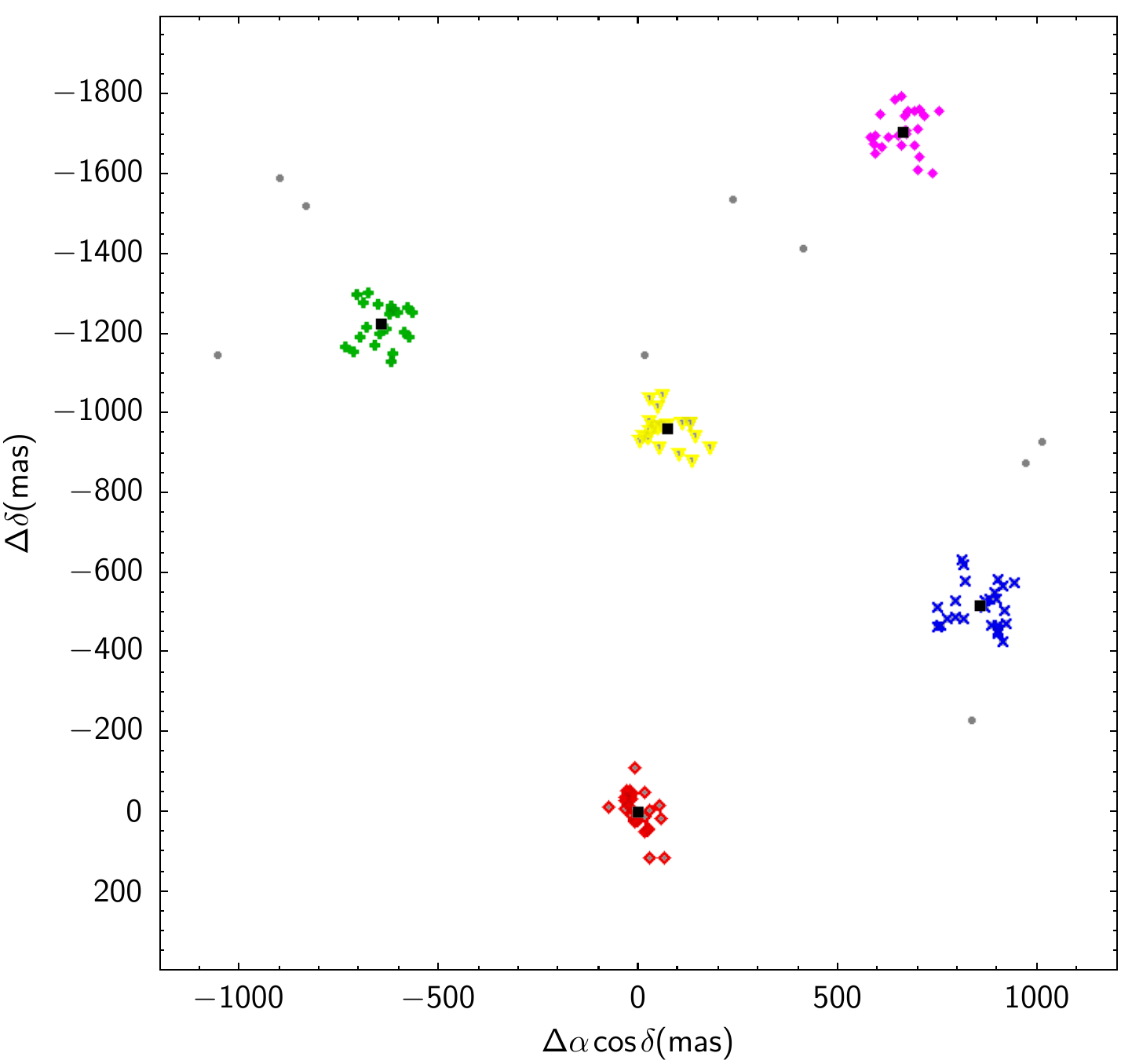}
  	\caption{GravLens results for the Einstein cross G2237+0305 (DR3Gaia224030.229+032130.03 in our output). The black dots represent the five components and the blue crosses, red diamonds, green crosses, and pink dots represent the four images of the quasar. The yellow triangles show the \gaia transits on the deflecting galaxy.
   The information is in the \texttt{ra\_obs}, \texttt{dec\_obs} fields of the \texttt{lens\_candidates} and \texttt{lens\_outlier} tables.}
    \label{Fig:DR3Gaia224030.229+032130.03}
\end{figure}

As an example of GravLens results, Fig.~\ref{Fig:DR3Gaia224030.229+032130.03} illustrates the application of the clustering algorithm on the well-known lens Einstein cross G2237+0305~\citep{1985AJ.....90..691H}. Five components are found by GravLens, corresponding to the four images of the quasar, and to the lensing galaxy that is also clearly detected.

\subsection{Clustering issues}\label{issues}
\begin{figure*}
    \begin{minipage}{0.45\textwidth}\centering \small
    \includegraphics[width=0.98\textwidth]{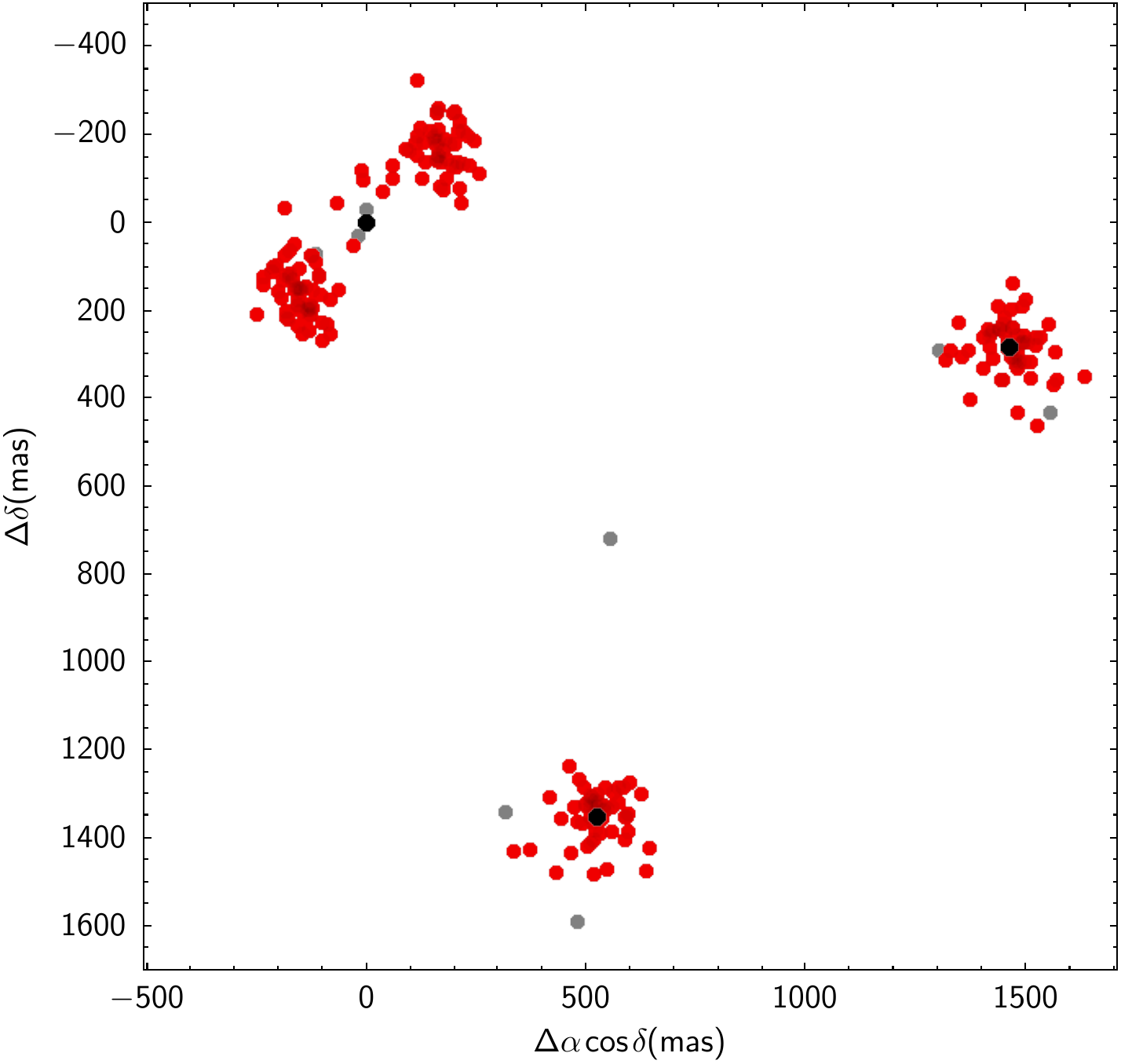} \\
    (a) DR3Gaia113100.075-441959.69
    \end{minipage}
    \begin{minipage}{0.45\textwidth}\centering \small
    \includegraphics[width=0.98\textwidth]{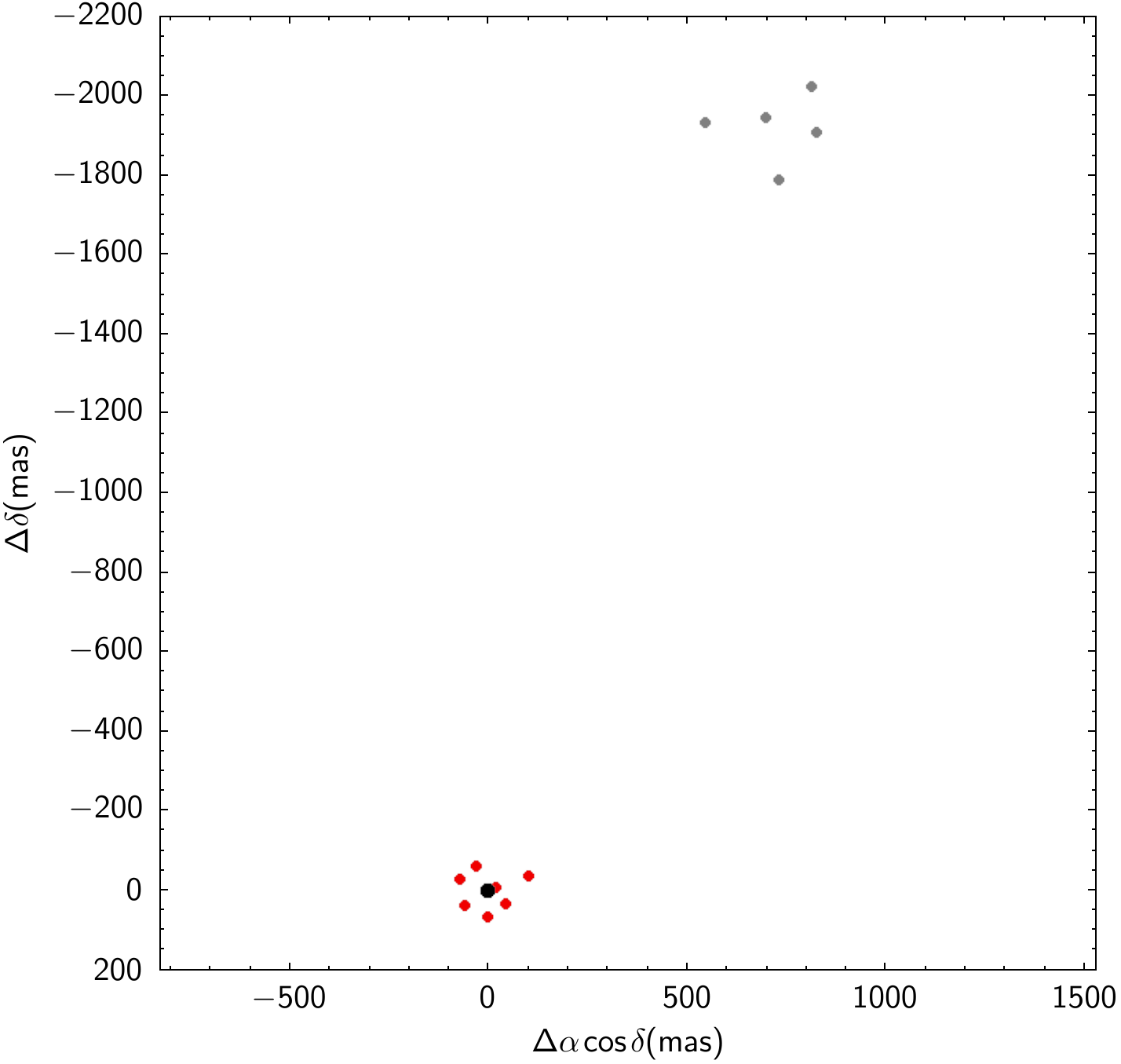} \\
    (b) DR3Gaia235007.548+365434.45
    \end{minipage}
    \\[0.5cm]
    \begin{minipage}{0.45\textwidth}\centering \small
    \includegraphics[width=0.98\textwidth]{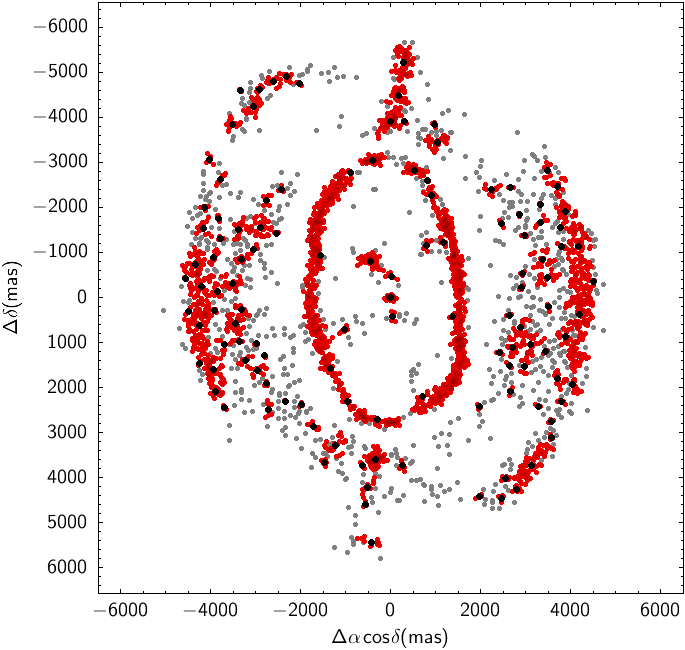} \\
    (c) DR3Gaia034732.982+350248.6
    \end{minipage}
    \begin{minipage}{0.45\textwidth}\centering \small
    \includegraphics[width=0.98\textwidth]{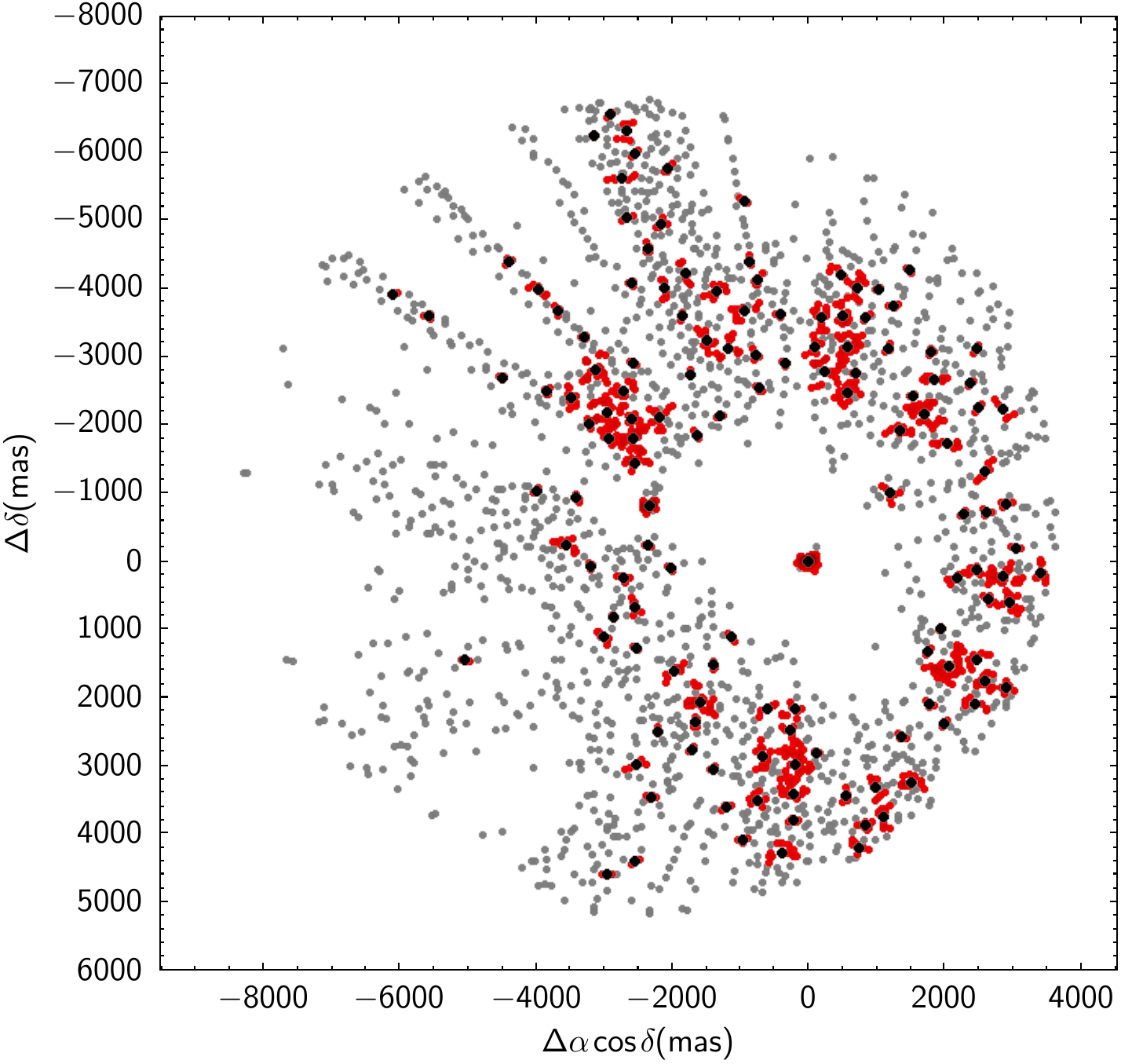} \\
    (d) DR3Gaia082523.532+241524.53
    \end{minipage}    
  	\caption{Examples of known issues. Black dots are the mean positions of the components, red points correspond to individual observations no matter the component and  gray dots are outliers. In (d) a planetary nebula (IC 351) that unduly entered in the quasar catalogue is decomposed by the algorithm into numerous sources. as well as in (d) for the halo of a bright star.}
    \label{Fig:clusteringIssues}
\end{figure*}

The GravLens algorithm is efficient and, in most lensing configurations reaches an optimal solution. However, occasionally, it converges to sub-optimal solutions. Fig.~\ref{Fig:clusteringIssues} illustrates some examples of known issues. 

We show in Fig.~\ref{Fig:clusteringIssues} (a) the known quadruply imaged lens \citep{2018Krone-Martins,2019A&A...628A..17W}, that corresponds to the multiplet DR3Gaia113100.075-441959.69. \gdr3 identifies four distinct sources with their own \texttt{source\_id}. In the figure, all \gaia detections are plotted, and the two known components in the top left were merged by GravLens, which only outputs three components. The two components are connected by detections closer than 100 mas which causes the method to group the two sources; this is a major drawback of the DBSCAN algorithm. The central point, identified as an outlier, could even bring some physical information about the lens. The end user of the tables of this FPR should be aware that useful information may be present in the outliers table.

We show another example of a clustering issue in Fig~\ref{Fig:clusteringIssues} (b). This figure corresponds to the source DR3Gaia235007.548+365434.45. 
~This source is a known doubly image quasar, but GravLens identified only one of its components. The component is $2.9$\arcsec away from the quasar and the individual detections at the quasar position were labeled as outliers. Therefore, the information is not completely lost, but present in the \texttt{lens\_outlier} table. Five transits are near the quasar but the distance between the transits is $> \epsilon=100$~mas so they are not considered to be a component. The astrometry may have been perturbed by the deflecting galaxy and the \gdr3 astrometry is unreliable, presenting errors of $\sim 17$~mas.

We illustrate the case of a large planetary nebula (IC 351) decomposed into many components in Fig.~\ref{Fig:clusteringIssues} (c). This figure represents the source DR3Gaia034732.982+350248.6. There are 3\,508 \gaia observations in this field. GravLens found 120 components with 2\,768 detections, and 740 outliers. This example highlights one of the causes of the large number of components in certain fields when an extended object is decomposed. 

Figure~\ref{Fig:clusteringIssues} (d) shows another odd example of a large number of detections (DR3Gaia082523.532+241524.53). In this case, the central source is a very bright object (magnitude $\sim$ 9.7). GravLens detects 136 components in radial spokes from the central source, based on 2\,481 observations, of which 1\,424 are considered outliers. A halo of outliers is present around this source. 

\subsection{Post-processing}\label{postproc}
After the GravLens processing, we perform a post-processing stage. The post-processing can handle specific situations and flag problematic sources or sources to be discarded. In particular, we observed an excess of doublets separated by less than 300 mas and with $\leq$5 observations of one of the components. A small fraction of these are probably real sources but the majority of them result from the excessive decomposition of single sources into doublets by the clustering algorithm. The post-processing gathered these nearby components into single sources for $\sim$200\,000 doublets. 

The post-processing aims at raising flags to indicate problematic multiplets or multiplets which are clearly not lensed quasars. The flags are raised at the quasar level \texttt{flag} and/or at the component level \texttt{component\_flag}. 

The \texttt{flag} is a two-bit binary flag. The first bit is set to 1 if the maximum difference of magnitude within the multiplet is larger than 5 mag which indicates that it is very improbable that this is a lensed quasar. The second bit is set to 1 if there are more outliers than clustered observations, such as could be the case for a galaxy and other extended objects, see, e.g. Fig.~\ref{Fig:clusteringIssues} (c).

The \texttt{component\_flag} is also a two-bit binary flag. The first bit is set to 1 if the standard deviation in right ascension or declination of a component is larger than $100$ \mas.  A point source should yield $\sigma_{RA, Dec}\approx 60$~mas, at the order of the uncertainty of the RA/Dec of the SM position. The second bit is set to 1 if the standard deviation of the raw mean magnitude is larger than 0.4 mag.  In both cases, \texttt{component\_flag} points to unusually high measurement uncertainty, possibly resulting from a bright nearby source or several very nearby sources which are considered a single component.
There are 4 444 145 components with both flags set to 00, indicating no alert is raised. This represents 93\% of all components.

\section{The catalogue of sources around quasars}\label{catalogue}
%
GravLens has analysed 183\,368\,062 transits matched to the $3\,760\,480$ quasars from our list during the first three years of Gaia operations. It attributed 171\,545\,519 transits to components and rejected 11\,822\,543 as outliers. GravLens did not converge in 448 cases. Within 6" of the $3\,760\,032$ quasars, $4\,760\,920$ sources were found (see Table~\ref{nombre}), including the quasars.

These results are included in the \texttt{lens\_candidates} table. The data model of the catalogue is presented in appendix \ref{datamodel}. Additional information can be found in the table \texttt{lens\_catalogue\_name} 
(see Sect.~\ref{inputlist}). The individual observations of each component and the 
outliers are found in the 
\texttt{lens\_observation} and \texttt{lens\_outlier} tables (see \red{link to the FPR documentation} for a detailed description of all tables and fields). 

\subsection{General properties}
\begin{figure*}
\centering
\includegraphics[width=0.33\textwidth]{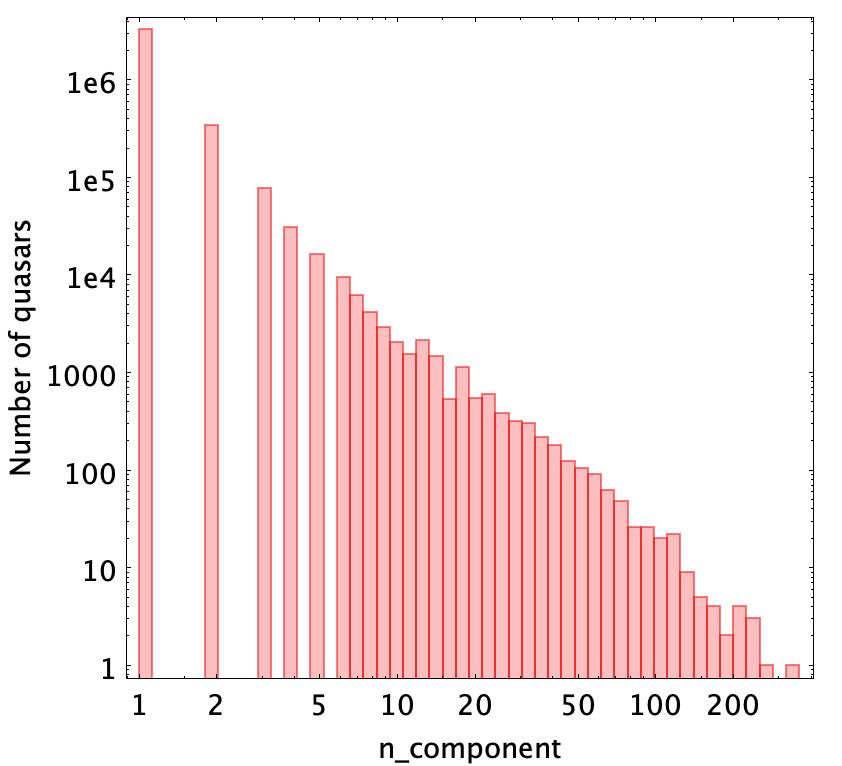}
\includegraphics[width=0.33\textwidth]{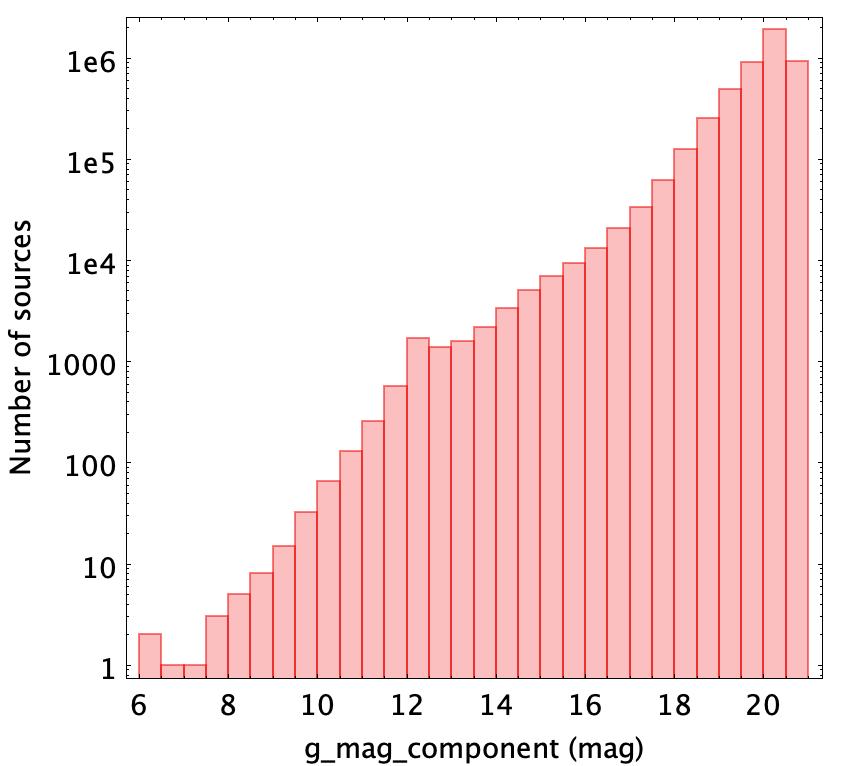}
\includegraphics[width=0.33\textwidth]{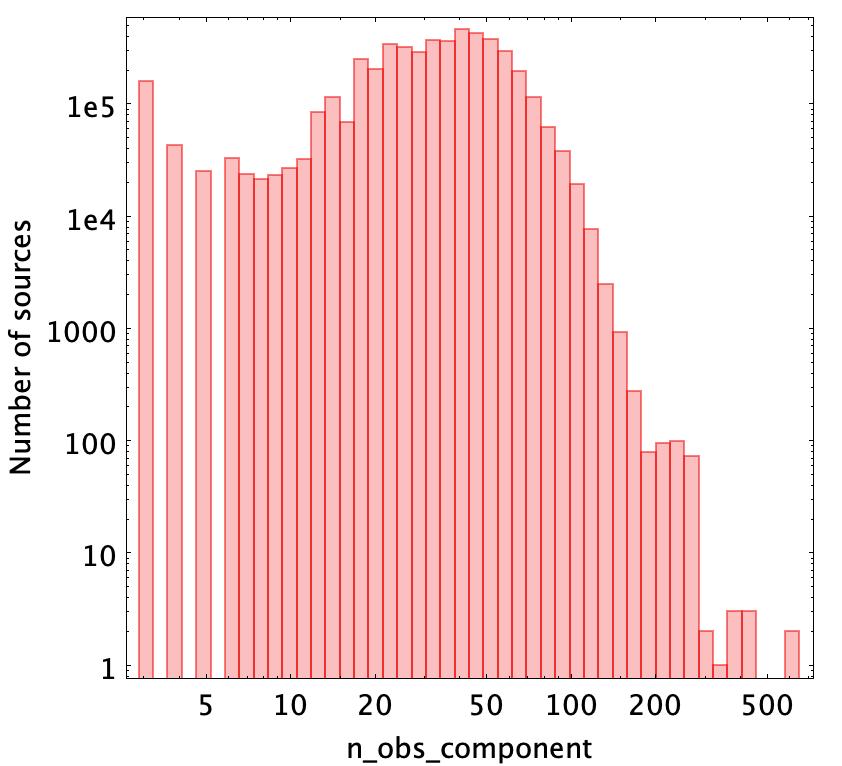} \\
\includegraphics[width=0.33\textwidth]{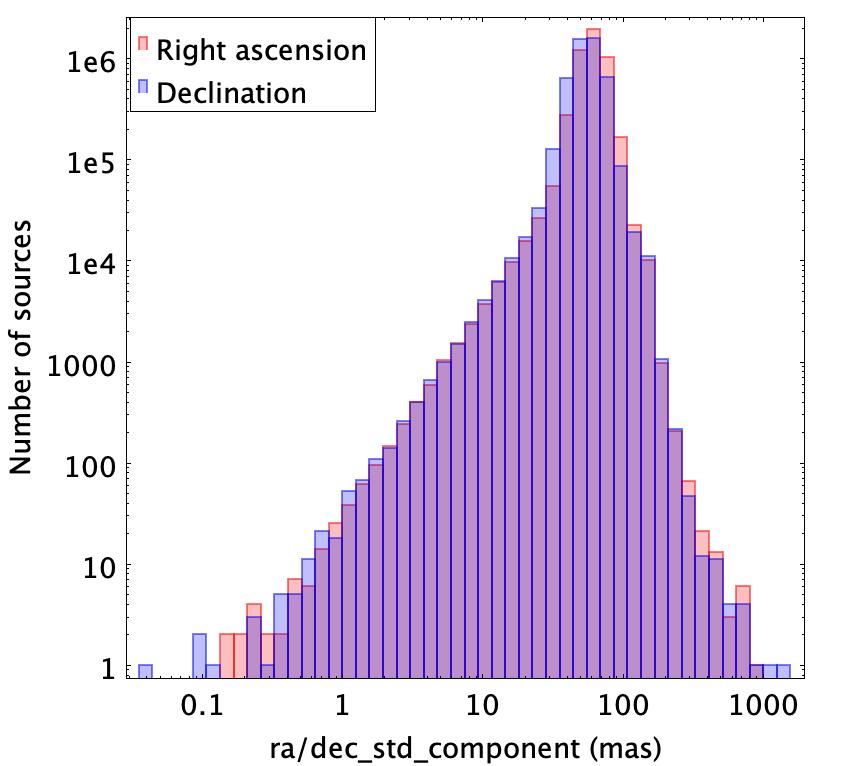}
\includegraphics[width=0.33\textwidth]{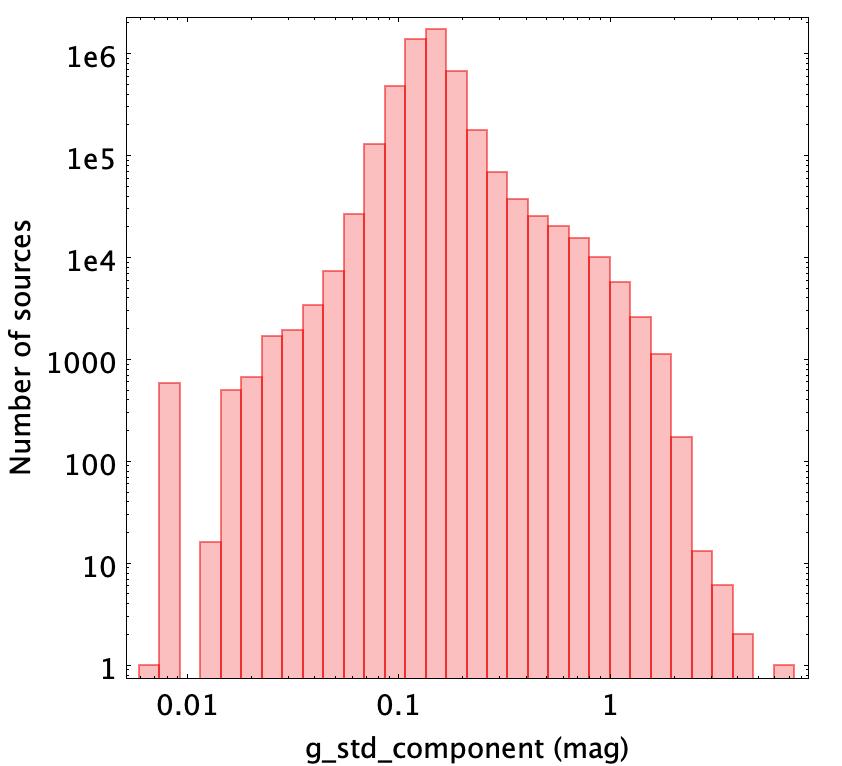}
\caption{Distributions of components' main features provided in the \texttt{lens\_candidates} table - (a): number of components found in the fields analysed, (b): mean $G$ magnitude of the components of all fields. (c): number of observations of the components. (d): standard deviation of mean coordinates (\texttt{ra\_component}, \texttt{dec\_component}) and (e): standard deviation of mean $G$ magnitude.}
\label{comp_nobs}
\end{figure*}

The distribution of the main properties of the components published in the \texttt{lens\_candidates} table is given in Fig.~\ref{comp_nobs} and the number of sources detected in the fields of the quasars along with the number of components in the fields is given in Table~\ref{nombre}.

\begin{table}[t!]
\centering
\caption{\label{nombre} Source counts in the \texttt{lens\_candidates} table according to the number of components in the field.} 
\begin{tabular}{lrrrr}
\hline
Selection   & Nb of quasars &  Nb of components\\
\hline
All             &$3\,760\,032$  & $4\,760\,920$\\	 
1 component     &$3\,258\,647$  & $3\,258\,647$\\
2 components    &$341\,551$     &   $683\,102$\\
3-10 components &$149\,953$     &   $618\,838$\\   	
11+ components  &  $9\,881$     &   $200\,333$\\ 
\hline
\end{tabular}
\end{table}

The vast majority of the quasars (87\%) have no neighbour within 6\arcsec ~and 9\% are doublets. There are $\sim159\,000$ multiplets with more than 2 components (4\%). The search for quadruply imaged quasars will therefore focus on this sample of multiplets. There are $\sim 9\,000$ multiplets containing a large number of components (> 10). They generally correspond to large galaxies decomposed in many sources as seen in Sect.~\ref{issues}. 

The distribution of magnitudes follows that of the quasars in the input list (see Fig.~\ref{qso_G}), reflecting the fact that the majority of the sources in our catalogue are not multiply-imaged.

Components have a median of $36$ observations, ranging from three to $630$ observations, and are time-resolved. Sources with a very low number of observations should be considered with caution and generally correspond to the faintest sources detected.

\subsection{Astrometry and photometry}
Gravlens astrometry and photometry are meant to complement the information from the current \gaia Data Releases, especially for the sources that are not present in the latter. GravLens magnitudes and fluxes use uncalibrated onboard magnitudes, for instance. For many GravLens sources not published in \gdr3 the measurements are poor, as these sources are usually faint. The mean standard deviations are $62$ mas and $57$ mas, respectively, for right ascension and declination and 0.15 mag for the magnitudes. 

\begin{figure}
    \includegraphics[width=0.45\textwidth]{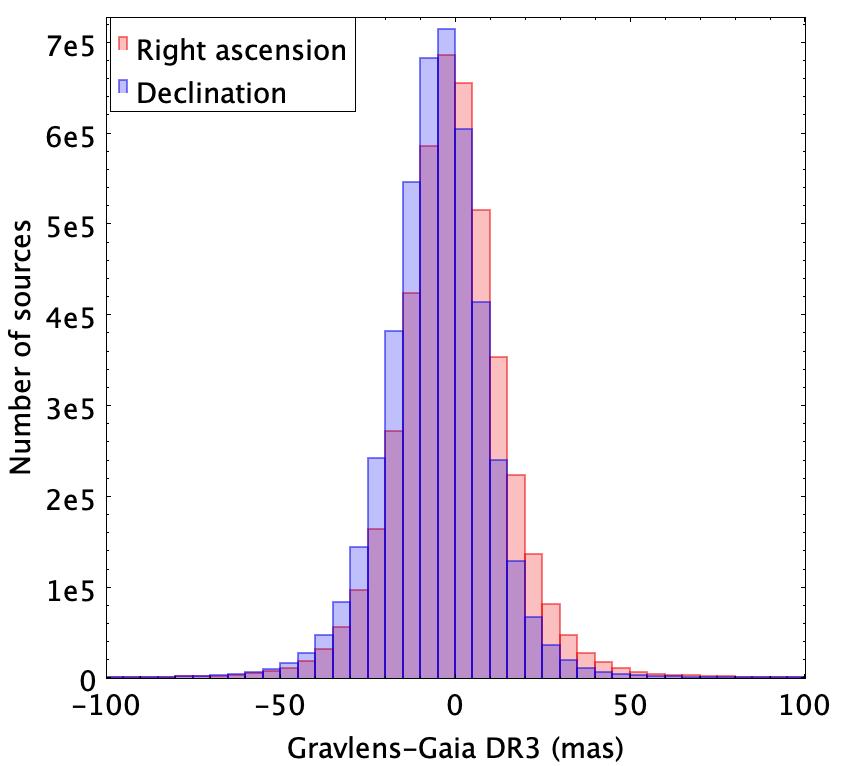} \\
    \includegraphics[width=0.45\textwidth]{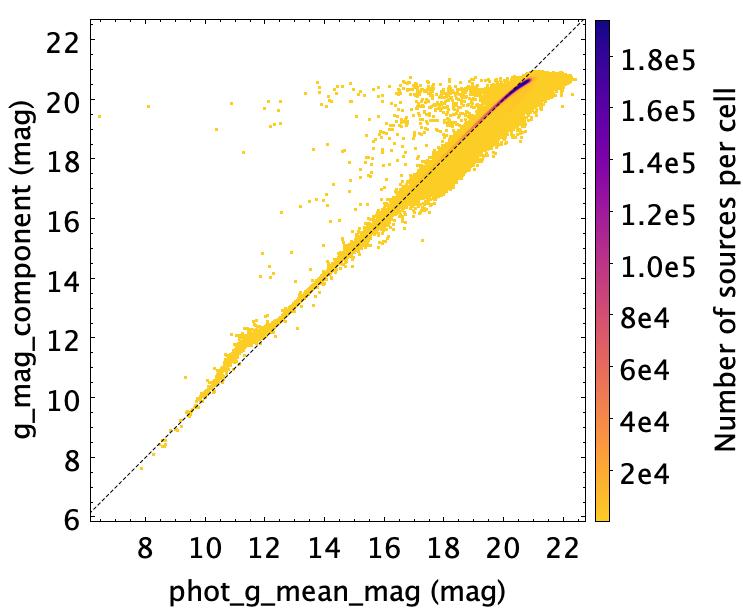}
  \caption{Comparison between coordinates and magnitudes from GravLens and \gdr3. (Top) Comparison of the coordinates (ra, dec) derived by GravLens and by \gdr3. $\Delta \text{ra}$ includes the cos(dec) factor. (Bottom) Comparison of Gravlens magnitudes (table \texttt{lens\_candidates} field \texttt{g\_mag\_component})\ with \gdr3 magnitudes (table \texttt{gaiadr3.gaia\_source} field \texttt{phot\_g\_mean\_mag}).}
    \label{compareDR3}
\end{figure}

The GravLens and \gdr3 positions and magnitudes for common sources are compared in Fig.~\ref{compareDR3}. A slight asymmetry is present in RA and Dec, with $(\mathrm{RA}_{GL}-\mathrm{RA}_{DR3}) \approx -1.33$mas and $\mathrm{Dec}_{GL}-\mathrm{Dec}_{DR3} \approx -5.4$mas, with dispersions of $\sim 13$mas. 

The GravLens magnitudes agree well with \gaia magnitudes with a mean difference of -0.06 mag and a standard deviation of 0.15 mag. Around $G=12\pm0.5$ mag, the Gravlens magnitudes are slightly higher. This is a well-known effect of the uncalibrated onboard magnitudes \citep{2018A&A...616A...3R}, and for \gdr3 magnitudes $\ga21$, the GravLens magnitudes are quite dispersed and generally lower (i.e. the magnitudes can be overestimating the true brightness of the source). 

The astrometry and photometry of the GravLens components are based on the Gaia onboard detections, and should be much improved in the future Data Releases solutions when the individual components are properly handled.

\subsection{New sources not in \gdr3}\label{new_sources_not_in_dr3}
\begin{figure}
    \includegraphics[width=0.45\textwidth]{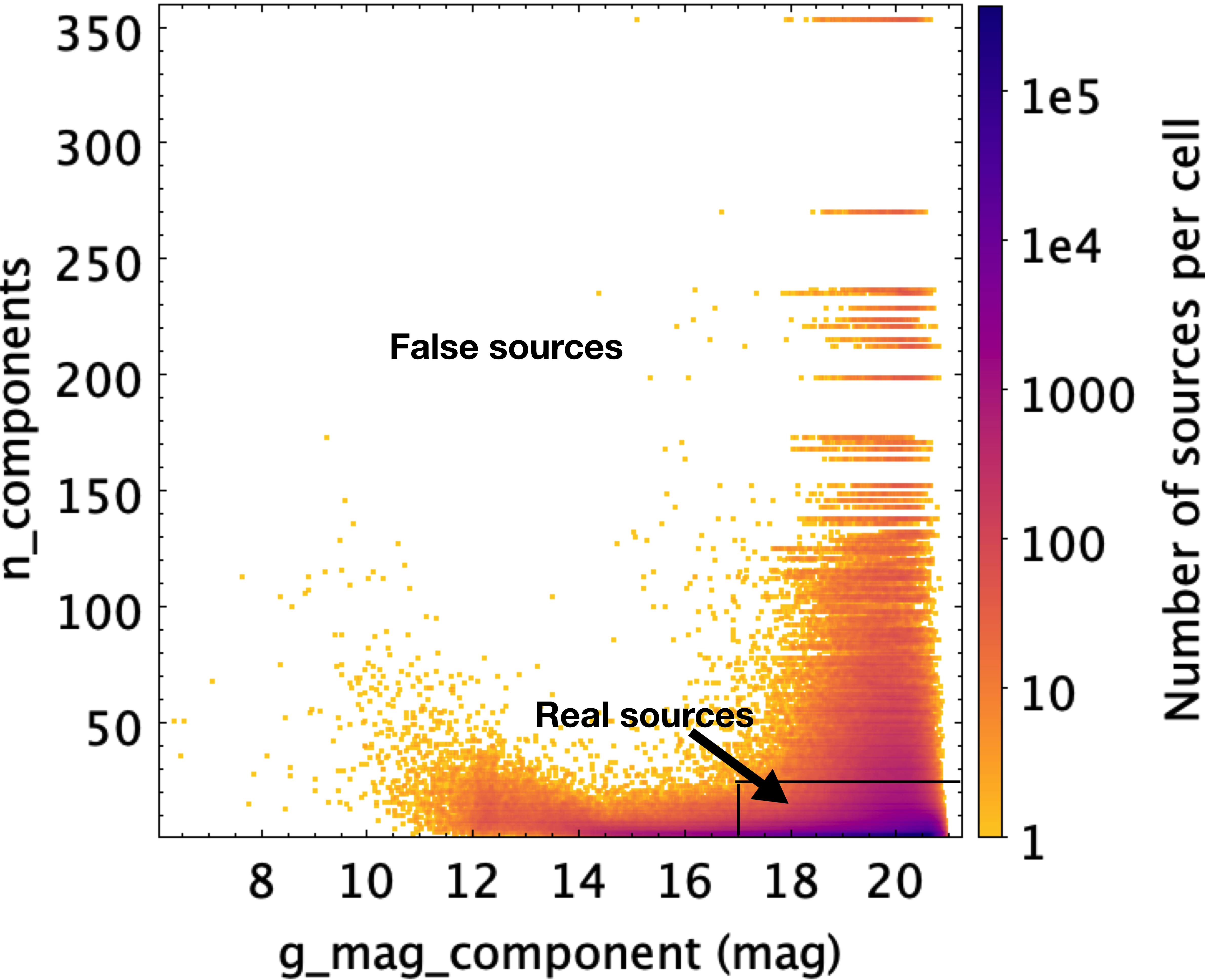}
  \caption{Density plot of the magnitudes of the GravLens components not present in \gdr3 along with the number of components in the multiplet. Coulour scale is logarithmic.
  }
    \label{new}
\end{figure}

There are ${\sim}10\,500$ \gdr3 sources in the vicinity of analysed quasars ($6$\arcsec) that are not among the GravLens components, representing less than $0.2\%$ of all GravLens components. Meanwhile, there are $306\,970$ new sources that are not in \gdr3 among the $4\,760\,920$ GravLens components. About ${\sim}200$\,000 new sources are either bright with \texttt{g\_mag\_component}<17.5 mag or in crowded fields with n\_components>20  (Fig.~\ref{new}). The bright new sources correspond to problems illustrated in Fig.~\ref{Fig:clusteringIssues} (c and d).
They are generally flagged either at the \textit{quasar} level (\texttt{flag}) or at the \textit{component} level (\texttt{component\_flag}) (see Sect. \ref{postproc} for a description of the flags). 
After this process, ${\sim}103\,000$ new sources remain which are not flagged and believed to be bona-fide sources.

\subsection{Known lenses}\label{known_lenses_dr3}
We first compare the GravLens results to known lenses. The GravLens catalogue includes $\sim$450 known or candidate lenses published in the literature, 76 with 4 images (quads) and the rest being doublets. For 67 quads out of the 76 quads, GravLens complements the existing measures from \gdr3 by measuring one or more additional components or the deflecting galaxy. 
In total GravLens measured 1\,293 components in the fields of known lenses while 1\,207 are present in \gdr3. The 86 newly detected components in the fields of known lenses are mostly faint real components lying around lenses with a previously small number of \gdr3 counterparts.

\begin{figure*}
    \includegraphics[width=0.49\textwidth]{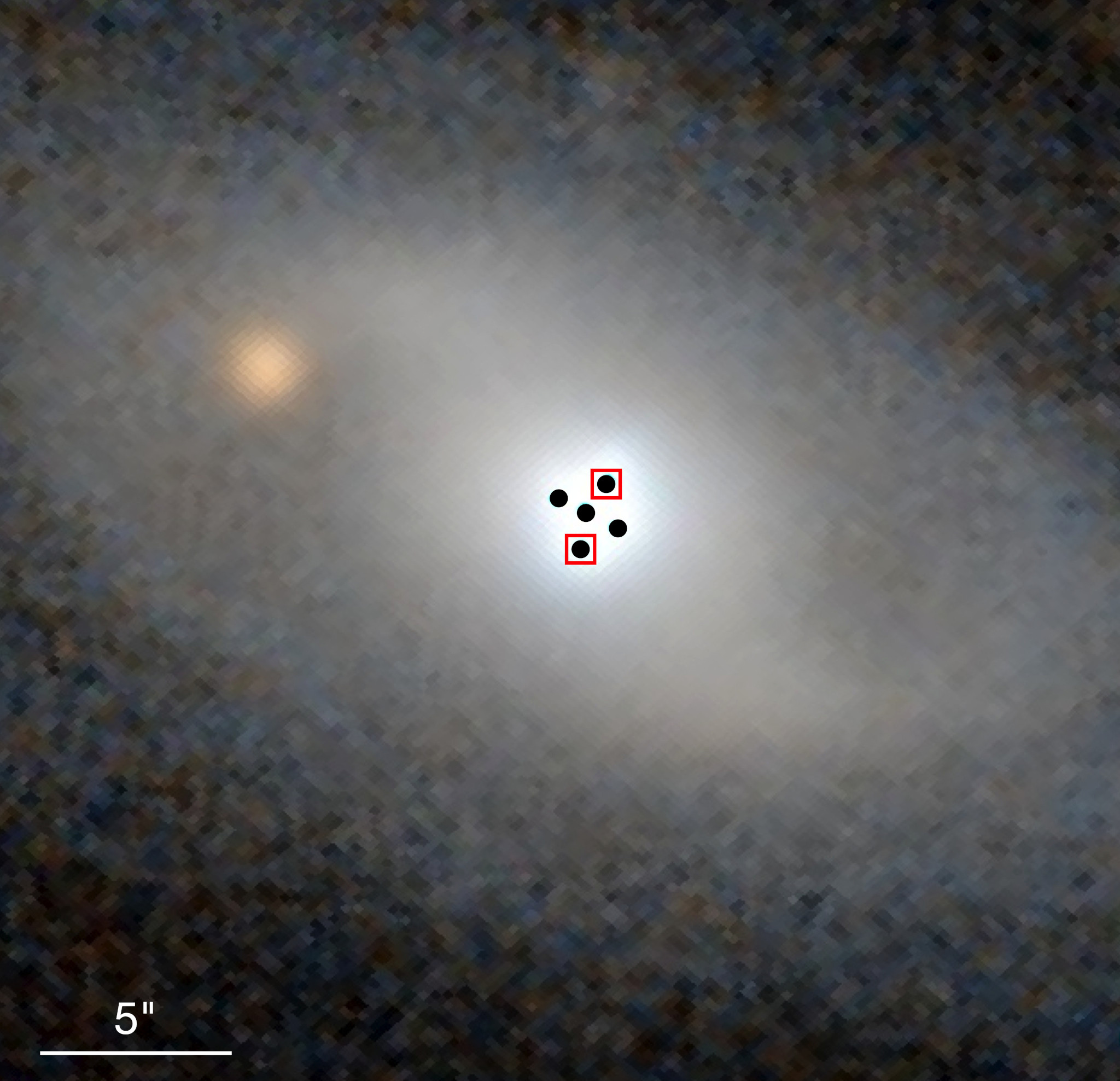}
    \includegraphics[width=0.49\textwidth]{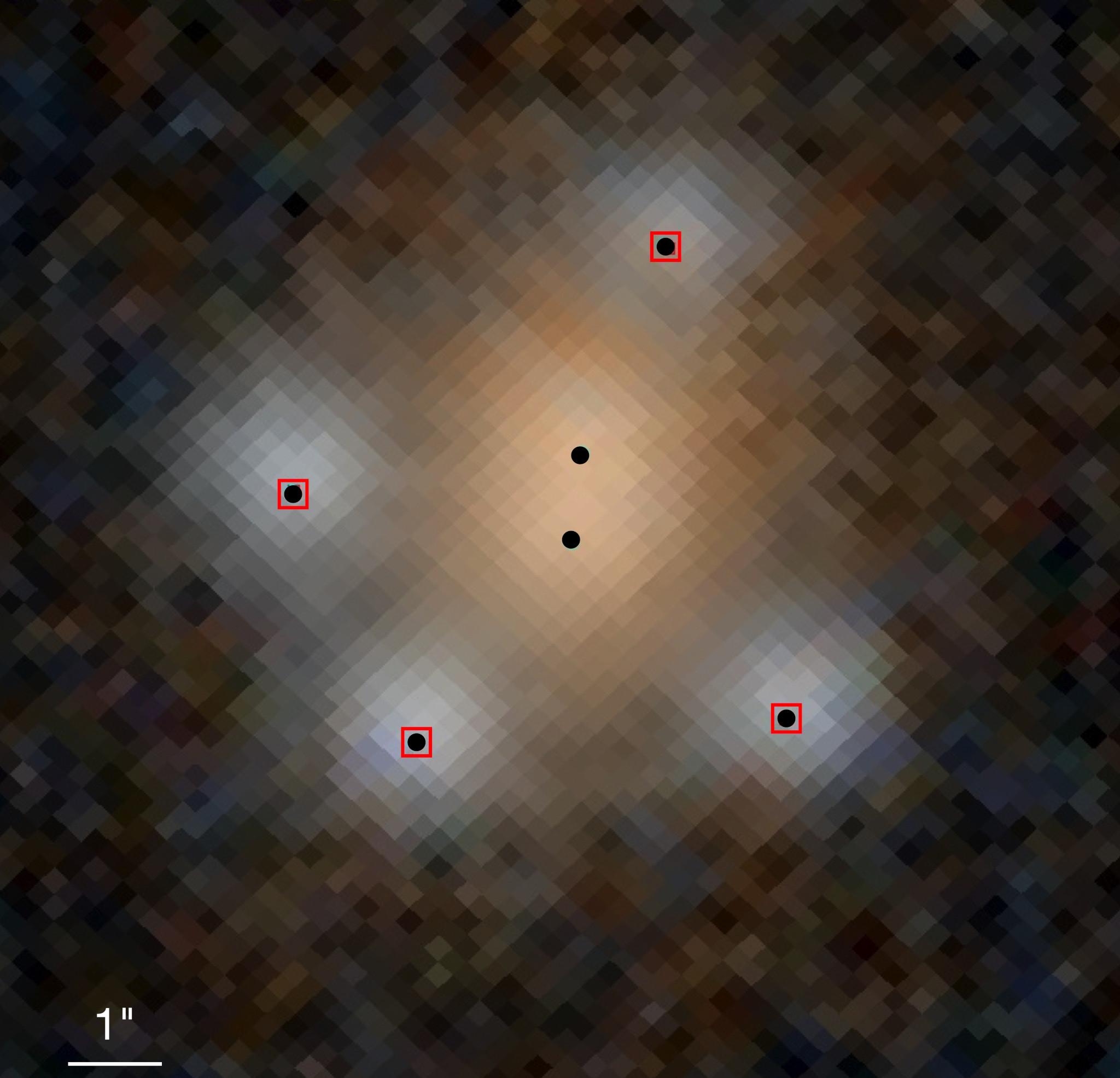}
  \caption{Pan-STARRS images \citep{panstarrs} of two known gravitational lenses with an indication of GravLens components in black (filled circles) and entry in \gdr3 in red (squares). Left: the Einstein cross (G2237+0305). Right: 2MASSJ13102005-1714579. The central sources in 2MASSJ13102005-1714579 encompasses two lensing galaxies recovered as  GravLens components.}
    \label{panstarrs}
\end{figure*}

We show two examples in Fig.~\ref{panstarrs}. This figure shows Pan-STARRS images \citep{panstarrs} of two known lenses: the Einstein cross G2237+0305 \citep{1985AJ.....90..691H} where GravLens detects all 4 images of the quasar and the deflecting galaxy while \gdr3 only contains two entries, and 2MASSJ13102005-1714579 \citep{2018Lucey} where GravLens detects the four images of the quasar and two central deflecting galaxies that had no entry in \gdr3.

Some of the presently known lensed quasars have been targeted by the Hubble Space Telescope (HST), and new structures can be found in the HST source catalogue version 3 \citep{HST}. However, of the 476 known lenses, only 69 have space-borne measurements from the HST catalogue. 
\section{Search for new lenses}\label{search}
To look for new lenses and help guiding the users of this FPR, we developed two methods using artificial intelligence: an outlier scoring algorithm (Sect. \ref{hesiod}) and the application of Extremely Randomised Trees \citep[][and Sect.~\ref{ert}]{2019A&A...622A.165D}.
When \gaia spectra are available, we also make use of this information, and in Sect. \ref{spectra} we explain the method that compares the mean BP and RP-spectra using chi-squares and Wasserstein distances. High-scoring multiplets are then visually inspected.

\subsection{The Hesiod score for the input list of quasars}\label{hesiod}
Only a small fraction of components near quasars are expected to be quasar images. So, analysing the quasars from the list presented in Sect.~\ref{inputlist} to identify good lens candidates can be seen as an outlier detection or a one-class classification problem. Accordingly, we can use these techniques to produce a lens score.

In these methods, distances, densities, and, in some instances, labeled data are used to train to identify a class called the positive class~\citep[see, e.g.][]{10.1145/1401890.1401920}. 
The supervised or semi-supervised training assumes (a) that the learning method has access to a reliable subset of positive examples such as spectroscopically confirmed lenses and (b) that the data contains positive and unknown examples (i.e. new lenses and other objects). 

To increase the reliability of the scores, photometric and astrometric indicators from the \gdr3 data~\citep{2022arXiv220605681G} and public unWISE data~\citep{2014AJ....147..108L} were calculated based on sources within $6\arcsec$ of each quasar. Missing data patterns appear when sources lack \gdr3 or unWISE data or when there is no unWISE counterpart for a \gaia source. Missing data is a serious problem that prevents the adoption of many approaches readily available in the literature \cite[e.g.][]{10.1007/s10994-020-05877-5}. So, to produce a score in this situation, we developed a simple heuristic method that we call \texttt{HESIOD} for Heuristical Ensemble Splitting Imputation and Organization of Data that can be applied to large datasets as it is embarrassingly parallel.

\texttt{HESIOD} 
assumes that the dataset can be described by a single matrix $\mat{D}$ of $n$ rows by $d$ columns. Each row corresponds to an astronomical source, and each column corresponds to a physical parameter (e.g. astrometric and photometric measurements from \gaia and unWISE, maximum and minimum angular distances and color differences between sources, etc.).
$\mat{D}$ can be incomplete in that not all elements $D_{ij}$ are filled (i.e. data for one or more column $j$ can be missing in any row $i$). 
The binary vector $\mathbf{c} \in \{0,1\}^n$ encodes the class of the $i-$th row ($i-$th source); $c_i=1$ if the source belongs to the {\it positive} class, here equivalent to a known lens, and $c_i = 0$ if the class is unknown.
Only $k$ components of the vector $\mathbf{c}$ are equal to one, with $k\ll n$.
\texttt{HESIOD} is a method $\mathscr{H}$ to estimate a vector $\mathbf{o} \in R^n | 0 \le o_i \le 1, \forall i \in [1,n]$, from $\mat{D}$ and $\mathbf{c}$ (i.e. $\mathscr{H}(\mat{D}, \mathbf{c}) \to \mathbf{o}$), such that $\mathbf{o}$ 
contains a score $o_i$ for all $n$ rows (sources) of $\mat{D}$ 
to indicate if the $i-$th source can belong to a different class than the {\it positive} class (the lenses). \texttt{HESIOD} thus starts with the known lenses $\mathbf{c}$ and ends with a new real-valued  vector $\mathbf{o}$. The vector $\mathbf{o}$ is initially an outlier score, that is, a score for the source not being a lens, which we complement (i.e. $1-\mathbf{o}$) to obtain a lens score. 

Informally, \texttt{HESIOD} solves this problem by creating ensembles of smaller problems that are easier to solve. It has two steps, an initial `inner' phase followed by an `outer' phase, as in UPMASK~\citep{2014A&A...561A..57K}. The inner phase randomly splits the matrix $\mat{D}$ into a set of $m$ smaller $p\times d$ sub-matrices $\{\mat{S}_j|\forall j \in [1,m]\}$, without replacement. This corresponds to random partitions of the catalogue into random samplings of sources (keeping all the associated data). Then, an imputation method $\mathscr{J}$ solves the less complex imputation problem for each sub-matrix ($\mathscr{J}(\mat{S}_j)\to\mat{\tilde{S}}_j)$. Afterward, a $\mat{\tilde{D}}$ matrix is reassembled from the results of the imputations on the $\mat{S}$ sub-matrices, and multiple outlier or one-class classification methods $\mathscr{C}$ produce scores from $\mat{\tilde{D}}$ and the vector $\mathbf{c}$, resulting in the matrix $\mat{\tilde{O}}$ containing the scores for each source (i.e. $\mathscr{C}(\mat{\tilde{D},c})\to\mat{\tilde{O}}$). 

The outer phase of \texttt{HESIOD} executes the inner phase $q$ times, resulting in a set of matrices $\{\mat{\tilde{O}}_l|\forall l \in [1,q]\}$. This ensures diversity in the imputation process due to the random splitting of $\mat{D}$.  Then it runs the final scoring method $\mathscr{O}$ over the matrix $\mat{O}$, where $\mat{O}=\mat{\tilde{O}}_1 | ... |\mat{\tilde{O}}_l$ (i.e. this matrix is the concatenation of the individual matrices resulting from the $q$ runs of the inner phase), producing a final score for each source (i.e. $\mathscr{O}(\mat{O})\to\mathbf{o}$), where $\mathbf{o}$ is a score for the source to be an outlier (i.e. not a lens), and $1-\mathbf{o}$ is the \texttt{HESIOD} score.

Since here the {\it positive class} is composed of lenses, which correspond to a small number of rows of the total dataset $\mat{D}$, all known lenses are concatenated with each sub-matrix $\mat{S}$. This is important to avoid significantly biasing the imputation process against the lenses.

\begin{figure}
    \centering
    \includegraphics[width=0.49\textwidth]{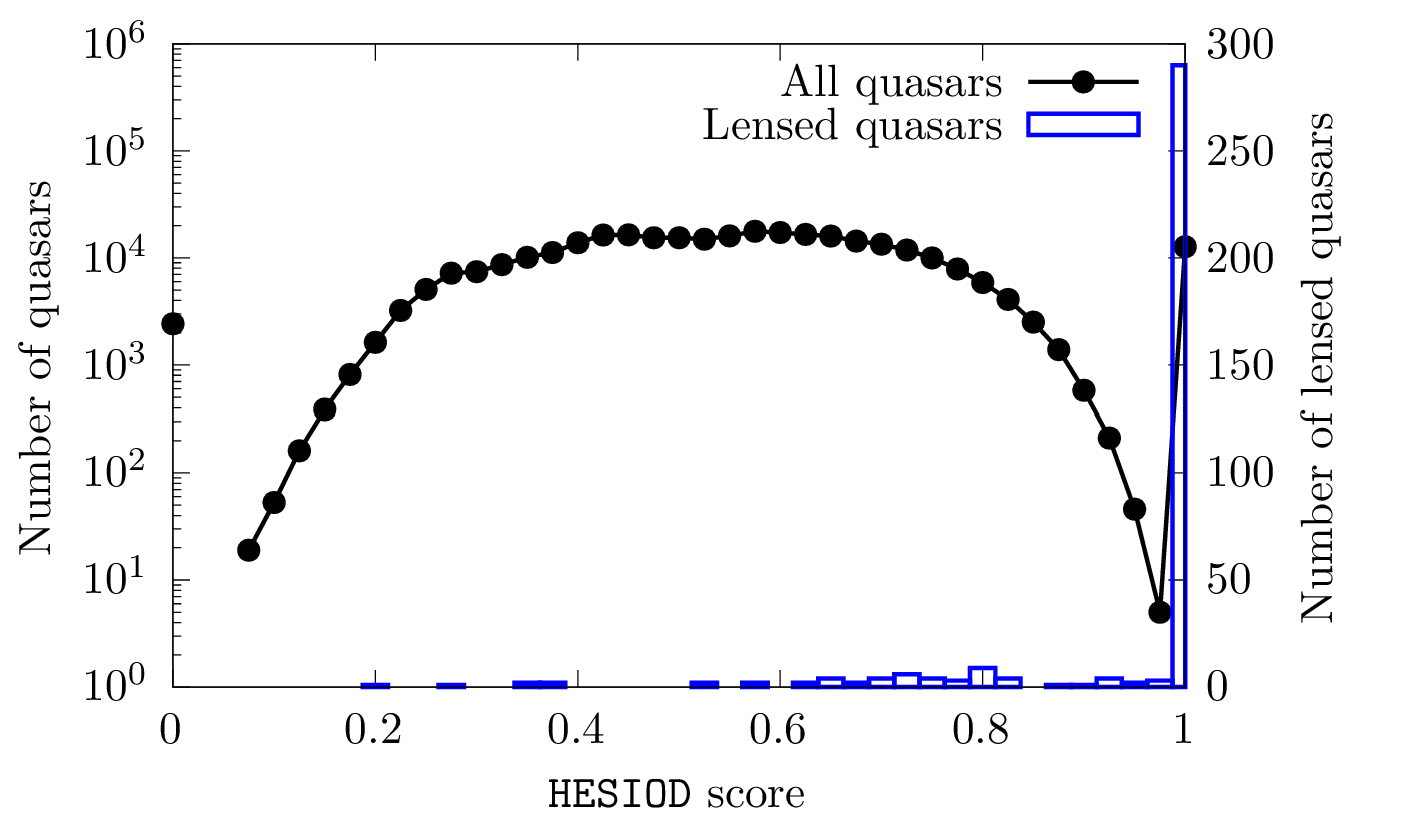}
    \caption{Distribution of the {\tt HESIOD} score for the 319\,296 quasars for which it was computed. We note that the left axis corresponds to all the quasars and is represented in logarithmic scale, while the right axis corresponds to the known lenses and is represented in linear scale.}
    \label{fig:hesiod_score}
\end{figure}

The \texttt{HESIOD} results depend on the choice of the ensembles of methods used for imputation $\mathscr{J}$ and classifications $\mathscr{C}$ and  $\mathscr{O}$. In this work, we adopted different methods based on ensembles of decision trees. We adopted \texttt{miceRanger}~\citep{miceRanger} for the imputation $\mathscr{J}$. This is a version of the Multiple Imputation by Chained Equations method~\citep{doi:10.1177/0962280206074463} that uses a random forest regression for individual imputations~\citep{10.1093/bioinformatics/btr597}, as in~\cite{2017A&A...597A..90D}. For the outlier scoring $\mathscr{C}$ we use three types of ensembles of decision trees as there is usually no optimal method for all problems~\citep[e.g.][]{6795940,585893,2021arXiv211013402C}: the classic Isolation Forest method~\citep{4781136}, SciForests~\citep{sciforests2010}, and Fair-Cut Forests~\citep{2019arXiv191106646C}. 
The final outlier score $\mathscr{O}$ also uses Fair-Cut Forests. The resulting distribution of the {\tt HESIOD} scores is presented in Fig. \ref{fig:hesiod_score}. Known lenses were iteratively used during the training, so it is expected that their scores would peak at high values, which indeed happens. The results on all $319\,296$ sources for which the method produced results show a central peak indicating more uncertain sources and two sharp peaks at the low and high score extremes.

The {\tt HESIOD} method seems effective in the present lens candidate scoring application since it was designed to deal with large datasets with missing data and, moreover, to consider a parameter space constructed from combinations of the measurements of all sources around the analysed quasar. The parameter space (i.e. columns of the matrix $\mat{D}$) is constructed from summary information about distributions of the measurements of all sources around the quasar, such as minimum and maximum differences in color in all possible \gdr3, unWISE, \gaia-unWISE W1 and W2 bands, astrometric errors, angular distances, between the images, as well as global properties of the parameter distributions as the minimum, mean and maximum astrometric and photometric errors, astrometric excess noise, RUWE, BPRP excesses and signal to noise ratios (e.g. fluxes, positions, proper motions and parallaxes over their errors) for all sources. As such, the {\tt HESIOD} score is assigned for the entire candidate system, composed of multiple \gdr3 sources. This parameter space also enables {\tt HESIOD} to deal with more challenging lensing cases. For instance, although gravitational lenses are achromatic, one or more of the quasar images can be superposed with parts of the lensing galaxy, and in the most extreme cases, the lensing galaxy can be completely unresolved and mixed with one or more images due to finite spatial resolution. In such cases, property gradients (such as color, and astrometry) could be expected to exist within the candidate system, and {\tt HESIOD} can deal with such non-textbook lensing cases as long as there are similar examples among the {\it positive} class sample in $\mat{D}$.

\subsection{Extremely Randomised Trees}\label{ert}
\begin{figure}
    \includegraphics[width=0.49\textwidth]{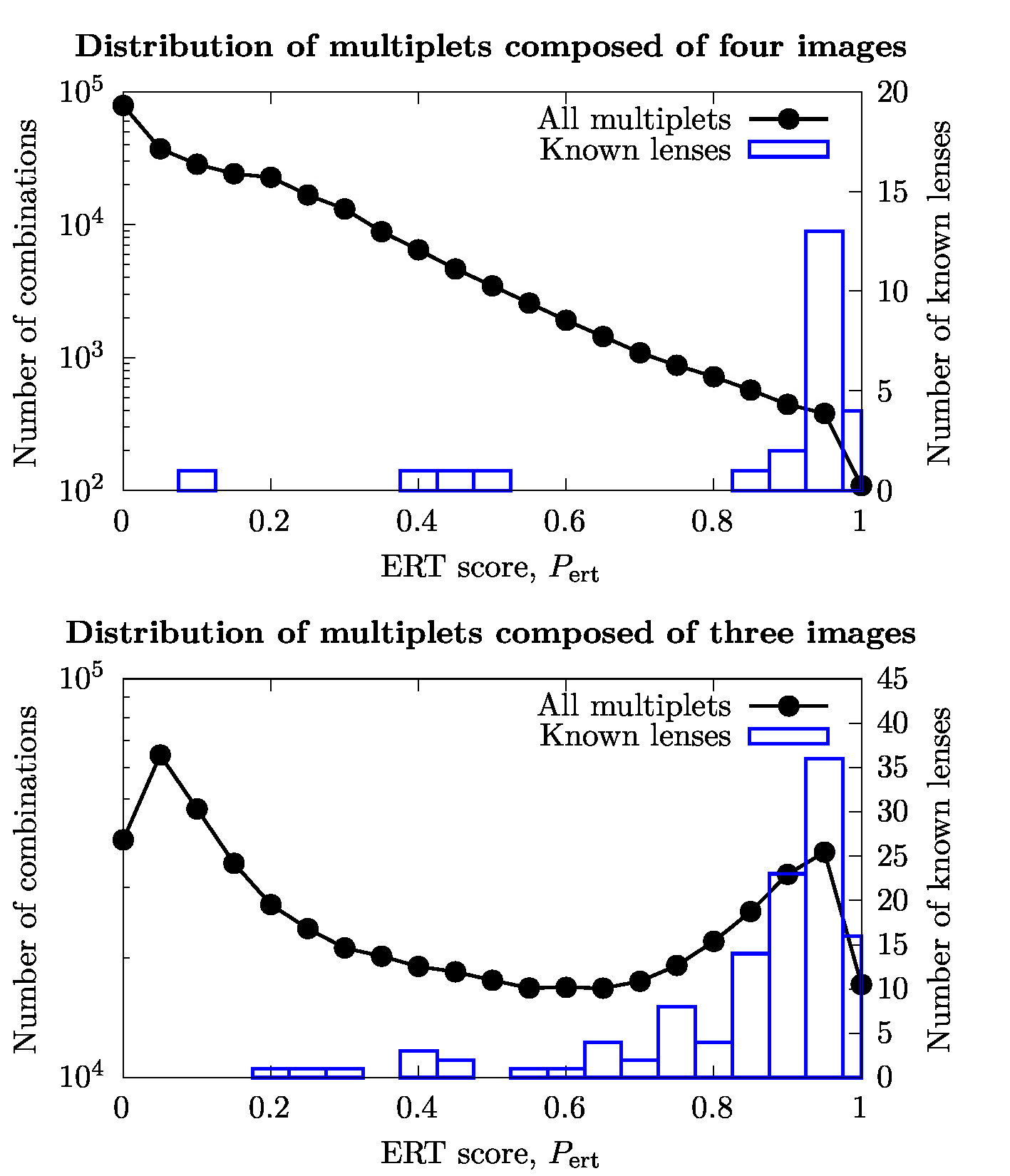}
    \caption{Distribution of the ERT scores for the 56\,398 multiplets (24 known lenses) composed of at least four components (top) and 134\,656 multiplets (45 known lenses) composed of at least three components (bottom). If multiplets are composed of more than three or four components, respectively, then all combinations of three and four components are considered for computing the ERT scores.}
    \label{fig:Pert_hist}
\end{figure}

Another technique for identifying strong gravitational lenses assumes that multiplets whose positions and magnitudes can be modelled by a singular isothermal ellipsoid in the presence of an external shear \citep[hereafter SIE$\gamma$ lens model]{Kormann1994} are good lens candidates. Whereas doublets do not yield a sufficient number of constraints to properly assess if their component positions and magnitudes can be reproduced through a SIE$\gamma$ lens model, those composed of three or four images do. Classical lens modelling tools, such as those from \cite{Keeton2001} or \cite{Birrer2015}, are based on a sampling of log-posterior distributions, which efficiently provide estimates of the lens model parameters along with a thorough estimate of their uncertainties. As we are not interested in those parameters but only in the ability the reproduce the multiplet positions and fluxes, we choose to simulate the relative positions and magnitudes of quadruple lenses using a SIE$\gamma$ lens model, then simulate random multiplets and train a supervised machine learning model to identify the simulated lenses from the random multiplets. 

For this purpose, we use an updated version of the method described in \cite{2019A&A...622A.165D}, which is based on Extremely Randomised Trees \citep[hereafter ERT]{Geurts2006}. The training used 112 784 simulated quadruple lenses drawn from a SIE$\gamma$ model. The simulations use random values of the ellipticity and shear drawn from the distributions provided in \cite{2022Petit} and in \cite{Holder2003}, respectively. A Gaussian noise with a standard deviation of $0.3$ mag was also added to the simulated magnitudes in order to deal with the imperfection of the SIE$\gamma$ lens model (galaxy substructures, micro-lensing, time delays, ...), see \cite[Section 3.2]{2019A&A...622A.165D} for details. We also simulated a similar number of random multiplets using $G$ magnitudes drawn from  the empirical \gdr{3} distribution. 
Cross-validation tests, where 20\% of our simulations are kept as a test set and 80\% of our simulations are used for training, show that 90.4\% of our simulated lenses are recovered by the method if four images are present while 0.7\% of the random multiplets are falsely classified as lenses. These numbers become 90.3\% and 12.5\% if triplets are considered\footnote{The areas under the receiver operating characteristic curve are equal to 0.9958 and 0.9554 for the cases of four images and three images, respectively.}. When only three out of the four lensed images are observed, we do not know -- a priori -- which lensed image is not detected (not necessarily the faintest). We test the four possibilities and keep the highest score. Similarly, if a multiplet is composed of more than 3 components, we consider all combinations of 3 and 4 components out of this multiplet. This allows us to identify quadruply imaged quasars having a contaminating star or (lensing) galaxy in their vicinity.

Figure~\ref{fig:Pert_hist} shows the ERT scores, $P_\text{ert}$, for all combinations of three and four images of the GravLens multiplets. We can see that 20/24 (83\%) of the known lenses have $P_\text{ert} > 0.8$ if four components are available, while only 0.75\% of the combinations of four components from the GravLens multiplets have $P_\text{ert} > 0.8$.

Regarding the combinations of three components, 91/117 (78\%) of the combinations from known lenses have $P_\text{ert} > 0.8$ compared to 22.7\% of all multiplets. This is in good agreement with the identification performance estimated from cross-validation tests. The differences for three components are mostly explained by the fact that we keep the maximal score out of the four ERT models (and explains the peak at $P_\text{ert} \approx 0.95$ in the bottom panel of Fig.~\ref{fig:Pert_hist}). 
Misclassified lenses can either be due to the inability of the SIE$\gamma$ model to reproduce the observed fluxes or positions of the lens (e.g. if two lensing galaxies are present), to extreme values of the eccentricity or shear (i.e. not covered by our simulations) or to microlensing \citep[see][for further discussions]{2019A&A...622A.165D}.

\subsection{Comparison of mean BP/RP spectra}\label{spectra}
\begin{figure*}
    \centering
    \includegraphics[width=0.9\textwidth]{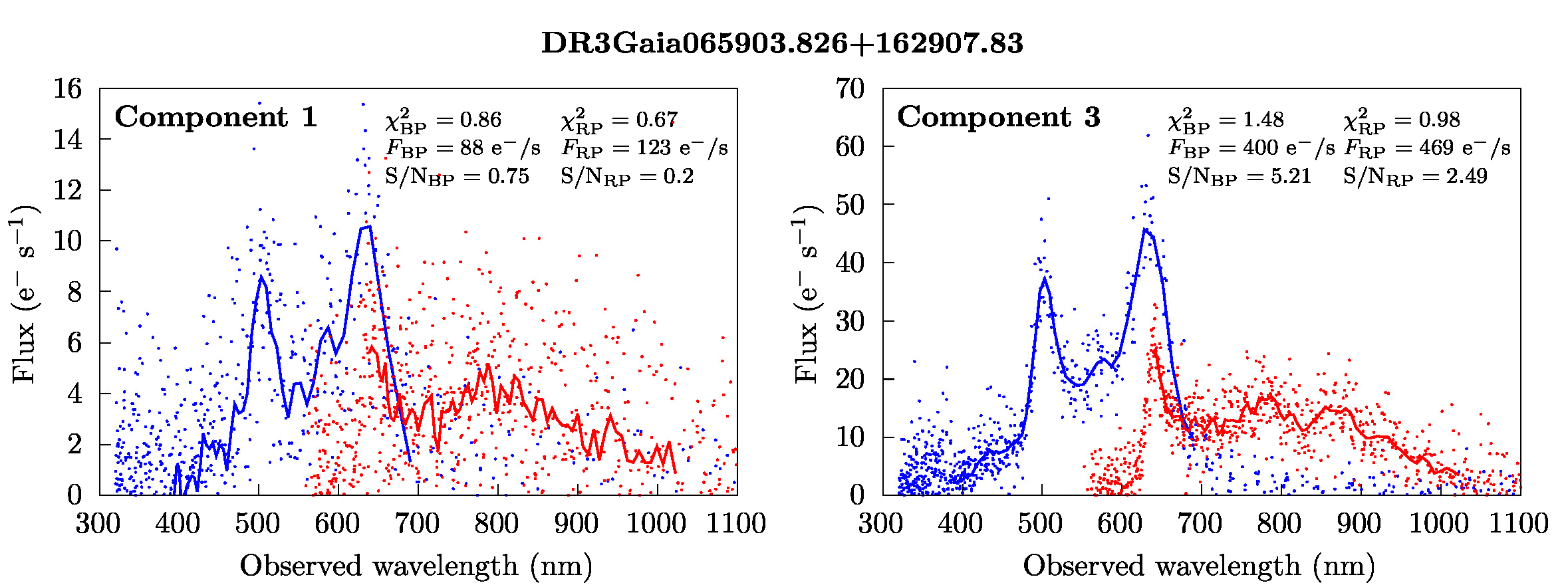}
    \caption{Epoch and resampled BP and RP spectra of the first and third components of quadruple lens system GraL J065904.1+162909 \citep{2019A&A...622A.165D,2021Stern}, corresponding to the GravLens multiplet DR3Gaia065903.826+162907.83. Points are the epoch BP/RP spectra of each of the components (blue for BP, red for RP) while solid lines are the resampled BP/RP spectra, as described in Sect.~\ref{spectra}. We also provide the additional parameters derived during the resampling phase: mean resampling chi-square ($\chi_\text{XP}^2$), integrated flux ($F_\text{XP}$) and signal-to-noise ratio (S/N$_\text{XP}$) where $\text{XP}$ stands for $\text{BP}$ and $\text{RP}$. Both component spectra show strong Ly$\alpha$ and \ion{C}{iv} emission lines that allow us to unambiguously identify this multiplet as a lensed quasar.}
    \label{fig:0659+1629}
\end{figure*}

The most secure way of identifying strong gravitational lenses is to compare the spectral energy distributions (SEDs) of their images. Indeed, as the background quasar is unique and the lensing phenomenon achromatic, all of the lensed images should have similar SEDs (except for any absorption by the deflecting galaxy; intervening gas and lens time delays). \gaia provides epoch spectro-photometry in the blue (300--700 nm, resp. BP) and in the red (600--1100 nm, resp. RP) part of the optical domain \citep{2016Prusti}, and hence should be a very powerful tool to identify lenses. \gaia's spectral resolution of $R = \lambda / \Delta \lambda \leq 100$ with a full width at half maximum between $10$ nm and $40$ nm \citep{2022Montegriffo}, can however hardly differentiate strongly lensed quasars from quasar pairs, although both have important applications in cosmology \citep[see][for examples]{2022Mannucci}.

To compare components of a GravLens multiplet, we use the \gaia epoch BP/RP spectra associated with each of the components since we cluster components at the transit level. Each of these epoch spectra has 60 fluxes, associated uncertainties, and pixel positions in the along-scan (AL) direction. Pixel positions are converted into wavelength positions using dedicated dispersion functions\footnote{Available at the \href{https://github.com/gaia-dpci/GaiaXPy/blob/main/gaiaxpy/config/TabulatedDispersionFunction-DR3-v375wiv142r_20200610.csv}{GaiaXPy github}.}. The spectra are not sampled on the same pixel scale due to the geometric and flux calibrations that minimise the discrepancies between otherwise similar spectra but acquired over different CCD rows, CCD columns, or TDI gates \citep[see][for details]{2022DeAngeli}. We resample the epoch spectra on a uniform pixel grid with $\vec{x}_\text{BP} = \lbrace 13, 13.5, \dots, 36\rbrace$ in BP and $\vec{x}_\text{RP} = \lbrace 13, 13.5, \dots, 49\rbrace$ in RP. These cover the wavelength regions 394--690 nm in BP and 638--1022 nm in RP. For each $x \in \vec{x}_\text{BP}$ or $x \in \vec{x}_\text{RP}$, we first isolate epoch BP or RP fluxes falling in the pixel range $[ x-0.5, x+0.5 [$ and reject those for which the distance to the median flux in this range is larger than $7.5 \sigma$. We then fitted a line to the remaining fluxes and take its value at $x$ as the value of the resampled flux, along with its associated uncertainty. Since the resampling bins overlap, correlations exist between the noise on the resampled fluxes that should be taken in account. During resampling, we estimate the total fluxes of each component, $F_\text{BP}$ and $F_\text{RP}$; their signal-to-noise ratio, S/N$_\text{BP}$ and S/N$_\text{RP}$; and a mean chi-square for the fit of the lines to the epoch fluxes in each of the resampling bin, $\chi_\text{BP}^2$ and $\chi_\text{RP}^2$. High $\chi_\text{BP}^2$ or $\chi_\text{RP}^2$ 
are indicative of the inability of our resampling to fully model the variance seen in the epoch spectra. This could be due to multiple effects, such as blended sources, unfiltered cosmic rays, border effects, and high intrinsic variability of the sources, to cite a few examples. The procedure is illustrated in Fig.~\ref{fig:0659+1629} for the case of the known lens GraL J065904.1+162909 \citep{2019A&A...622A.165D,2021Stern}.

The resampled spectra of $N$ components are compared using the method described in the appendices of \cite{2022arXiv220605681G}\footnote{The Octave/Matlab source code is publicly available at \url{https://github.com/ldelchambre/gls_mean}.}. If $\vec{f_i}$ is the resampled BP or RP spectrum of the $i$th component of the multiplet, then we aim to find a mean vector, $\vec{m}$, and linear coefficients, $s_i$, that minimise the reduced chi-square defined by
\begin{equation}
\chi_\nu^2 = \frac{1}{\nu} \sum_{i=1}^N \left\| \, \mat{W_i} \, \left[ \, \vec{f_i} - \vec{m} \, s_i \, \right] \, \right\|^2
\label{eq:chi2spec}
\end{equation}
where $\nu$ are degrees of freedom of the problem and $\mat{W_i}$ is the inverse of the Cholesky decomposition of the covariance matrix associated with $\vec{f_i}$, $\mat{C_i}$, such that $\mat{W_i}^T \mat{W_i} = \mat{C_i}^{-1}$. Absorption of quasar light by the lens affects the colour of the lensed images so Equation \ref{eq:chi2spec} was evaluated separately for BP and RP before producing a single reduced chi-square. Multiplets composed of components having similar spectra thus have $\chi_\nu^2 \approx 1$. Finally, to ease the comparison of the $\chi_\nu^2$, we use the well-known cubic root transformation \citep{1931Wilson},
\begin{equation}
\texttt{gof} = \sqrt{\frac{9 \nu}{2}} \; \left( \, \sqrt[3]{\chi_\nu^2} + \frac{2}{9 \nu} - 1 \, \right),
\label{eq:gaussianised_chi_square}
\end{equation}
which approximately follows a standard Gaussian distribution that is independent of the degrees of freedom, $\nu$, once $\nu$ is large (here the mode of $\nu$ is equal to $118$).

We complement this chi-square approach by a comparison based on the Wasserstein distance \citep{Kantorovich1942,Kantorovich2006OnTT}, which is potentially more robust to outliers. Intuitively, the Wasserstein distance corresponds to the minimal `effort', or optimal transport cost \citep[e.g.][]{villani2003optimal,villani2016optimal, 2019-Peyre-computational-ot}, that is needed in order to convert a pile of earth into another pile,
hence the reason why it is often called the earth mover's distance. Given two set of epoch spectra, $\vec{f}$ and $\vec{g}$, and their  
linear interpolations
in pixel space, $f(x)$ and $g(x)$, we define the 1-Wasserstein distance between $\vec{f}$ and $\vec{g}$ as
\begin{equation}
W_d = \int_\mathbb{R} \left| \frac{F_z}{F_\infty} - \frac{G_z}{G_\infty} \right| dz
\label{eq:wasserstein_dist}
\end{equation}
where $F_z = \int_{-\infty}^z f(x) \; dx$ and $G_z = \int_{-\infty}^z g(x) \; dx$. Two components with similar SEDs then have $W_d \ll 1$. No resampling is needed here and we do not use the uncertainties on the epoch spectra as the comparison is done on the overall shape of the epoch spectra only.

\subsection{Selection of the lens candidates}\label{candidates}

\begin{figure}
    \centering
    \includegraphics[width=0.49\textwidth]{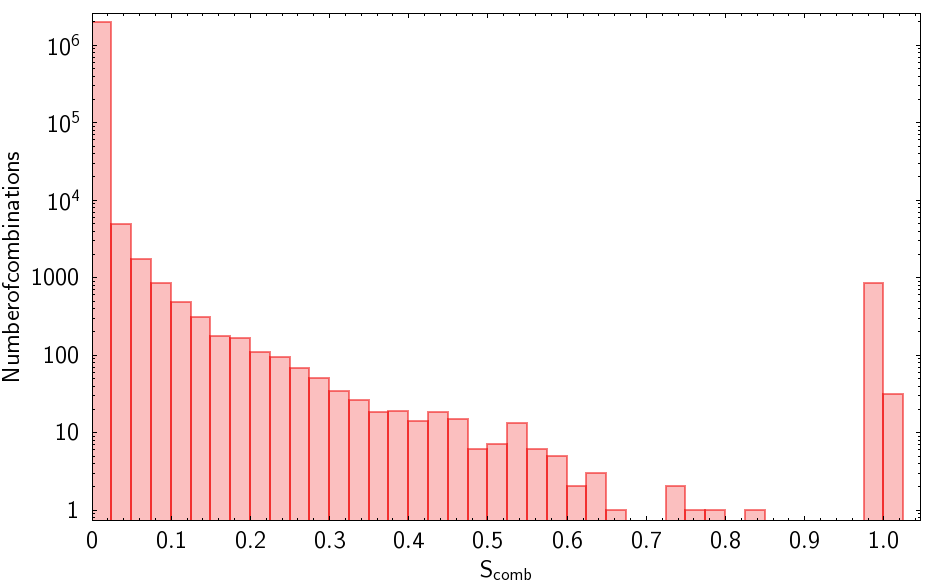}
    \caption{Distribution of the combined score, $S_\text{comb}$.}
    \label{fig:Scomb}
\end{figure}

Focusing on providing a first non-exhaustive list of lens candidates based on the \gaia FPR data, we applied the methods described above to the 491\,504 multiplets with less than 7 components (Table \ref{nombre}) to provide scores quantifying if a multiplet is likely to be a strongly lensed quasar. As the scoring methods have different limitations, not all scores are available for all multiplets.  The selection of the most promising candidates is done by isolating regions in the space defined by the parameters: $P_\text{ert}$ score; minimal {\tt HESIOD} score ($\mathscr{O}_\text{min}$); $\chi_\nu^2$; {\tt gof} and $W_d$, with the addition of the galactic latitude, $b$; maximal separation between pair of components; $G$ magnitudes; S/N$_\text{XP}$; integrated flux $F_\text{XP}$ and mean resampling chi-square, $\chi_\text{XP}^2$. Instead of performing cuts manually, we use machine learning to compute a combined score, $S_\text{comb}$, that reflects the similarities between the multiplets and the set of known lenses. To do so, we use a cross-validation procedure where we split the set of 1\,957\,559 combinations of 2--6 components from the 491\,504 multiplets into 100 subsets of approximately equal size. For each subset, we run a Random Forest classifier \citep{RFBreiman} built on the set of known lenses and on the combinations from the 99 other subsets. The combinations from these 99 subsets that are closer than 10\arcsec from one of the combinations in the selected subset are discarded, as the combinations from multiplets share input parameters (e.g. $G$ magnitudes, S/N$_\text{XP}$, \ldots). Fig.~\ref{fig:Scomb} shows the distribution of the combined score for all the 1\,957\,559 combinations. The 869 combinations at $S_\text{comb} > 0.9$ correspond to combinations from known lenses. As the known lenses are always included in the Random Forest training sample, these will automatically have $S_\text{comb} \approx 1$. Most of the combinations, however, have $S_\text{comb} \ll 1$, and sorting by this score can drastically reduce the number of combinations the user has to explore. The set of known lenses is limited in number, and thus the score combination is naturally biased due to the lack of coverage of the parameter space. Thus, it is expected that some yet-to-be-discovered lenses may have low $S_\text{comb}$ values. We hence encourage users of this FPR to explore alternative selections.

The scores and discriminators we use to isolate the lens candidates are provided in Table~\ref{tbl:cds_table}. We concentrate our search on multiplets having $S_\text{comb} > 0.01$ but also consider each of the discriminators from Table \ref{tbl:cds_table} separately. Finally, for deeper cleaning, we cross-match the GravLens components with the \gdr{3} in order to obtain information on proper motions, parallaxes, redshifts, and source classification; and with the CatWISE2020 catalogue \citep{2021ApJS..253....8M} to obtain W1-W2 colours. The use of these two public catalogues allows us to discard some obvious contaminants with large proper motions, large parallaxes, or low W1-W2 colours as well, as to select interesting candidates based on components with nearly equal redshifts or magnification biases \citep{1980ApJ...242L.135T}.
 
We finally selected 1 307 candidates from Table~\ref{tbl:cds_table} using loose cuts on our discriminators so as to favour completeness. This selection consists in a compilation of various subsets that were independently drawn by the main authors of this paper and is consequently very heterogeneous. These candidates however share some common characteristics: $|b| > 5\deg$, W1-W2$>0.275$, maximal component separation$<6.5\arcsec$, \texttt{gof} $< 3$, $W_d < 6$ and $S_\text{comb} > 0.01$; although not all candidates satisfy all these characteristics at the same time. These candidates were then visually ranked from A to D, where A corresponds to the most promising candidates, where the lensing hypothesis is the most probable. Out of these candidates, 621 were ruled out (ranked D) because of one or more components exhibiting very large proper motions, large parallaxes, low W1-W2 WISE colors, or because they are spectroscopically confirmed as stars, nearby galaxies, or AGN. 305 candidates are given a rank of C, because the visual inspection tends to support the stars, QSO+star or QSO+galaxy hypothesis, or fortuitous alignment of QSOs. 332 candidates have interesting lens-like features and are classified as plausible candidates (rank B). Rank A is further subdivided into two subcategories: A+ if all components have similar spectra, image(s) that support the lensing hypothesis while exhibiting a potential deflector and are ranked A- otherwise. Rank B candidates are similarly split into B+ and B-, depending of the degree of confidence we put on the observed lens-like features. The 381 candidates A and B are presented in Table \ref{list_candidates} (available online), highlighting the 49 candidates ranked A (see also Appendix \ref{sec:lens_candidates}). The rank A candidates have angular sizes from 1.03\arcsec to 5.97\arcsec, reaching minimal image separations of 0.41 \arcsec.

We note that depending on the involved redshifts, lensing galaxies, and image separations, the lensing galaxies can be hard to detect from the currently available ground-based imaging surveys. Thus, good candidates for lensed quasars can present no detectable lensing galaxy based on stamps from current ground-based survey archives, such as PS1. However, the lensing galaxy can later be identified in higher resolution and/or deeper images or via absorption lines directly in spatially unresolved, slit-based, follow-up spectra. Moreover, some effects can effectively bias the eye-detection of the lensing galaxy: first, we use a default color scale to display the DESI and PanSTARRS images in Appendix \ref{sec:lens_candidates}, while sharper cuts or a more detailed inspection of the individual g, r, i, z, y images can reveal hints of a lensing galaxy in several cases (e.g. DR3Gaia014718.509-465709.04 or DR3Gaia020209.884-431922.09). Secondly, for more compact lenses, the lensing galaxy is often blended with the lensed images, and the {\tt HESIOD} method then reports a high score, likely due to a blue-red color gradient in the system that does not appear immediately by the eye (e.g. DR3Gaia115352.588-252027.70).

\begin{table*}
\centering
\caption{Extract of the table containing 1\,957\,559 combinations of 2--6 components out of the 491\,504 multiplets including the scores calculated in this work. All combinations of components from the multiplets are considered here to enable the detection of cases consisting of multiple images of a lens system, plus contaminants (e.g. deflecting galaxy or nearby star). Comp. are the component\_id's (from Table \ref{datamodeltab}) of the components in each combination.\label{tbl:cds_table}} 
{\footnotesize
\begin{tabular}{rllrrrrrrr}
  \multicolumn{1}{c}{\#} &
  \multicolumn{1}{c}{name} &
  \multicolumn{1}{c}{Comp.} &
  \multicolumn{1}{c}{$\alpha$} &
  \multicolumn{1}{c}{$\delta$} &
  \multicolumn{1}{c}{$P_\text{ert}$} &
  \multicolumn{1}{c}{$\mathscr{O}_\text{min}$} &
  \multicolumn{1}{c}{\tt gof} &
  \multicolumn{1}{c}{$W_d$} &
  \multicolumn{1}{c}{$S_\text{comb}$} \\
\hline \hline
  1 & DR3Gaia052025.426+331443.46 & 1/2 & 80.1057322 & 33.2448383 &  & 0.615 & -2.770 & 2.448 & 0.000 \\
  2 & DR3Gaia084114.124+223614.72 & 1/2 & 130.3082652 & 22.6042005 &  & 0.521 & -3.070 & 3.035 & 0.000 \\
  3 & DR3Gaia102800.048+803922.91 & 1/2 & 157.0020369 & 80.6567414 &  & 0.628 & 0.577 & 4.126 & 0.000 \\
  4 & DR3Gaia235819.841+482027.63 & 1/2 & 359.5817555 & 48.3413855 &  &  & 0.597 & 5.803 & 0.000 \\
  5 & DR3Gaia192635.516+405005.98 & 1/2 & 291.6473954 & 40.8348653 &  & 0.642 & 1.247 & 4.707 & 0.000 \\
  6 & DR3Gaia062751.407-253223.83 & 1/2 & 96.9646288 & -25.5393739 &  & 0.374 & 6.391 & 4.224 & 0.000 \\
  7 & DR3Gaia062751.407-253223.83 & 1/3 & 96.9642328 & -25.5398715 &  &  & 5.097 & 4.224 & 0.000 \\
  8 & DR3Gaia062751.407-253223.83 & 2/3 & 96.9638001 & -25.5404539 &  & 0.374 & 1.278 & 4.224 & 0.000 \\
  9 & DR3Gaia062751.407-253223.83 & 1/2/3 & 96.9642205 & -25.5398998 & 0.060 & 0.374 & 5.646 & 4.224 & 0.000 \\
  10 & DR3Gaia191906.417-110304.27 & 1/2 & 289.7765560 & -11.0504709 &  &  & 0.377 & 21.999 & 0.000 \\
  \multicolumn{1}{r}{\vdots} & \multicolumn{1}{c}{\vdots} & \multicolumn{1}{l}{\vdots} & \multicolumn{1}{r}{\vdots} & \multicolumn{1}{r}{\vdots} & \multicolumn{1}{r}{\vdots} & \multicolumn{1}{r}{\vdots} & \multicolumn{1}{r}{\vdots} & \multicolumn{1}{r}{\vdots} & \multicolumn{1}{r}{\vdots} \\
  1957557 & DR3Gaia190225.249-311341.45 & 2/3/4 & 285.6043694 & -31.2287627 & 0.580 & 0.705 & -3.967 & 2.688 & 0.000 \\
  1957558 & DR3Gaia190225.249-311341.45 & 1/2/3/4 & 285.6046004 & -31.2289471 & 0.000 & 0.705 & 5.939 & 131.139 & 0.000 \\
  1957559 & DR3Gaia201938.091-142206.89 & 1/2 & 304.9092924 & -14.3684893 &  & 0.759 & 3.843 & 1.984 & 0.000 \\
\hline
\end{tabular}
\tablefoot{The full table is available at the CDS in electronic form via anonymous ftp to \href{http://cdsarc.u-strasbg.fr}{cdsarc.u-strasbg.fr} (\href{http://130.79.128.5}{130.79.128.5}) or via http://cdsarc.u-strasbg.fr/viz-bin/qcat?J/A+A/\red{???}/\red{???}.}
}
\end{table*}

\begin{table*}
\centering
\caption{\label{list_candidates} List of the most promising 49 lens candidates out of the 381 selected in this work (the full is available online). Candidates with $P_\text{ert}$ are quad candidates, others are double candidates. The resampled spectra and Dark Energy Survey or Pan-STARRS1 images for this selection can be found in Appendix \ref{sec:lens_candidates}. The candidate identifier (\#) corresponds to the row identifier from Table \ref{tbl:cds_table}.}
{ \footnotesize 
\begin{tabular}{rllrrrrrrr}
  \multicolumn{1}{c}{\#} &
  \multicolumn{1}{c}{name} &
  \multicolumn{1}{c}{grade} &
  \multicolumn{1}{c}{$\alpha$} &
  \multicolumn{1}{c}{$\delta$} &
  \multicolumn{1}{c}{$P_\text{ert}$} &
  \multicolumn{1}{c}{$\mathscr{O}_\text{min}$} &
  \multicolumn{1}{c}{\tt gof} &
  \multicolumn{1}{c}{$W_d$} &
  \multicolumn{1}{c}{$S_\text{comb}$} \\
  \hline \hline
  1186594 & DR3Gaia010120.807-494324.36 & A+ & 15.3367775 & -49.7228775 &  & 0.739 & 1.062 & 1.194 & 0.000\\
  1410355 & DR3Gaia015426.227-440213.66 & A+ & 28.6086993 & -44.0372670 &  & 0.844 & -1.701 & 1.501 & 0.003\\
  1209020 & DR3Gaia015739.213-683707.60 & A+ & 29.4137683 & -68.6187615 &  & 1.000 & -1.220 & 2.797 & 0.178\\
  482036 & DR3Gaia020209.884-431922.09 & A+ & 30.5410743 & -43.3235545 &  &  & -0.477 & 0.540 & 0.000\\
  400875 & DR3Gaia020501.994-323348.59 & A+ & 31.2589438 & -32.5640445 &  &  & -1.904 & 0.189 & 0.000\\
  506340 & DR3Gaia044652.260-310219.85 & A+ & 71.7179854 & -31.0383946 &  & 1.000 & -0.963 & 2.369 & 0.003\\
  748543 & DR3Gaia050613.596-253047.45 & A+ & 76.5561851 & -25.5134752 &  & 0.621 & -0.623 & 0.955 & 0.000\\
  1879078 & DR3Gaia060216.151-433540.97 & A+ & 90.5670711 & -43.5945222 &  & 0.866 & 5.328 & 3.712 & 0.004\\
  50830 & DR3Gaia105221.613-195238.39 & A+ & 163.0900708 & -19.8771667 &  & 0.754 & 8.572 & 7.774 & 0.088\\
  1572115 & DR3Gaia121504.295-200556.84 & A+ & 183.7681509 & -20.0992440 &  & 0.997 & -0.912 & 4.616 & 0.153\\
  48388 & DR3Gaia151030.678-791857.87 & A+ & 227.6266420 & -79.3157777 &  & 0.521 & -3.740 & 1.952 & 0.000\\
  1734022 & DR3Gaia151723.117-241848.13 & A+ & 229.3462903 & -24.3139543 &  & 0.796 & -2.028 & 4.264 & 0.002\\
  147896 & DR3Gaia170842.333+064614.31 & A+ & 257.1758007 & 6.7705216 &  & 0.679 & -2.583 & 4.762 & 0.000\\
  1654705 & DR3Gaia172201.867+201920.75 & A+ & 260.5075498 & 20.3222709 & 0.740 & 0.391 & -2.150\tablefootmark{a} & 5.784\tablefootmark{a} & 0.000\\
  294425 & DR3Gaia193647.137-320217.79 & A+ & 294.1957241 & -32.0385626 &  & 0.833 & -2.110 & 1.282 & 0.000\\
  602038 & DR3Gaia210752.320-161131.67 & A+ & 316.9684944 & -16.1922981 & 0.880 & 0.782 & 1.747\tablefootmark{b} & 4.827\tablefootmark{b} & 0.003\\
  1466139 & DR3Gaia221540.110-520404.66 & A+ & 333.9167767 & -52.0676474 &  & 0.998 & 0.143 & 0.877 & 0.176\\
  700931 & DR3Gaia230405.819-802805.72 & A+ & 346.0281735 & -80.4686247 &  &  & -0.419 & 1.415 & 0.000\\
  141767 & DR3Gaia014718.509-465709.04 & A- & 26.8266523 & -46.9530303 &  & 0.799 & -0.124 & 1.573 & 0.000\\
  884711 & DR3Gaia021120.383+210749.64 & A- & 32.8343751 & 21.1299320 & 0.090 & 0.634 & 0.474\tablefootmark{c} & 3.008\tablefootmark{c} & 0.000\\
  559412 & DR3Gaia031013.747+352414.86 & A- & 47.5561827 & 35.4045044 & 0.320 & 0.278 & -2.067\tablefootmark{d} & 4.703\tablefootmark{d} & 0.000\\
  1002146 & DR3Gaia033001.688-441335.60 & A- & 52.5063646 & -44.2268054 &  & 0.675 & 1.658 & 1.575 & 0.000\\
  9023 & DR3Gaia045755.331+124238.67 & A- & 74.4805842 & 12.7104176 &  & 0.603 & -2.458 & 1.522 & 0.000\\
  1420154 & DR3Gaia055409.442-234754.13 & A- & 88.5387470 & -23.7982264 &  & 0.610 & 3.493 & 1.395 & 0.000\\
  771791 & DR3Gaia070020.352+132813.68 & A- & 105.0851179 & 13.4707837 &  & 0.458 & -1.776 & 2.121 & 0.000\\
  1113851 & DR3Gaia092321.265-020554.21 & A- & 140.8399647 & -2.0980878 &  &  & -4.474 & 3.419 & 0.023\\
  496839 & DR3Gaia110527.117-391343.61 & A- & 166.3628841 & -39.2284930 &  & 1.000 & -2.382 & 1.397 & 0.122\\
  287077 & DR3Gaia111221.158-201111.55 & A- & 168.0885649 & -20.1865341 &  & 0.600 & -1.452 & 1.381 & 0.010\\
  1204565 & DR3Gaia114934.110-172651.95 & A- & 177.3923073 & -17.4478355 &  & 0.709 & 1.989 & 4.780 & 0.036\\
  1147299 & DR3Gaia115352.588-252027.70 & A- & 178.4693002 & -25.3410790 &  & 0.747 & 2.111 & 2.093 & 0.051\\
  13466 & DR3Gaia124708.184-092332.50 & A- & 191.7842239 & -9.3921238 &  & 0.793 & -0.925 & 6.682 & 0.057\\
  399786 & DR3Gaia125238.119-270906.98 & A- & 193.1589992 & -27.1514771 &  & 0.503 & 1.112 & 1.845 & 0.000\\
  1938680 & DR3Gaia133741.153-132524.24 & A- & 204.4217226 & -13.4235215 &  & 0.630 & 0.855 & 6.103 & 0.029\\
  950559 & DR3Gaia134839.786+002343.29 & A- & 207.1655325 & 0.3946152 &  &  & -3.192 & 3.462 & 0.000\\
  787035 & DR3Gaia150826.916+670544.68 & A- & 227.1123665 & 67.0955375 &  & 0.469 & -0.758 & 2.561 & 0.023\\
  468388 & DR3Gaia160508.549+024739.44 & A- & 241.2857807 & 2.7943234 &  & 0.779 & -1.242 & 3.677 & 0.052\\
  651231 & DR3Gaia161135.764+515346.43 & A- & 242.8993897 & 51.8965852 &  & 0.766 & 0.509 & 12.918 & 0.019\\
  726741 & DR3Gaia173144.453+250232.26 & A- & 262.9352928 & 25.0424432 &  & 0.639 & 1.918 & 5.289 & 0.026\\
  1741359 & DR3Gaia175323.439+144702.74 & A- & 268.3482034 & 14.7845840 & 0.280 & 0.587 & -2.918\tablefootmark{e} & 3.794\tablefootmark{e} & 0.000\\
  675406 & DR3Gaia180734.677+475943.60 & A- & 271.8948565 & 47.9953497 &  & 1.000 & & & 0.021\\
  820377 & DR3Gaia190007.256-624734.16 & A- & 285.0303752 & -62.7924480 &  & 0.822 & -0.214 & 22.185 & 0.001\\
  1157981 & DR3Gaia201951.245-062931.96 & A- & 304.9638007 & -6.4925101 &  & 0.367 & 1.034 & 1.331 & 0.000\\
  628816 & DR3Gaia202042.974-265023.86 & A- & 305.1794142 & -26.8399389 &  & 0.780 & 6.270 & 2.452 & 0.009\\
  173950 & DR3Gaia202627.737+161850.69 & A- & 306.6157010 & 16.3144130 &  & 0.595 & -1.076 & 3.107 & 0.000\\
  1381292 & DR3Gaia202710.607+060438.30 & A- & 306.7943588 & 6.0773550 &  & 0.650 & -0.260 & 2.325 & 0.033\\
  937383 & DR3Gaia204449.725-040357.87 & A- & 311.2067405 & -4.0665789 &  & 0.441 & -1.300 & 0.831 & 0.000\\
  1401461 & DR3Gaia220231.754-800425.40 & A- & 330.6355963 & -80.0735089 &  & 0.696 & 0.181 & 0.633 & 0.000\\
  1557618 & DR3Gaia222638.124-521519.18 & A- & 336.6584846 & -52.2557299 &  & 0.518 & 2.052 & 1.633 & 0.000\\
  379838 & DR3Gaia235506.238-455335.44 & A- & 358.7762969 & -45.8929542 &  & 0.469 & -1.552 & 3.957 & 0.001\\
  \multicolumn{10}{c}{\vdots} \\
\hline\end{tabular}

\tablefoot{
\tablefoottext{a}{Taken from components 1 and 2 of DR3Gaia172201.867+201920.75 (\#1654702).}
\tablefoottext{b}{Taken from components 1 and 2 of DR3Gaia210752.320-161131.67 (\#602035).}
\tablefoottext{c}{Taken from components 1 and 2 of DR3Gaia021120.383+210749.64 (\#884708).}
\tablefoottext{d}{Taken from components 1 and 3 of DR3Gaia031013.747+352414.86 (\#559410).}
\tablefoottext{e}{Taken from components 2 and 3 of DR3Gaia175323.439+144702.74 (\#1741358).}
The full table is available at the CDS in electronic form via anonymous ftp to \href{http://cdsarc.u-strasbg.fr}{cdsarc.u-strasbg.fr} (\href{http://130.79.128.5}{130.79.128.5}) or via http://cdsarc.u-strasbg.fr/viz-bin/qcat?J/A+A/\red{???}/\red{???}.}
}
\end{table*}

\section{Conclusions}\label{conclusion}
The \gaia satellite has all-sky coverage with an angular resolution of $\sim 0.18$". This is unprecedented for an astronomical survey operating in optical wavelengths. In this article we describe the \gaia Focused Product Release (FPR) aimed at detecting strongly lensed quasars and the results of the DPAC \textit{GravLens} processing. We developed novel methods to analyse the \gaia detections near quasars and produce a list of secondary sources that complement the current Gaia Data Releases. The methods produce a series of scores that can guide the user in the selection of promising new lensed quasar candidates.

A list of 3\,760\,480 quasar candidates from well-known catalogues (Sect.~\ref{inputlist}) was input to our \textit{GravLens} pipeline. \textit{GravLens} uses the DBSCAN unsupervised clustering algorithm to produce a list of sources within a 6\arcsec radius of each quasar. It identifies clusters of \gaia detections, referred to as components, around the quasar and labels anomalous ones as outliers. A list of point sources with mean positions, fluxes, and magnitudes of the components are computed and stored in the table \texttt{lens\_candidates}. \textit{GravLens} has analysed 183~368~062 transits around quasars obtained during the first three years of \gaia operations, and produced a catalogue of 4 760 920 sources of which $\sim$103\,000 are new sources complementing those from \gdr3. 87\% of the quasars were identified as single sources, while $501\,385$ resulted in multiplets (doublets or more).

We developed scoring methods to guide the selection of the best candidates for new lenses of different types, quads, and doubles. Two of these are the {\tt HESIOD} score (Sect.~\ref{hesiod}), an outlier detection algorithm, and an Extremely Randomised Tree algorithm (ERT, Sect.~\ref{ert}). These methods use astrometric and photometric data. When available, \gaia spectrophotometry was used to ascertain whether a component was a probable image of the quasar (Sect.~\ref{spectra}). The outlier detection methods were trained on real data from a set of known lenses, while the ERT method was trained on a large number ($\sim10^5$) of simulated lenses. The methods are complementary as the ERT score works best for quads and triplets, whereas {\tt HESIOD} is particularly effective for doublets. The scores accompany this Focused Product Release.

Finally, we use our scores complemented by visual inspection to derive a refined, non-exhaustive, list of 381 lensed quasar candidates, each assigned quality grades. Among these candidates, 49 are particularly promising. 

The spatial resolution and all-sky coverage make \gaia data a treasure for lensing studies. This Focused Product Release provides a first list of new lens candidates and data beyond the \gdr3 to establish an all-sky catalogue of multiply-imaged quasars at the full \gaia angular resolution. We anticipate that the data products from this FPR and the upcoming \gaia Data Releases can contribute to various realms of cosmology. After identification and confirmation of lensed quasars through spectroscopic analysis, these lenses can help to progress on the elusive topics of dark matter and dark energy, and potentially offer insights into the tension surrounding the determination of the Hubble constant.

\begin{acknowledgements}
We thank the anonymous referee for providing valuable comments that helped improve this paper. This work is part of the reduction of the \gaia satellite observations (https://www.cosmos.esa.int/gaia).
The \gaia space mission is operated by the European Space Agency, and the data are being processed by the \gaia Data
Processing and Analysis Consortium (DPAC, https://www.cosmos.esa.int/web/gaia/dpac/consortium)). The \gaia archive website is https://archives.esac.esa.int/gaia. Funding for the DPAC is provided by national institutions, in particular, the
institutions participating in the \gaia Multi-Lateral Agreement (MLA).
We acknowledge the French “Centre National d’Etudes Spatiales” (CNES), the French national program PN-GRAM, and Action Sp\'ecifique \gaia as well as Observatoire Aquitain des Sciences de l'Univers (OASU) for financial support along the years. We also acknowledge funding from the Brazilian Fapesp institution as well as from the Brazilian-French cooperation institution CAPES/COFECUB. 
Our work was eased by the use of the data handling and visualisation software TOPCAT \citep{2005Taylor}.
This research has made use of the "Aladin sky atlas" developed at CDS, Strasbourg Observatory, France \citep{Aladin2014ASPC..485..277B, Aladin2000A&AS..143...33B}.
This research has made use of the VizieR catalogue access tool, CDS, Strasbourg, France.
This research was eased by the use of The Pan-STARRS1 Surveys (PS1) and the Dark Energy Survey for a visual check of lens candidates.

\end{acknowledgements}


%
\bibliographystyle{aa}
\bibliography{main}
\begin{appendix} 

\section{Catalogue data model}\label{datamodel}
The data model of the catalogue of sources in the vicinity of quasars is described in Table \ref{datamodeltab}.

\begin{table*}[ht]
\centering
\caption{\label{datamodeltab} \texttt{lens\_candidates} table that presents the content of the table of all sources found in the fields of the quasars analysed. For more information about the data model please refer to https://gaia.esac.esa.int/dpacsvn/DPAC/docs/ReleaseDocumentation/FPR/FPR\_master.pdf} 
\begin{tabular}{ll}
\hline
Name   & Content\\
\hline
\hline
solution\_id        & Solution Identifier \\
source\_id          & Unique source identifier of the quasar analysed \\
name                & Name of the multiplet corresponding to the coordinates of the quasar analysed\\
flag                & Flag at the quasar level (see \ref{postproc} for detailed description)\\
n\_components       & Number of components found in the field of the quasar analysed\\
component\_id       & Index of the component for this quasar field\\
n\_obs\_component   & Number of valid observations used for this component\\
component\_flag     & flag of this component\\
ra\_component       & Mean right ascension of the component\\
ra\_std\_component  & Standard deviation of the right ascension of the component\\
dec\_component      & Mean declination of the component\\
dec\_std\_component & Standard deviation of the declination of the component\\
g\_flux\_component  & Mean $G$ flux of the component\\
g\_flux\_component\_error & Uncertainty of the mean flux value for this component\\
g\_mag\_component   & Mean onboard $G$ magnitude of the component\\
g\_mag\_std\_component & Standard deviation of the onboard $G$ magnitude of the component\\
\hline
\end{tabular}
\end{table*}

\section{Lens candidates}
\label{sec:lens_candidates}

This section compares the resampled spectra of the components from some of the most promising lens candidates in Table \ref{list_candidates} and displays the associated Dark Energy Survey \citep{2019AJ....157..168D} or Pan-STARRS1 \citep{panstarrs} images. Candidates composed of three components either have spectra for two components only (DR3Gaia021120.383+210749.64) or we decided to discard one of the spectrum for clarity purpose. The discarded spectrum is either the faintest (DR3Gaia031013.747+352414.86 and DR3Gaia210752.320-161131.67) or the most contaminated (DR3Gaia172201.867+201920.75 and DR3Gaia175323.439+144702.74). None of these discarded spectra allows to rule out the lensing hypothesis.

\begin{figure*}\centering\includegraphics[height=6cm]{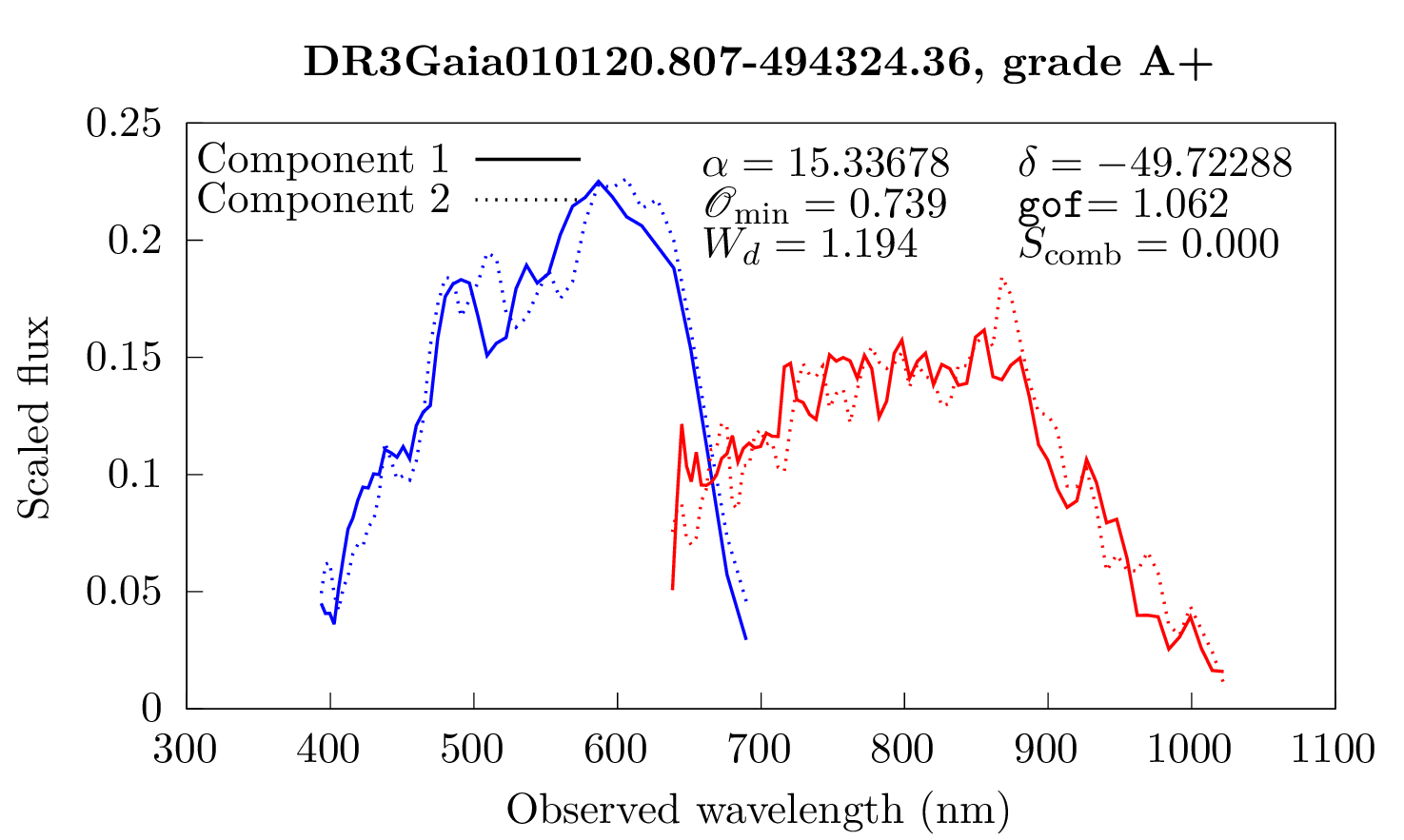}\hspace{0.5cm}\includegraphics[height=5.5cm]{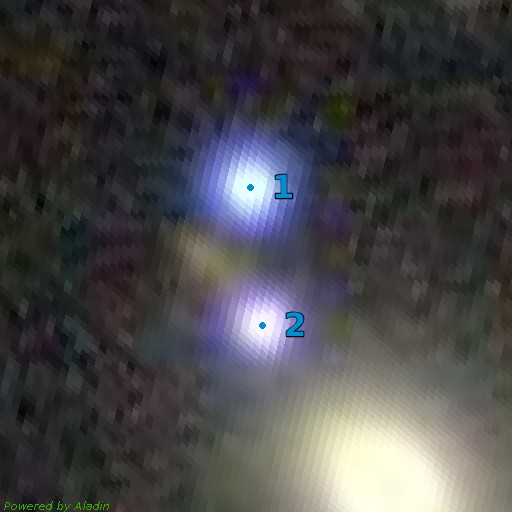}\caption{\label{fig:DR3Gaia010120.807-494324.36} Comparison of the resampled spectra of the DR3Gaia010120.807-494324.36 multiplet (Left) and associated Dark Energy Survey image (Right) \citep{2019AJ....157..168D}. Blue dots correspond to the GravLens components. Cutout size is $15.0 \arcsec \times 15.0 \arcsec$, north is up, east is left.} \end{figure*}
\begin{figure*}\centering\includegraphics[height=6cm]{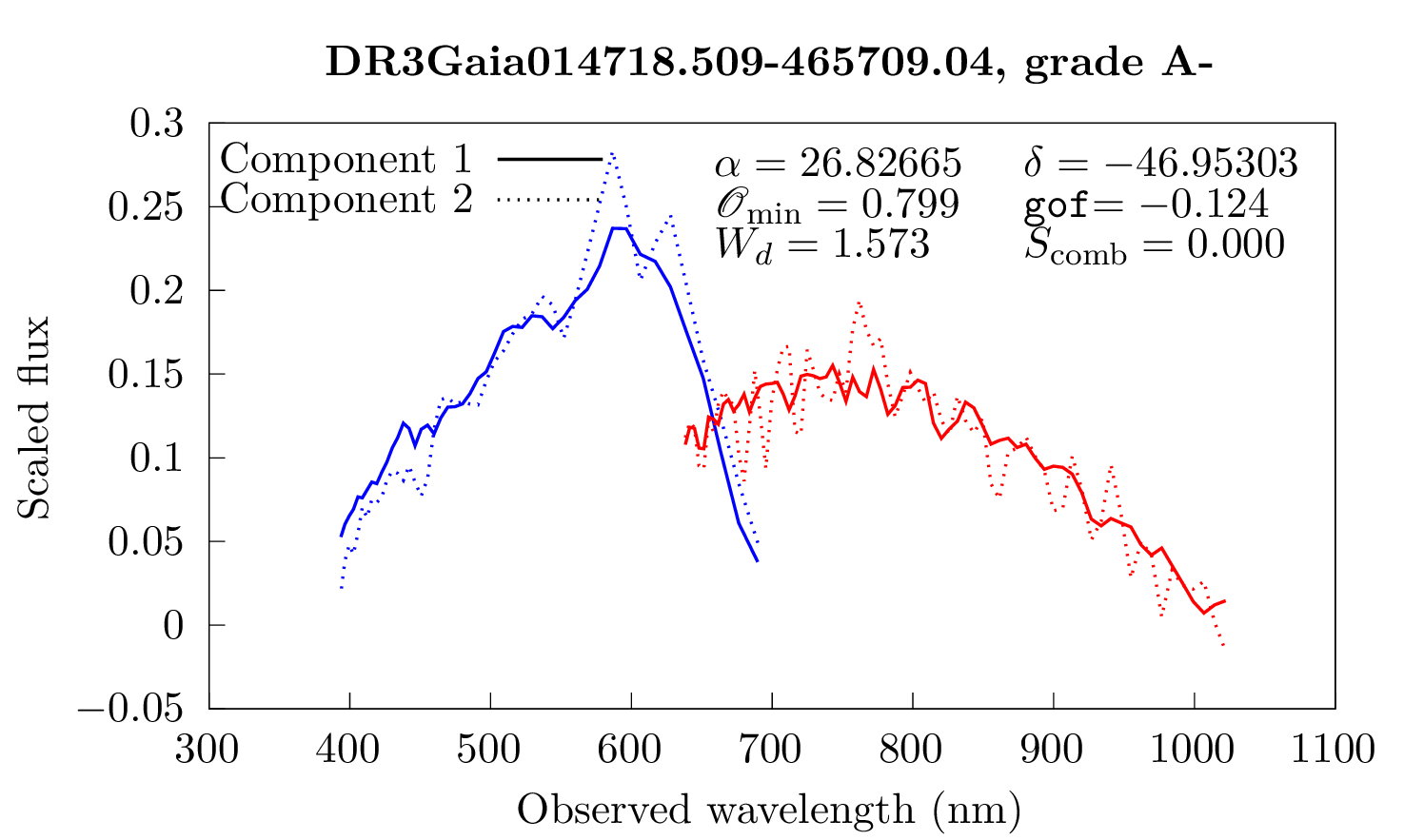}\hspace{0.5cm}\includegraphics[height=5.5cm]{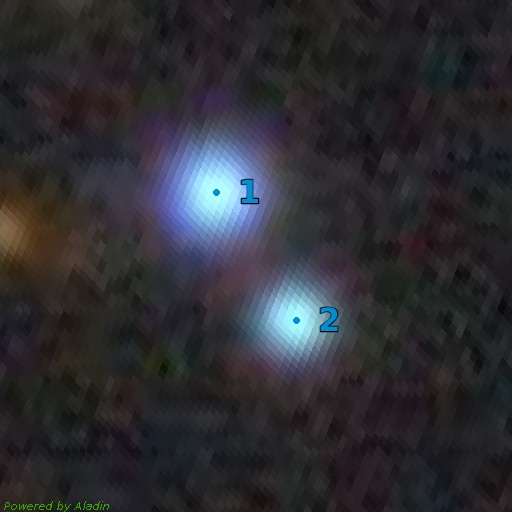}\caption{\label{fig:DR3Gaia014718.509-465709.04} Comparison of the resampled spectra of the DR3Gaia014718.509-465709.04 multiplet (Left) and associated Dark Energy Survey image (Right) \citep{2019AJ....157..168D}. Blue dots correspond to the GravLens components. Cutout size is $15.0 \arcsec \times 15.0 \arcsec$, north is up, east is left.} \end{figure*}
\begin{figure*}\centering\includegraphics[height=6cm]{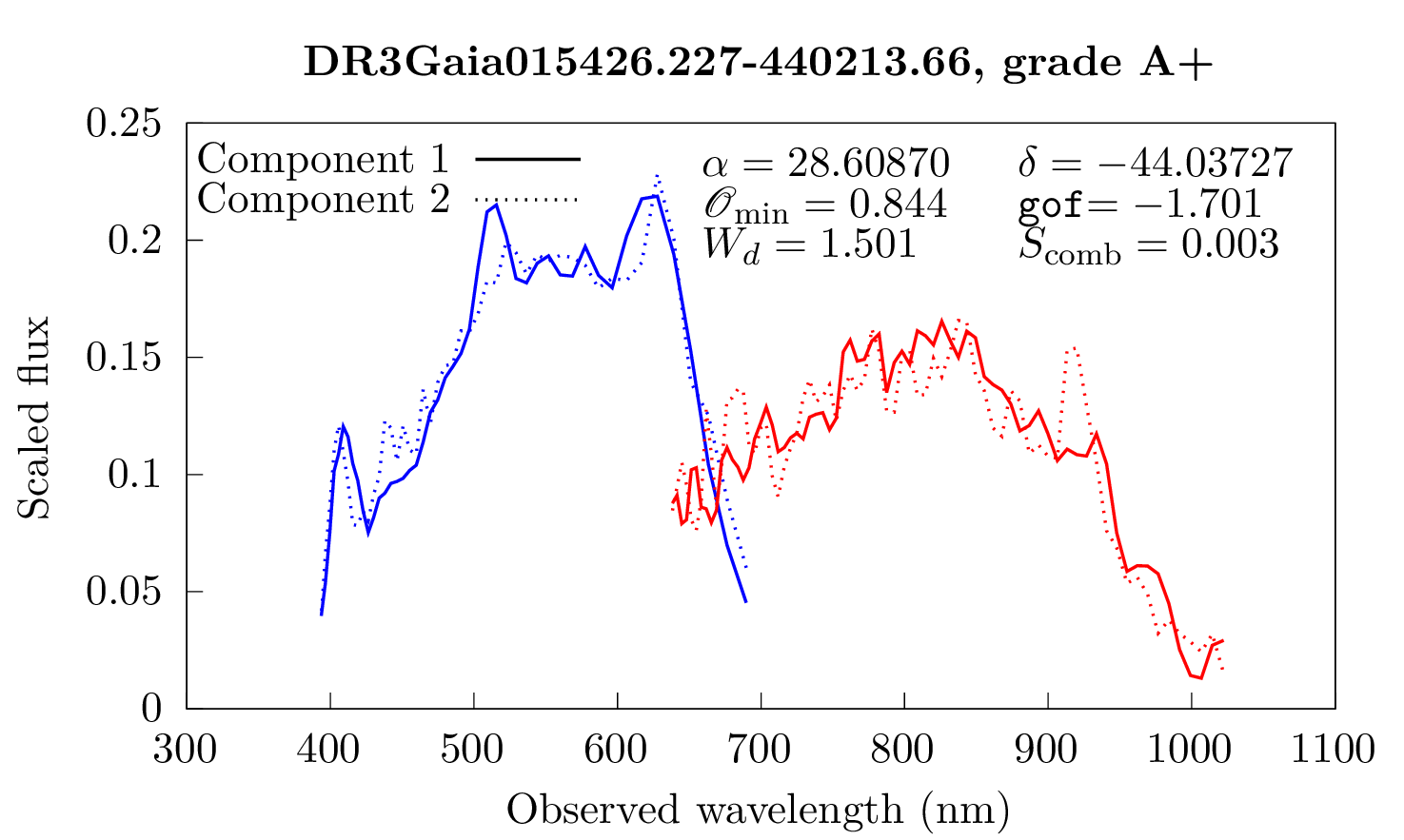}\hspace{0.5cm}\includegraphics[height=5.5cm]{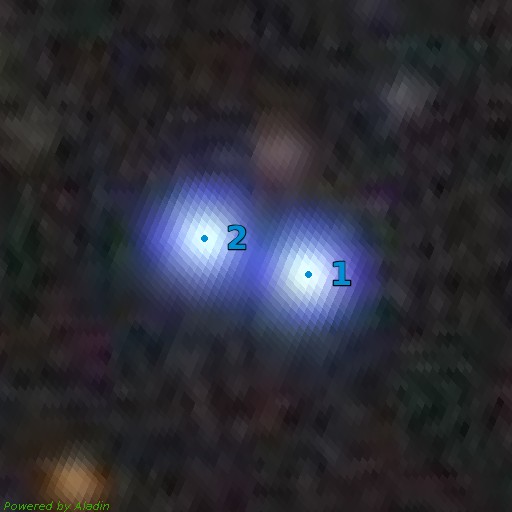}\caption{\label{fig:DR3Gaia015426.227-440213.66} Comparison of the resampled spectra of the DR3Gaia015426.227-440213.66 multiplet (Left) and associated Dark Energy Survey image (Right) \citep{2019AJ....157..168D}. Blue dots correspond to the GravLens components. Cutout size is $15.0 \arcsec \times 15.0 \arcsec$, north is up, east is left.} \end{figure*}
\begin{figure*}\centering\includegraphics[height=6cm]{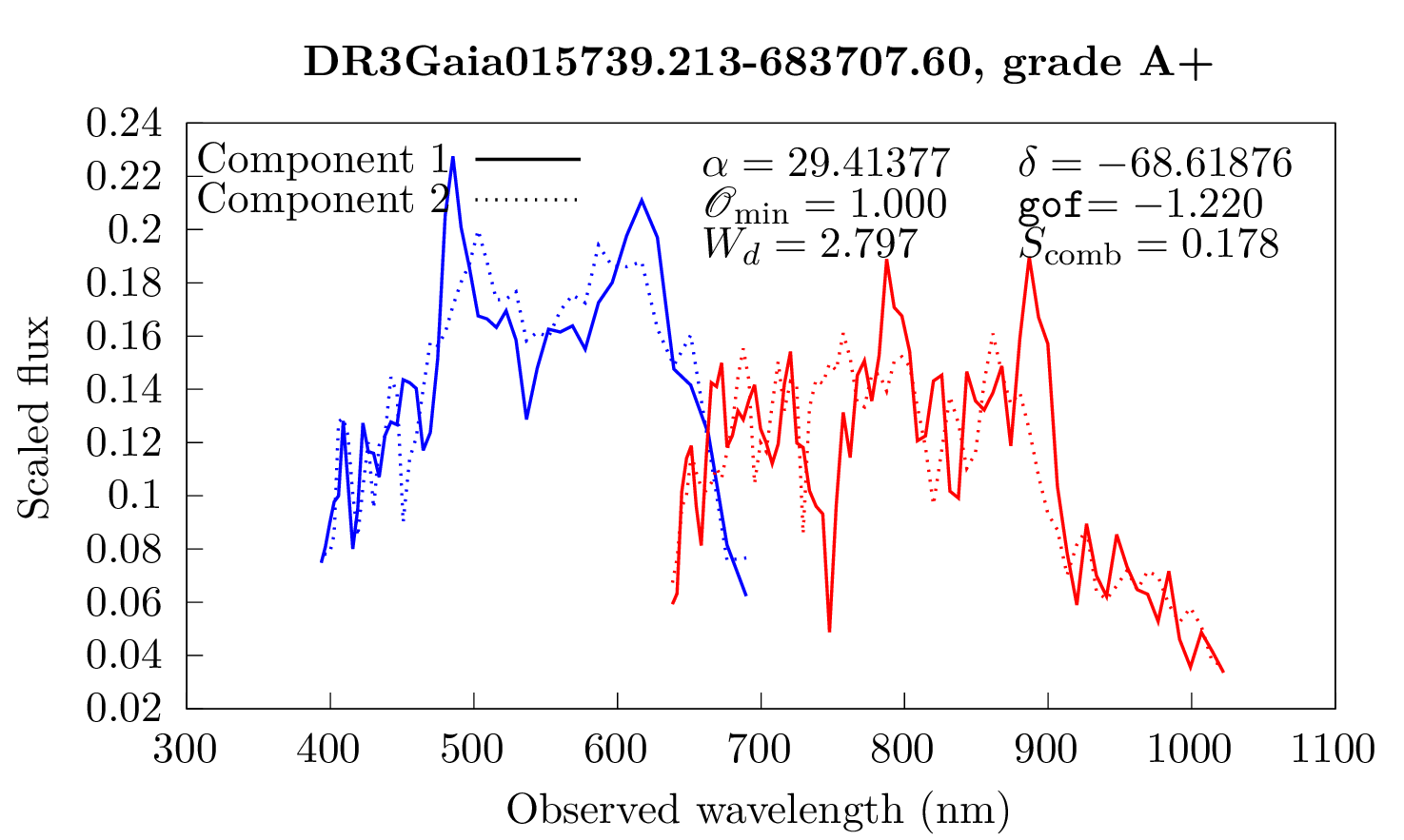}\hspace{0.5cm}\includegraphics[height=5.5cm]{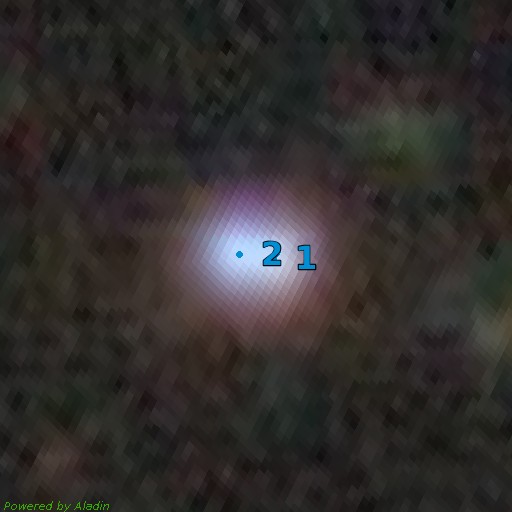}\caption{\label{fig:DR3Gaia015739.213-683707.60} Comparison of the resampled spectra of the DR3Gaia015739.213-683707.60 multiplet (Left) and associated Dark Energy Survey image (Right) \citep{2019AJ....157..168D}. Blue dots correspond to the GravLens components. Cutout size is $15.0 \arcsec \times 15.0 \arcsec$, north is up, east is left.} \end{figure*}
\begin{figure*}\centering\includegraphics[height=6cm]{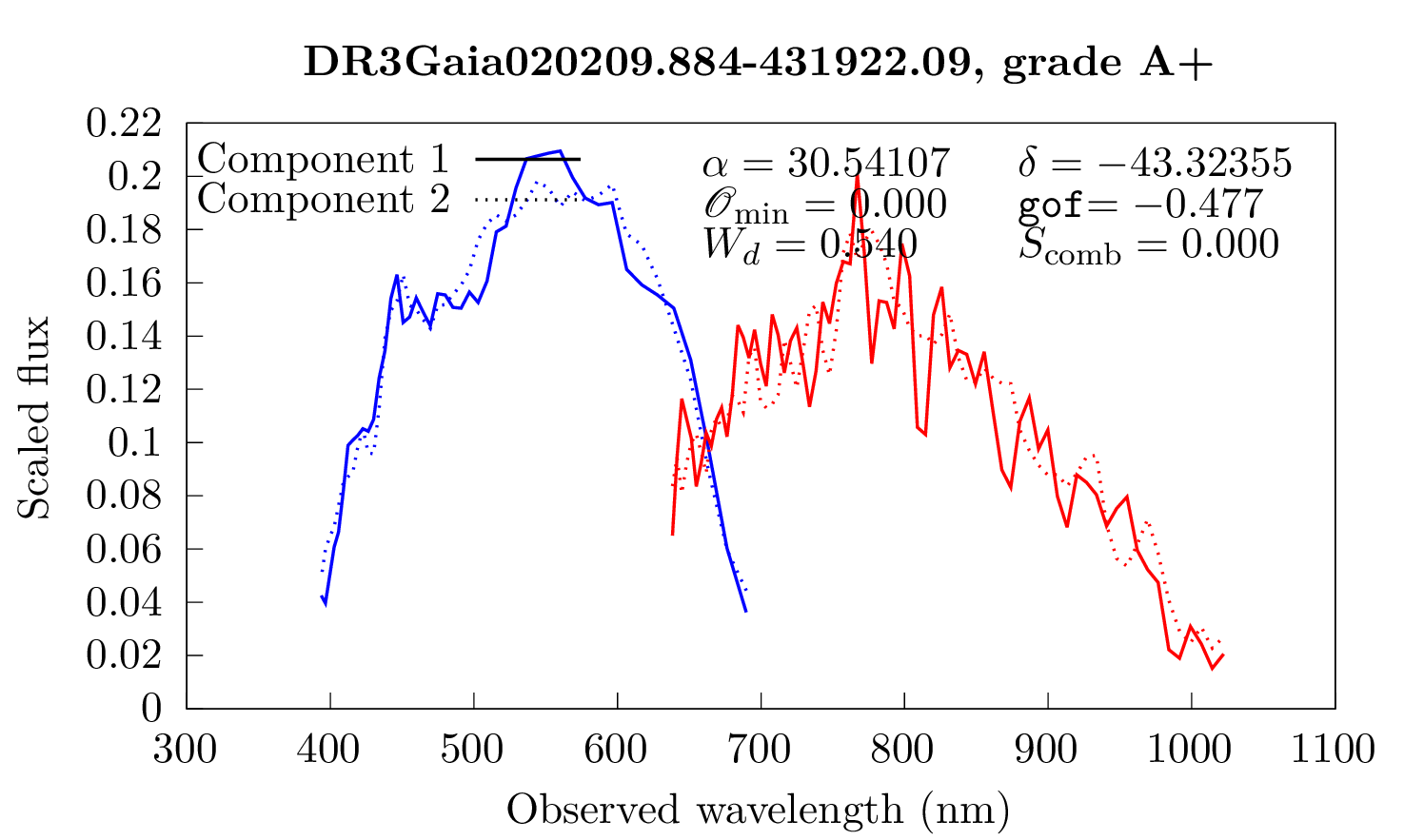}\hspace{0.5cm}\includegraphics[height=5.5cm]{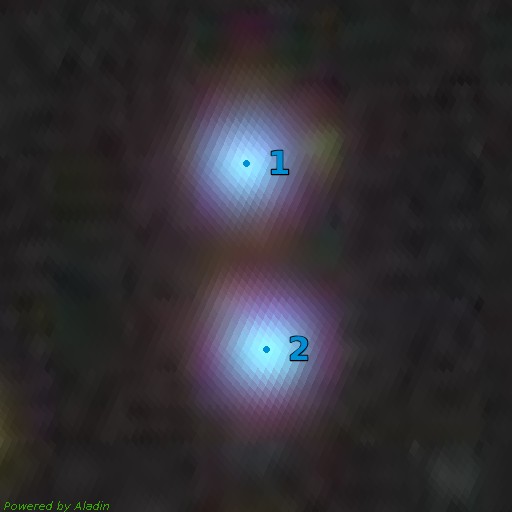}\caption{\label{fig:DR3Gaia020209.884-431922.09} Comparison of the resampled spectra of the DR3Gaia020209.884-431922.09 multiplet (Left) and associated Dark Energy Survey image (Right) \citep{2019AJ....157..168D}. Blue dots correspond to the GravLens components. Cutout size is $15.0 \arcsec \times 15.0 \arcsec$, north is up, east is left.} \end{figure*}
\begin{figure*}\centering\includegraphics[height=6cm]{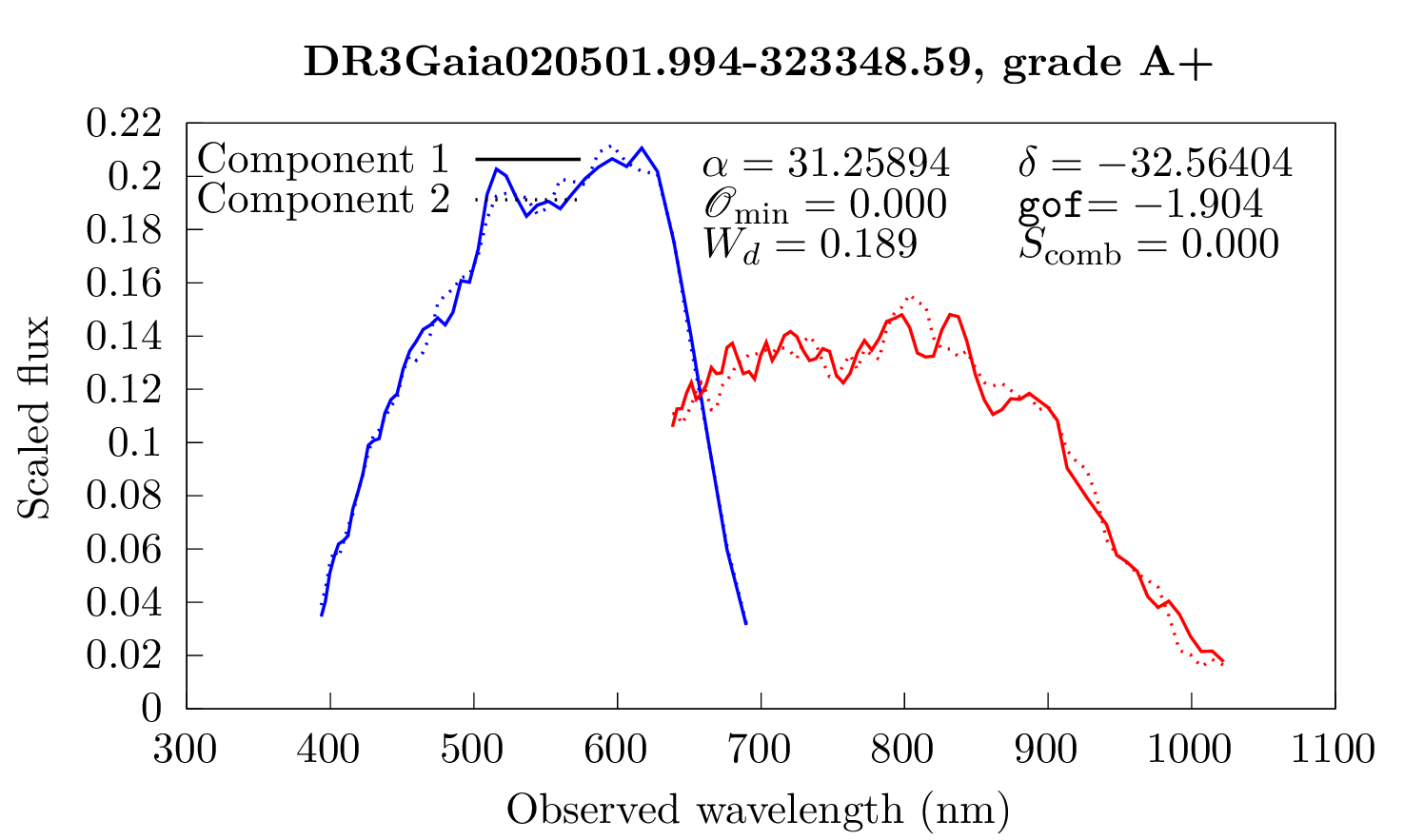}\hspace{0.5cm}\includegraphics[height=5.5cm]{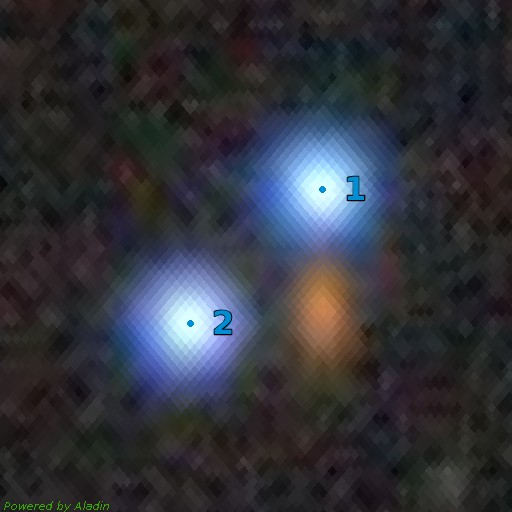}\caption{\label{fig:DR3Gaia020501.994-323348.59} Comparison of the resampled spectra of the DR3Gaia020501.994-323348.59 multiplet (Left) and associated Dark Energy Survey image (Right) \citep{2019AJ....157..168D}. Blue dots correspond to the GravLens components. Cutout size is $15.0 \arcsec \times 15.0 \arcsec$, north is up, east is left.} \end{figure*}
\begin{figure*}\centering\includegraphics[height=6cm]{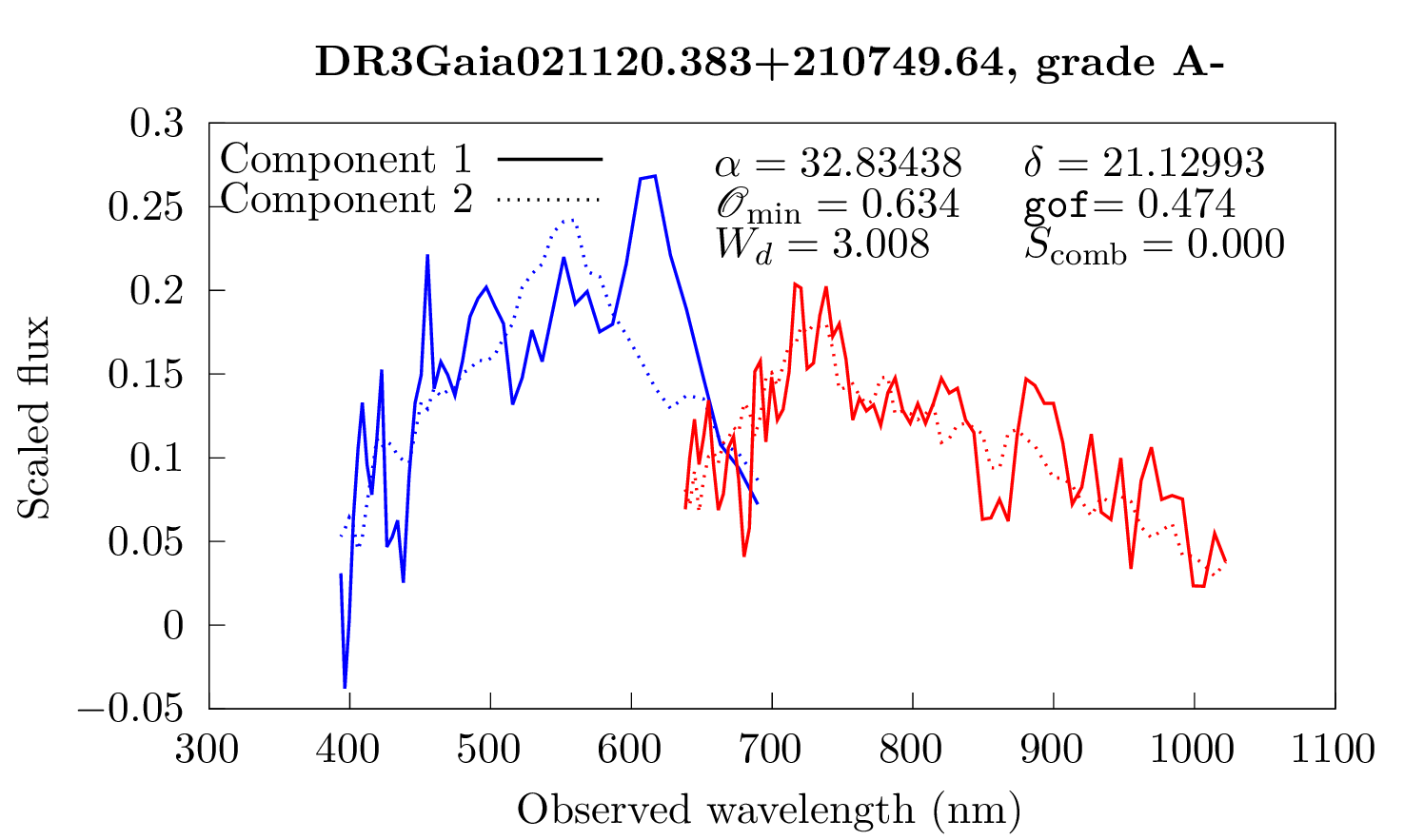}\hspace{0.5cm}\includegraphics[height=5.5cm]{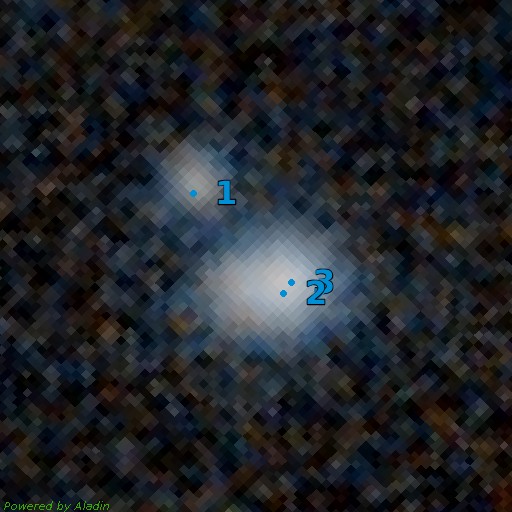}\caption{\label{fig:DR3Gaia021120.383+210749.64} Comparison of the resampled spectra of the DR3Gaia021120.383+210749.64 multiplet (Left) and associated Pan-STARRS1 image (Right) \citep{panstarrs}. Blue dots correspond to the GravLens components. Cutout size is $15.0 \arcsec \times 15.0 \arcsec$, north is up, east is left.} \end{figure*}
\begin{figure*}\centering\includegraphics[height=6cm]{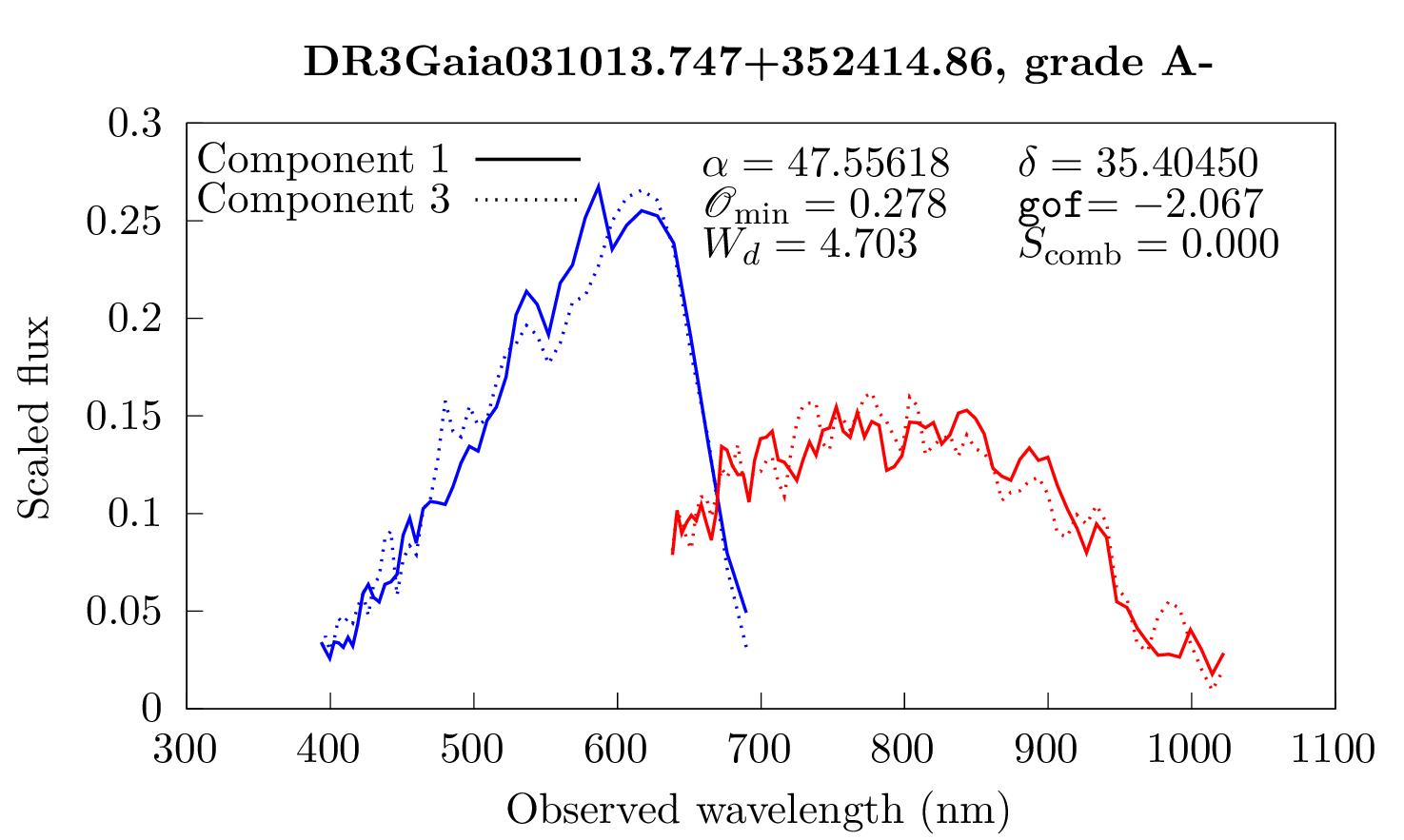}\hspace{0.5cm}\includegraphics[height=5.5cm]{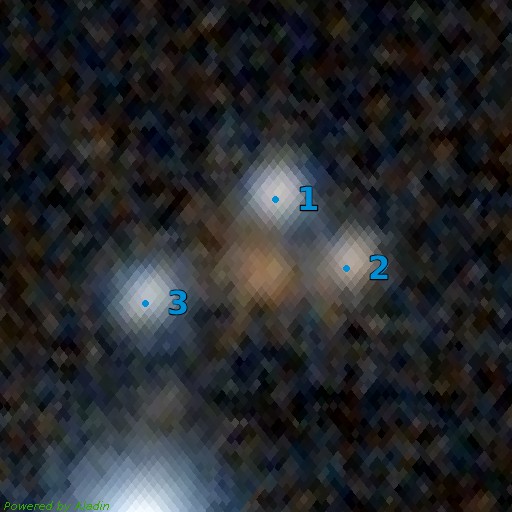}\caption{\label{fig:DR3Gaia031013.747+352414.86} Comparison of the resampled spectra of the DR3Gaia031013.747+352414.86 multiplet (Left) and associated Pan-STARRS1 image (Right) \citep{panstarrs}. Blue dots correspond to the GravLens components. Cutout size is $15.0 \arcsec \times 15.0 \arcsec$, north is up, east is left.} \end{figure*}
\begin{figure*}\centering\includegraphics[height=6cm]{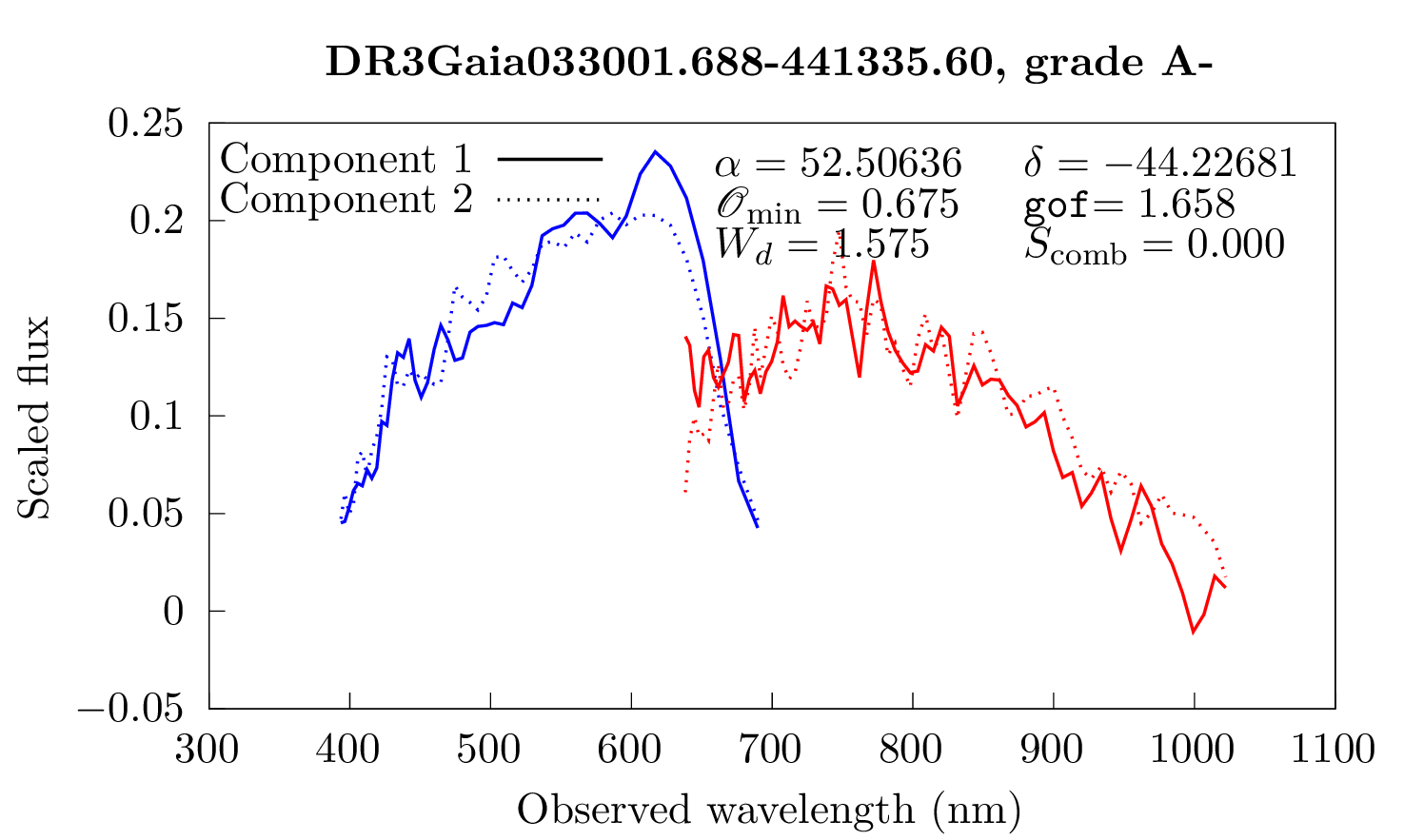}\hspace{0.5cm}\includegraphics[height=5.5cm]{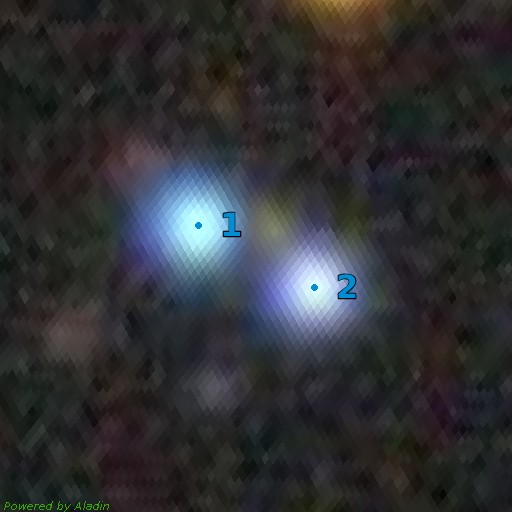}\caption{\label{fig:DR3Gaia033001.688-441335.60} Comparison of the resampled spectra of the DR3Gaia033001.688-441335.60 multiplet (Left) and associated Dark Energy Survey image (Right) \citep{2019AJ....157..168D}. Blue dots correspond to the GravLens components. Cutout size is $15.0 \arcsec \times 15.0 \arcsec$, north is up, east is left.} \end{figure*}
\begin{figure*}\centering\includegraphics[height=6cm]{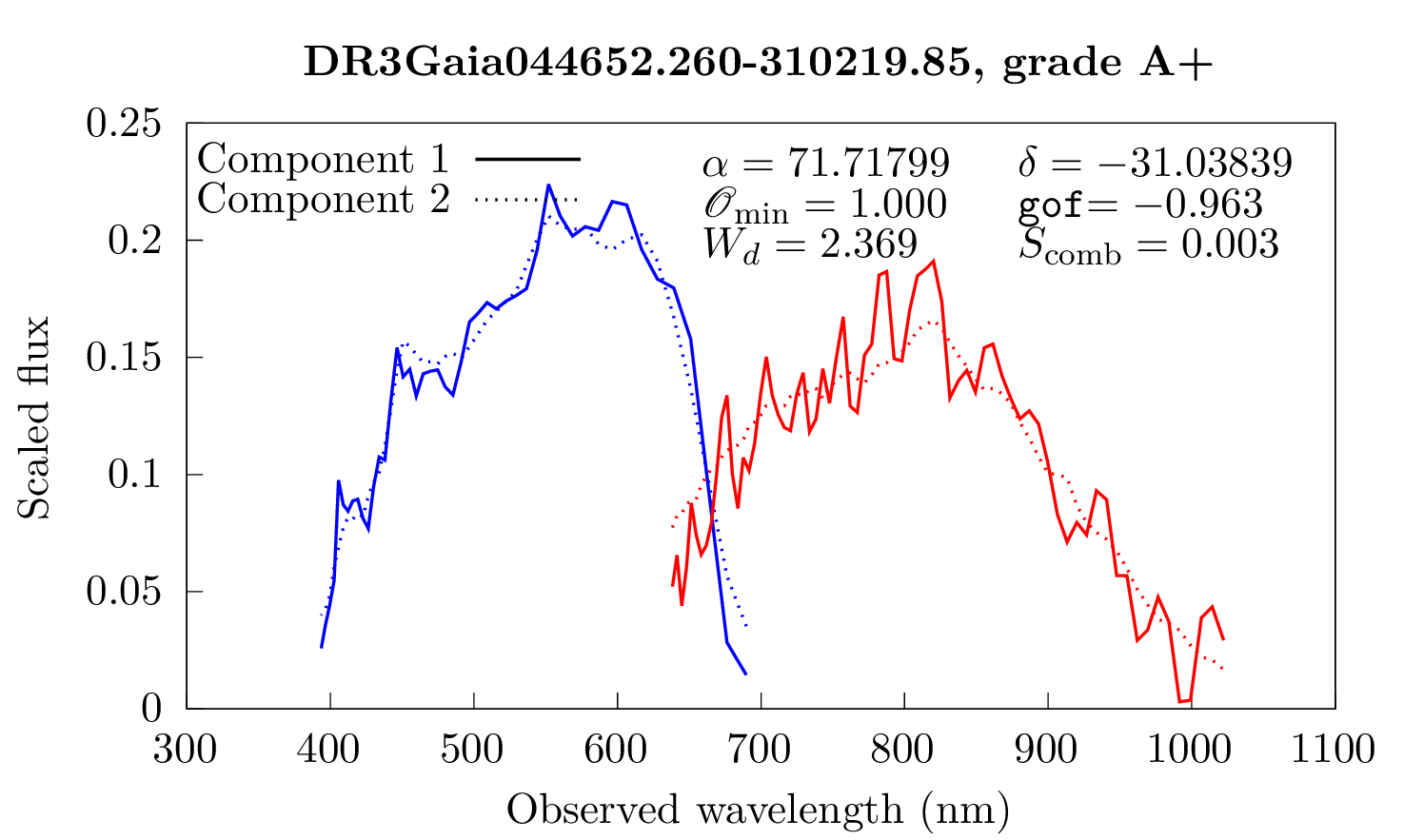}\hspace{0.5cm}\includegraphics[height=5.5cm]{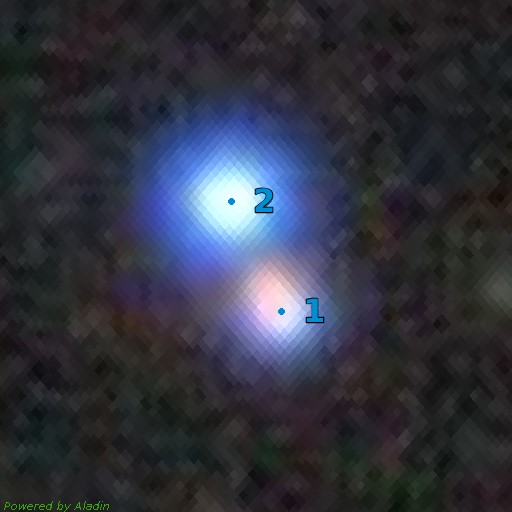}\caption{\label{fig:DR3Gaia044652.260-310219.85} Comparison of the resampled spectra of the DR3Gaia044652.260-310219.85 multiplet (Left) and associated Dark Energy Survey image (Right) \citep{2019AJ....157..168D}. Blue dots correspond to the GravLens components. Cutout size is $15.0 \arcsec \times 15.0 \arcsec$, north is up, east is left.} \end{figure*}
\begin{figure*}\centering\includegraphics[height=6cm]{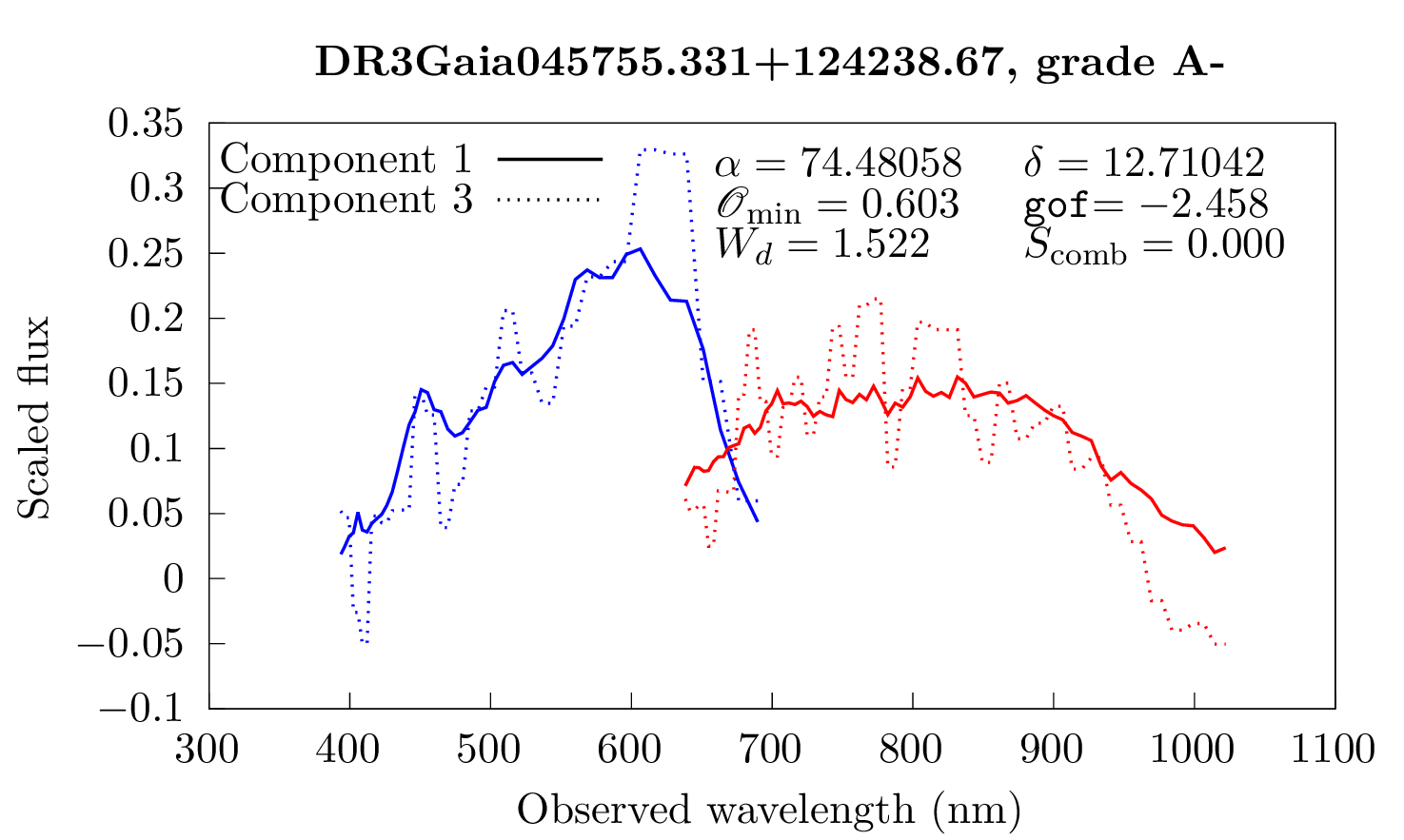}\hspace{0.5cm}\includegraphics[height=5.5cm]{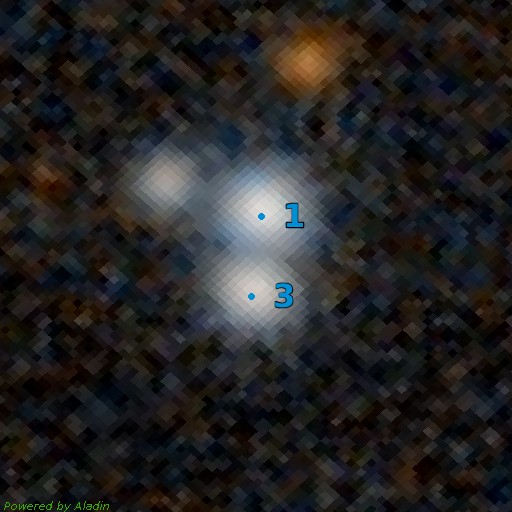}\caption{\label{fig:DR3Gaia045755.331+124238.67} Comparison of the resampled spectra of the DR3Gaia045755.331+124238.67 multiplet (Left) and associated Pan-STARRS1 image (Right) \citep{panstarrs}. Blue dots correspond to the GravLens components. Cutout size is $15.0 \arcsec \times 15.0 \arcsec$, north is up, east is left.} \end{figure*}
\begin{figure*}\centering\includegraphics[height=6cm]{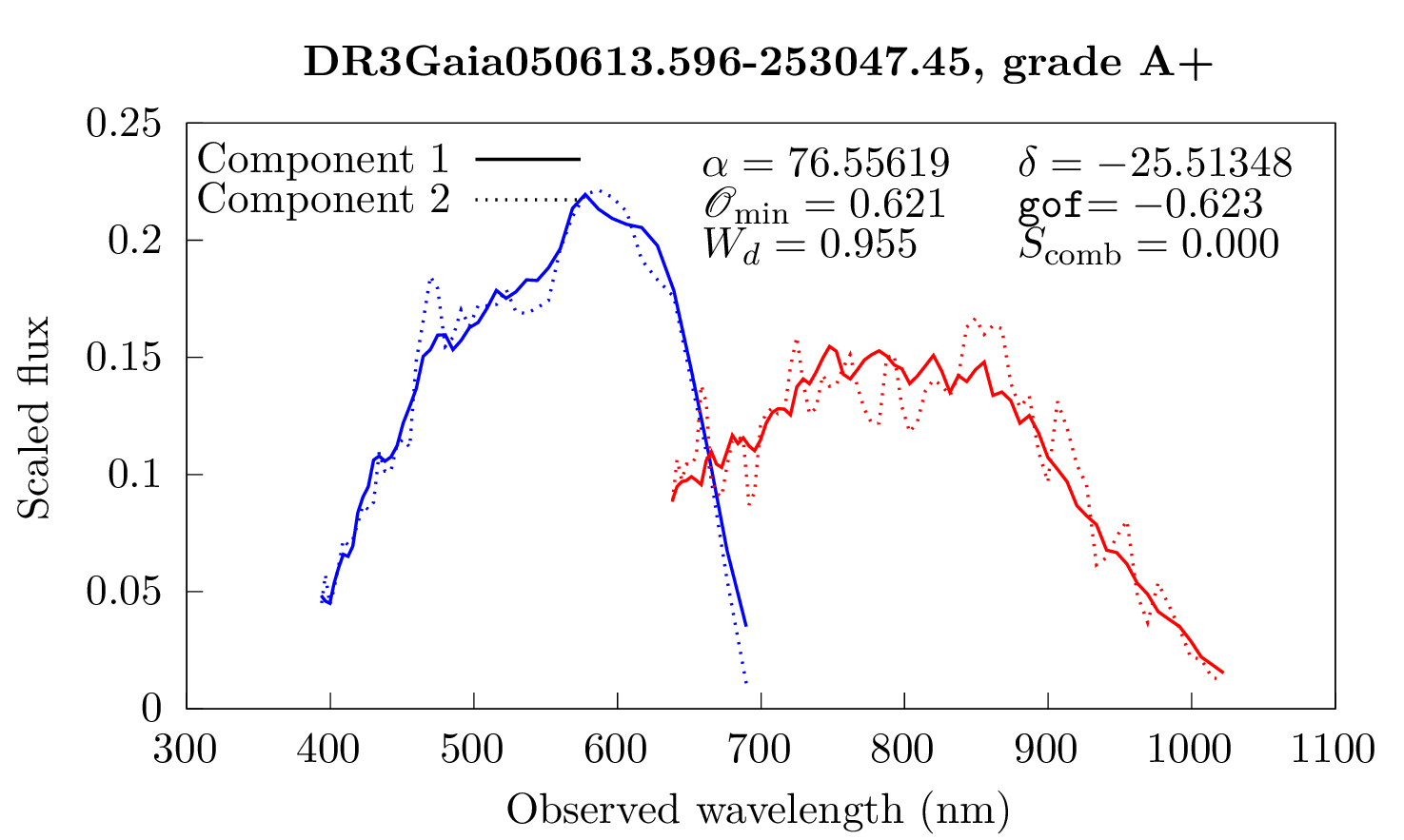}\hspace{0.5cm}\includegraphics[height=5.5cm]{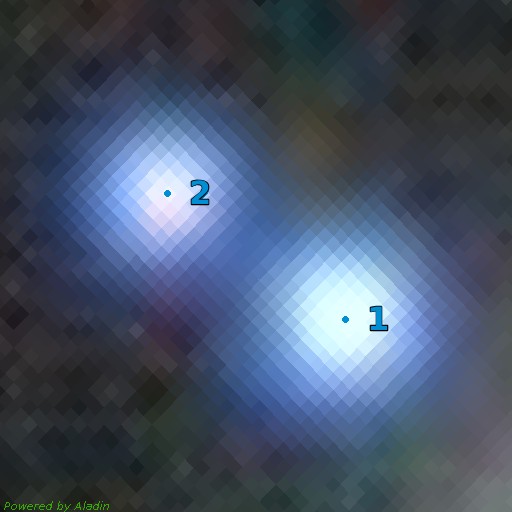}\caption{\label{fig:DR3Gaia050613.596-253047.45} Comparison of the resampled spectra of the DR3Gaia050613.596-253047.45 multiplet (Left) and associated Dark Energy Survey image (Right) \citep{2019AJ....157..168D}. Blue dots correspond to the GravLens components. Cutout size is $15.0 \arcsec \times 15.0 \arcsec$, north is up, east is left.} \end{figure*}
\begin{figure*}\centering\includegraphics[height=6cm]{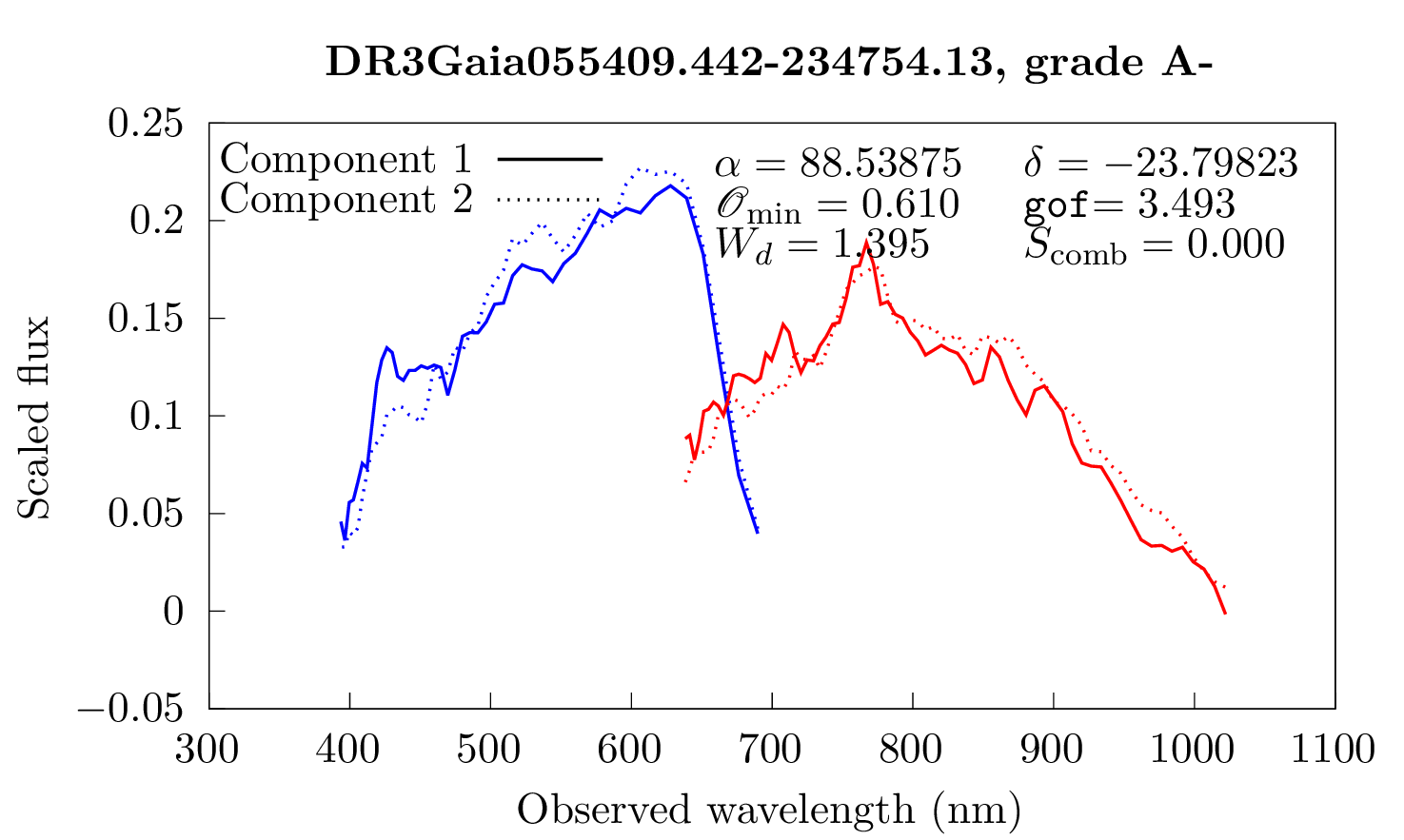}\hspace{0.5cm}\includegraphics[height=5.5cm]{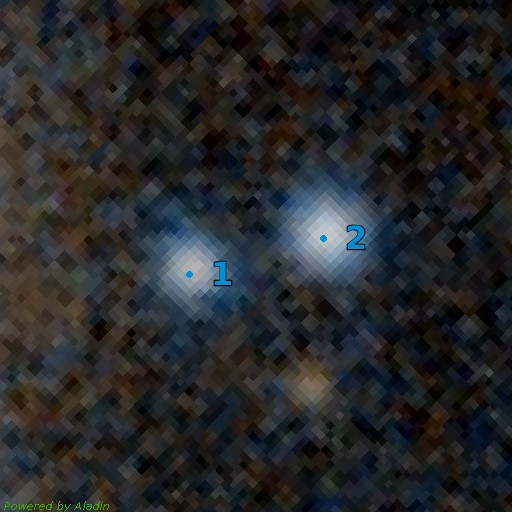}\caption{\label{fig:DR3Gaia055409.442-234754.13} Comparison of the resampled spectra of the DR3Gaia055409.442-234754.13 multiplet (Left) and associated Pan-STARRS1 image (Right) \citep{panstarrs}. Blue dots correspond to the GravLens components. Cutout size is $15.0 \arcsec \times 15.0 \arcsec$, north is up, east is left.} \end{figure*}
\begin{figure*}\centering\includegraphics[height=6cm]{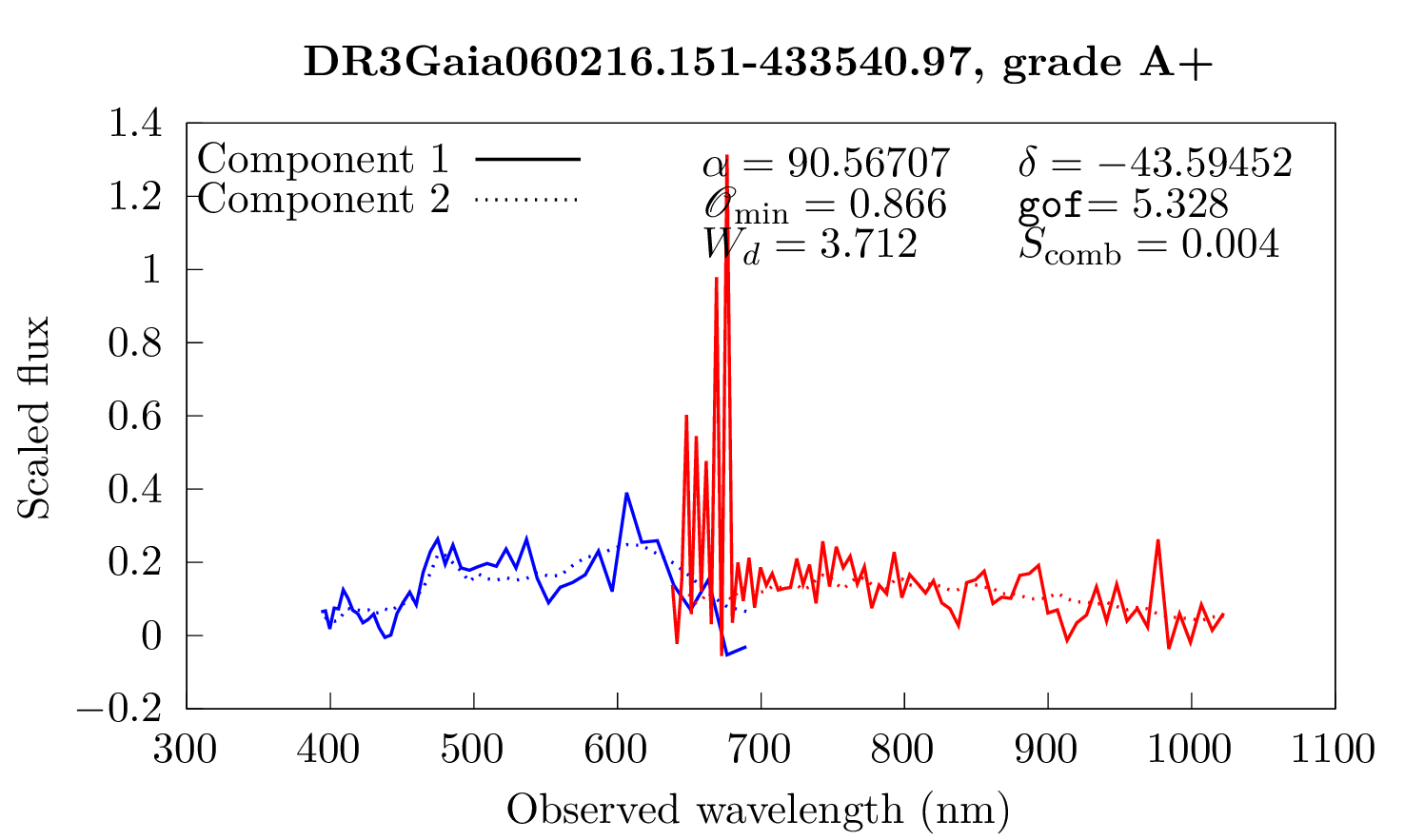}\hspace{0.5cm}\includegraphics[height=5.5cm]{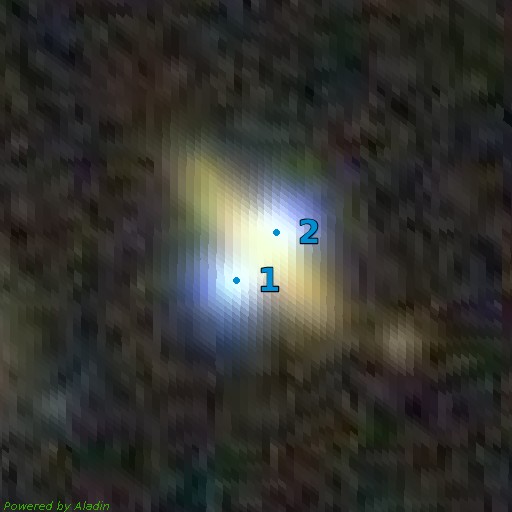}\caption{\label{fig:DR3Gaia060216.151-433540.97} Comparison of the resampled spectra of the DR3Gaia060216.151-433540.97 multiplet (Left) and associated Dark Energy Survey image (Right) \citep{2019AJ....157..168D}. Blue dots correspond to the GravLens components. Cutout size is $15.0 \arcsec \times 15.0 \arcsec$, north is up, east is left.} \end{figure*}
\begin{figure*}\centering\includegraphics[height=6cm]{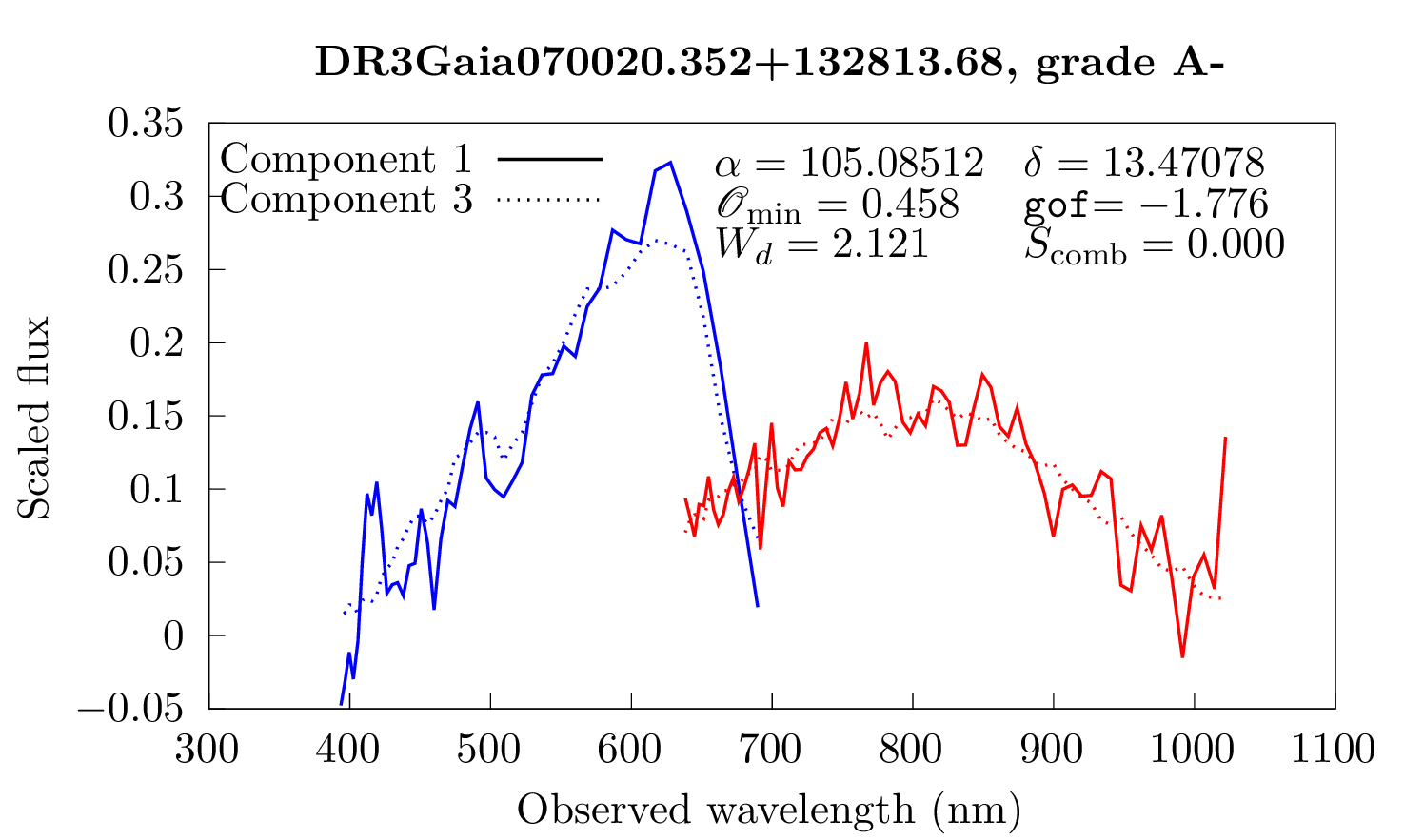}\hspace{0.5cm}\includegraphics[height=5.5cm]{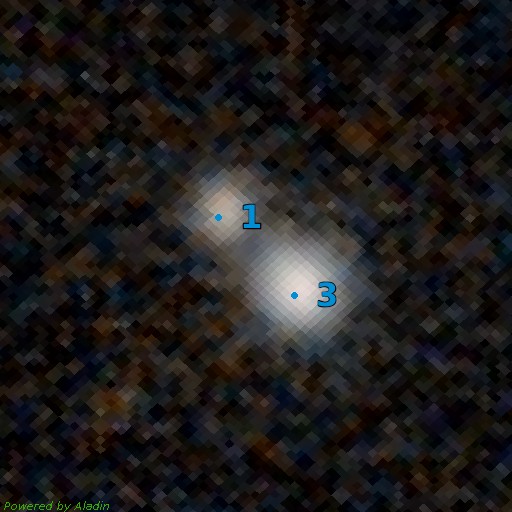}\caption{\label{fig:DR3Gaia070020.352+132813.68} Comparison of the resampled spectra of the DR3Gaia070020.352+132813.68 multiplet (Left) and associated Pan-STARRS1 image (Right) \citep{panstarrs}. Blue dots correspond to the GravLens components. Cutout size is $15.0 \arcsec \times 15.0 \arcsec$, north is up, east is left.} \end{figure*}
\begin{figure*}\centering\includegraphics[height=6cm]{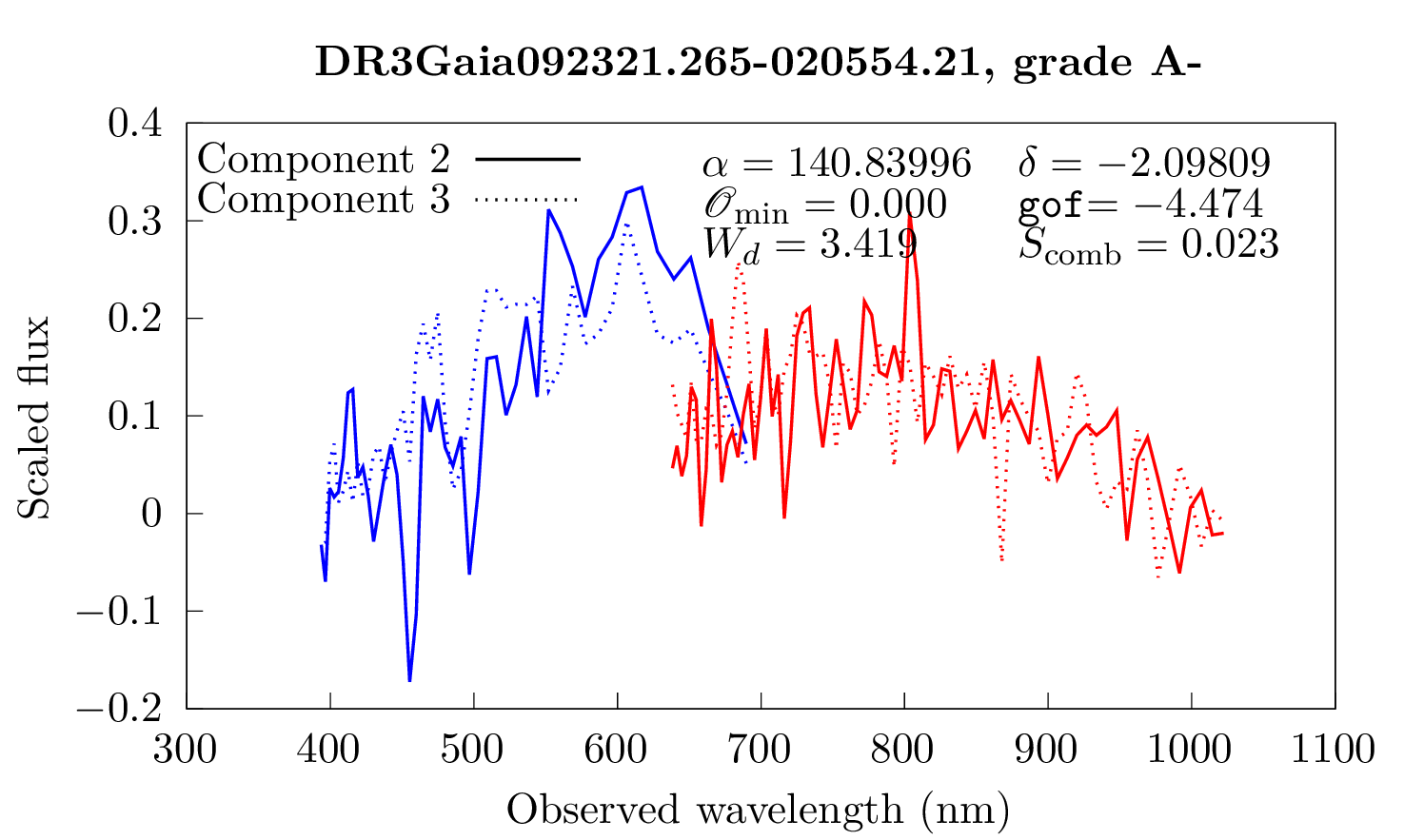}\hspace{0.5cm}\includegraphics[height=5.5cm]{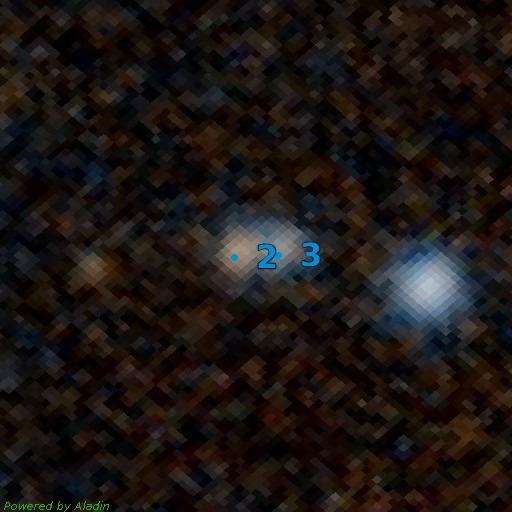}\caption{\label{fig:DR3Gaia092321.265-020554.21} Comparison of the resampled spectra of the DR3Gaia092321.265-020554.21 multiplet (Left) and associated Pan-STARRS1 image (Right) \citep{panstarrs}. Blue dots correspond to the GravLens components. Cutout size is $15.0 \arcsec \times 15.0 \arcsec$, north is up, east is left.} \end{figure*}
\begin{figure*}\centering\includegraphics[height=6cm]{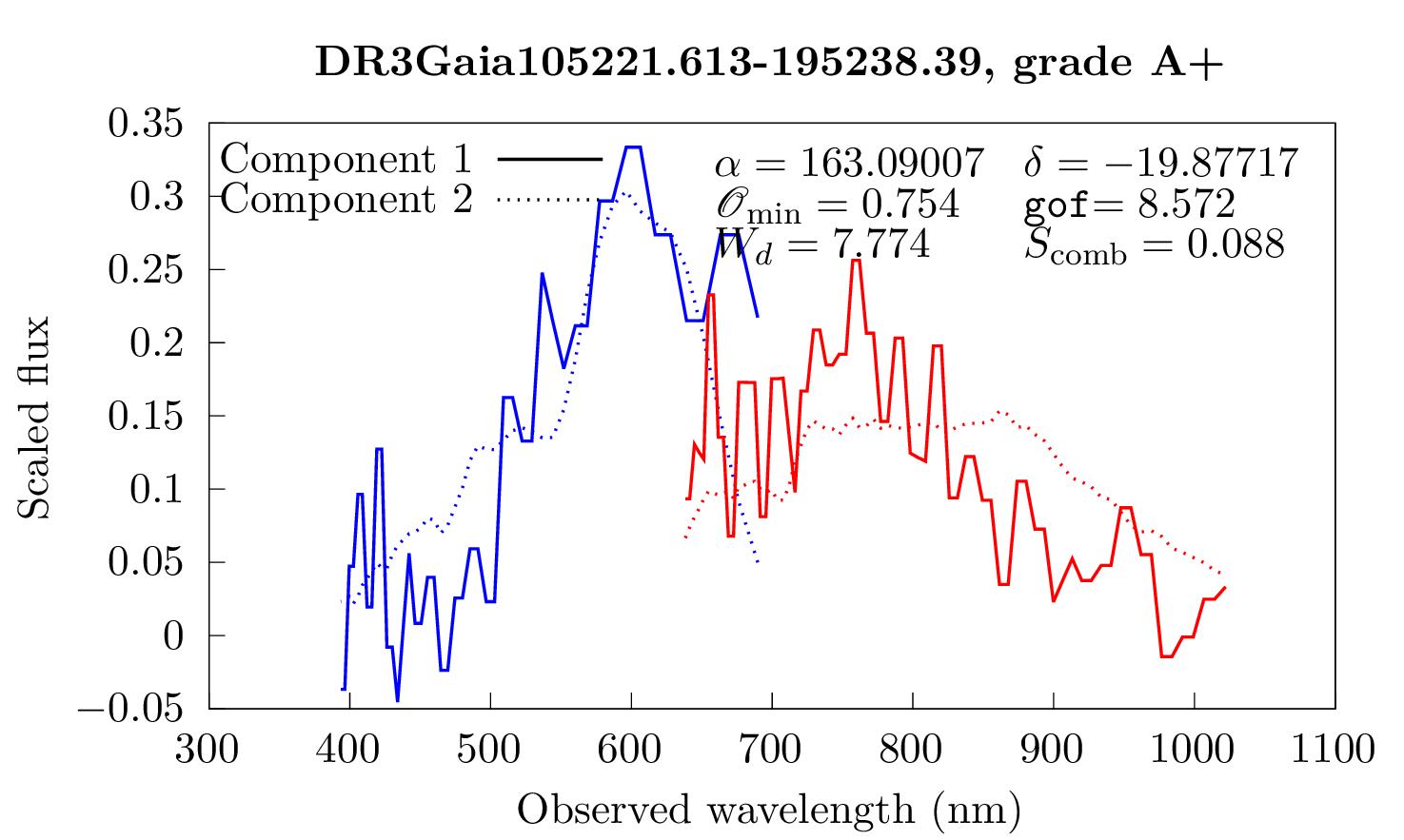}\hspace{0.5cm}\includegraphics[height=5.5cm]{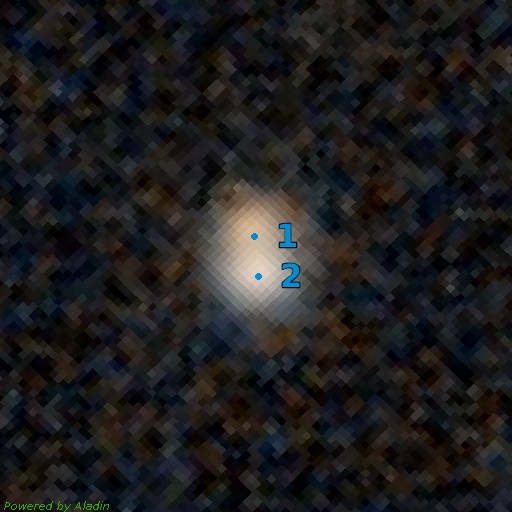}\caption{\label{fig:DR3Gaia105221.613-195238.39} Comparison of the resampled spectra of the DR3Gaia105221.613-195238.39 multiplet (Left) and associated Pan-STARRS1 image (Right) \citep{panstarrs}. Blue dots correspond to the GravLens components. Cutout size is $15.0 \arcsec \times 15.0 \arcsec$, north is up, east is left.} \end{figure*}
\begin{figure*}\centering\includegraphics[height=6cm]{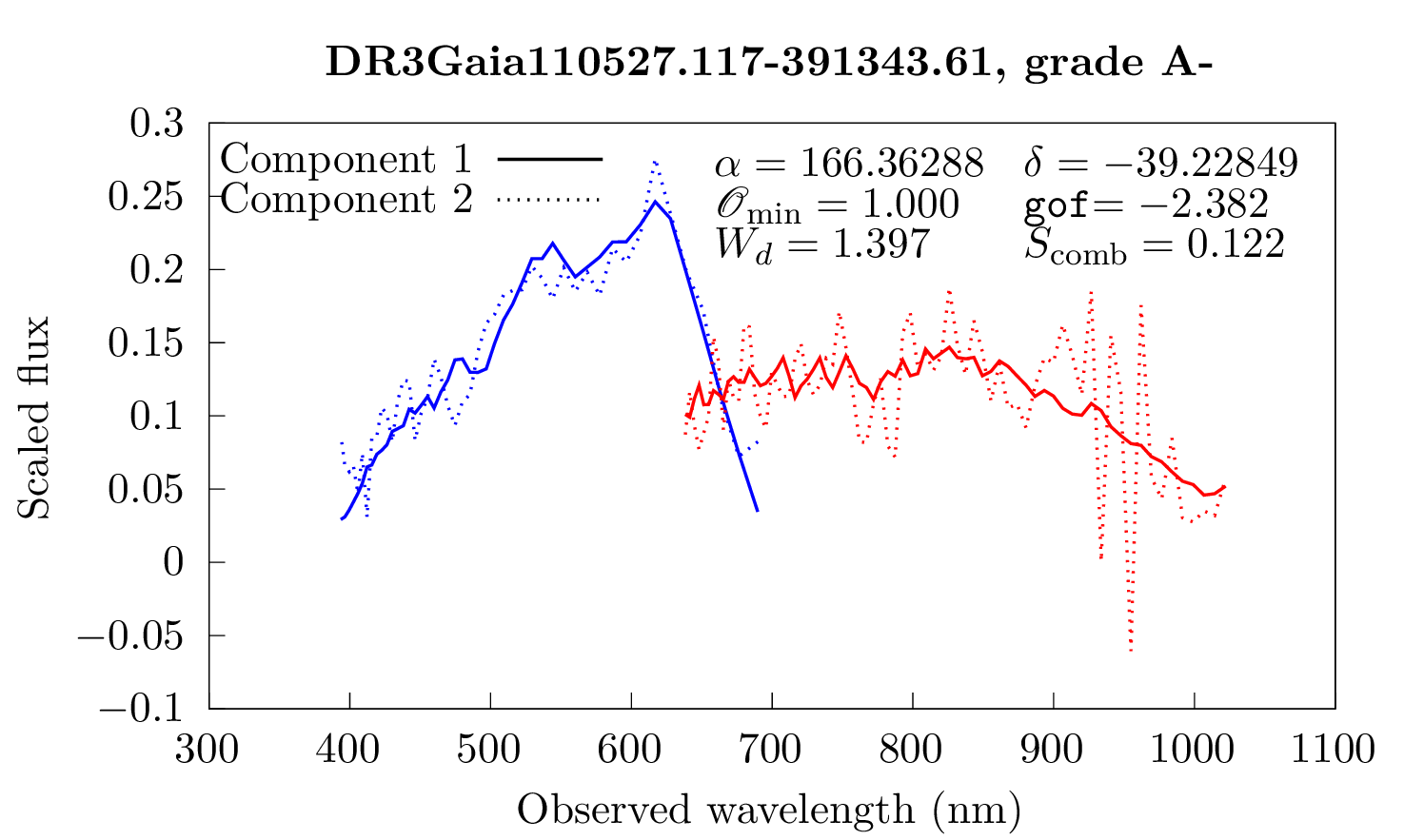}\hspace{0.5cm}\includegraphics[height=5.5cm]{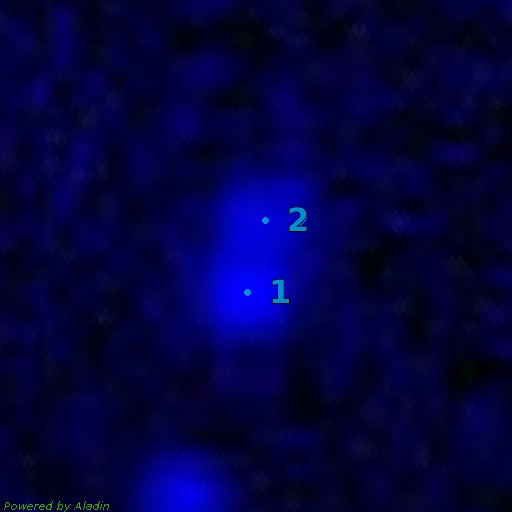}\caption{\label{fig:DR3Gaia110527.117-391343.61} Comparison of the resampled spectra of the DR3Gaia110527.117-391343.61 multiplet (Left) and associated Dark Energy Survey image (Right) \citep{2019AJ....157..168D}. Blue dots correspond to the GravLens components. Cutout size is $15.0 \arcsec \times 15.0 \arcsec$, north is up, east is left.} \end{figure*}
\begin{figure*}\centering\includegraphics[height=6cm]{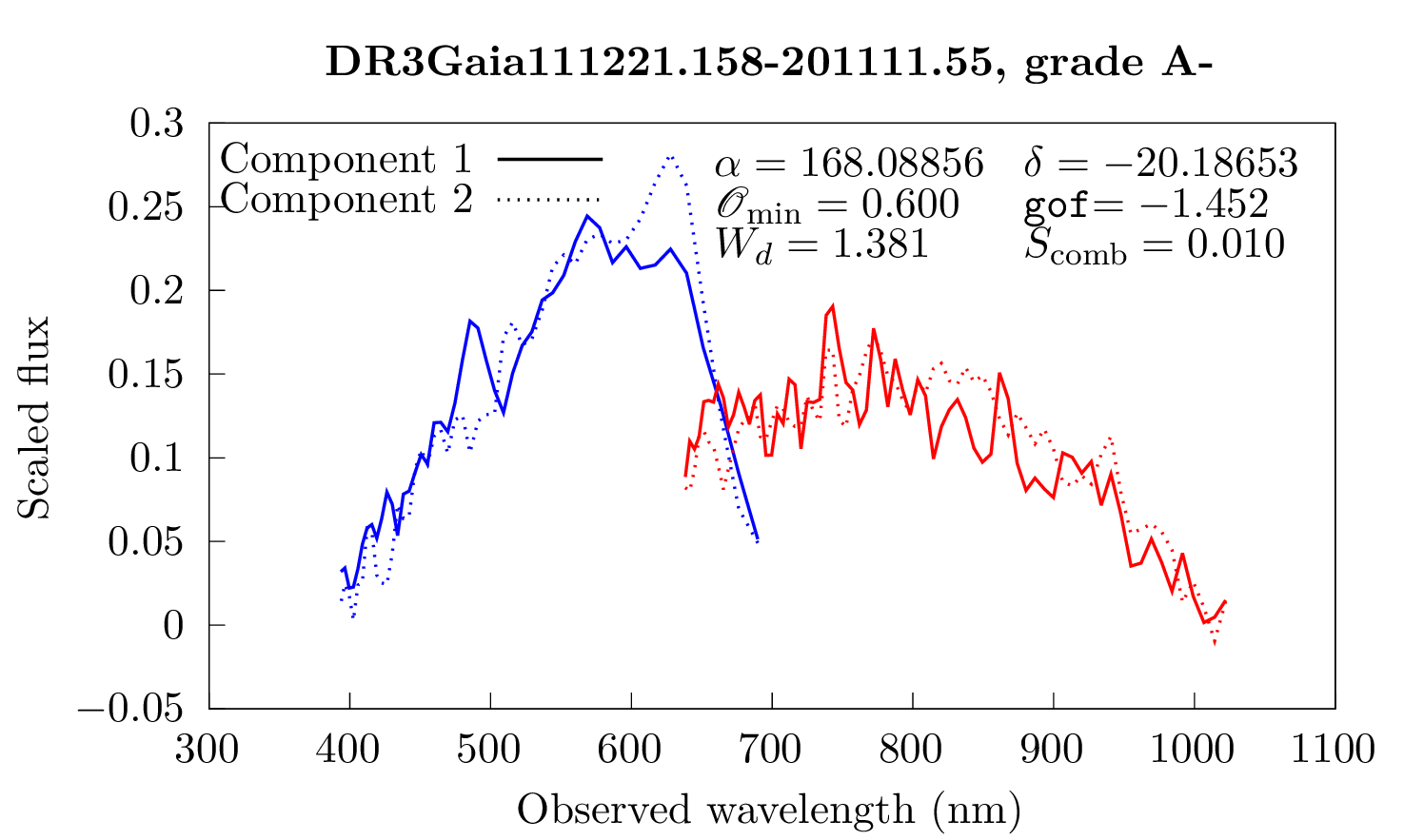}\hspace{0.5cm}\includegraphics[height=5.5cm]{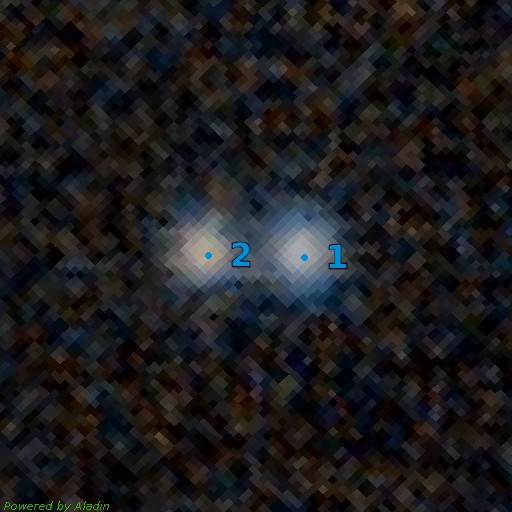}\caption{\label{fig:DR3Gaia111221.158-201111.55} Comparison of the resampled spectra of the DR3Gaia111221.158-201111.55 multiplet (Left) and associated Pan-STARRS1 image (Right) \citep{panstarrs}. Blue dots correspond to the GravLens components. Cutout size is $15.0 \arcsec \times 15.0 \arcsec$, north is up, east is left.} \end{figure*}
\begin{figure*}\centering\includegraphics[height=6cm]{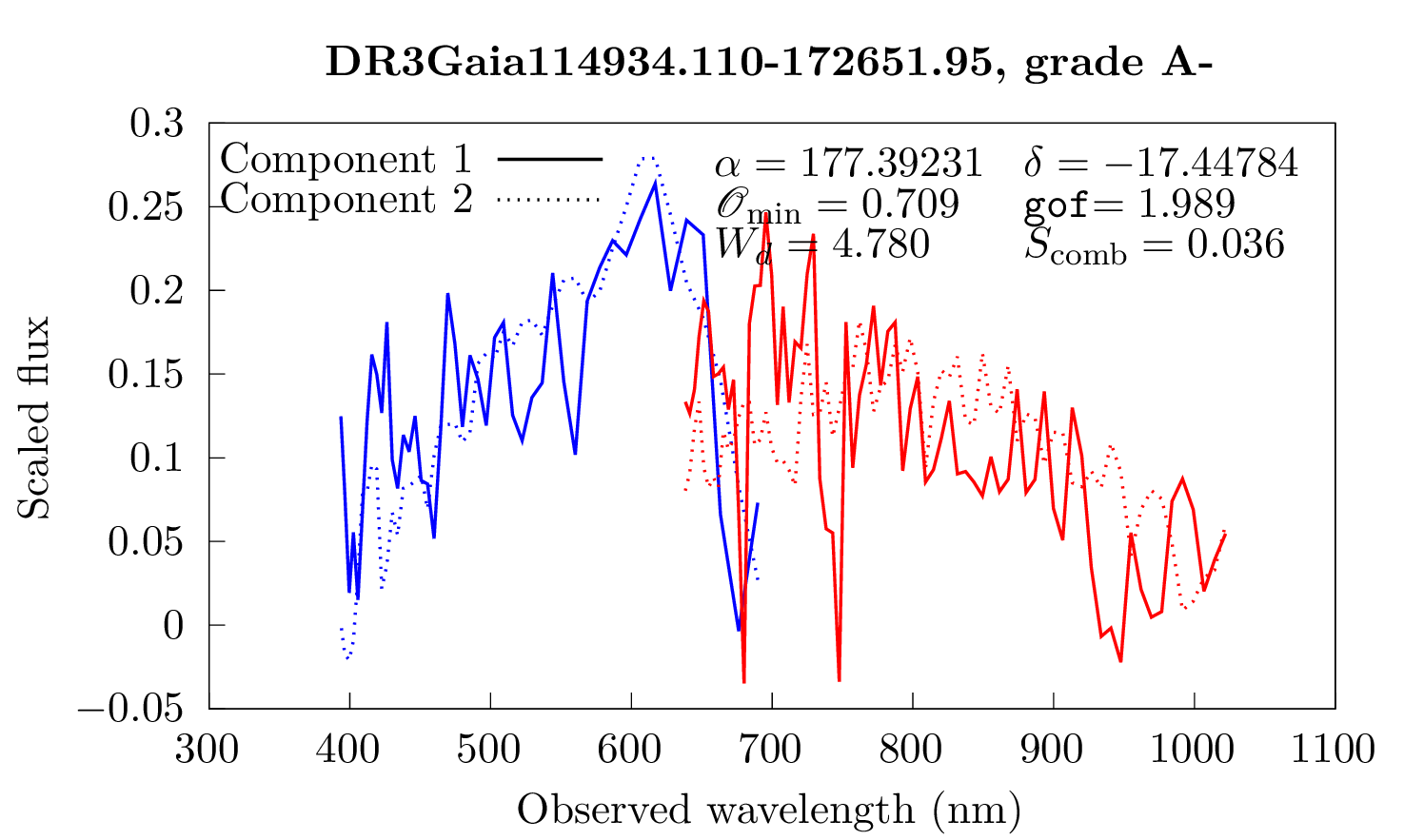}\hspace{0.5cm}\includegraphics[height=5.5cm]{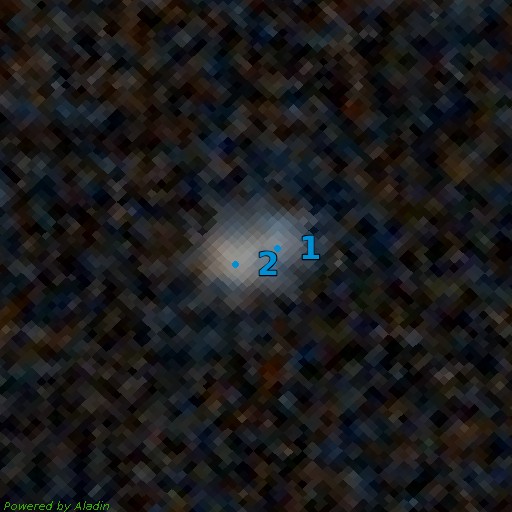}\caption{\label{fig:DR3Gaia114934.110-172651.95} Comparison of the resampled spectra of the DR3Gaia114934.110-172651.95 multiplet (Left) and associated Pan-STARRS1 image (Right) \citep{panstarrs}. Blue dots correspond to the GravLens components. Cutout size is $15.0 \arcsec \times 15.0 \arcsec$, north is up, east is left.} \end{figure*}
\begin{figure*}\centering\includegraphics[height=6cm]{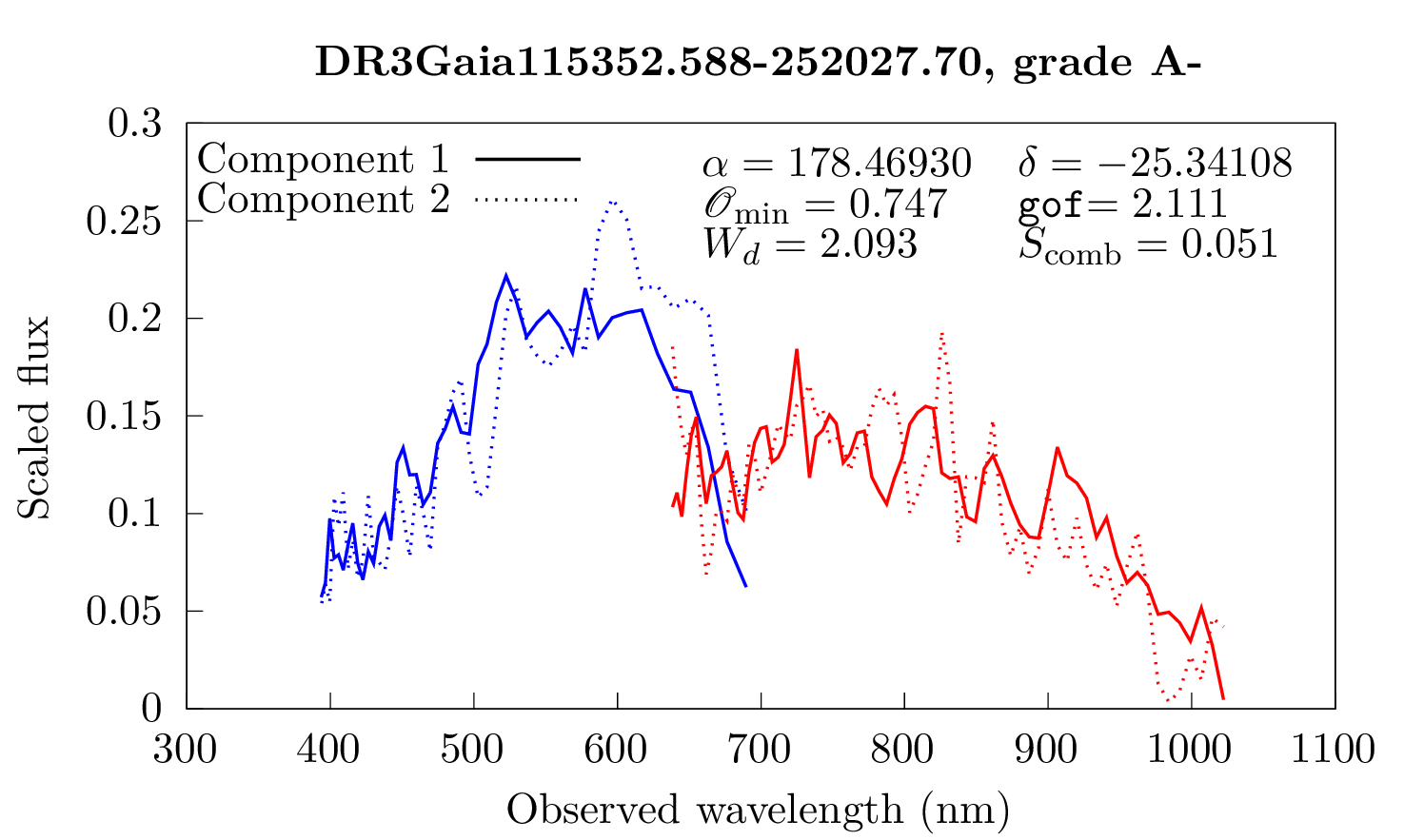}\hspace{0.5cm}\includegraphics[height=5.5cm]{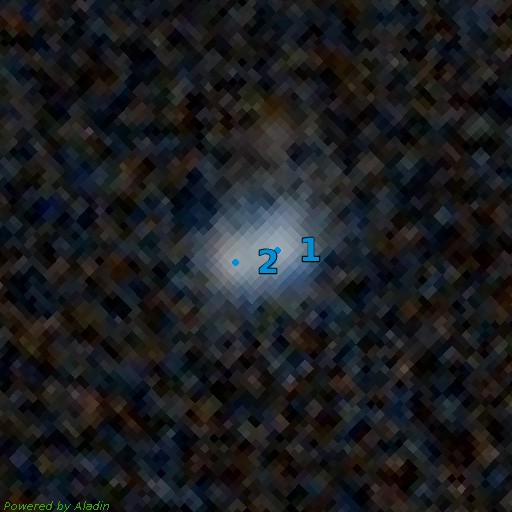}\caption{\label{fig:DR3Gaia115352.588-252027.70} Comparison of the resampled spectra of the DR3Gaia115352.588-252027.70 multiplet (Left) and associated Pan-STARRS1 image (Right) \citep{panstarrs}. Blue dots correspond to the GravLens components. Cutout size is $15.0 \arcsec \times 15.0 \arcsec$, north is up, east is left.} \end{figure*}
\begin{figure*}\centering\includegraphics[height=6cm]{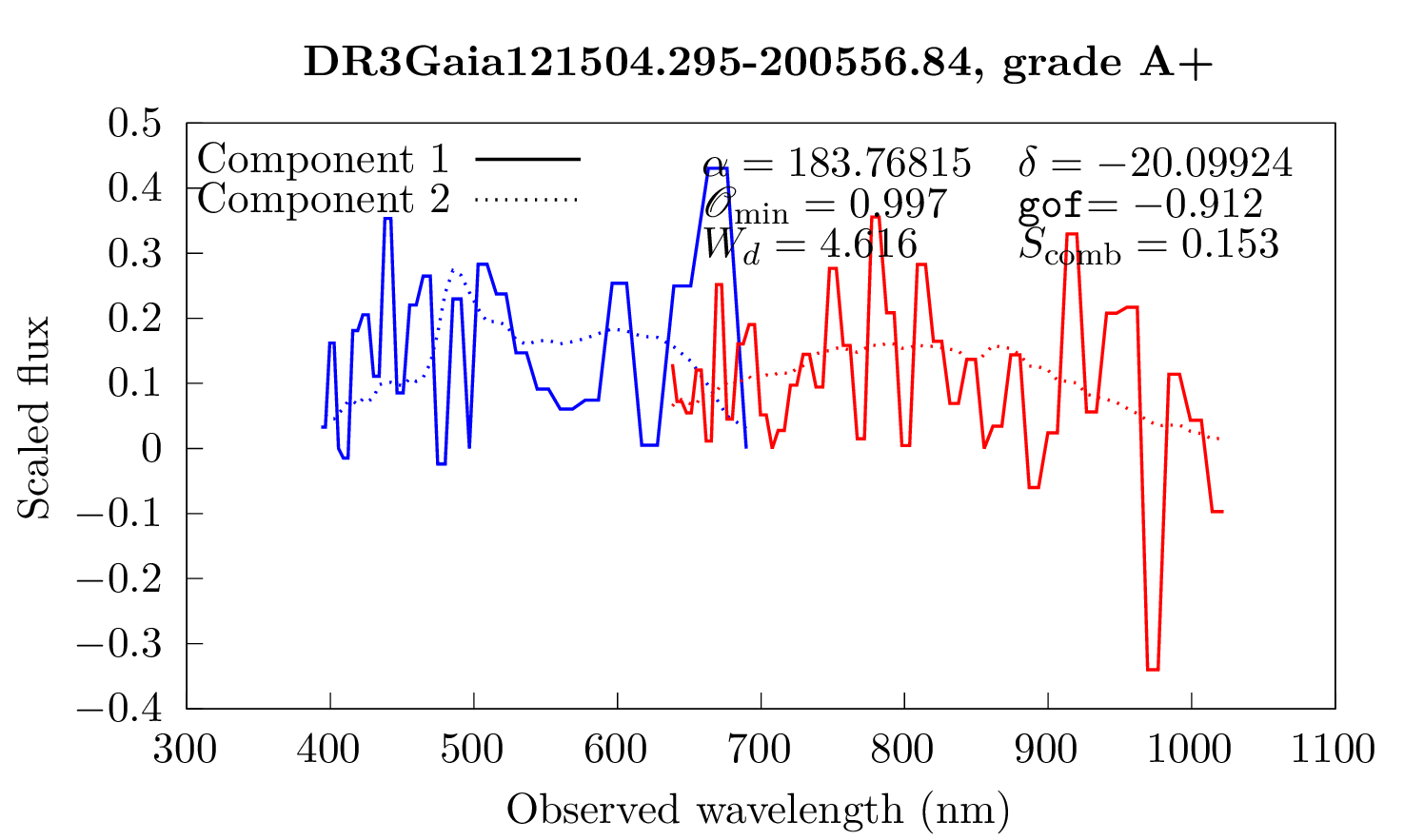}\hspace{0.5cm}\includegraphics[height=5.5cm]{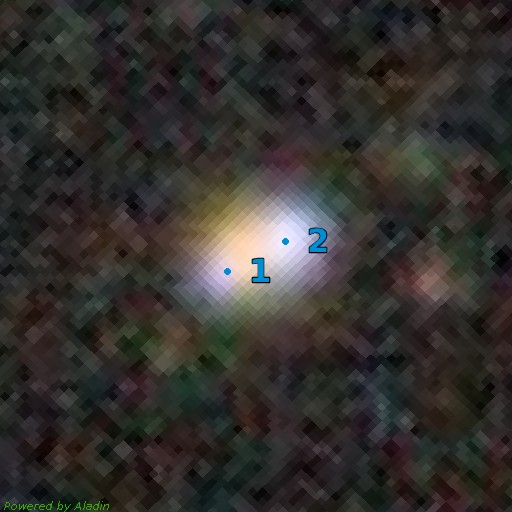}\caption{\label{fig:DR3Gaia121504.295-200556.84} Comparison of the resampled spectra of the DR3Gaia121504.295-200556.84 multiplet (Left) and associated Dark Energy Survey image (Right) \citep{2019AJ....157..168D}. Blue dots correspond to the GravLens components. Cutout size is $15.0 \arcsec \times 15.0 \arcsec$, north is up, east is left.} \end{figure*}
\begin{figure*}\centering\includegraphics[height=6cm]{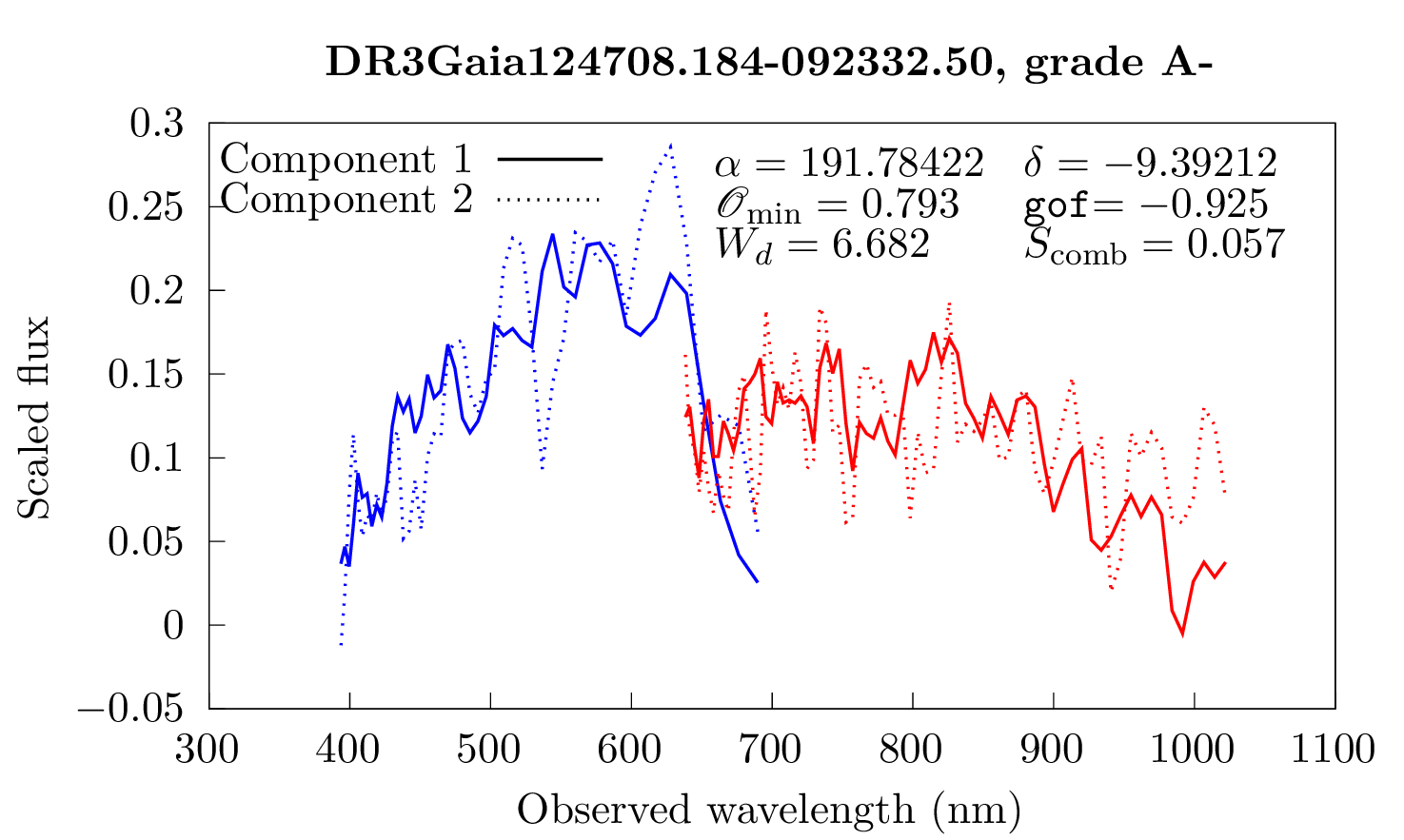}\hspace{0.5cm}\includegraphics[height=5.5cm]{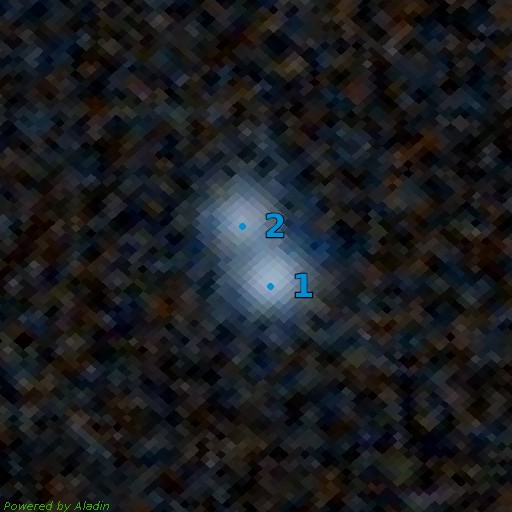}\caption{\label{fig:DR3Gaia124708.184-092332.50} Comparison of the resampled spectra of the DR3Gaia124708.184-092332.50 multiplet (Left) and associated Pan-STARRS1 image (Right) \citep{panstarrs}. Blue dots correspond to the GravLens components. Cutout size is $15.0 \arcsec \times 15.0 \arcsec$, north is up, east is left.} \end{figure*}
\begin{figure*}\centering\includegraphics[height=6cm]{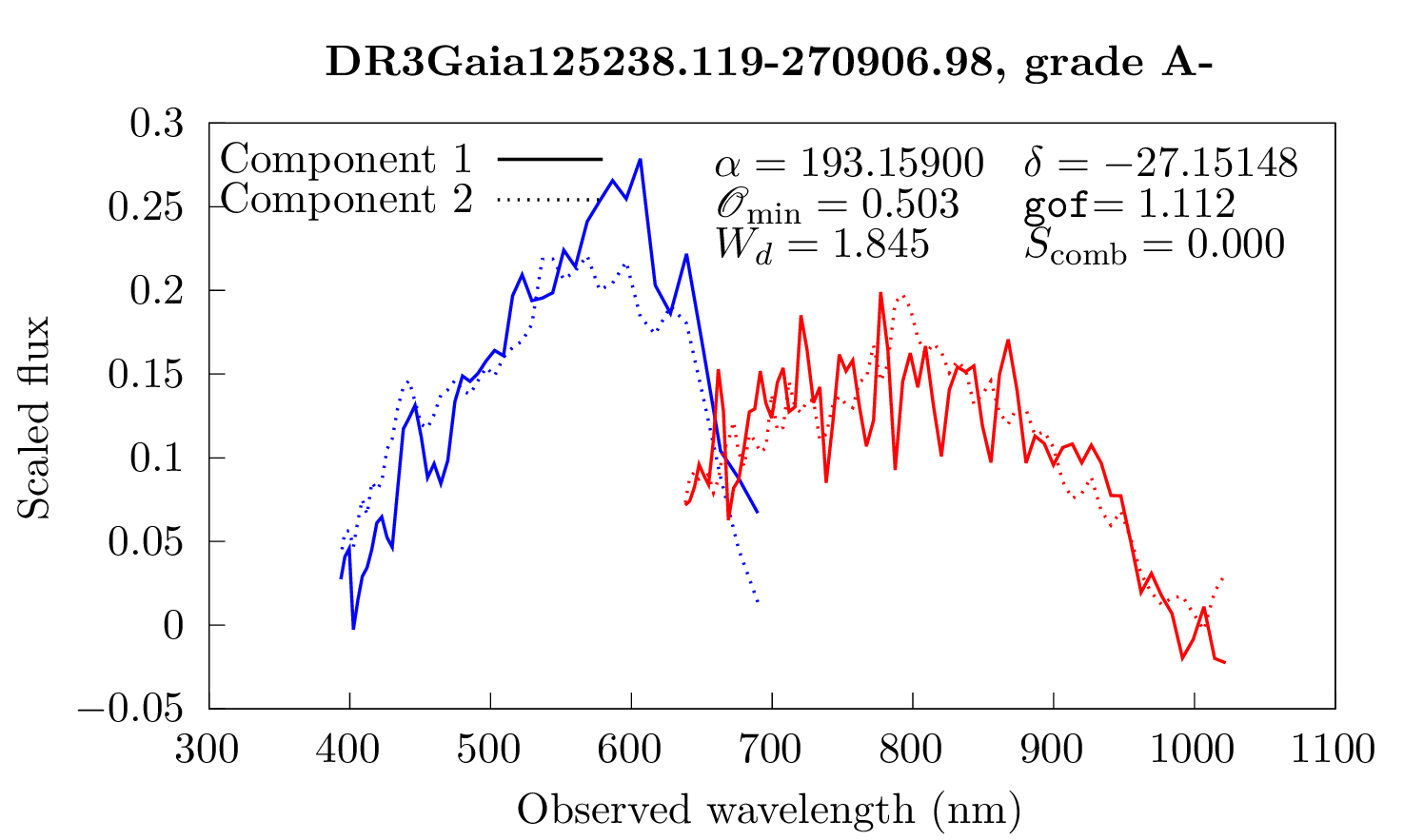}\hspace{0.5cm}\includegraphics[height=5.5cm]{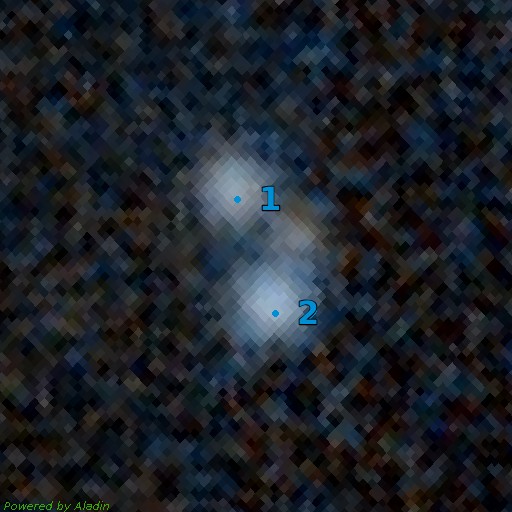}\caption{\label{fig:DR3Gaia125238.119-270906.98} Comparison of the resampled spectra of the DR3Gaia125238.119-270906.98 multiplet (Left) and associated Pan-STARRS1 image (Right) \citep{panstarrs}. Blue dots correspond to the GravLens components. Cutout size is $15.0 \arcsec \times 15.0 \arcsec$, north is up, east is left.} \end{figure*}
\begin{figure*}\centering\includegraphics[height=6cm]{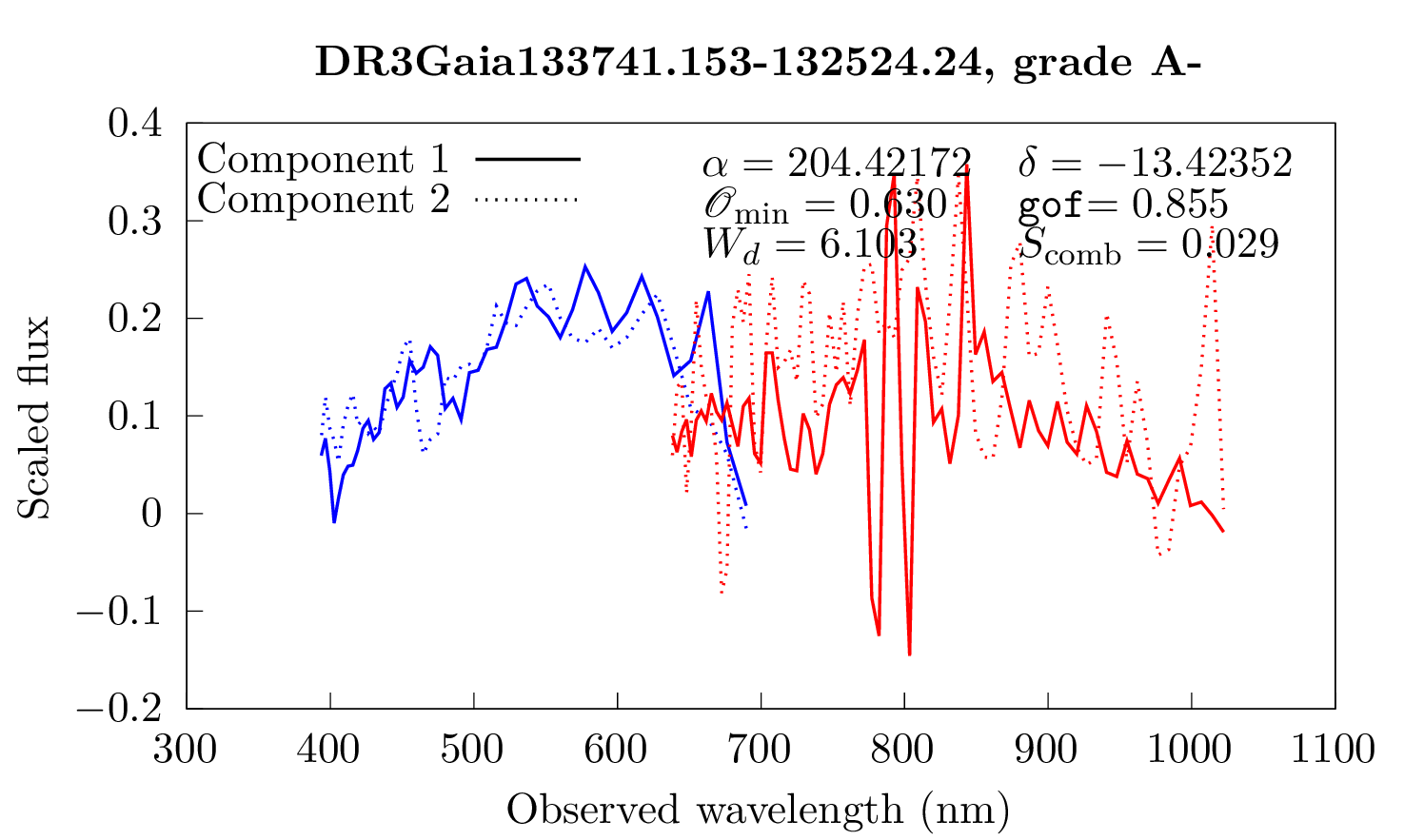}\hspace{0.5cm}\includegraphics[height=5.5cm]{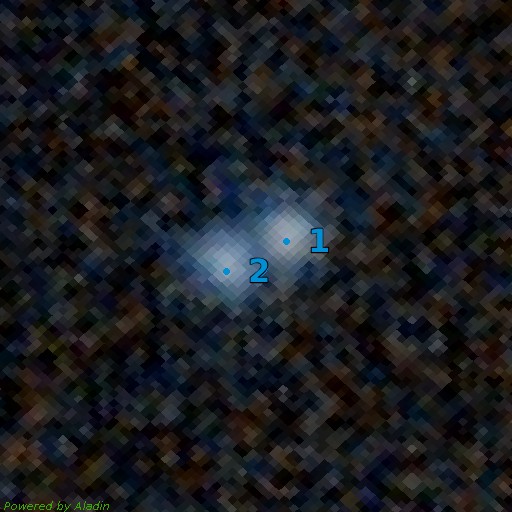}\caption{\label{fig:DR3Gaia133741.153-132524.24} Comparison of the resampled spectra of the DR3Gaia133741.153-132524.24 multiplet (Left) and associated Pan-STARRS1 image (Right) \citep{panstarrs}. Blue dots correspond to the GravLens components. Cutout size is $15.0 \arcsec \times 15.0 \arcsec$, north is up, east is left.} \end{figure*}
\begin{figure*}\centering\includegraphics[height=6cm]{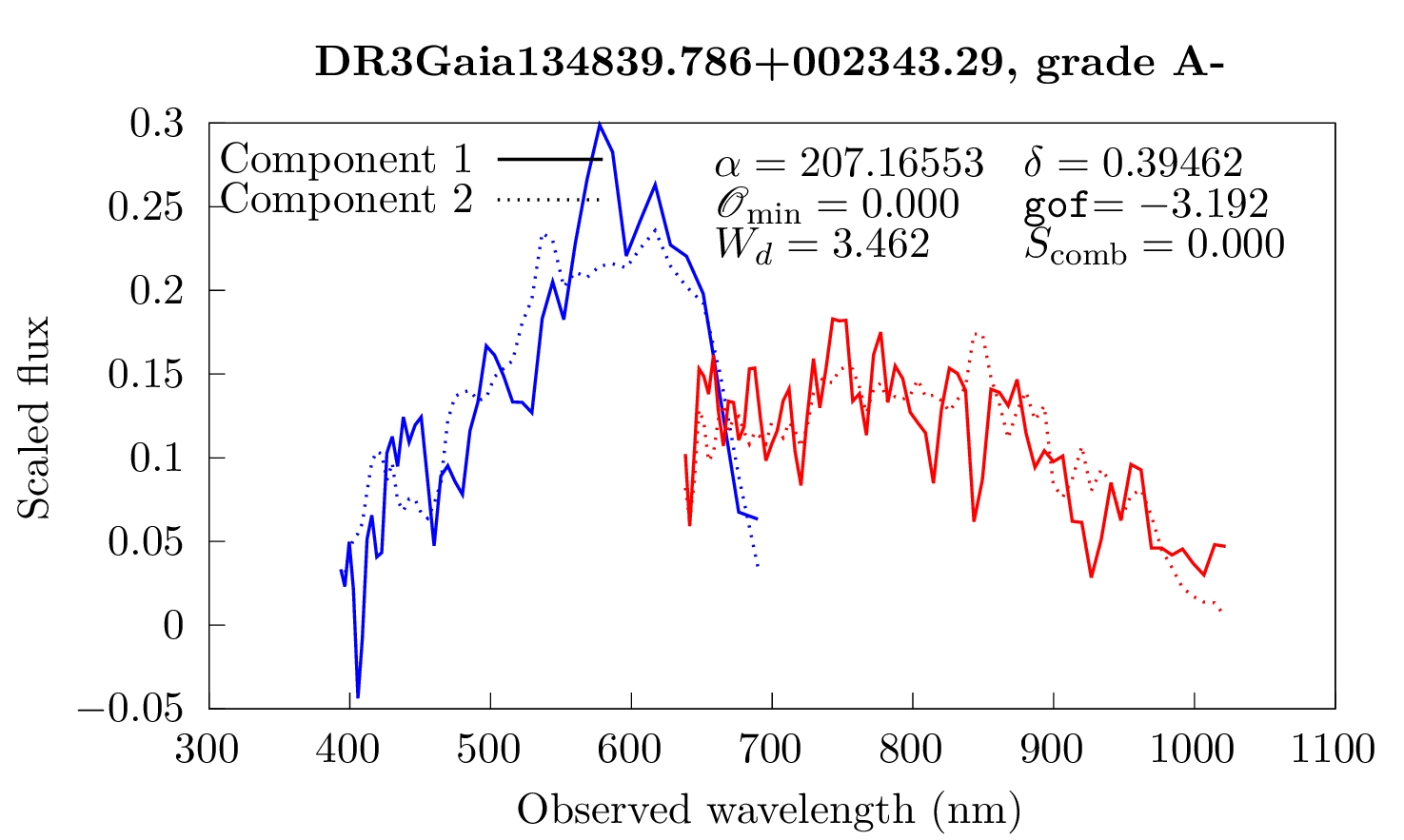}\hspace{0.5cm}\includegraphics[height=5.5cm]{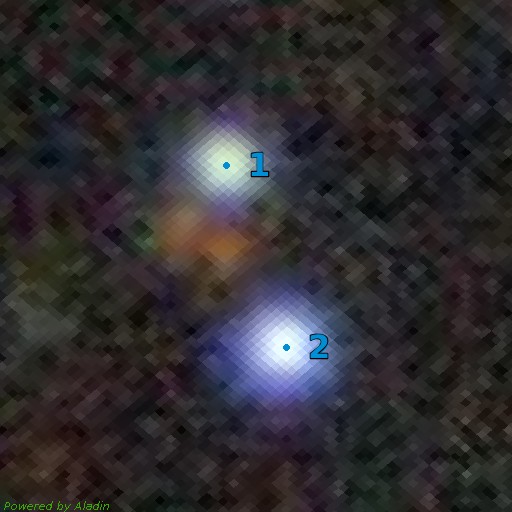}\caption{\label{fig:DR3Gaia134839.786+002343.29} Comparison of the resampled spectra of the DR3Gaia134839.786+002343.29 multiplet (Left) and associated Dark Energy Survey image (Right) \citep{2019AJ....157..168D}. Blue dots correspond to the GravLens components. Cutout size is $15.0 \arcsec \times 15.0 \arcsec$, north is up, east is left.} \end{figure*}
\begin{figure*}\centering\includegraphics[height=6cm]{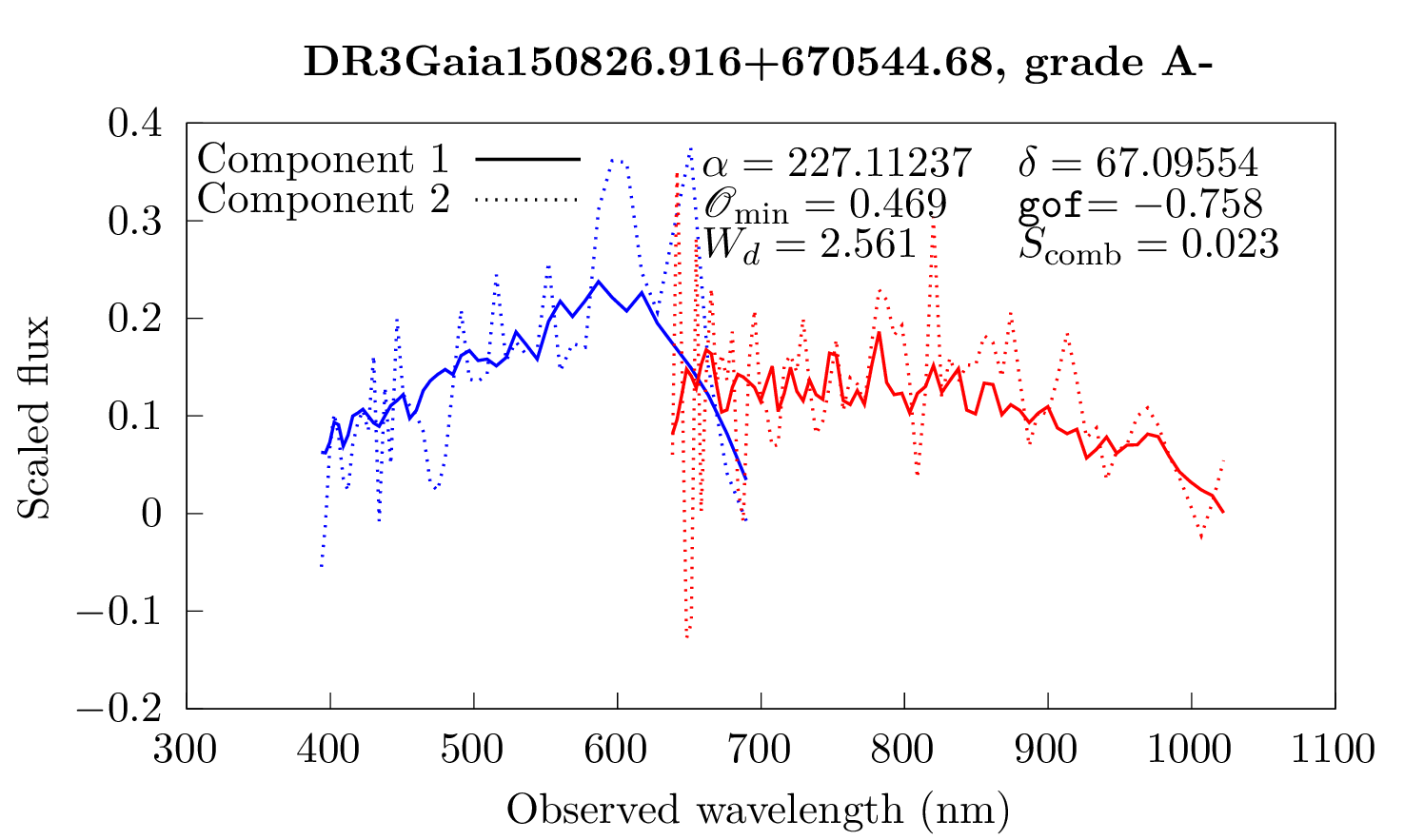}\hspace{0.5cm}\includegraphics[height=5.5cm]{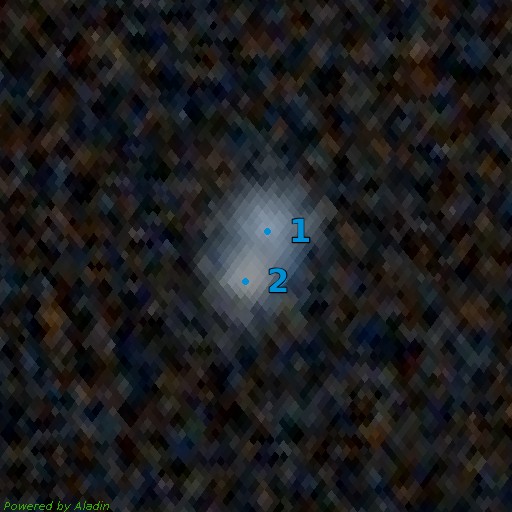}\caption{\label{fig:DR3Gaia150826.916+670544.68} Comparison of the resampled spectra of the DR3Gaia150826.916+670544.68 multiplet (Left) and associated Pan-STARRS1 image (Right) \citep{panstarrs}. Blue dots correspond to the GravLens components. Cutout size is $15.0 \arcsec \times 15.0 \arcsec$, north is up, east is left.} \end{figure*}
\begin{figure*}\centering\includegraphics[height=6cm]{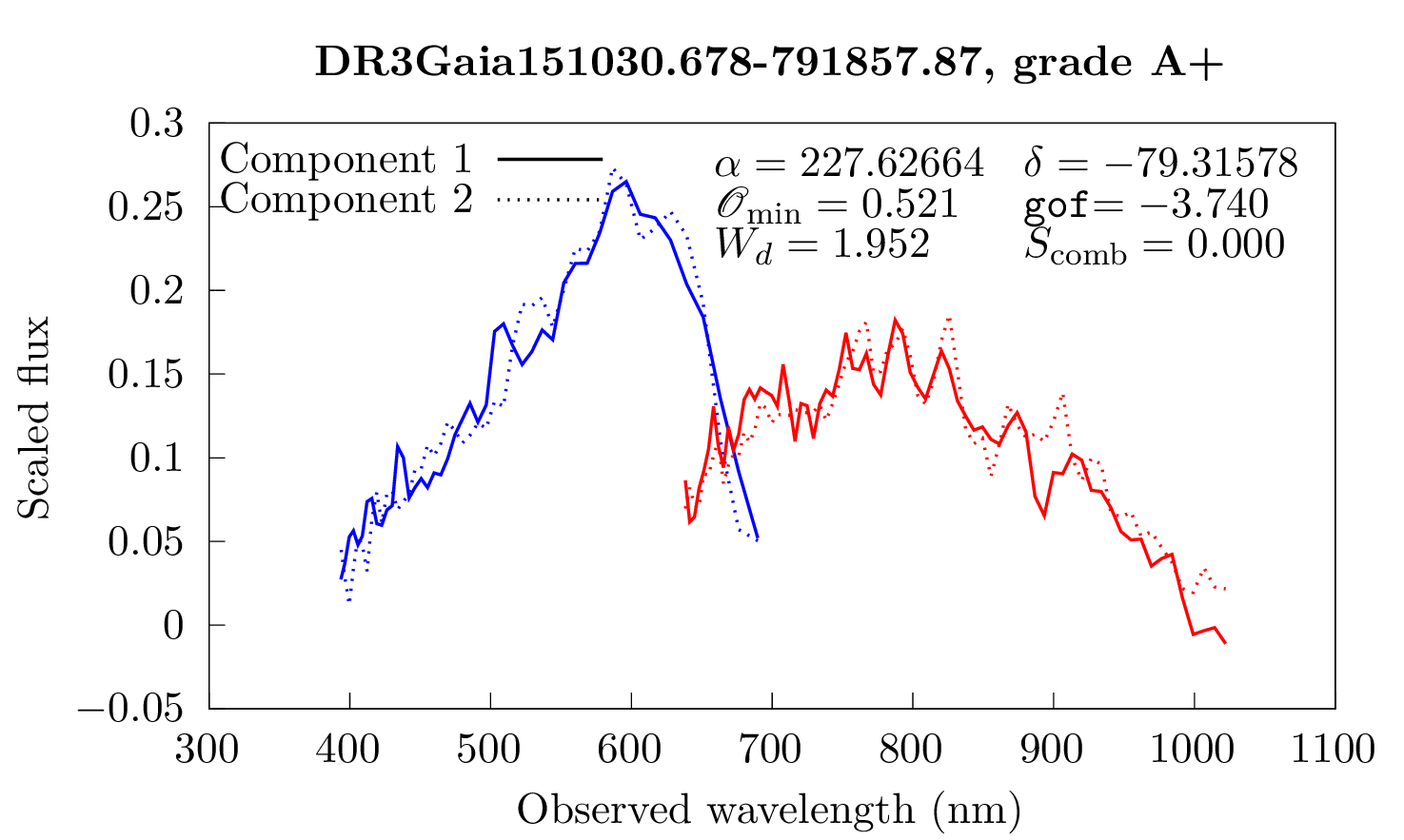}\hspace{0.5cm}\includegraphics[height=5.5cm]{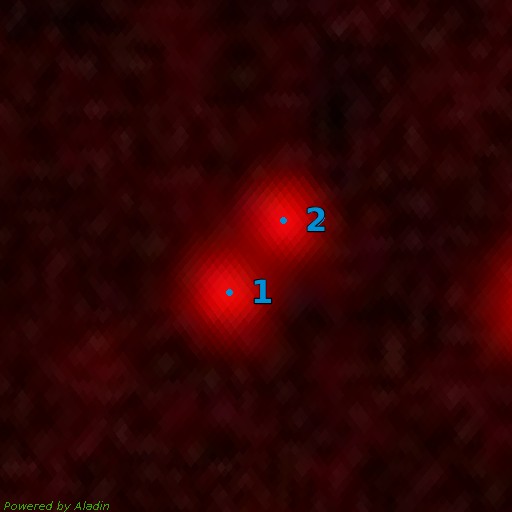}\caption{\label{fig:DR3Gaia151030.678-791857.87} Comparison of the resampled spectra of the DR3Gaia151030.678-791857.87 multiplet (Left) and associated Dark Energy Survey image (Right) \citep{2019AJ....157..168D}. Blue dots correspond to the GravLens components. Cutout size is $15.0 \arcsec \times 15.0 \arcsec$, north is up, east is left.} \end{figure*}
\begin{figure*}\centering\includegraphics[height=6cm]{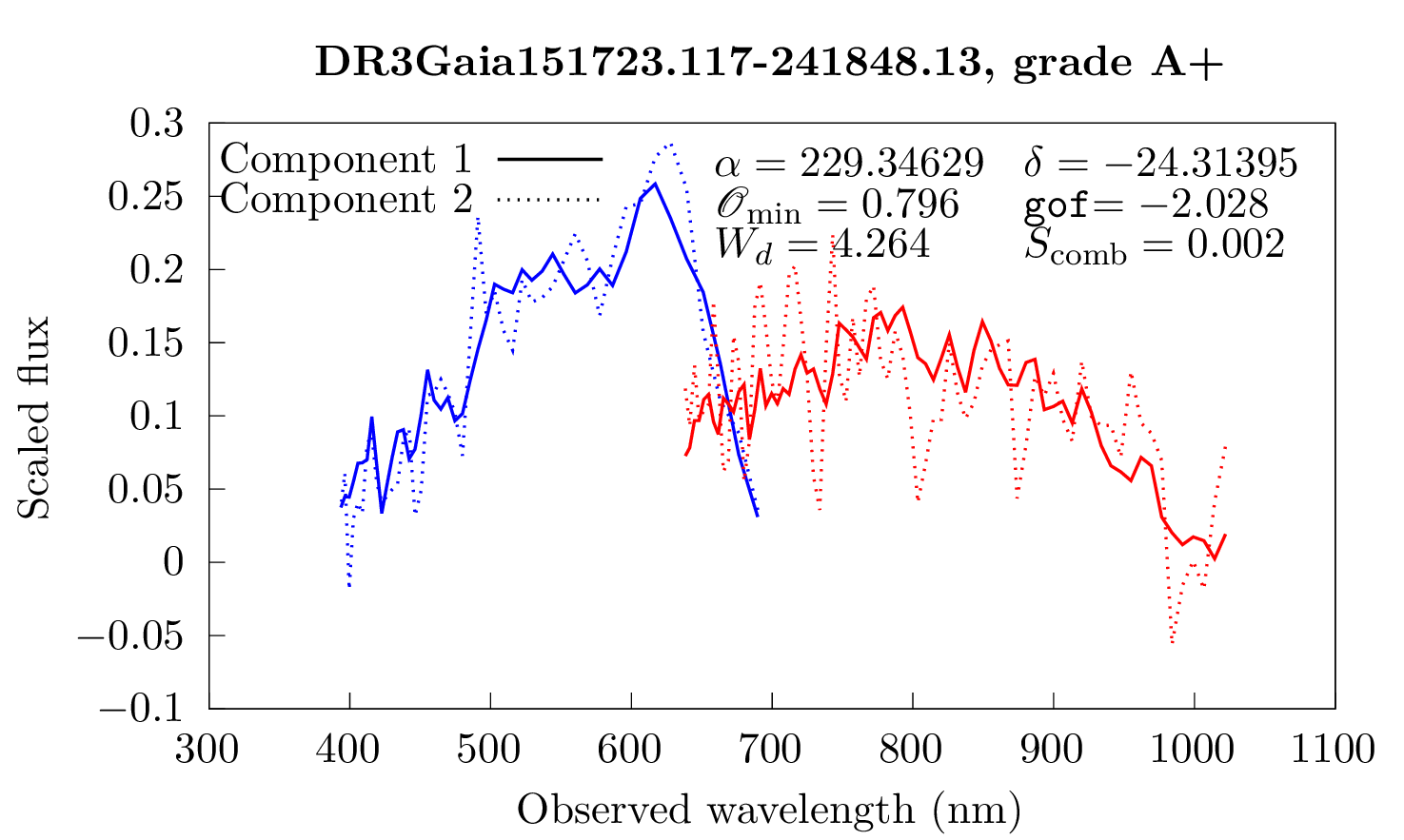}\hspace{0.5cm}\includegraphics[height=5.5cm]{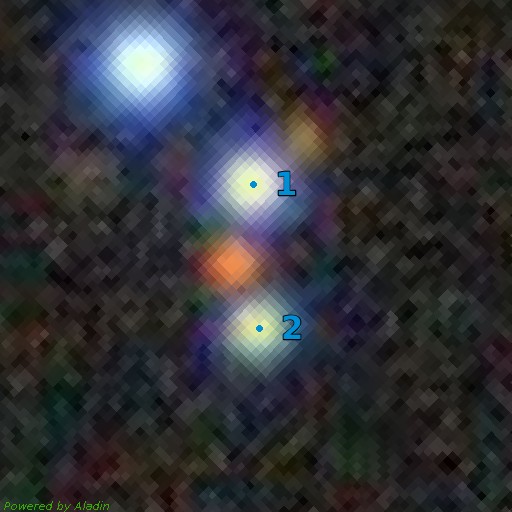}\caption{\label{fig:DR3Gaia151723.117-241848.13} Comparison of the resampled spectra of the DR3Gaia151723.117-241848.13 multiplet (Left) and associated Dark Energy Survey image (Right) \citep{2019AJ....157..168D}. Blue dots correspond to the GravLens components. Cutout size is $15.0 \arcsec \times 15.0 \arcsec$, north is up, east is left.} \end{figure*}
\begin{figure*}\centering\includegraphics[height=6cm]{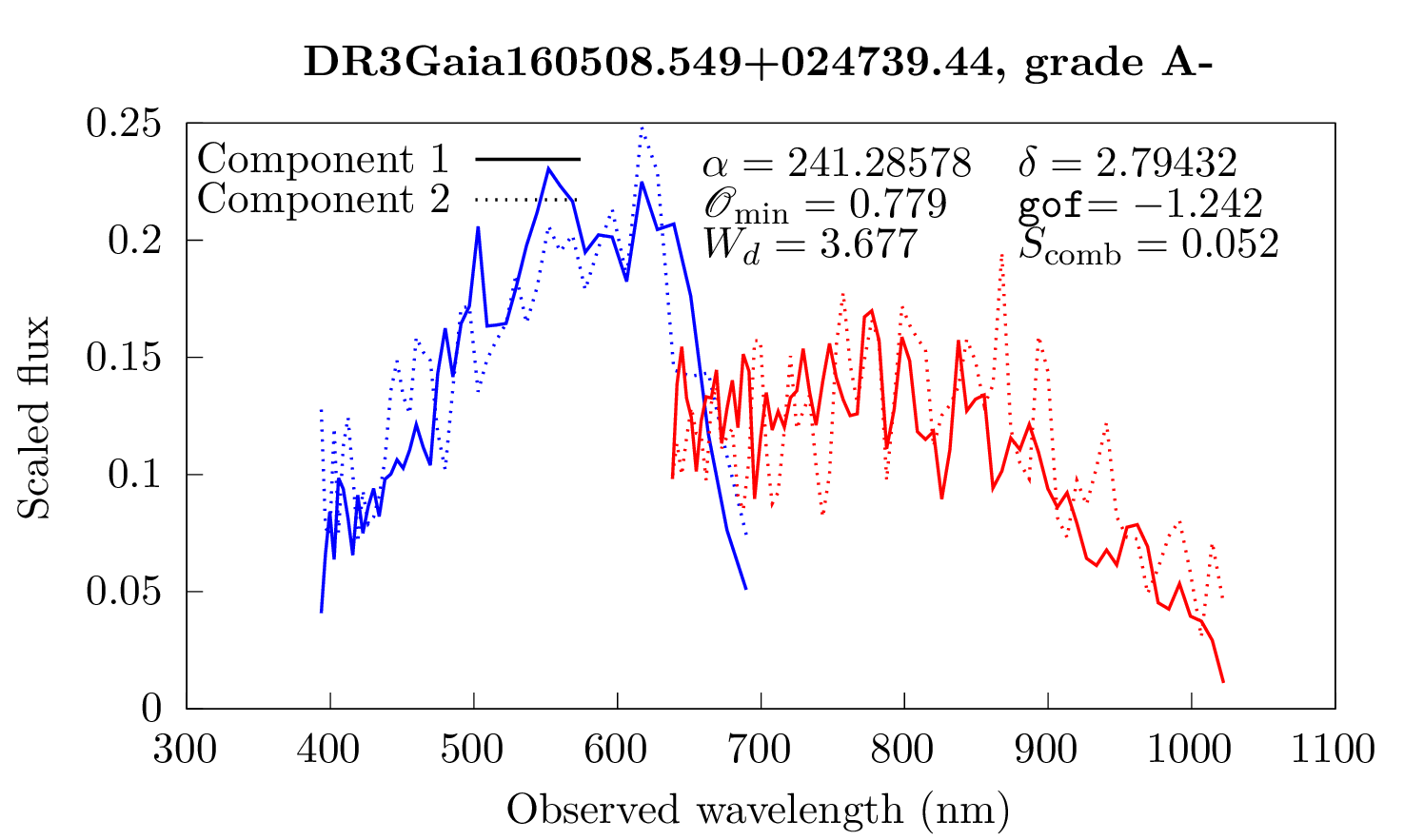}\hspace{0.5cm}\includegraphics[height=5.5cm]{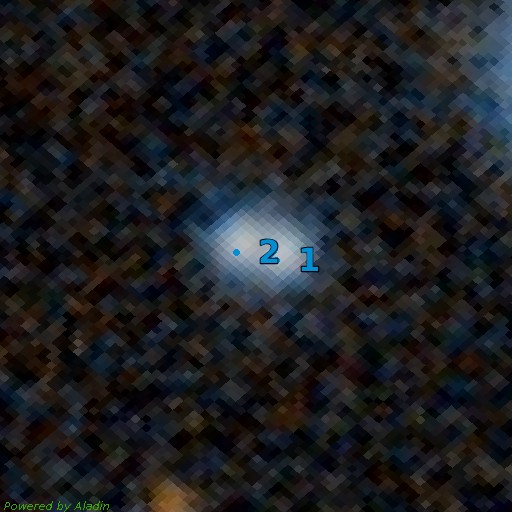}\caption{\label{fig:DR3Gaia160508.549+024739.44} Comparison of the resampled spectra of the DR3Gaia160508.549+024739.44 multiplet (Left) and associated Pan-STARRS1 image (Right) \citep{panstarrs}. Blue dots correspond to the GravLens components. Cutout size is $15.0 \arcsec \times 15.0 \arcsec$, north is up, east is left.} \end{figure*}
\begin{figure*}\centering\includegraphics[height=6cm]{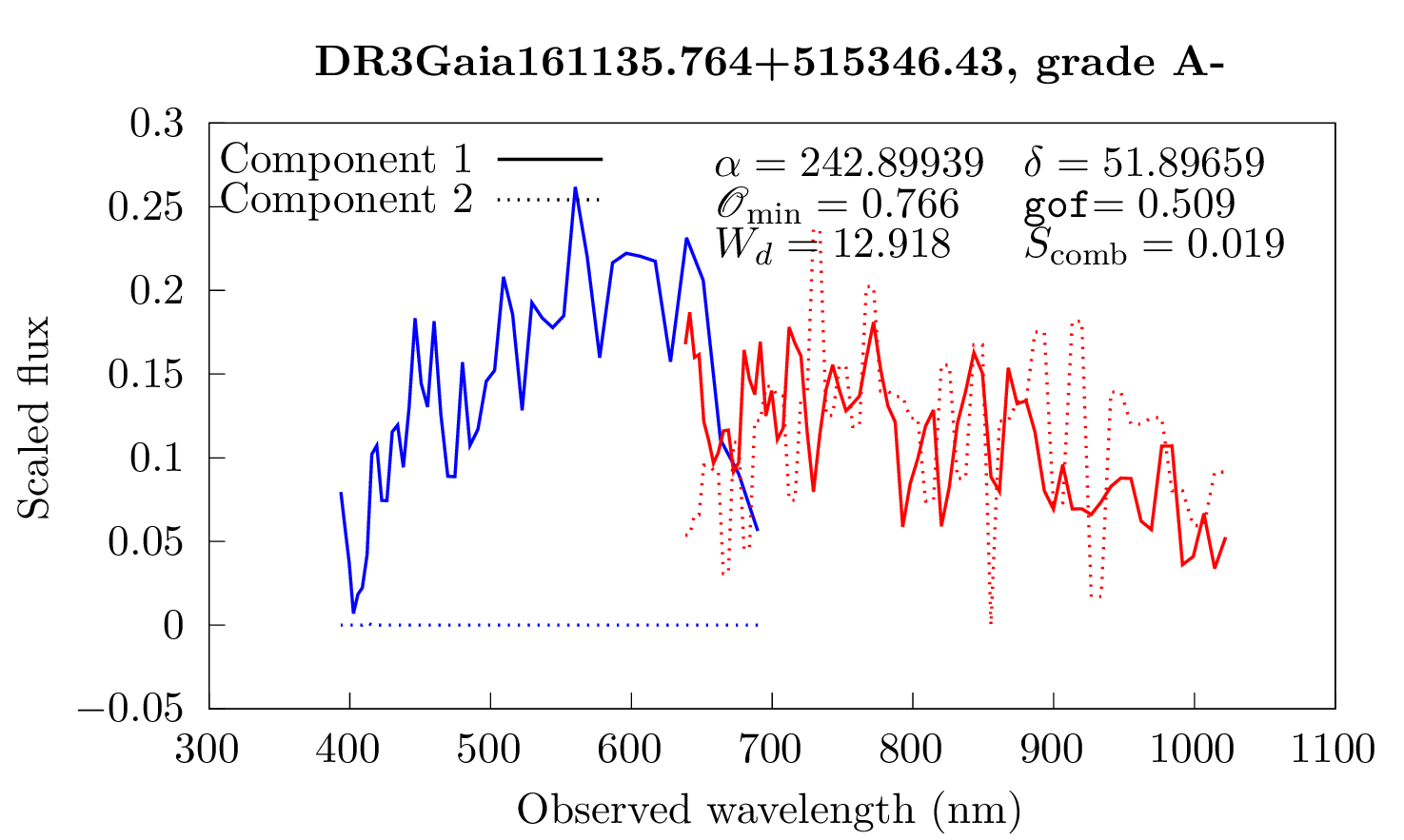}\hspace{0.5cm}\includegraphics[height=5.5cm]{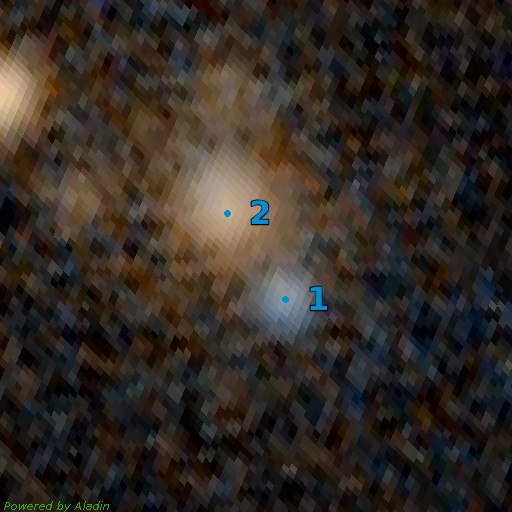}\caption{\label{fig:DR3Gaia161135.764+515346.43} Comparison of the resampled spectra of the DR3Gaia161135.764+515346.43 multiplet (Left) and associated Pan-STARRS1 image (Right) \citep{panstarrs}. Blue dots correspond to the GravLens components. Cutout size is $15.0 \arcsec \times 15.0 \arcsec$, north is up, east is left.} \end{figure*}
\begin{figure*}\centering\includegraphics[height=6cm]{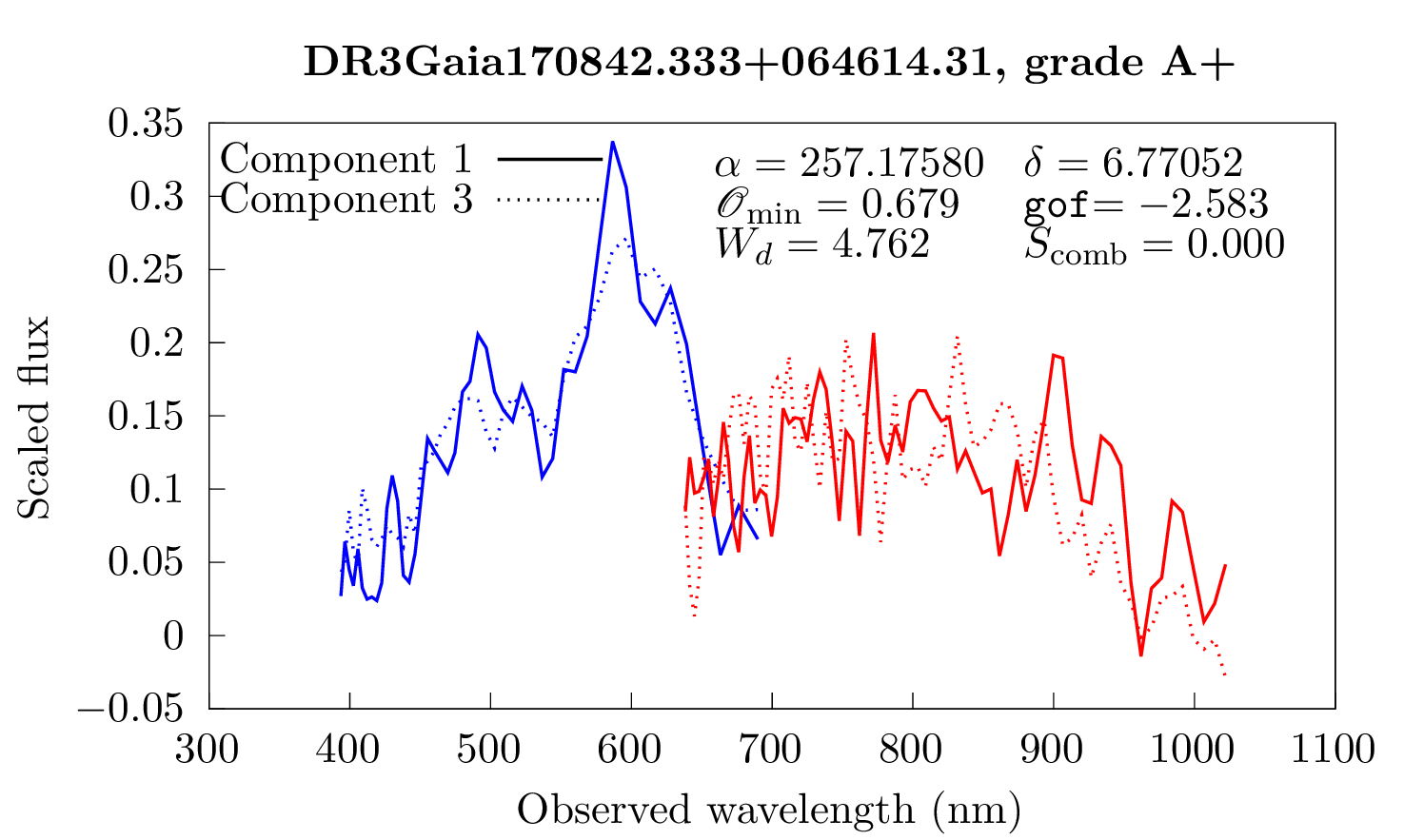}\hspace{0.5cm}\includegraphics[height=5.5cm]{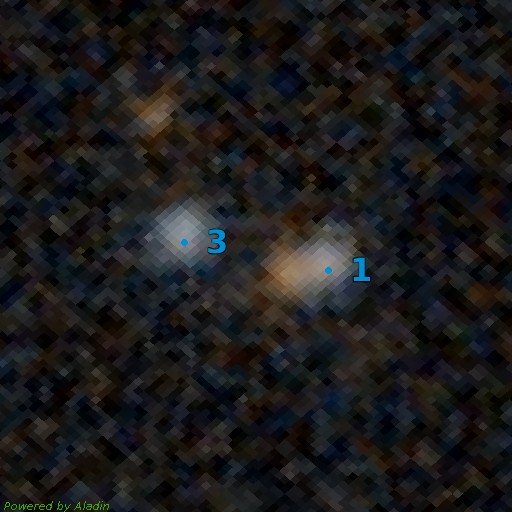}\caption{\label{fig:DR3Gaia170842.333+064614.31} Comparison of the resampled spectra of the DR3Gaia170842.333+064614.31 multiplet (Left) and associated Pan-STARRS1 image (Right) \citep{panstarrs}. Blue dots correspond to the GravLens components. Cutout size is $15.0 \arcsec \times 15.0 \arcsec$, north is up, east is left.} \end{figure*}
\begin{figure*}\centering\includegraphics[height=6cm]{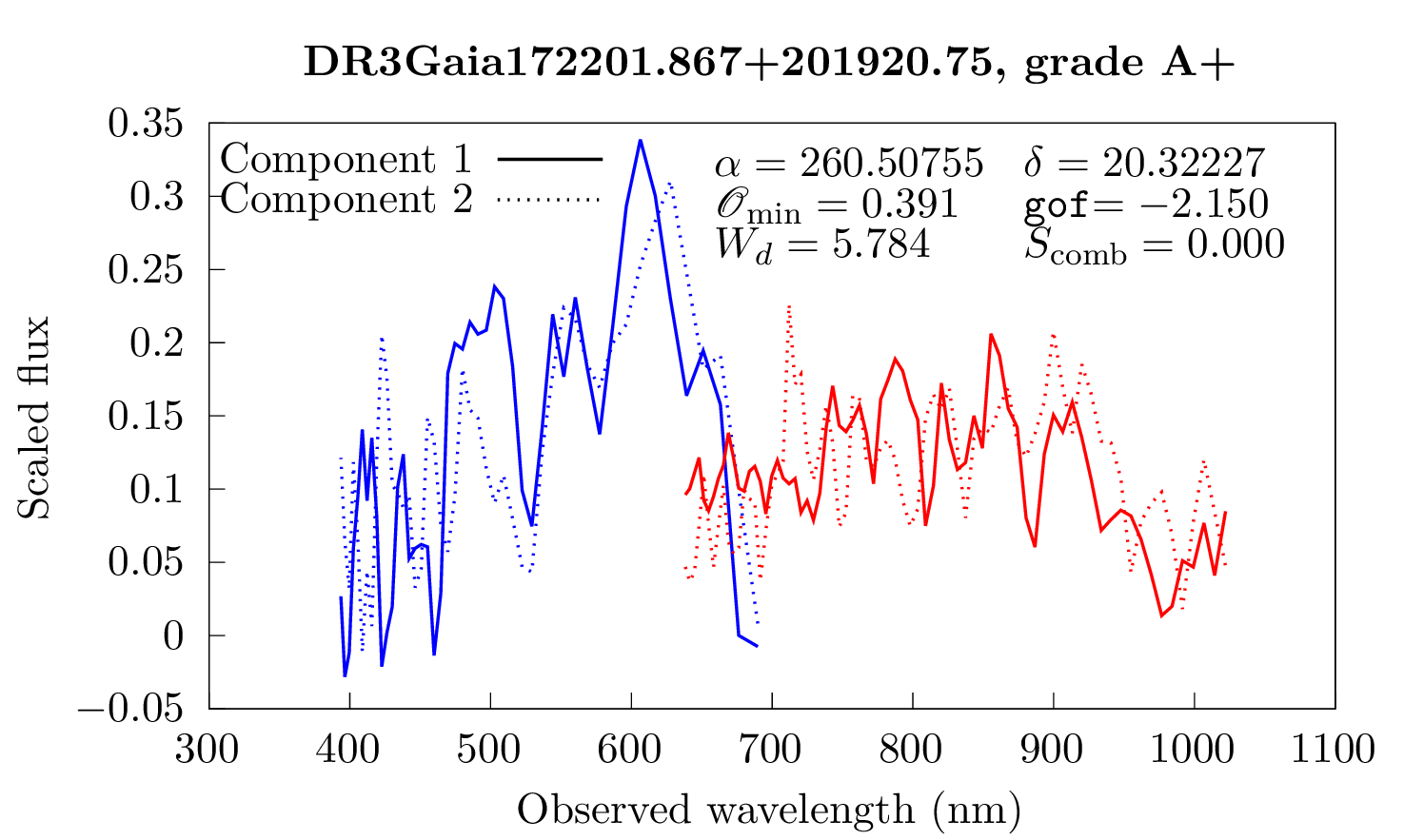}\hspace{0.5cm}\includegraphics[height=5.5cm]{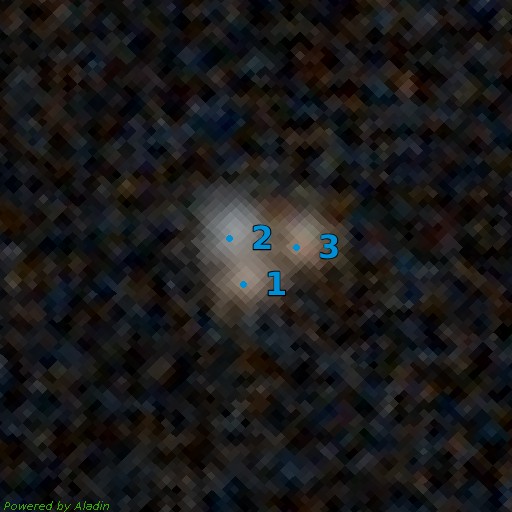}\caption{\label{fig:DR3Gaia172201.867+201920.75} Comparison of the resampled spectra of the DR3Gaia172201.867+201920.75 multiplet (Left) and associated Pan-STARRS1 image (Right) \citep{panstarrs}. Blue dots correspond to the GravLens components. Cutout size is $15.0 \arcsec \times 15.0 \arcsec$, north is up, east is left.} \end{figure*}
\begin{figure*}\centering\includegraphics[height=6cm]{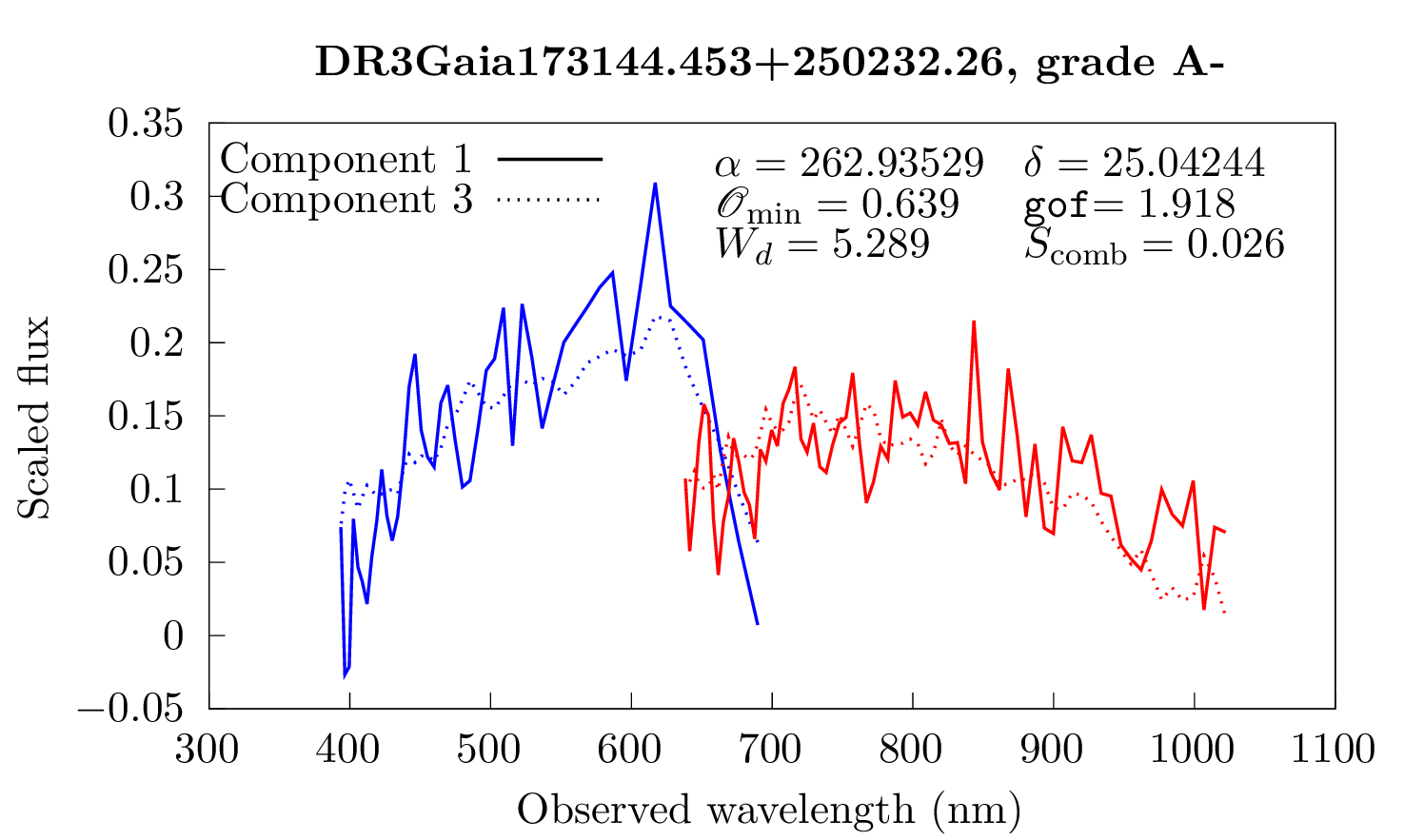}\hspace{0.5cm}\includegraphics[height=5.5cm]{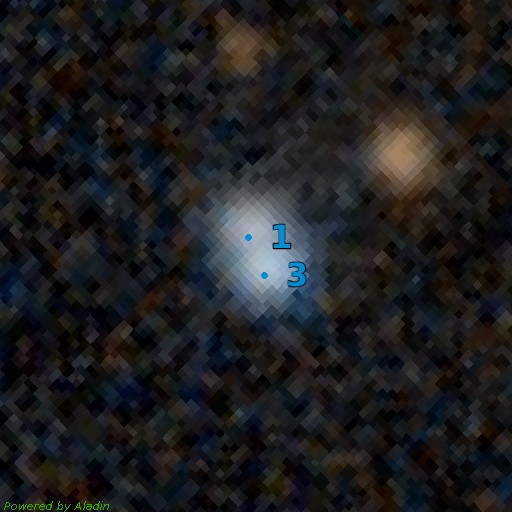}\caption{\label{fig:DR3Gaia173144.453+250232.26} Comparison of the resampled spectra of the DR3Gaia173144.453+250232.26 multiplet (Left) and associated Pan-STARRS1 image (Right) \citep{panstarrs}. Blue dots correspond to the GravLens components. Cutout size is $15.0 \arcsec \times 15.0 \arcsec$, north is up, east is left.} \end{figure*}
\begin{figure*}\centering\includegraphics[height=6cm]{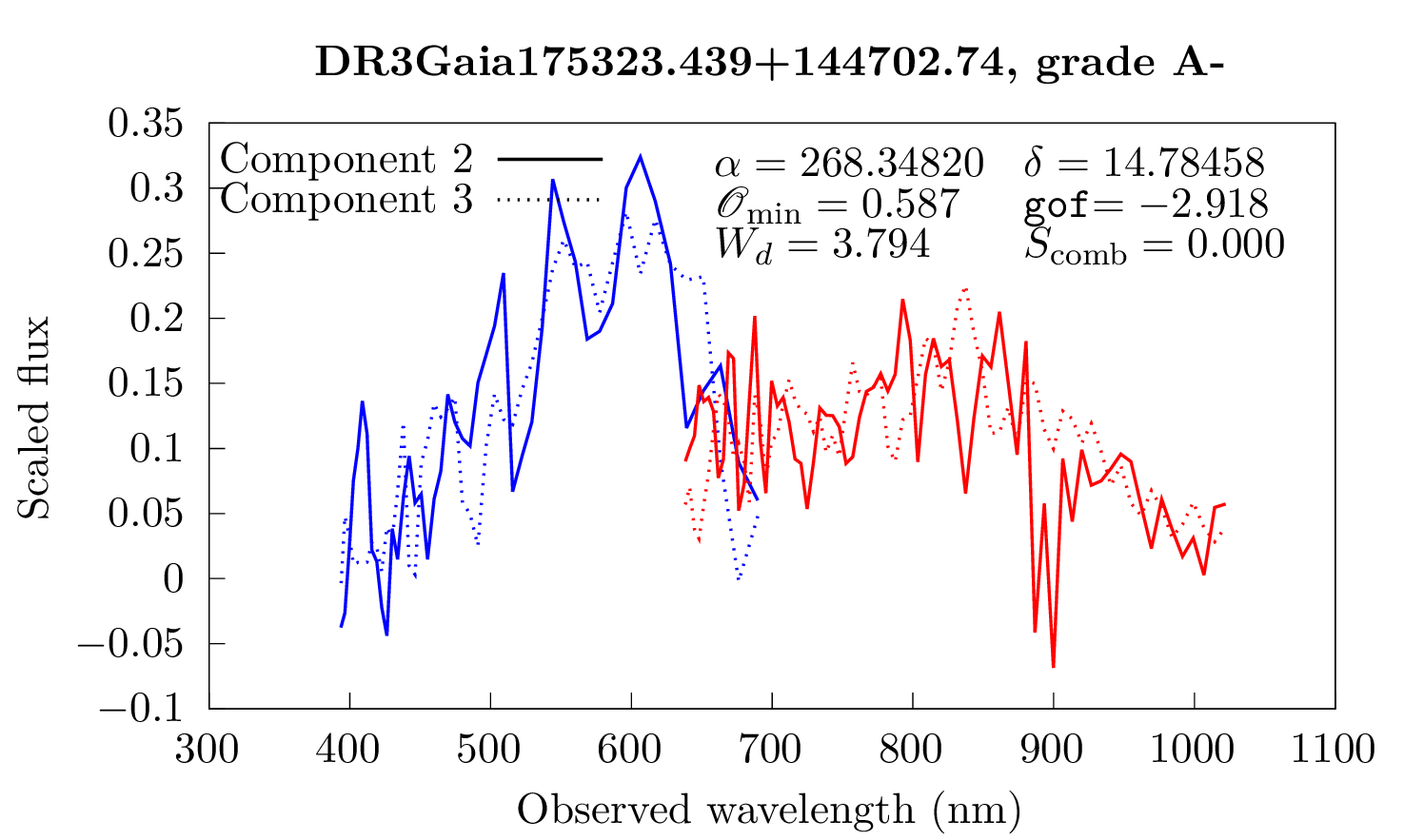}\hspace{0.5cm}\includegraphics[height=5.5cm]{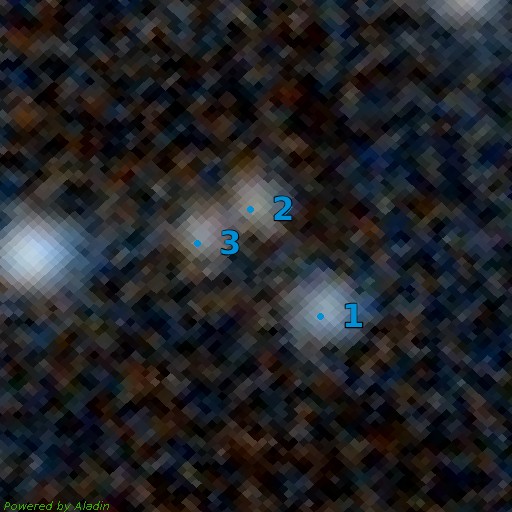}\caption{\label{fig:DR3Gaia175323.439+144702.74} Comparison of the resampled spectra of the DR3Gaia175323.439+144702.74 multiplet (Left) and associated Pan-STARRS1 image (Right) \citep{panstarrs}. Blue dots correspond to the GravLens components. Cutout size is $15.0 \arcsec \times 15.0 \arcsec$, north is up, east is left.} \end{figure*}
\begin{figure*}\centering\includegraphics[height=6cm]{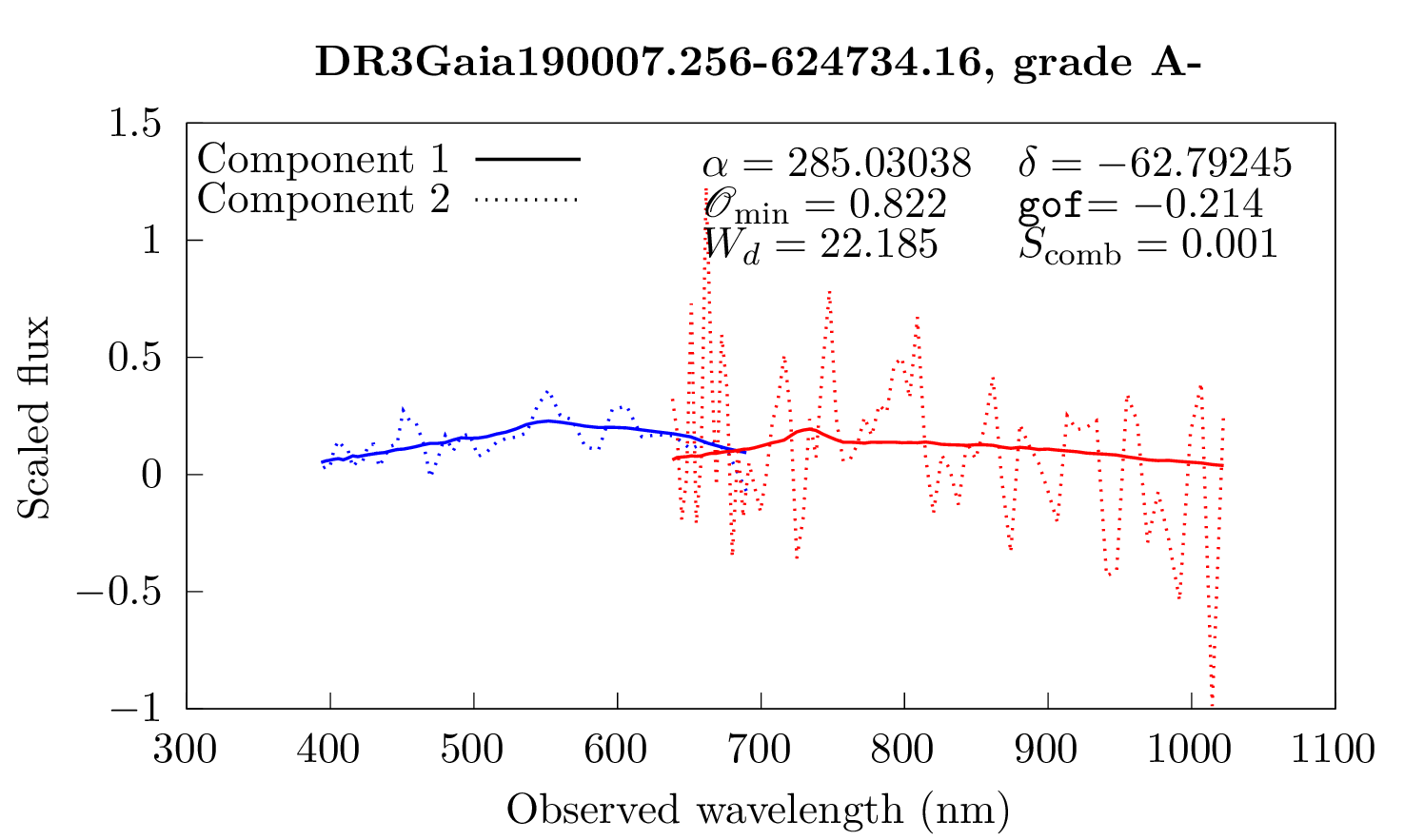}\hspace{0.5cm}\includegraphics[height=5.5cm]{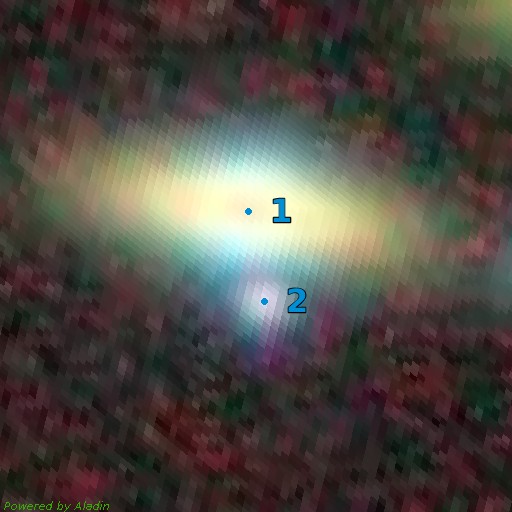}\caption{\label{fig:DR3Gaia190007.256-624734.16} Comparison of the resampled spectra of the DR3Gaia190007.256-624734.16 multiplet (Left) and associated Dark Energy Survey image (Right) \citep{2019AJ....157..168D}. Blue dots correspond to the GravLens components. Cutout size is $15.0 \arcsec \times 15.0 \arcsec$, north is up, east is left. {\bf Component 2 is presumably a contaminating source.}} \end{figure*}
\begin{figure*}\centering\includegraphics[height=6cm]{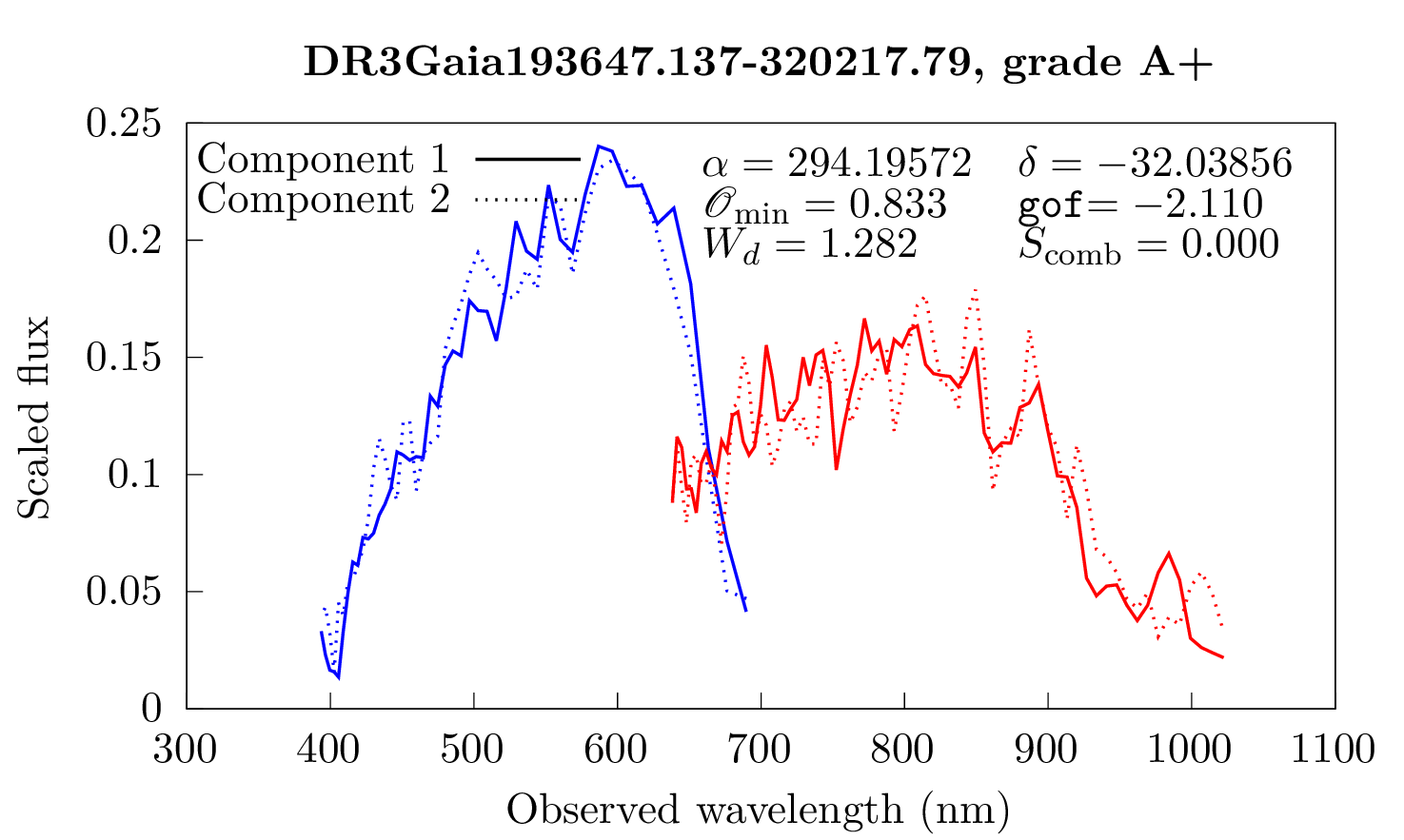}\hspace{0.5cm}\includegraphics[height=5.5cm]{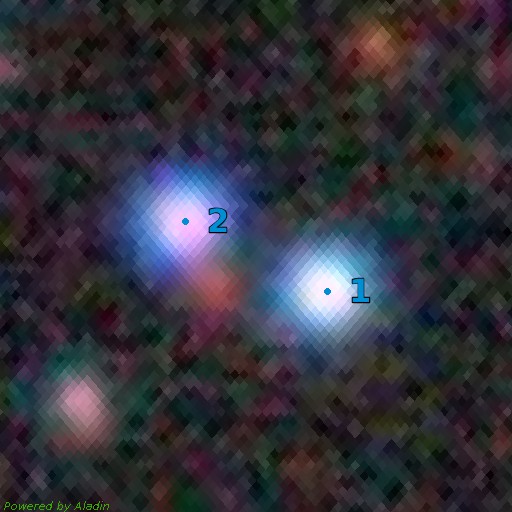}\caption{\label{fig:DR3Gaia193647.137-320217.79} Comparison of the resampled spectra of the DR3Gaia193647.137-320217.79 multiplet (Left) and associated Dark Energy Survey image (Right) \citep{2019AJ....157..168D}. Blue dots correspond to the GravLens components. Cutout size is $15.0 \arcsec \times 15.0 \arcsec$, north is up, east is left.} \end{figure*}
\begin{figure*}\centering\includegraphics[height=6cm]{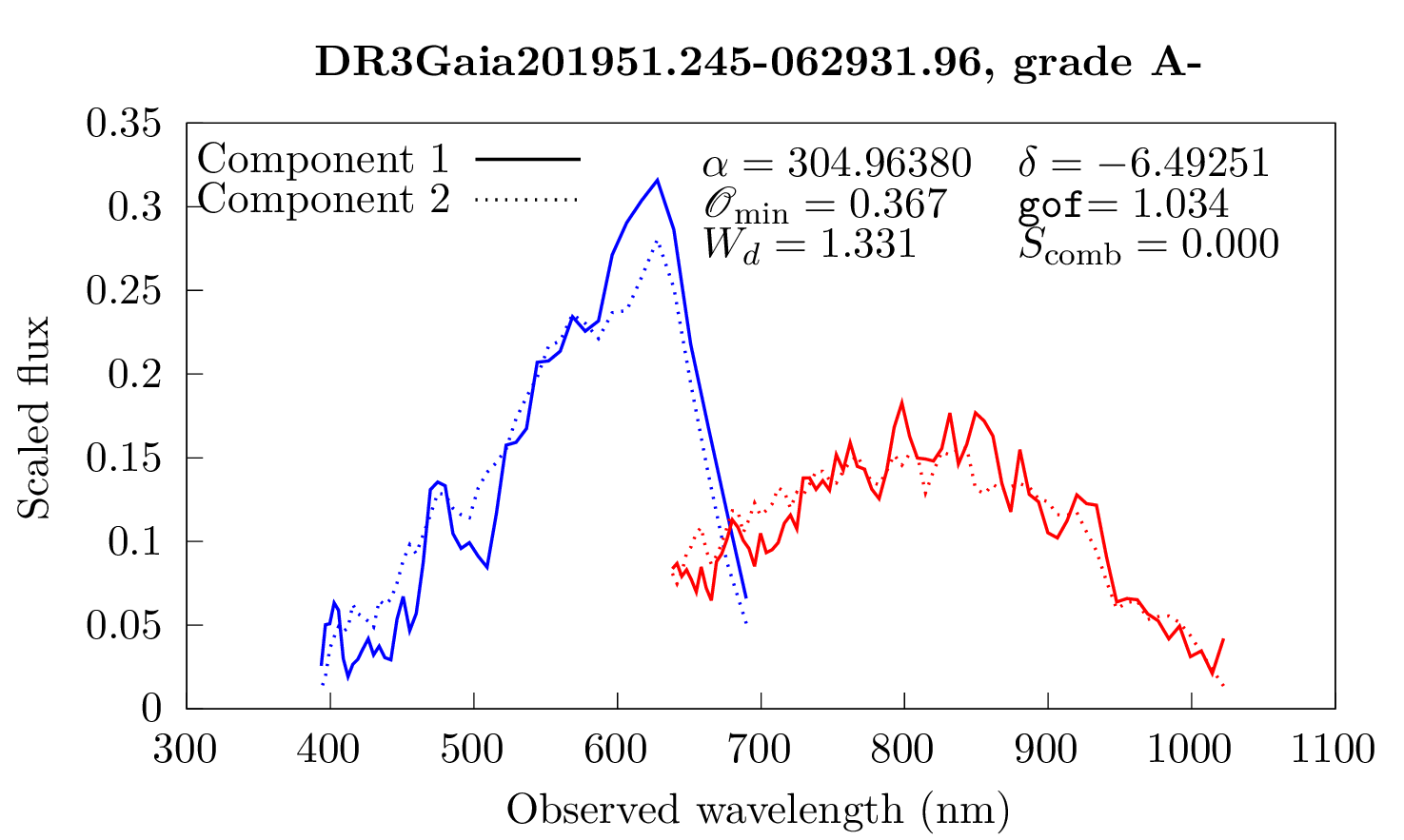}\hspace{0.5cm}\includegraphics[height=5.5cm]{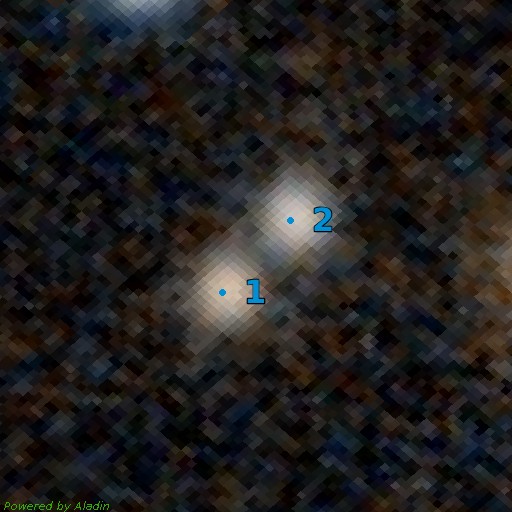}\caption{\label{fig:DR3Gaia201951.245-062931.96} Comparison of the resampled spectra of the DR3Gaia201951.245-062931.96 multiplet (Left) and associated Pan-STARRS1 image (Right) \citep{panstarrs}. Blue dots correspond to the GravLens components. Cutout size is $15.0 \arcsec \times 15.0 \arcsec$, north is up, east is left.} \end{figure*}
\begin{figure*}\centering\includegraphics[height=6cm]{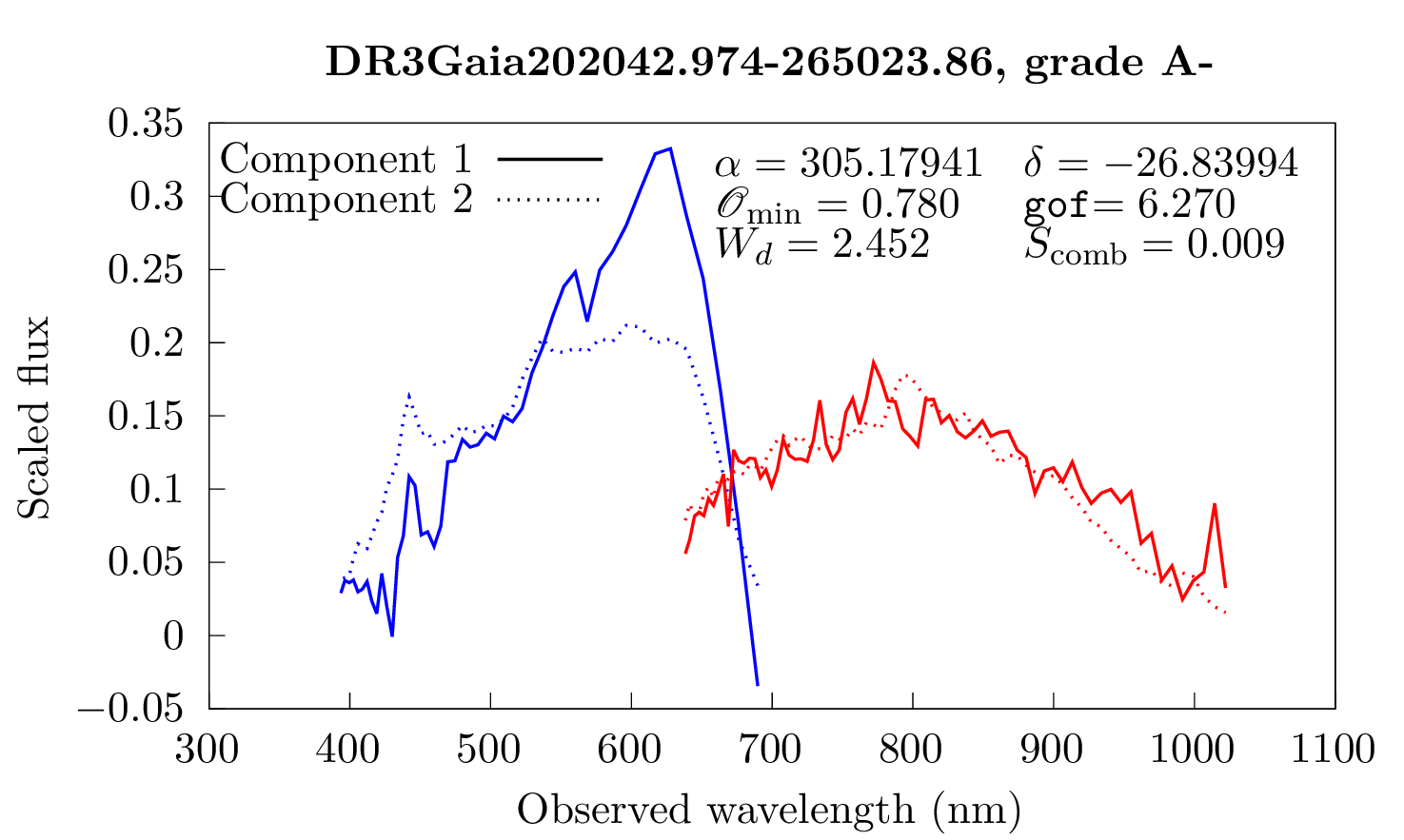}\hspace{0.5cm}\includegraphics[height=5.5cm]{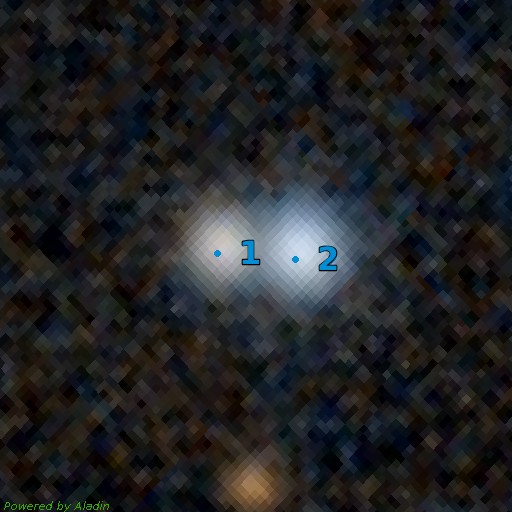}\caption{\label{fig:DR3Gaia202042.974-265023.86} Comparison of the resampled spectra of the DR3Gaia202042.974-265023.86 multiplet (Left) and associated Pan-STARRS1 image (Right) \citep{panstarrs}. Blue dots correspond to the GravLens components. Cutout size is $15.0 \arcsec \times 15.0 \arcsec$, north is up, east is left.} \end{figure*}
\begin{figure*}\centering\includegraphics[height=6cm]{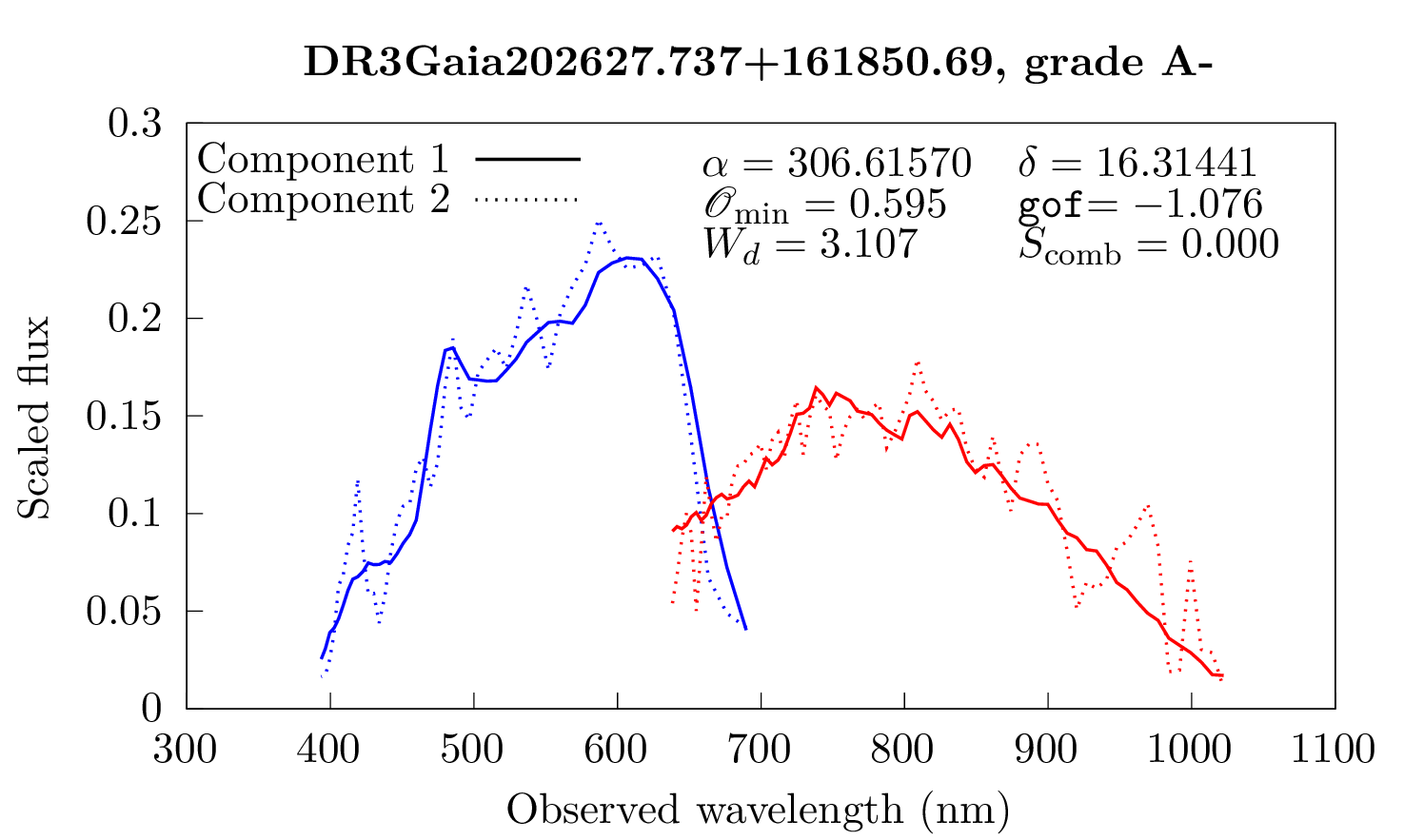}\hspace{0.5cm}\includegraphics[height=5.5cm]{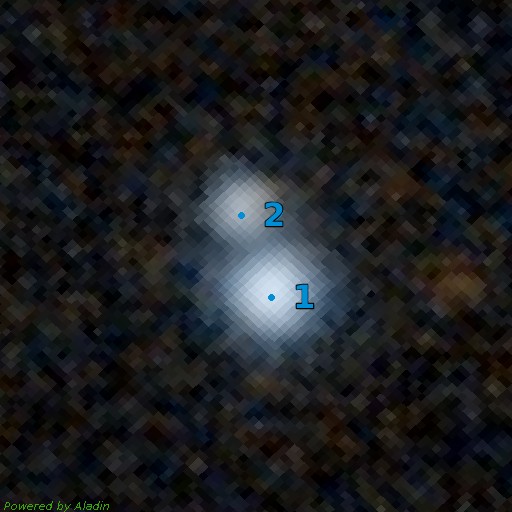}\caption{\label{fig:DR3Gaia202627.737+161850.69} Comparison of the resampled spectra of the DR3Gaia202627.737+161850.69 multiplet (Left) and associated Pan-STARRS1 image (Right) \citep{panstarrs}. Blue dots correspond to the GravLens components. Cutout size is $15.0 \arcsec \times 15.0 \arcsec$, north is up, east is left.} \end{figure*}
\begin{figure*}\centering\includegraphics[height=6cm]{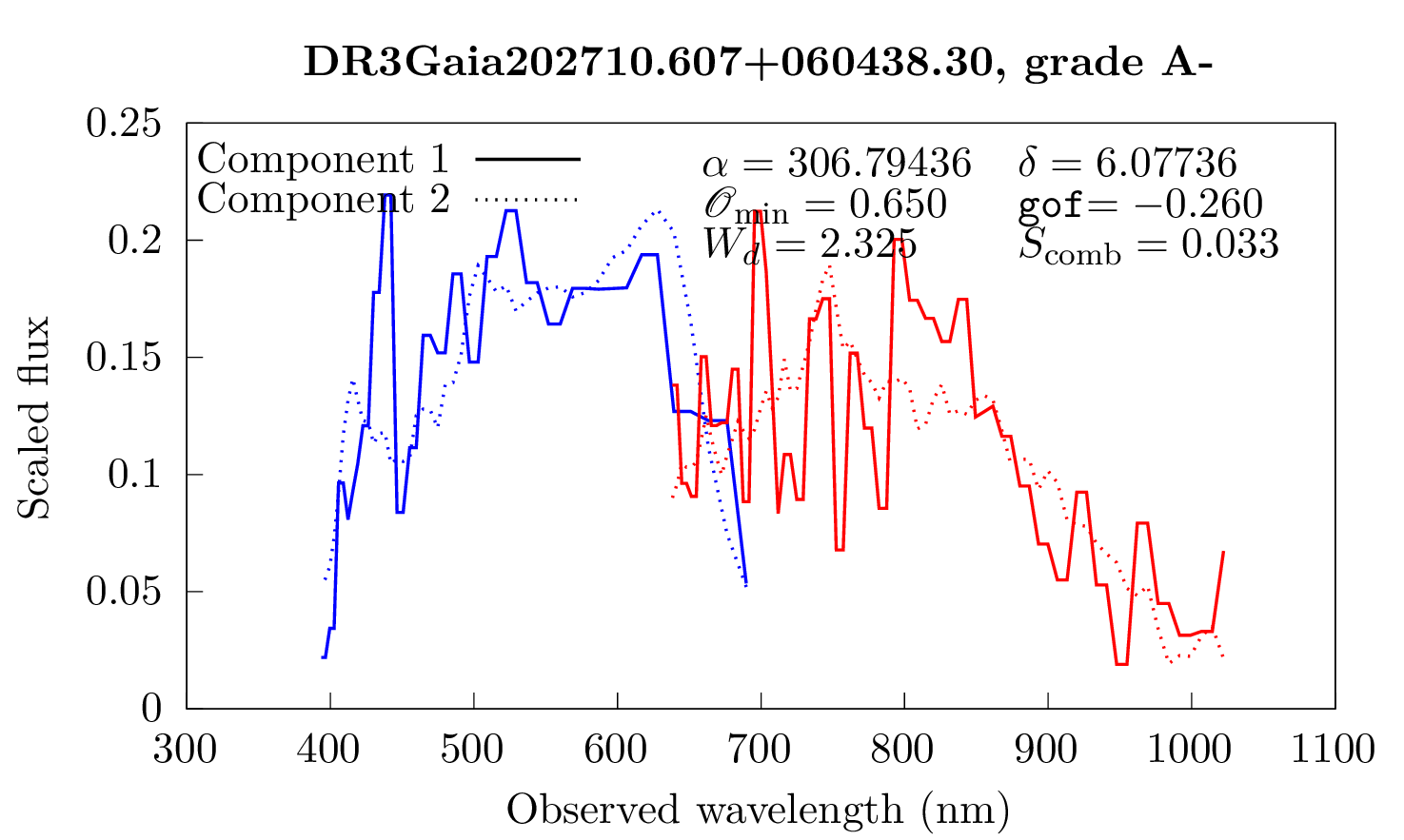}\hspace{0.5cm}\includegraphics[height=5.5cm]{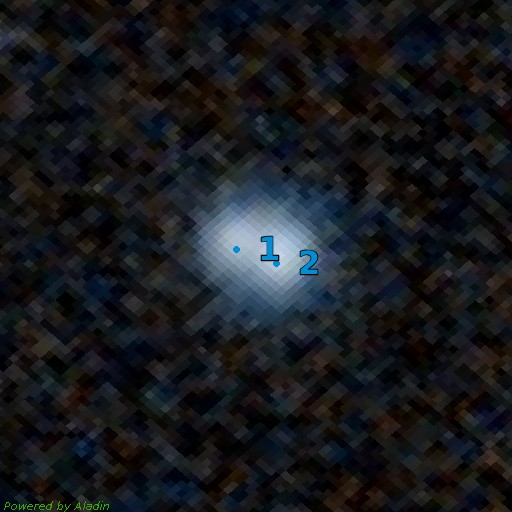}\caption{\label{fig:DR3Gaia202710.607+060438.30} Comparison of the resampled spectra of the DR3Gaia202710.607+060438.30 multiplet (Left) and associated Pan-STARRS1 image (Right) \citep{panstarrs}. Blue dots correspond to the GravLens components. Cutout size is $15.0 \arcsec \times 15.0 \arcsec$, north is up, east is left.} \end{figure*}
\begin{figure*}\centering\includegraphics[height=6cm]{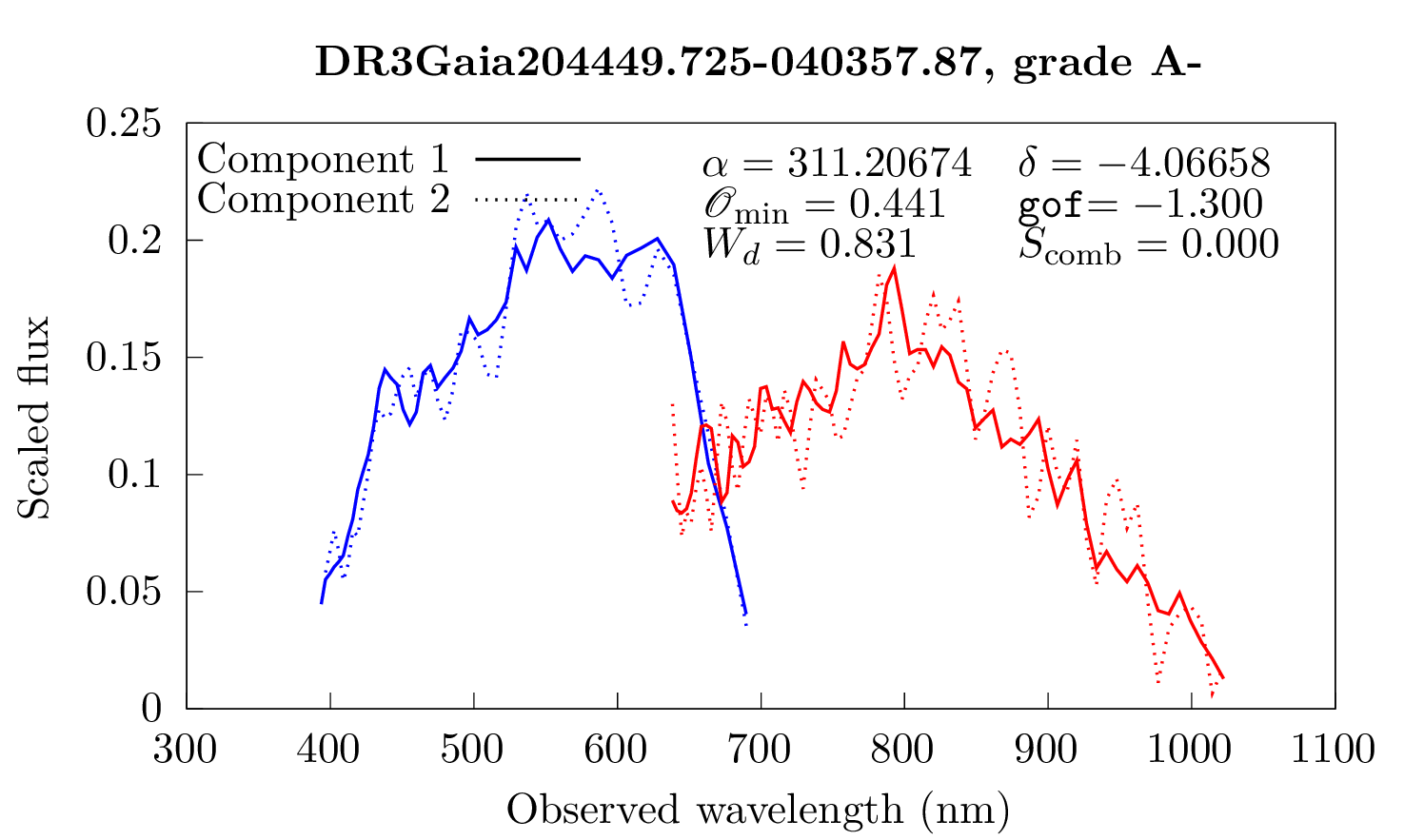}\hspace{0.5cm}\includegraphics[height=5.5cm]{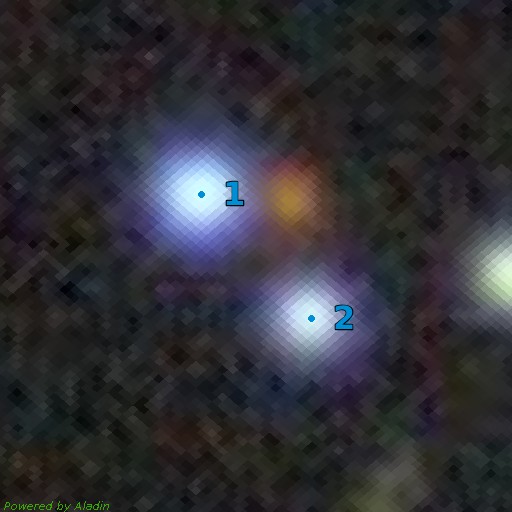}\caption{\label{fig:DR3Gaia204449.725-040357.87} Comparison of the resampled spectra of the DR3Gaia204449.725-040357.87 multiplet (Left) and associated Dark Energy Survey image (Right) \citep{2019AJ....157..168D}. Blue dots correspond to the GravLens components. Cutout size is $15.0 \arcsec \times 15.0 \arcsec$, north is up, east is left.} \end{figure*}
\begin{figure*}\centering\includegraphics[height=6cm]{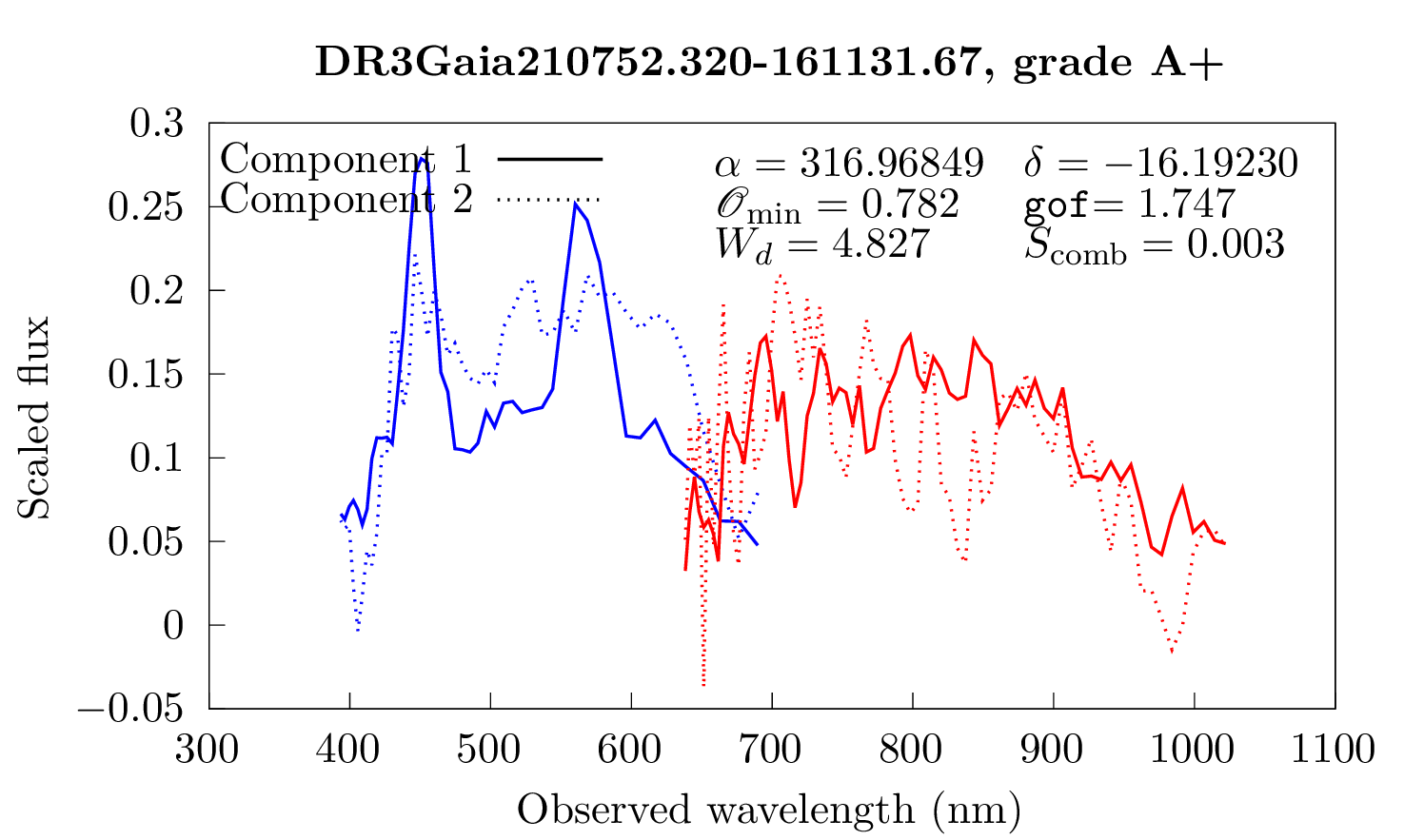}\hspace{0.5cm}\includegraphics[height=5.5cm]{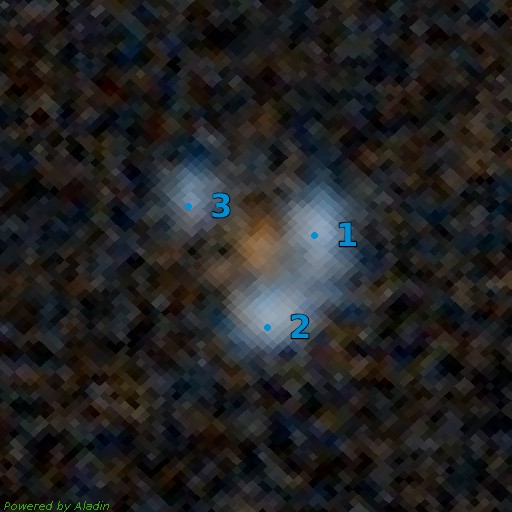}\caption{\label{fig:DR3Gaia210752.320-161131.67} Comparison of the resampled spectra of the DR3Gaia210752.320-161131.67 multiplet (Left) and associated Pan-STARRS1 image (Right) \citep{panstarrs}. Blue dots correspond to the GravLens components. Cutout size is $15.0 \arcsec \times 15.0 \arcsec$, north is up, east is left.} \end{figure*}
\begin{figure*}\centering\includegraphics[height=6cm]{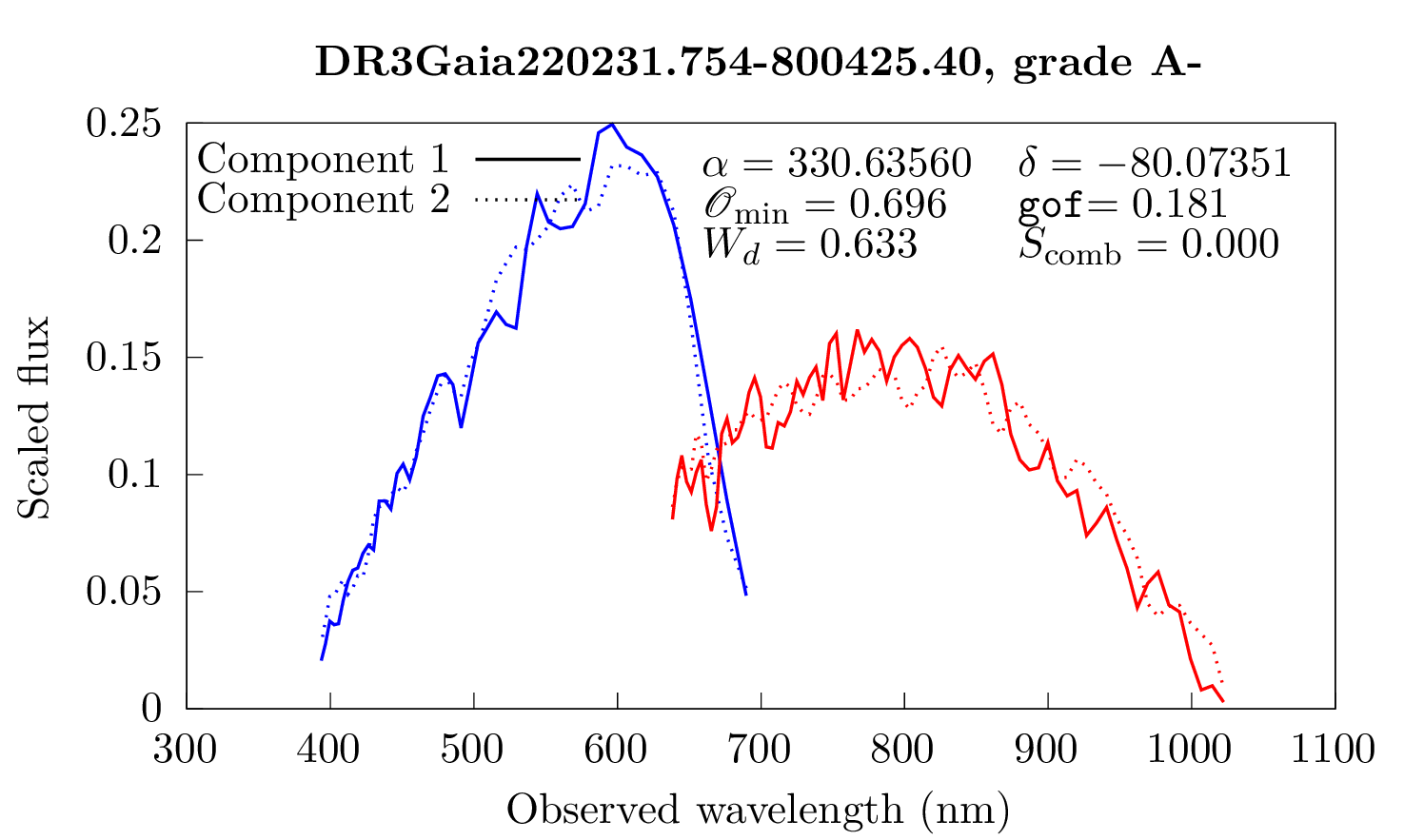}\hspace{0.5cm}\includegraphics[height=5.5cm]{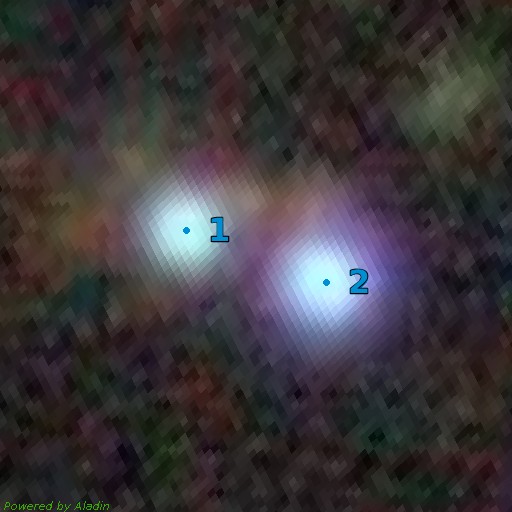}\caption{\label{fig:DR3Gaia220231.754-800425.40} Comparison of the resampled spectra of the DR3Gaia220231.754-800425.40 multiplet (Left) and associated Dark Energy Survey image (Right) \citep{2019AJ....157..168D}. Blue dots correspond to the GravLens components. Cutout size is $15.0 \arcsec \times 15.0 \arcsec$, north is up, east is left.} \end{figure*}
\begin{figure*}\centering\includegraphics[height=6cm]{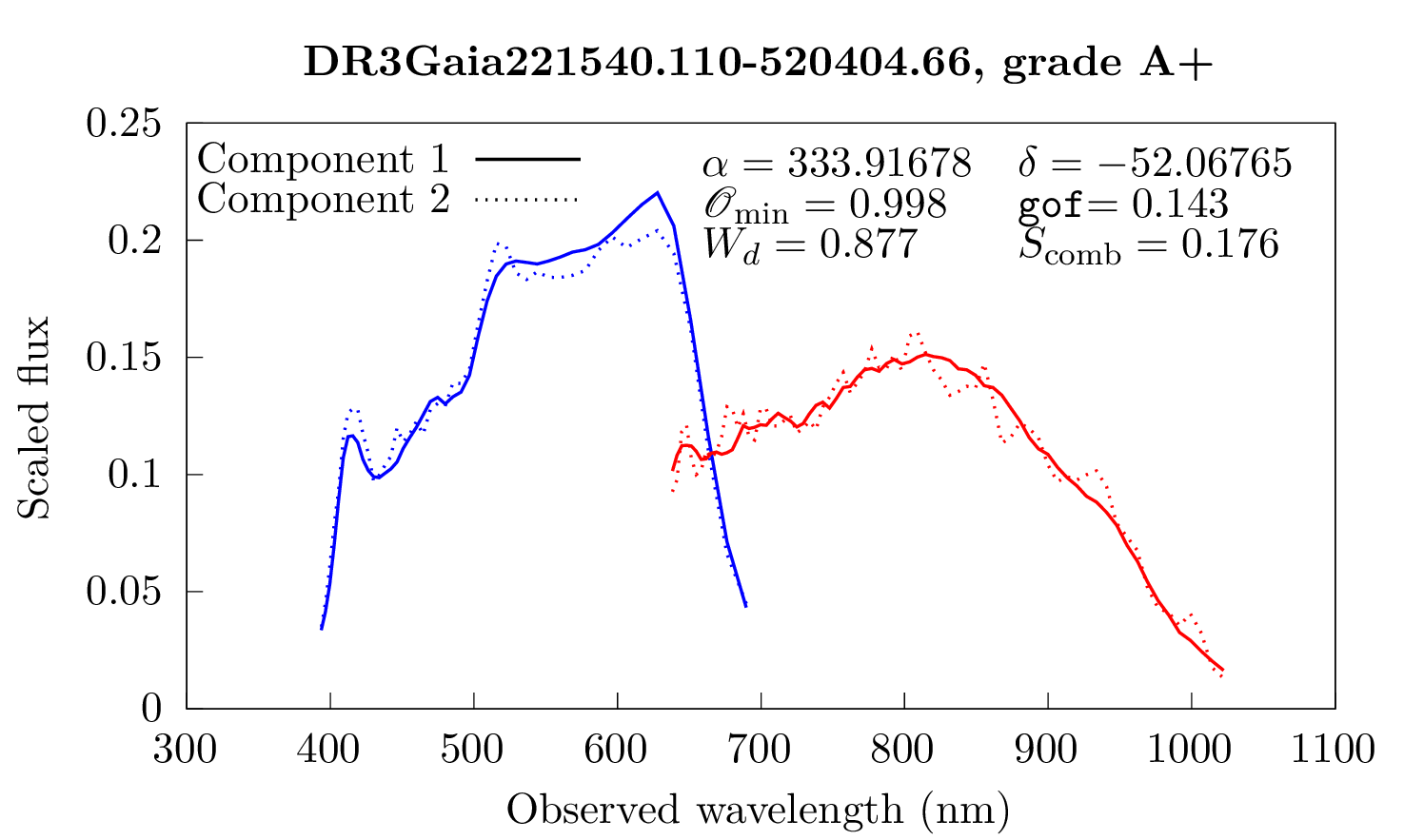}\hspace{0.5cm}\includegraphics[height=5.5cm]{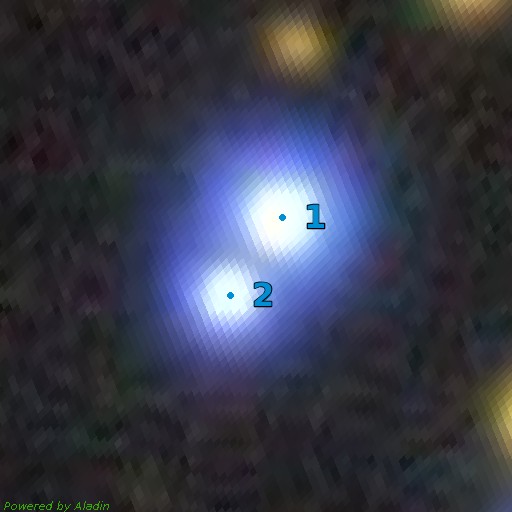}\caption{\label{fig:DR3Gaia221540.110-520404.66} Comparison of the resampled spectra of the DR3Gaia221540.110-520404.66 multiplet (Left) and associated Dark Energy Survey image (Right) \citep{2019AJ....157..168D}. Blue dots correspond to the GravLens components. Cutout size is $15.0 \arcsec \times 15.0 \arcsec$, north is up, east is left.} \end{figure*}
\begin{figure*}\centering\includegraphics[height=6cm]{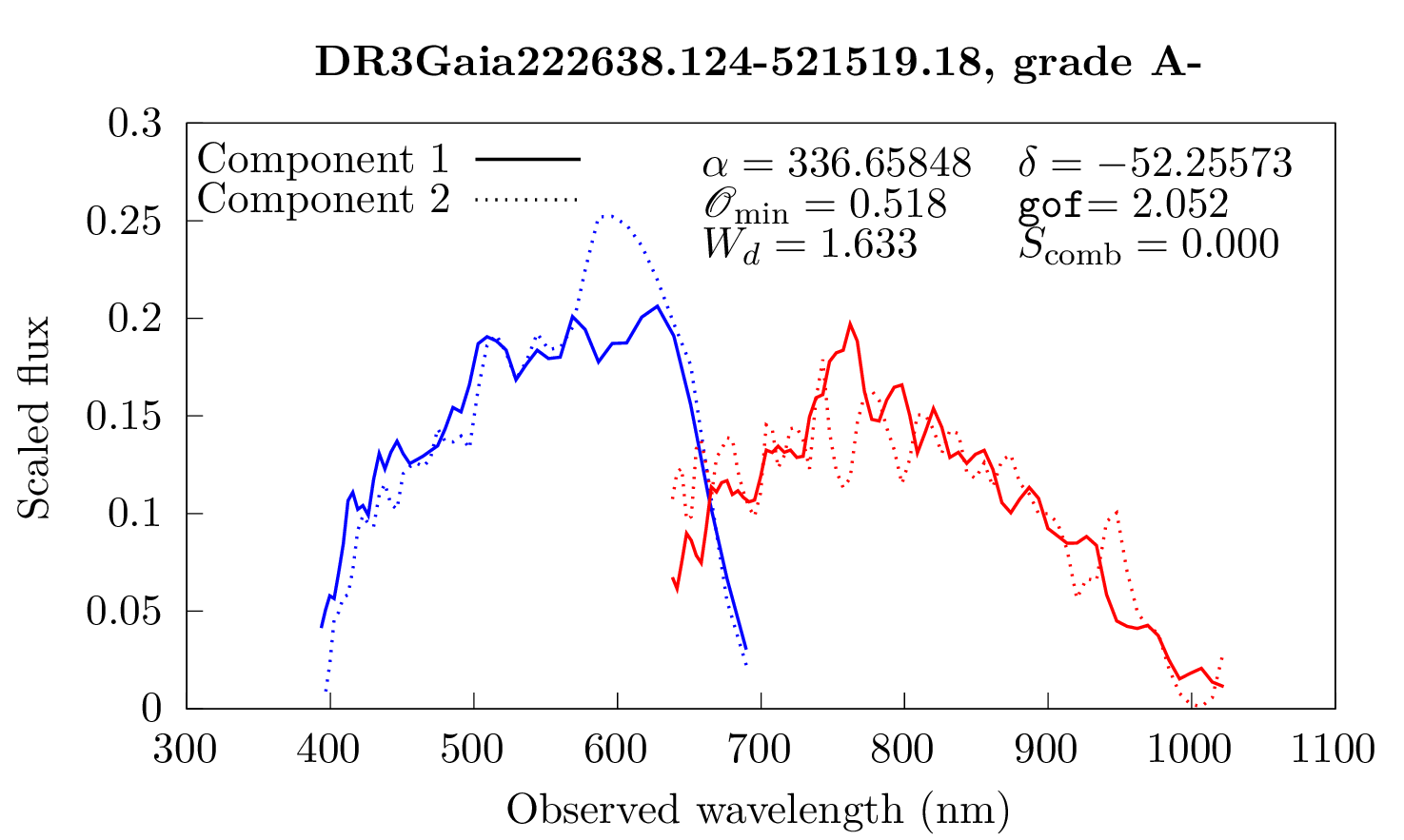}\hspace{0.5cm}\includegraphics[height=5.5cm]{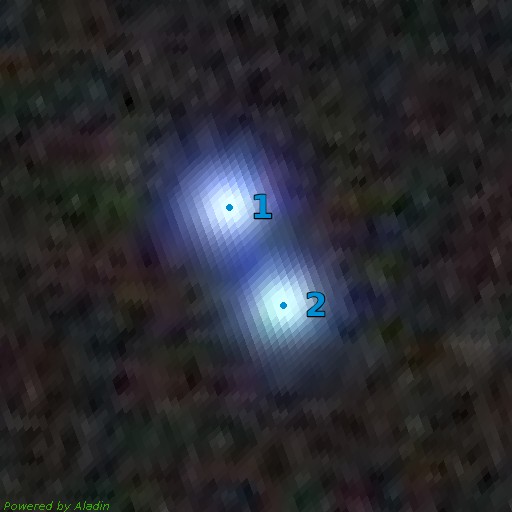}\caption{\label{fig:DR3Gaia222638.124-521519.18} Comparison of the resampled spectra of the DR3Gaia222638.124-521519.18 multiplet (Left) and associated Dark Energy Survey image (Right) \citep{2019AJ....157..168D}. Blue dots correspond to the GravLens components. Cutout size is $15.0 \arcsec \times 15.0 \arcsec$, north is up, east is left.} \end{figure*}
\begin{figure*}\centering\includegraphics[height=6cm]{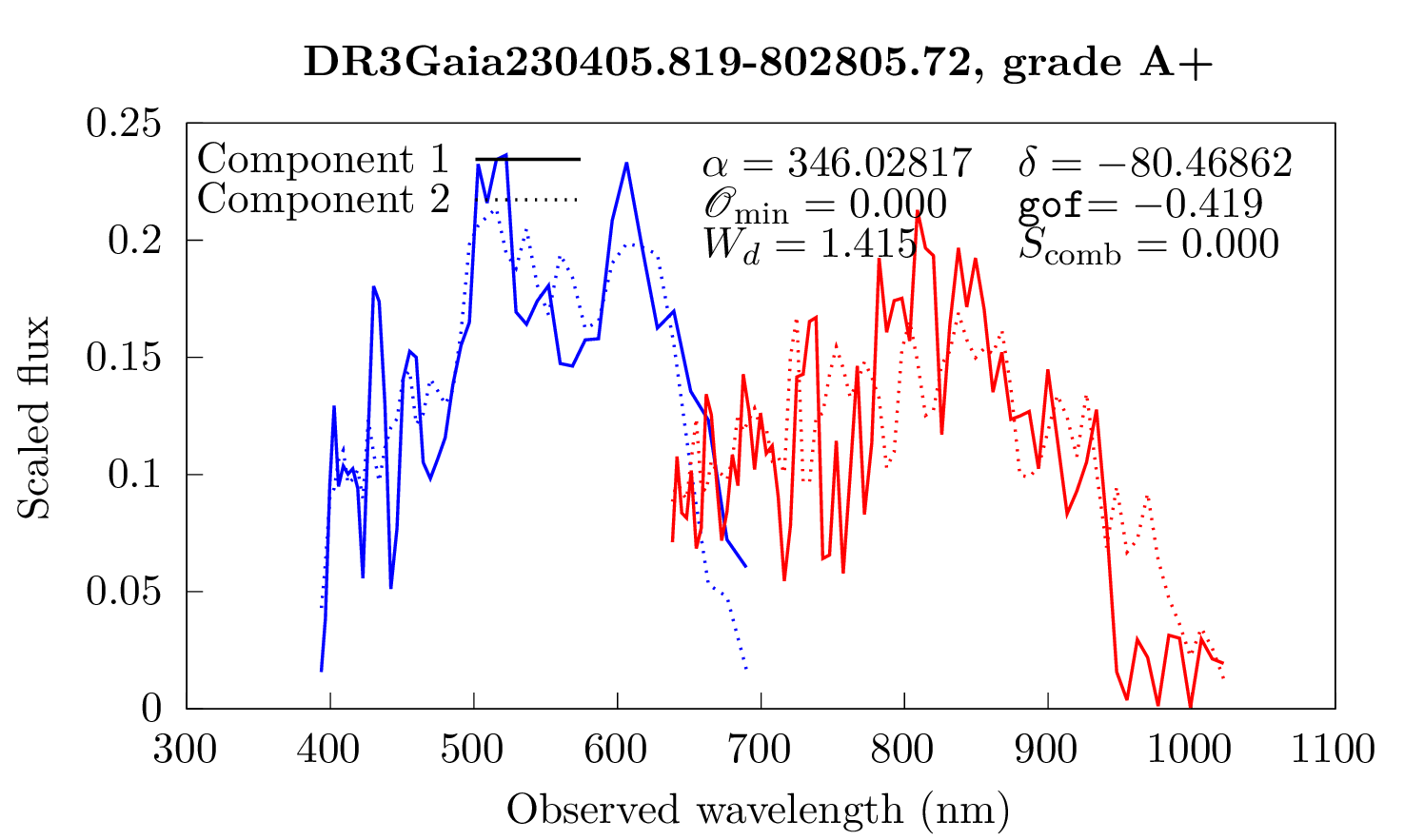}\hspace{0.5cm}\includegraphics[height=5.5cm]{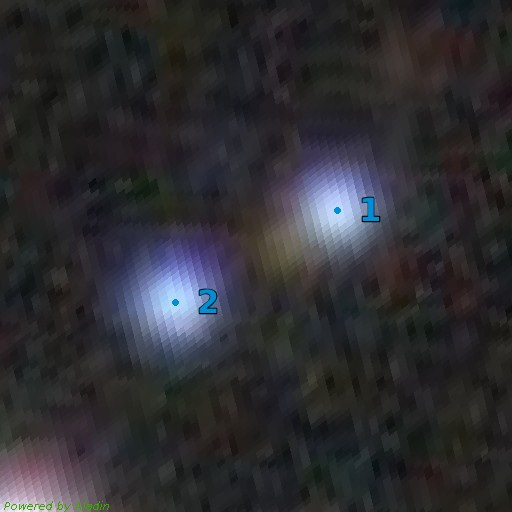}\caption{\label{fig:DR3Gaia230405.819-802805.72} Comparison of the resampled spectra of the DR3Gaia230405.819-802805.72 multiplet (Left) and associated Dark Energy Survey image (Right) \citep{2019AJ....157..168D}. Blue dots correspond to the GravLens components. Cutout size is $15.0 \arcsec \times 15.0 \arcsec$, north is up, east is left.} \end{figure*}
\begin{figure*}\centering\includegraphics[height=6cm]{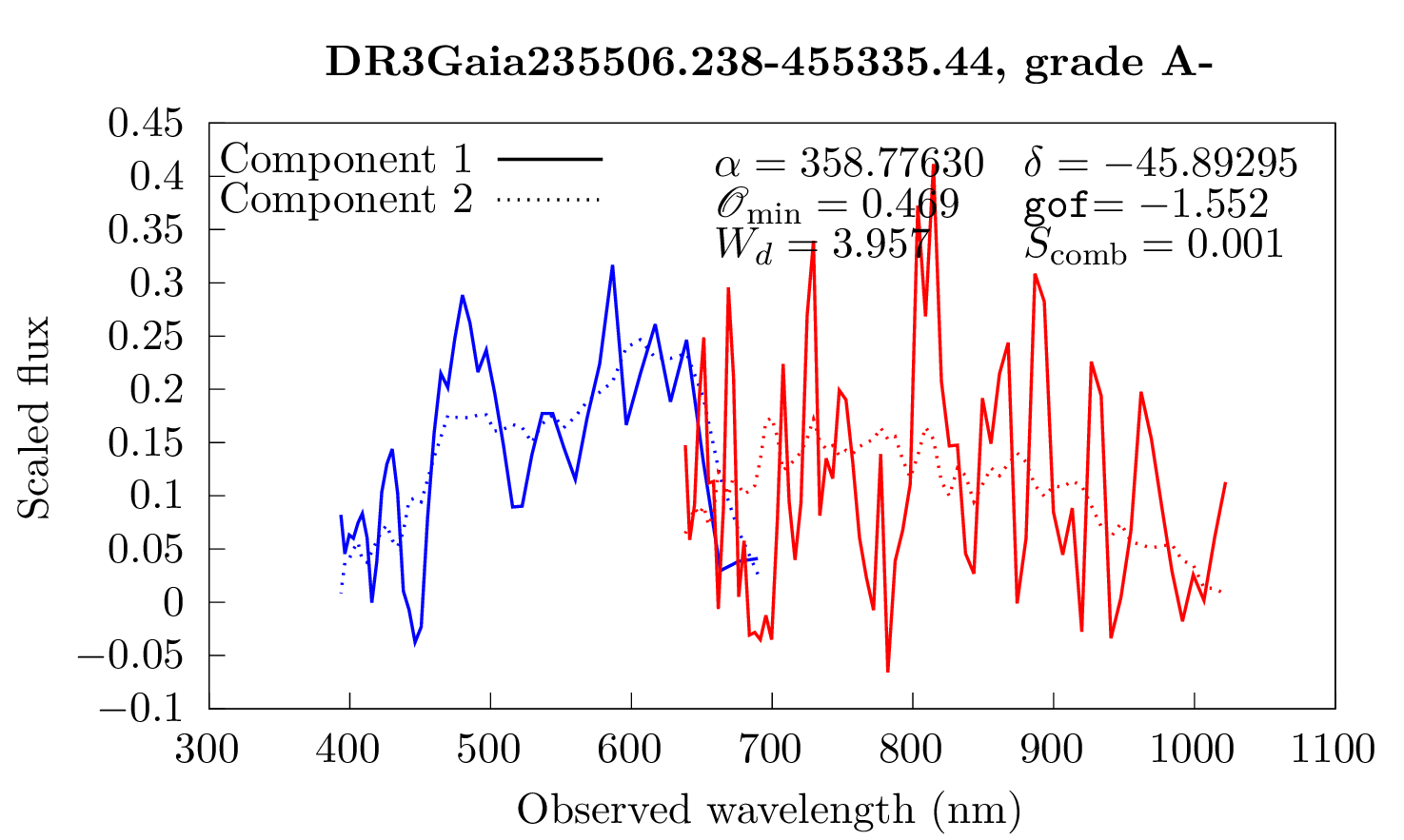}\hspace{0.5cm}\includegraphics[height=5.5cm]{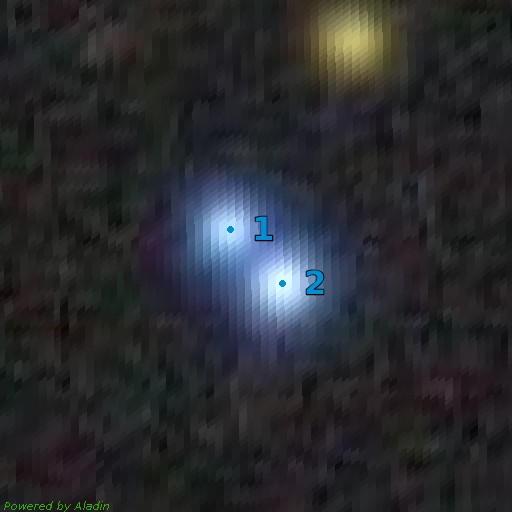}\caption{\label{fig:DR3Gaia235506.238-455335.44} Comparison of the resampled spectra of the DR3Gaia235506.238-455335.44 multiplet (Left) and associated Dark Energy Survey image (Right) \citep{2019AJ....157..168D}. Blue dots correspond to the GravLens components. Cutout size is $15.0 \arcsec \times 15.0 \arcsec$, north is up, east is left.} \end{figure*}

\end{appendix}
\end{document}